\documentclass[binding=0.1cm]{sapthesis}

\usepackage{microtype}
\usepackage{hyperref}
\usepackage{wasysym}
\usepackage{float}
\usepackage[title]{appendix}

\usepackage[
backend=biber,
style=alphabetic,
sorting=nyt
]{biblatex}
\AtEveryBibitem{\clearfield{issue}}

\usepackage{amssymb}
\usepackage{mathrsfs}
\usepackage{bm}
\usepackage{bbold}

\usepackage{comment}
\usepackage{svg}
\usepackage{soul,xcolor}
\setstcolor{blue}
\usepackage{cancel}

\DeclareFontEncoding{LS1}{}{}
\DeclareFontSubstitution{LS1}{stix}{m}{n}
\DeclareSymbolFont{symbols4}{LS1}{stixbb}{m}{it}
\DeclareMathSymbol{\hexagonblack}{\mathord}{symbols4}{"DE}

\newcommand{\de}{\text{d}}    
\newcommand{\mH}{\mathcal{H}} 
\newcommand{\mZ}{\mathcal{Z}} 
\newcommand{\mG}{\mathcal{G}} 
\newcommand{\mD}{\mathcal{D}} 
\newcommand{\mQ}{\mathcal{Q}}
\newcommand{\msQ}{\mathscr{Q}}
\newcommand{\mfq}{\mathfrak{q}}
\newcommand{\mR}{\mathcal{R}}
\newcommand{\mM}{\mathcal{M}}
\newcommand{\mA}{\mathcal{A}}
\newcommand{\msD}{\mathscr{D}}
\newcommand{\rvline}{\hspace*{-\arraycolsep}\vline\hspace*{-\arraycolsep}}
\DeclareMathOperator{\Tr}{Tr}
\DeclareMathOperator{\erf}{erf}

\title{Realistic Model for Random Lasers \\
from Spin-Glass Theory}

\author{Jacopo Niedda}
\IDnumber{1563260}
\authoremail{jacopo.niedda@uniroma1.it}

\courseorganizer{Scuola di Dottorato in Scienze Astronomiche, \\
Chimiche, Fisiche e Matematiche “Vito Volterra”}
\course{Dottorato di Ricerca in Fisica}
\cycle{XXXV}

\submitdate{2022/2023}
\copyyear{2023}

\advisor{Prof. Luca Leuzzi}
\coadvisor{Prof. Giacomo Gradenigo}

\reviewerlabel{Thesis Reviewers}
\reviewer{Prof. Markus M\"{u}ller}
\reviewer{Prof. Juan Jesus Ruiz Lorenzo}

\examdate{12 May 2023}
\examiner{Prof. Chiara Cammarota} 
\examiner{Prof. Marc Mezard} 
\examiner{Prof. Juan Jesus Ruiz Lorenzo}  
\examiner{Prof. Prof. Adriano Barra}  

\thesistype{PhD}

\begin{document}

\frontmatter
\maketitle
\dedication{A Claudia, \\ compagna}
\begin{abstract}
This work finds its place in the statistical mechanical approach to light
amplification in disordered media, namely Random Lasers (RLs). The problem of
going beyond the standard mean-field Replica Symmetry Breaking (RSB) theory
employed to find the solution of spin-glass models for RLs is addressed,
improving the theory towards a more realistic description of these optical
systems.

The leading model of the glassy lasing transition is considered, justifying the
emergence of the 4-body interaction term in the context of RL semiclassical
theory. In the slow amplitude basis, the mode-couplings are selected by a
Frequency Matching Condition (FMC) and the Langevin equation for the
complex amplitude dynamics has a white noise, leading to an effective
equilibrium theory for the stationary regime of RLs. The spin-glass 4-phasor
Hamiltonian is obtained by taking disordered couplings, as induced by the
randomness of the mode spatial extension and of the nonlinear optical
response. A global constraint on the overall intensity is implemented to ensure
the system stability.

Standard mean-field theory requires the model to be defined on the fully-connected
interaction graph, where the FMC is always satisfied. This approximation allows one
to use standard RSB techniques developed for mean-field spin glasses, but only applies to
a very special regime, the narrow-bandwidth limit, where the emission
spectrum has a width comparable to the typical linewidth of the modes. This
prevents the theory from being applied to generic experimental situations, e.g.,
hindering the reproduction of the central narrowing in RL empirical spectra. It is
of great interest, then, to investigate the model on the Mode-Locked (ML)
diluted interaction graph.

To address the problem, both a numerical and an analytical approach are
followed. A major result is the evidence of a mixed-order ergodicity breaking
transition in the ML 4-phasor model, as revealed by exchange Monte Carlo
numerical simulation. The joint study of the specific-heat divergence at the 
critical point and of the low temperature behavior of the Parisi overlap
distribution reveals both the second and the first-order nature of the transition.
This feature, already analytically predicted on the fully-connected model, seems quite solidly
preserved in the diluted model. However, in numerical simulations preceding
this work, the transition is found not to be compatible with mean-field theory,
according to the estimated value of the scaling exponent of the critical region,
which appears to be outside the boundaries corresponding to a mean-field
universality class. We derive these bounds through a general argument for
mean-field second order transitions.

New results from numerical simulations show how the previous ones were
haunted by strong finite-size effects, as expected in simulations of a dense
model such as the ML RL: the number of connections in the graph requires a
number of operations which scales as the cube of the system size, thus
forbidding the simulation of large enough sizes. To reduce these effects, we
develop a simulation strategy based on periodic boundary conditions on the
frequencies, for which the simulated model at a given size can be regarded as
the bulk of the model with free boundaries pertaining to a larger size. By means
of this strategy, we assess that the scaling of the critical region is actually compatible
with mean-field theory. However, the universality class of the model seems not
to be the same as its fully connected counterpart, suggesting that the ML RL
needs a different mean-field solution.

The possibility of a localization transition in the ML RL is also investigated. In this
context, localization - else termed \emph{power condensation} - is the
phenomenon whereby a finite number of modes carries an extensive amount
of light intensity. The presence of localization, as the global constraint on the
overall intensity is tuned above a given threshold, is only theoretically possible
in presence of dilution with respect to the fully-connected case, where the high
connectivity of the model guarantees equipartition of the constraint among all
degrees of freedom. From the finite-size study of the localization order
parameter, we assess that, despite some evidence of incipient localization, the
glassy phase of light is not strictly speaking localized. Moreover, the study of the
spectral entropy reveals that the low temperature phase of the model is
characterized by intensity equipartition breaking. We have termed “pseudo-
localization” the transition to this hybrid phase, where light intensity is not
completely localized and at the same time is not equipartitioned among the modes. 
One of the most relevant aspects revealed by the numerical results is 
that the critical temperature of the glass and of the pseudo-localization transitions is the
same. This occurrence makes the ML RL an interesting problem where
ergodicity breaking manifests itself in a twofold way: replica-symmetry breaking
and condensation. The opportunity given by this model is to study both
transitions at the same time, opening the way to more general studies for
arbitrary nonlinearities and degrees of dilution.

Supported by the numerical evidence that the ML RL is, indeed, a mean-field
model, we address its analytical solution. Our approach is based on a technique
developed for the Merit Factor problem, which has the same topology of the
ML network. This is an ordered model, which due to antiferromagnetic
couplings, exhibits a frustrated glassy phenomenology. The presence of a glass transition
is investigated through the replica method applied to the model in the space
where the spin variables are mapped by a random unitary matrix. 
We call this version of the model Random Unitary Model (RUM).
A careful study of the saddle-point self-consistency equations of the RUM, both in the replica symmetric and in the one step replica-symmetry breaking scheme reveals the absence of a phase transition for this model and leads us to
question whether the mapping between the original deterministic (though frustrated) model and the RUM is under control.

The technique is then applied to the ML RL, where after averaging over the
disordered couplings we pass to a generalized Fourier space by transforming the
local overlaps with a random unitary matrix. The major difficulty of defining a
global order parameter for the model and finding closed equations to determine
it as function of temperature is successfully addressed, with the introduction of
a new order parameter, a \emph{superoverlap}, which is a measure of the correlations
among local overlaps. However, the solution suffers the same problem of the RUM for the Merit Factor problem.
To the best of our knowledge, this represents the first
tentative solution ever attained of a spin-glass model out of the fully-connected or sparse
graph cases.
\end{abstract}

\begin{acknowledgments}
First of all, I would like to thank my advisors, Luca Leuzzi and Giacomo Gradenigo.
Thanks to Luca for introducing me to this fascinating research topic,
for being patient and always taking me seriously, for his humanity and availability,
qualities that are rare to find.
Thanks to Giacomo for helping me in some difficult moments, for the many physics conversations,
for giving me esteem and trust, and for the many pieces of advice.

I also thank the referees of this thesis work, Juan Jesus Ruiz Lorenzo and Markus Muller,
for their positive and encouraging reports. In particular, thanks to Markus
for carefully correcting the thesis and for raising some interesting questions.

I thank Daniele Ancora for supporting me in a difficult part of this work,
providing his expertise. Thanks for his listening and friendship.

I thank the Chimera group for welcoming me in a stimulating and pleasant environment,
and, no less important, for funding all my research.

Thanks to Giorgio Parisi for being the moral inspiration of this work and
for finding the time to listen to us: it is an honor to feel like adding a small piece to a great story.

A special thanks goes to Matteo Negri and Pietro Valigi, colleagues and friends, because it is nice
to spend time with you talking about everything.

I thank all the people who have crossed our dear $\pi$-room: thank you from the bottom of my heart for
sharing the hardships of research and for making the toughest days easier.

Finally, I would also like to thank Silvio Franz and Ada Altieri for hosting me in Paris
for two visiting periods during my PhD. Thanks in particular to Silvio for teaching me so much
and for showing me that research can be done by getting lost in long and passionate discussions,
without the hurry of commitments.
\end{acknowledgments}
\tableofcontents

\mainmatter
\chapter{Introduction} 
Statistical mechanical models for spin glasses were first introduced in the '70s by Edwards and Anderson \cite{Edwards75} for the study of certain magnetic alloys displaying an intriguing low temperature behavior, which significantly differed from ferromagnetism. In such systems, lowering the temperature did not lead to the onset of long-range order in terms of global magnetization, but rather to the freezing of the material in apparently random configurations. The problem of dealing with this structural and athermal kind of randomness proved to be hard also in the mean-field approximation \cite{Sherrington75}. It took almost a decade and a remarkable series of papers by Parisi \cite{Parisi79a,Parisi79b,Parisi80a,Parisi80b,Parisi80c} to lay the foundations of the mean-field theory of spin glasses and to deepen the knowledge about the spin-glass phase transition. The effort required the development of new mathematical techniques, such as the algebraic replica-symmetry breaking method and the probabilistic cavity approach. The physical scenario coherently revealed by these techniques is that at least at the mean-field level some kind of magnetic order arises at low temperature, where the system exhibits a behavior compatible with ergodicity breaking in multiple pure states non related by a symmetry operation and organized in a highly nontrivial structure \cite{Parisi83,Mezard87}. 

Spin-glass theory, then, took the shape of the ideal settlement to rigorously frame the physical meaning of complexity and describe a number of out-of-equilibrium phenomena, including weak ergodicity breaking and aging, i.e.~the phenomenon by which the relaxation of a system depends on its history \cite{Bouchaud92}. As new spin-glass models with nonlinear interactions were considered \cite{Derrida80,Gross84}, it was soon understood that spin glasses could represent a powerful tool to describe a much larger class of systems spreading over many different fields of research, such as condensed matter physics, biophysics and computer science. The first and probably most studied applications can be traced in structural glasses \cite{Kirkpatrick87,Kirkpatrick89}, the amorphous state reached by many supercooled liquids, when cooled fast enough to avoid crystallization, and neural networks \cite{Amit89}, the prototype of learning systems, which mimic the interactions among neurons in the brain. Nowadays, the list of systems and problems where spin-glass models and techniques have been applied is quite long, ranging from colloids \cite{Dawson01} to granular materials \cite{Metha94}, from protein folding \cite{Bryngelson87} to optimization and constrained satisfaction problems in computer science \cite{Mezard87,Mezard16} and theoretical ecology \cite{Altieri19}. All those systems, ubiquitous in science, where frustration leads to a complex structure of states may be described as spin glasses.

If spin-glass theory represents the perfect framework for a large number of systems, it is also true that new insights on the theory have been acquired from many applications, such as the Random First Order Transition \cite{Lubchenko07,Leuzzi08}, developed in the context of structural glasses, to describe the glass transition and the theory of the jamming transition for the packing of hard spheres \cite{Parisi10,Parisi19}. For this reason, spin glasses can be fairly regarded as one of the most interdisciplinary line of research in statistical mechanics. 

However, despite the number of applications, the mean-field theory of spin glasses has not yet found a clear correspondence in experiments on physical systems. In particular, its most prominent feature, replica-symmetry breaking, has been a long debated issue, leading to question whether it is just an artifact of long-range interactions, rather than an actual physical mechanism \cite{Berthier11}. One would be naturally interested in understanding what of the mean-field picture remains true in finite dimension, that is in the case of the vast majority of physical systems described within the framework of spin-glass theory, which are characterized by rapidly decaying interactions. Unfortunately, unlike the case of ferromagnetism, an approach based on a renormalizable field theory is still missing for spin glasses, albeit very hardly investigated \cite{DeDominicis06} (see also Ref.~\cite{Altieri17,Angelini17,Angelini20} for more recent approaches). However, among the applications of spin-glass theory there is a fortunate one, to which the present work is devoted, which is very promising as an experimental benchmark of replica-symmetry breaking: the study of optical waves in disordered media with gain, namely \emph{random lasers}. Indeed, recently the order parameter of the replica symmetry breaking theory has been experimentally measured in these optical systems, for which the mean-field theory is exact~\cite{Ghofraniha15,Pincheira16,Gomes16,Tommasi16}.

\subsubsection{Random Lasers}
A Random Laser (RL) is made of an optically active medium with randomly placed scatterers \cite{Wiersma96}. As in standard lasers, the optical activity\footnote{With optical activity of the medium we refer to the inversion of the atomic level population of the material by means of external energy injection, which is necessary for stimulated emission. In optics, optical activity also stands for the ability of a substance to rotate the polarization plane of light passing through it.} of the medium provides the gain, whose specific relation with the frequency of the radiation depends on the material. However, random lasers differ from their ordered counterpart both in the inhomogeneity of the medium and in the absence of a proper resonating cavity, which accounts for feedback in standard lasers. In order to have lasing without a cavity some other mechanism at least for light confinement must exist, which manages to overcome the strong leakages of these systems. Since Letokhov's groundbreaking work \cite{Letokhov68}, where light amplification in random media was first theoretically predicted, the trapping action for light has been attributed to the multiple scattering with the constituents of the material. The nature of the feedback, instead, whether it was resonant or non-resonant\footnote{Non-resonant or incoherent feedback leads to amplified spontaneous emission (ASE) or superluminescence \cite{Beenakker98}, which is light produced by spontaneous emission optically amplified by stimulated emission. In this case, interference effects are neglected, and the laser output is only determined by the gain curve of the active medium.}, remained intensely debated for a long time and with it the nature of the modes of RLs \cite{Andreasen11}.

In the original theory by Letokhov only light intensity was considered, with phases and interference not playing any role in mode dynamics. A key finding obtained by means of a diffusion equation with gain is that there is a threshold for amplification, when the volume of the medium is sufficiently large with respect to the gain length. The diffusive limit applies to the case when the mean free path of the photon with respect to scattering is much larger than its wavelength and, at the same time, much smaller than the average dimension of the region occupied by the medium \cite{Letokhov68}. In this approach, above the threshold the emission spectrum is predicted to be continuous and peaked in the frequency corresponding to the maximum gain. These features were observed in early experiments \cite{Markushev86,Gouedard93}, fueling the idea that the notion of modes looses its meaning in RLs. 

Later experiments, based on more accurate techniques and spectral refinement, revealed the emergence of sharp peaks in the emission spectra of RLs on top of the global narrowing as the external pumping was increased \cite{Frolov99,Cao05}. The observation of highly structured and heterogeneous spectra brought evidence in favor of the existence of many coupled modes with random frequencies. Studies on photon statistics \cite{Cao01,Polson01} confirmed this idea, by showing that the intensity of light emitted at the peak frequencies exhibits a Poisson photon count distribution, as in the case of standard multimode lasers. In view of these experiments, it was generally accepted that random lasing is, in fact, characterized by a resonant feedback mechanism, which induces the existence of well-defined cavity modes. This idea is also supported by more recent results drawn from numerical simulations based on the semiclassical theory of RLs \cite{Andreasen11}.


The physical picture that one has to bear in mind is the following: the multiple scattering of light with the randomly placed scatterers not only confines part of the spectrum inside the medium, but also allows for the existence of cavity modes with a lifetime long enough to compete for amplification. The key role of scattering in random lasing is quite remarkable, especially if one thinks that in laser theory scattering is usually considered to be deleterious to the lasing action, since it is responsible for losses disturbing the intensity and directionality of the output. The modes of RLs are many, they are characterized by a complex spatial profile of the electromagnetic field and in most RLs they are extended\footnote{Just a note on the use of some words, which may be misleading: the modes of RLs are \emph{confined} in the medium, in the sense that their spatial extension is comparable with the characteristic length of the sample material, but not properly \emph{localized} in the sense of Anderson localization (see the next paragraph); they are \emph{extended} over the whole volume of the sample, as if the sample itself represents a cavity.} and coupled. In a fascinating way, one can say that random lasers are ``mirror-less'' systems, but not ``mode-less'' \cite{Wiersma08}. 

What is not yet completely understood is the physical source of the oscillating modes and of the corresponding peaks in RLs spectra. Some attempts to explain the occurrence of well defined resonances have been made in terms of light localization, the counterpart for the photons of Anderson localization of electrons \cite{Anderson58}, which was claimed to have been experimentally revealed in Refs.~\cite{Wiersma97,Storzer06,Sperling13}. The presence of localization was inferred from measurements of the deviation from diffusion theory, e.g., through the study of photon time of flight. However, it has been later theoretically proved \cite{Skipetrov14,vanTiggelen21} that light localization can not take place in 3D, due to the vectorial nature of light, as revealed by a comparison between the spectra of the random Green matrix describing the propagation of light from one atom to another in the vector case and in the scalar approximation\footnote{The difference with respect to electron localization lies in the different role played by polarization with respect to electron spin: in the case of light, elementary excitations from one atom to another can be mediated not only by the transverse electromagnetic waves but also by the direct interaction of atomic dipole moments, which is accounted for by the longitudinal component of the electromagnetic field~\cite{Skipetrov14}. 
While the former phenomenon would be reduced by increasing the number density of
atoms, the latter becomes more and more efficient as the typical distance between neighboring atoms decreases.}. This is coherent with the observation that in many materials the modes, though confined inside the sample, are extended all over its volume. The deviations from diffusion theory mistaken for Anderson localization were then traced back to experimental effects, such as delays in fluorescence \cite{Sperling16}. Therefore, though in less than 3D it may truly be observed in particular random lasers \cite{Kumar21}, Anderson localization can not be taken as a general feedback mechanism for these systems.

Whatever the physical mechanism leading to the existence of cavity modes in RLs may be, a multimode theory of RLs based on quantum mechanics principles has to include the openness of the cavity which leads to a nonperturbative effect of the leakages and the inhomogeneity of the medium, which causes the irregular spatial structure of the electromagnetic field. Though a complete quantum theory of light amplification in random media is still missing, when treated in a semiclassical perspective \cite{Viviescas03,Hackenbroich05,Tureci08,Zaitsev10b}, random lasers display two basic features of complex disordered systems: nonlinear interactions and disorder. 

Evidence that random lasing may be a complex phenomenon comes from more recent experiments \cite{vanderMolen06,Mujumdar07,Papadakis07}, which have revealed a new feature. The positions of the spectral peaks were already known to change, if different parts of a sample were illuminated, as a clear consequence of medium heterogeneity. These experiments show a very peculiar behavior in the temporal and spectral response of RLs, when taking shots of the spectrum produced by exactly the same piece of sample at different times, each one corresponding to a pump pulse. The positions of the random scatterers as well as the external conditions are kept fixed all along the data acquisition. The intriguing result is that each shot shows a different pattern of the peaks (see Fig.~\ref{fig:spettriIntro}), meaning that, at variance with standard multimode lasers, there is no specific frequency which is preferred, but depending on the initial state, with the disorder kept fixed, the narrow emission peaks change frequency every time\footnote{Incidentally, this phenomenon may have also contributed to early observations of continuous RL spectra, where data were averaged over many shots, smoothing the spectral profile.}.

This behavior strongly resembles the freezing of magnetic alloys or supercooled liquids in random configurations, making the idea of a spin-glass theory of random lasing quite tempting. Moreover, statistical mechanics is not new to lasing systems: the so-called Statistical Light-mode Dynamics (SLD) proved to be a successful way to deal with standard multimodal lasers, where the number of modes is high enough and nonlinear effects are present \cite{Gordon02,Gordon03}. The main merit of SLD is to show that an effective thermodynamic theory of these photonic systems is possible, where noise, mainly due to spontaneous emission, can be treated in a non-perturbative way. It may seem inappropriate to develop an equilibrium theory for lasers, which are out-of-equilibrium systems by definition, being constantly subjected to external energy injection. However, a stationary regime is achieved in such systems thanks to gain saturation, a phenomenon connected to the fact that, as the power is kept constant, the emitting atoms periodically decade into lower states, saturating the gain of the laser. This justifies the introduction of an equilibrium measure, giving weights to steady lasing states. The extension of the SLD approach to RLs has led quite naturally to the development of a research line devoted to the theoretical modeling of optical waves in random media within the framework of spin-glass theory~\cite{Angelani06a,Angelani06b,Angelani07,Leuzzi09a,Conti11}.

\subsubsection{A Glassy Random Laser}
The two main goals of the spin-glass approach to RLs are the following: (\emph{i}) to provide a theoretical interpretation of the lasing phase of optically active random media in terms of glassy light, which can be regarded as the amorphous phase of light modes; (\emph{ii}) to create the opportunity of experimentally testing the theory of spin glasses, and in particular replica symmetry breaking, on systems in which the glassy state is much easier to access than in structural and spin glasses. Regarding (\emph{ii}), one reason why this is the case is that the dynamics of light modes is incomparably faster with respect to the dynamics of particles in liquids or condensed matter systems, so that an effective equilibrium state is easier to reach for RLs. Incidentally, this is also the reason why by glassy light, here, it is only meant that RLs seem to be characterized by a multi-valley landscape with many possible equilibrium states: phenomena like aging, memory and rejuvenation, which are typical of the dynamics of supercooled liquids, may not be observable on the short timescale in which a laser reaches the steady state. The other -- and maybe more important -- reason is that many RLs are naturally represented by a statistical mechanical system with long-range interactions as a consequence of the fact that the effective mode couplings are determined by the spatial overlap among the wave functions of the modes, which can be extended over the whole medium.

Another merit of this approach is to provide a theoretical framework for the analysis of the mode-locking process in multimode lasers, both standard and random. In standard lasers, mode-locking entails the formation of very short, regularly spaced pulses in the laser output \cite{Haus00}. To produce ultrafast multimode lasers, special devices are required which sustain the pulse formation through nonlinear couplings selected by a particular rule called the frequency-matching condition (FMC). Given four modes, they form an interacting quadruplet only if their frequencies satisfy the following relation 
\begin{align*}
    \text{FMC}: | \omega_1 -\omega_2 +\omega_3 -\omega_4 | \lesssim \gamma,
\end{align*}
where $\gamma$ represents the typical single mode linewidth. In random lasers, pulse formation is, in principle, hindered by the disordered spatial structure of the electromagnetic field and by the random frequency distribution. Indeed, it has never been observed in such systems. However, nonlinear interactions and FMC are intrinsic to a RL and do not require \emph{ad hoc} devices. Though the possibility of a pulsed random laser is still only hypothetical, evidence of a self-induced mode-locked phase has been recently found in Ref.~\cite{Antenucci21}. Within the statistical mechanics approach, the formation of a mode-locked phase is interpreted as a phase transition: while increasing the pump energy, the system leaves a random fluctuating regime to enter a \emph{locked} one, where the oscillation modes have different phases and intensities, but they are fixed, “locked” and “frozen”.

\begin{figure}[t]
\centering
\includegraphics[width=\textwidth]{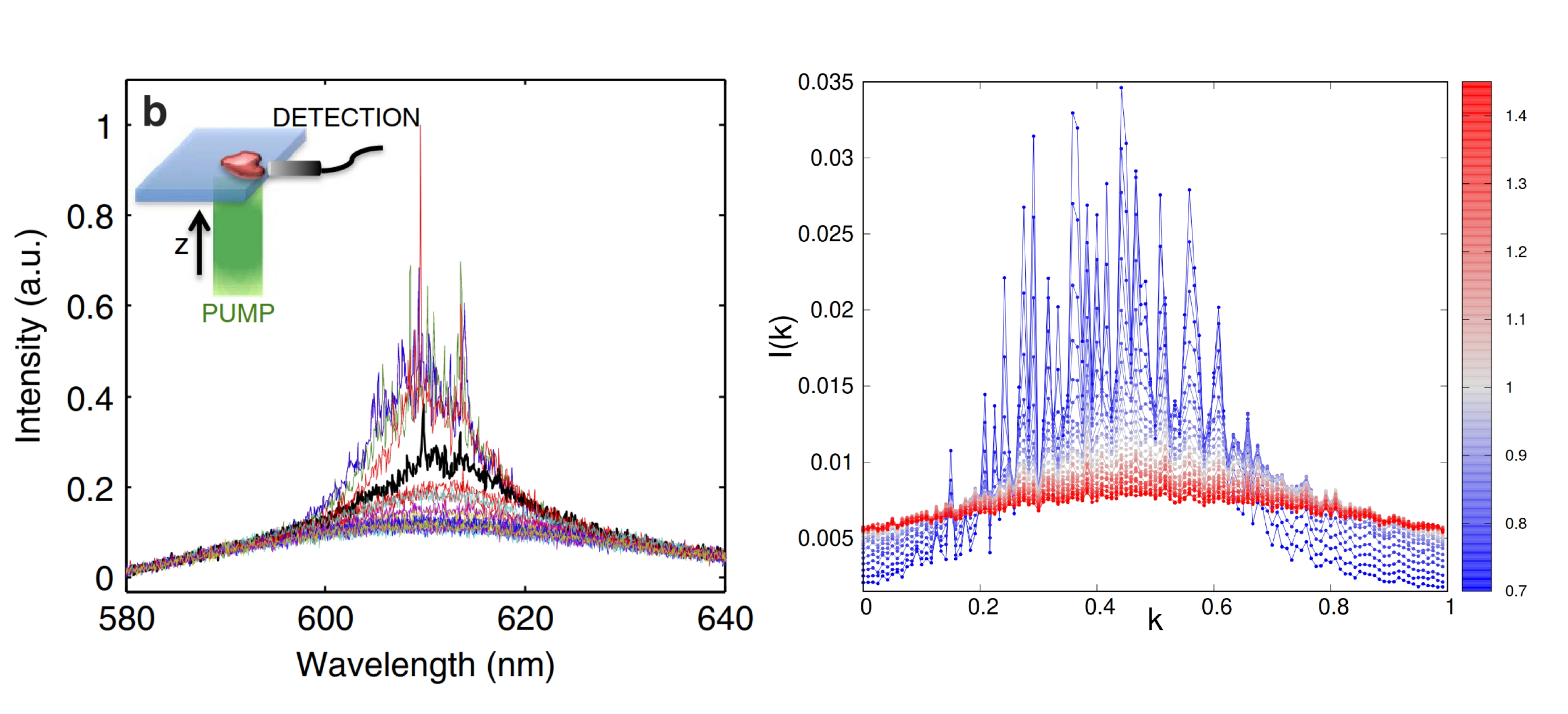}
\caption{Left panel: RL experimental single shot spectra from an amorphous solid sample of T5OCx, thienyl-S,S-dioxide quinquethiophene (reprinted from Ref.~\cite{Ghofraniha15}). Right panel: reproduction of a RL spectrum in numerical simulation of the Mode-Locked 4-phasor model with a frequency comb distribution. The mode index $k$ in abscissa, which pertains to the mode of frequency $\omega_k$, is normalized to the total number of modes, $N=120$. Increasing the pumping rate (from red to blue) the typical narrowing of the spectrum can be observed, together with the emergence of sharp peaks. The aim of this figure is only to show the ability of the spin-glass model for the mode-locked random laser to qualitatively reproduce experimental spectra in numerical simulations. Even if there is no precise mapping between the index $k$ of the right panel and the wavelength in the left panel, the resemblance of the two plots is remarkable.}
\label{fig:spettriIntro}
\end{figure} 

The statistical mechanics description of RLs has led to the definition of the Mode-Locked (ML) $p$-phasor model, a mixed $p$-spin model ($p=2$ and 4, i.e.~both two and four body interactions) with complex variables constrained on a $N$-dimensional sphere and quenched disordered couplings. In this framework, the oscillation modes of the electromagnetic field are represented by $N$ phasors placed on the nodes of the interaction graph and the total optical intensity of the laser is fixed by the spherical constraint on the amplitudes of the phasors. The model can be adapted to describe multimode lasers in the presence of an arbitrary degree of disorder and non-linearity, resulting in a comprehensive theory of the laser mode-locking transition in both random and standard lasers. The interaction graph is dense because the interactions among the modes are long range as a consequence of the evidence of extended modes. The specific topology of the graph is defined by the FMC, which yields a deterministic dilution of the interaction network. 

In Refs. \cite{Antenucci15a,Antenucci15b} the model has been analytically solved in a certain regime compatible with a fully connected graph of interaction, where standard mean-field techniques for spin-glass models can be applied. This particular regime is the \emph{narrow-bandwidth} limit, where the typical linewidth of the modes is comparable with the entire emission bandwidth of the laser. The replica solution of the fully-connected model already presents a very rich phenomenology, with various kinds of replica symmetry breaking, corresponding to nontrivial optical phases. In this context, shot-to-shot fluctuations of the emission spectra are shown to be compatible with an organization of mode configurations in cluster of states similar to the one occurring in spin glasses. Such correspondence relies on the equivalence between the distribution of the Intensity Fluctuation Overlap (IFO), which can be experimentally measured, and the distribution of the overlap between states, the order parameter of the spin-glass transition \cite{Antenucci15c}. Experimental evidence of replica symmetry breaking in the IFO probability distribution function has been found in Ref.~\cite{Ghofraniha15}. 

However, a complete understanding of the physics of RLs requires to go beyond the narrow-bandwidth limit and needs to incorporate in the description the FMC, which is an essential ingredient of the ML $p$-phasor model for the reproduction of the experimental spectra, see Fig.~\ref{fig:spettriIntro}. For combinatorial reasons modes at the center of the spectrum are frequently selected by the FMC, so that when the external pumping is increased and the nonlinear interactions become dominant, the spectrum develops a central narrowing on top of the gain profile curve (which, instead, prevails in the fluorescence regime). The inclusion of the FMC is the main goal of this work, where the problem of dealing with the diluted mode-locked graph is addressed both numerically and analytically. 

It is worth stressing that, besides being of interest for a more realistic description of RLs, the subject of this work is also fascinating from a purely theoretical point of view. In fact, we deal with a nonlinear (4-body) disordered model with complex spherical variables and couplings selected according to a deterministic rule. The presence of a 2-body interaction term, which takes into account the net gain profile of the medium and the radiation losses, allows for the competition between linear and nonlinear interactions, which is known to be responsible for mixed-order replica symmetry breaking. In fact, in the fully-connected case, the model is the generalization of the (2+$p$)-spin (with $p=4$) model \cite{Crisanti04,Crisanti06,Crisanti13} to complex variables, both magnitudes and phases. One of the most interesting features of the model is the dilution of the interaction graph, which is of the order of the system size $N$. This leaves the 4-body interaction network still dense, i.e. still $\mathcal{O}(N^2)$ connections per mode, which is an intermediate situation between the fully-connected ($\mathcal{O}(N^3)$ interactions per mode) and the sparse case (each mode participating in $\mathcal{O}(1)$ interaction terms). To the best of our knowledge, no spin-glass model has been analytically solved in this particular regime of dilution. Moreover, given the presence of a global quantity conserved through a hard constraint (i.e.~the total optical power) the model offers the possibility of studying the occurrence of a \emph{power condensation} transition in the space of the modes, especially in relation to the breaking of ergodicity, which is signaled by replica-symmetry breaking. The possibility of intensity localization is also suggested by the sharp peaks in the spectra of Fig.~\ref{fig:spettriIntro}, which are evidence of the fact that the total value of the intensity is not homogeneously parted among the modes and might be a precursor to a sharp condensation of the whole intensity on $\mathcal{O}(1)$ modes.

\subsubsection{Organization of the Thesis}
The Thesis is divided in two parts, a numerical and an analytical one, which are preceded by an introductory chapter on the mean-field theory of RLs. The organization of the chapters follows the natural development of the research: after acquiring confidence with the status of the art, the results of numerical simulations are presented and discussed in Part I; then, inspired by the physical insights obtained through the simulations, in Part II the analytical approach is developed. Three Appendices contain much of the technicalities of the computations. Each chapter opens with a brief introduction to the topic to which it is devoted. In what follows, we sketch the contents of each chapter.
\begin{itemize}
    \item \textbf{Chapter \ref{chap:IntroCap}} contains an Introduction to the spin-glass theory of RLs, where the main analytical results obtained within the mean-field fully-connected approximation are described in some detail. After introducing the reader to the statistical mechanics approach to standard (ordered) multimode lasers, the spin-glass model for random laser is derived starting from the semiclassical laser theory for open and disordered systems in the system-and-bath approach. The presence of off-diagonal linear terms of interactions among the modes is related to the openness of the system, while the presence of nonlinearity accounts for the light-matter interaction at the third order in perturbation theory in the mode amplitudes. The relevant approximations which are needed in order to obtain the mean-field fully-connected model are described. Then, the replica computation to derive the quenched average of the free energy is considered and the phase diagram of the model is described. The last section is devoted to an introduction to the IFO and how they are related to the Parisi overlap in the mean-field fully-connected theory.
    \item \textbf{Chapter \ref{chap:mixedorder}} deals with the first attempt at including the dilution effect due to the FMC in the theory through numerical simulations. The numerical technique used to simulate the model is described in detail and the results of Ref.~\cite{Gradenigo20} are carefully reviewed. The density of the model interaction graph represents an additional difficulty with respect to those already present in Monte Carlo simulations of finite-dimensional spin-glass models: not only the relaxation to equilibrium is hindered by the presence of local minima in the free energy landscape, but each attempt of changing configuration has a computational complexity which scales as the square of the system size. Moreover, especially for the study of non-self-averaging quantities, many samples corresponding to different realizations of the disordered couplings have to be simulated. In order to deal with these difficulties, a Parallel Tempering Monte Carlo algorithm has been developed and parallelized for Graphic Processing Units. From the simulations, the typical behavior of a Random First-Order Transition is revealed for the simulated model, though the results are plagued by strong finite-size effects, making the assessing of the universality class of the model a nontrivial task.
    \item \textbf{Chapter \ref{chap:Univ}} is devoted to a refinement of the finite-size scaling analysis of the glass transition for the ML 4-phasor model. Many of the results presented here are contained in Ref.~\cite{Niedda22a}. In order to reduce the finite size effects in numerical simulations of the mode-locked glassy random laser, two strategies have been exploited: first, simulations with larger sizes and a larger number of disordered samples have been performed; secondly, and more remarkably, a version of the model with periodic boundary conditions on the frequencies has been introduced in order to simulate the bulk spectrum of the model. The results obtained by a more precise finite-size scaling technique allow us to conclude that the ML 4-phasor model is indeed compatible with a mean-field theory, though it may be in a different universality class with respect to its fully-connected counterpart. This is the main output of this chapter; then, the study of the glass transition is completed by presenting results which pertain to various overlap probability distribution functions.
    \item \textbf{Chapter \ref{chap:Condens}} is devoted to the numerical study of the power condensation phenomenon in the ML 4-phasor model. The results presented here are contained in Ref.~\cite{Niedda22b}, where evidence of an emergent pseudo-localized phase characterizing the low-temperature replica symmetry breaking phase of the model is provided. A pseudo-localized phase corresponds to a state in which the intensity of light modes is neither equipartited among all modes nor really localized on few of them. Such a hybrid phase has been recently characterized in other models, such as the Discrete Non-Linear Schrödinger equation \cite{Gradenigo21a}, just as a finite size effect, while in the low temperature phase of the glassy random laser it seems to be robust in the limit of large size. The differences between such non-interacting models and generic $p$-body nonlinear interacting models are highlighted: in particular, the role played by the dilution of the interaction network is clarified.  
    \item \textbf{Chapter \ref{chap:MF}} is the first analytical chapter of this Thesis and the only one which is not directly dedicated to the ML 4-phasor model for the glassy random laser. The similarity between the topology of the mode-locked graph and the structure of the Hamiltonian of the Bernasconi model for the Merit Factor problem \cite{Bernasconi87}, has led us to devote our attention to this model first. Although it is a model with long-ranged ordered interactions, finite-size numerical studies, which have been replicated in this work, point in the direction of a glassy behavior at low temperature. The solution technique proposed in Ref.~\cite{Marinari94a}, which is based on quenched averaging over the unitary group of transformations of the spin variables, is carefully analyzed and completed through the study of the saddle-point equations with different ansatzes of solution. No evidence of phase transition at finite temperature has been found with one step of Replica Symmetry Breaking (RSB), up to the precision of our analysis; however, we believe that the solution technique may be the right tool to address the computation of the free energy in the mode-locked random laser. The three Appendices to this chapter deal respectively with the integration over the Haar measure of the unitary group and the Replica Symmetric (RS) and 1RSB details of the computation.
    \item \textbf{Chapter \ref{chap:ML}} is devoted to the proposal of a new mean-field theory for the mode-locked glassy random laser. The quenched average over the disordered couplings leads to a long-range ordered matrix field theory in the local overlap, which is characterized by a Hamiltonian formally similar to the one of the Merit Factor problem, but at the level of the local overlaps rather than of the spins. The technique developed for the Bernasconi model is then applied to the model of interest, allowing us, after averaging over the unitary group, to introduce a global order parameter, which we have called \emph{superoverlap}. As the global overlap usually represents a two-point correlation between spin variables, the superoverlap denotes a correlation between local overlaps. The RS and 1RSB self-consistency equations have been derived, and their study is in progress.
    \item \textbf{Chapter \ref{chap:disc}} contains the conclusions of this Thesis and a discussion on the research lines opened by the present work on the topic. Among them we mention the integration of the saddle point equations of the ML 4-phasor model, the numerical simulation of models with realistic frequency distributions and gain profiles, a detailed analysis of the comparison between the experimentally measured and the numerically computed overlap distributions, considering thermalization, size and time-averaging effects
\end{itemize}

\newpage

\chapter{Mean-Field Theory of the Glass Transition in Random Lasers} \label{chap:IntroCap}
In this chapter the most salient features of the mean-field spin-glass theory of random lasers are described. Before getting to the heart of the discussion, some background knowledge is provided about multimode lasing systems, in order to make the reader confident with the most relevant physical properties of these systems from a statistical mechanics point of view.

The case of standard multimode lasers is discussed first, since it represents a constant basis for comparison for the more general statistical theory of random lasers. Given the large number of modes ($10^2-10^9$ in long lasers) and the stabilizing effect of gain saturation, an effective thermodynamic theory can be developed for the stationary regime \cite{Gordon02,Gordon03}. The main outcome of the mean-field analysis of these systems is that the onset of the mode-locking regime \cite{Haus00}, can be interpreted as a noise driven first-order phase transition \cite{Gat04}. Then, we briefly review the system-and-bath approach to random laser theory developed in Refs. \cite{Hackenbroich02,Viviescas03,Viviescas04} to deal with the openness of the cavity and the light-matter interaction. In our perspective, the main merit of this approach is to provide reasonable explanations for the origin of all the essential elements of the general spin-glass model of a RL, starting from the semiclassical approximation to the quantum dynamics of the electromagnetic field in an open and disordered medium. 

In the second part of the chapter, the spherical (2+4)-phasor model \cite{Antenucci15a,Antenucci15b}, which represents the leading mean-field spin-glass model for RLs, is presented in connection with the semiclassical derivation. The particular regime where the theory applies is carefully described, by presenting all the approximations which make the model compatible with mean-field fully-connected theory. After a brief summary of the replica method for the solution of quenched disordered systems, the replica computation for the model of interest is reviewed in its main steps and the results are described. The general phase diagram of the model is presented, with particular attention to the glass transition, which will be studied in the rest of this work. The phenomenology of the model is very rich already in the fully-connected case, where the system exhibits four different phases corresponding to different regimes in the output of a laser depending on the amount of energy injected into the system and on the degree of disorder of the medium. Moreover, the breaking of replica symmetry occurs with three different kind of structures depending on the degree of non-linearity. In the last section, the theory is put in correspondence with experiments through the study of the overlap among intensity fluctuations \cite{Ghofraniha15} (i.e.~IFO), an experimentally measurable quantity whose analytical counterpart can be expressed in terms of the Parisi overlap in the fully-connected approximation \cite{Antenucci15c}.

\section{Statistical Light-mode Dynamics}
Though concepts borrowed from phase transition physics were already present in the seminal work of Lamb on multimode lasers \cite{Lamb64,Scully67,Haken84}, it is not until the early '00s that statistical mechanics methods were systematically applied to the study of optical systems. Statistical Light-mode Dynamics (SLD) is an approach developed by Gordon and Fisher \cite{Gordon02,Gordon03} to deal with open problems regarding the mode-locking phenomenon in multimode lasers. Mode-locking is a consequence of the fact that, unlike a conventional laser, a mode-locked laser oscillates among longitudinal modes whose frequencies are in a coherent relationship. In standard lasers the interaction among axial modes necessary for pulse formation is induced by \emph{ad hoc} devices: either the system is made time dependent by means of an amplitude modulator, or a suitable nonlinearity, as the one provided by a saturable absorber\footnote{Saturable absorption is the property of a material with a certain absorption loss for light, which is reduced at high optical intensities. Since the absorption coefficient depends on the light intensity, the absorption process is nonlinear.}, is added to the system dynamics. Between the two methods, which are commonly referred to as, respectively, active and passive mode-locking, only the latter is known to produce ultra-short pulses (of the order of femtoseconds). 


The mode-locking theory developed in the seventies \cite{Kuizenga70,Haus75,Haus00} has many merits, such as the prediction of the pulse shape and of its duration. However, the underlying mechanism to pulse formation remained unclear, until the SLD approach was formulated. It was already known that pulse formation may be achieved when the optical power reaches a certain threshold (besides the one needed for the onset of lasing) and that the emergence of pulses upon reaching this threshold is abrupt. Several hypotheses were put forward to explain this phenomenon, by identifying some mechanism which opposed to mode-locking \cite{Ippen90,Haus91,Chen95}, but no one was really satisfactory. In most of these approaches, the antagonist of optical power for the onset of pulse formation was correctly identified with noise, which however, was treated as a small perturbation. In fact, noise plays a central role in the dynamics of a laser: besides the usual sources of noise to which a physical system is subjected, in lasers a fundamental source of noise is represented by spontaneous emission, which can also be amplified due to optical activity. By treating noise in perturbation theory, many interesting features of the system can be missed when the noise is large.

The main novelty introduced by SLD is represented by the inclusion of noise in the theory in a non-perturbative way, as an effective temperature. This has lead to the first many-body thermodynamic theory of multimode lasers, where the onset of mode-locking is interpreted as a phase transition driven by the ratio between external pumping and noise. As the energy pumped into the system makes the interactions strong enough to overcome noise, then, global correlations arise among the phases of the modes, which sharply divide the unlocked and locked thermodynamic phases. In this framework, the difference between active and passive mode-locking becomes evident: when considered from the point of view of the interaction networks, the passive case corresponds to a long-range model \cite{Gat04}, where a global order can arise below a certain level of noise, whereas the active case corresponds to a one-dimensional short-range model \cite{Gordon04}, where a phase transition can in principle occur only at zero temperature\footnote{Amplitude modulation produces sidebands of the central frequency of the spectrum, say $\omega_0$, at the neighbor frequencies $\omega=\omega_0 \pm \delta\omega$, where the $\delta \omega$ is the frequency spacing, which lock the corresponding modes to the central one, and so on. This leads to nearest neighbors interactions on a linear chain.}. Hence, the fragility of active mode-locking ca be interpreted as a manifestation of the lack of global ordering at finite temperature in the one-dimensional spherical spin model \cite{Berlin52}: any weak noise breaks a bond between two modes, thus eliminating global ordering.

In the following, we focus on the theory of passive mode-locking, which is the most interesting one for the random laser case. In an ideal cavity, i.e.~by neglecting the leakages, the electromagnetic field can be expanded in $N$ normal modes 
\begin{align}
     \bm{E}(\bm{r},t) = \sum_{k=1}^N a_k(t) e^{-i \omega_k t} \bm{E}_k(\bm{r}) + \text{c.c.}
\end{align}
where the presence of nonlinearity makes the complex amplitudes $a_k(t)$ time dependent. In the physical situation, $N$ corresponds to the number of distinguishable resonances selected according to the distance between the mirrors. If the frequencies of adjacent modes are too close with respect to the spectral resolution, then the actual number of cavity modes is larger than the number of bins in the revealed spectrum. The frequency distribution of the modes is that imposed by a Fabry-Perot resonator, namely a linear comb:
\begin{align} \label{FreqComb}
    \omega_k=\omega_0 + k \delta\omega, ~~~~~~~~~ k\in \left[-\frac{N}{2},\frac{N}{2}\right]
\end{align}
where $\omega_0$ is the central frequency of the spectrum and $\delta\omega$ is the frequency spacing. In the high finesse limit, if we denote by $\Delta\omega$ the bandwidth of the entire spectrum and by $\gamma$ the typical linewidth of the modes\footnote{Even if photons are emitted exactly with the atomic frequency $\omega_{ij} = (E_i-E_j)/\hbar$, broadening effects give to the resonator modes a width $\gamma$. These effects can be homogeneous, like collision broadening, leading to a Lorentzian line-shape function or inhomogeneous, like Doppler broadening, leading to a Gaussian line-shape function: the Voigt profile takes into account both kinds of broadening \cite{Haken84}. In principle, then, each mode has a different linewidth. By neglecting these effects one would have a frequency distribution made of sharp delta peaks.} then we have $\gamma \ll \delta \omega \ll \Delta \omega$. Moreover, we consider the \emph{slow amplitude} mode basis, in which given a mode with frequency $\omega_k$, the time dependence of the amplitude of such a mode is on a time scale much larger than $\omega_k^{-1}$. Lasing modes are by definition slow amplitude modes, since their expression in the frequency domain must be approximately equal to a delta centered in their frequency:
\begin{align} \label{slowAmpl}
    a_k(t) e^{-i\omega_k t} \rightarrow \int \de t~a_k(t) e^{-i\omega_k t} e^{i\omega t}  \approx a_k(t) \delta(\omega-\omega_k),
\end{align}
where the Fourier transformation of the electromagnetic field basically reduces to a time average over the fast phase oscillations.

The SLD description of passive mode-locking can be obtained by considering the standard Langevin master equation \cite{Haus00} 
\begin{align} \label{HausME}
    \frac{\de a_l}{\de t} = (G_l + i D_l)a_l(t) + (\Gamma - i \Delta) \sum_{i-j+k=l} a_i(t) \overline{a}_j(t) a_k(t) + \eta_l(t),
\end{align}
where $G_l$ is the net gain profile, i.e. the gain minus the losses, $D_l$ is the group velocity dispersion coefficient, $\Gamma$ is the self-amplitude modulation coefficient resulting from saturable absorption and $\Delta$ is the self-phase modulation coefficient responsible of the Kerr lens effect. The nonlinear term in the dynamics is characterized by the selection rule $i-j+k=l$, which comes from averaging away the fast phase oscillations in the slow amplitude basis. This is actually a condition on the frequencies, the so-called \emph{frequency matching condition} (FMC), which reduces to a relation among indices because of Eq.~\eqref{FreqComb}. The noise $\eta_l(t)$, mainly due to spontaneous emission, is generally assumed Gaussian, white and uncorrelated:
\begin{align}
    <\overline{\eta}_l(t_1) \eta_k(t_2) > = 2 T \delta_{lk} \delta(t_1-t_2)  ~~~~~~~~~ <\eta_l(t_1) \eta_k(t_2) > = 0,
\end{align}
where $T$ is the spectral power of the noise, which has a dependency on the actual temperature of the sample (i.e.~the laboratory temperature in the experimental case). 

We distinguish two kinds of dynamics in Eq.~\eqref{HausME}: a dissipative one, which involves gain and saturable absorption and a dispersive one. In order to develop an effective thermodynamic approach, a necessary requirement is laser stability. The total optical intensity of the laser $\mathcal{E} = \sum_{k=1}^N |a_k|^2$ is a constant of motion only in the purely dispersive limit, i.e.~when $G_l=\Gamma=0$. However, in the general case the stability of the laser is ensured by gain saturation, the effect for which the gain decreases as the intensity increases \cite{Chen94}. In laser theory, gain saturation is usually implemented by assuming a time dependent gain, e.g. for a flat gain curve $G = G_0/(1+\mathcal{E}/E_{\text{sat}})$, where $E_{\text{sat}}$ is the saturation power of the amplifier and $G_0$ is the unsaturated gain. 

In Ref.~\cite{Gordon02} a simpler alternative has been proposed: one can assume that at each time $G_0$ takes exactly the value necessary to keep the optical power fixed to its original value. The precise value of $G_0$ can be obtained by imposing that $\partial \mathcal{E}/\partial t = 0$ and exploiting the equation of motion \eqref{HausME}. This way, the stabilizing effect of gain saturation can be modeled by considering a time independent gain profile $G_l$ and imposing a hard constraint on the total intensity, which forces the dynamics on the hypersphere $\mathcal{E}=E_0$. We refer to this choice as fixed-power ensemble \cite{Antenucci16}.
This hard constraint might be relaxed studying the dynamics of the overall total intensity under
saturation, evolving at a much larger time scale than the dynamics of the single mode phasors.
The fluctuations in a variable-power ensemble might, then, be studied, in a way that pretty much resembles the relation between ensembles in statistical mechanics \cite{Gat04}. We will come back to this topic in Chap.~\ref{chap:Condens}.

In the purely dissipative limit, i.e. $D_l = \Delta =0$, it has been shown in Ref.~\cite{Gordon03} that the stationary distribution of configurations, solution to the Fokker-Planck equation associated to Eq.~\eqref{HausME}, tends to a Gibbs-Boltzmann measure. Therefore, the equilibrium properties of the system can be investigated by studying the Hamiltonian
\begin{align} \label{OrderedHamilt}
    \mH = - \sum_{k=1}^N G_k |a_k|^2 - \frac{\Gamma}{2} \sum_{\text{FMC}(\bm{k})} a_{k_1} \overline{a}_{k_2} a_{k_3} \overline{a}_{k_4},
\end{align}
with the spherical constraint $\sum_k |a_k|^2 = E_0 = \epsilon N$. The model can be, then, studied as a statistical mechanical system in the canonical ensemble at equilibrium in a thermal bath at the effective temperature 
\begin{align} \label{E-TPHOTON}
    T_{\rm photonic} = \frac{T}{\epsilon^2} = \frac{1}{\mathcal P^2},
\end{align}
where $\mathcal{P}$ is the so-called \emph{pumping rate}. This temperature accounts for the competition between the optical power injected into the system, which favors the ordering action of the interactions, and the noise, which acts in the opposite direction. It is worth stressing that the effective temperature $T_{\rm photonic}$ reduces to the spectral power of noise $T$, if one considers $\epsilon=1$ (i.e.~if one fixes the spherical constraint to a specific value of the total intensity). 

In Ref.~\cite{Gat04}, the model has been solved in the \emph{narrow-bandwidth approximation}, in which the typical linewidth of the modes $\gamma$ is comparable to the total spectral bandwidth $\Delta \omega$. In this limit the FMC is always satisfied, by any quadruplet of modes, thus yielding a fully-connected graph of interactions. The mean-field analysis of the model reveals a first-order transition with respect to the value of $\mathcal{P}$ between two thermally disordered and ordered phases, characterized respectively by unlocked and locked phases of the mode amplitudes $a_k$. The former is the low-$\mathcal{P}$ (high temperature) phase corresponding to an incoherent output of the multimode laser (continuous wave - CW), the latter is the high-$\mathcal{P}$ (low temperature) phase corresponding to a coherent output, equivalent to pulses in the time domain (mode-locking - ML). The theory has found experimental confirmation in Ref. \cite{Vodonos04}. 

It is worth noting that in the narrow-bandwidth limit, the modes are locked in a trivial way: almost all of them are aligned in the same direction in the complex plane, i.e. they have the same value of the phase. In the language of magnetic systems (the analogy here is with the ordered XY model), the locking of the phasors to the same angle leads to the presence of global magnetization. Correspondingly, the output of the laser is a approximately a plain wave, which is equivalent to sharp delta-like pulses in the time domain. Therefore, spectra are not possible in this case, in the sense that they reduce to a single spectral line plus some noise. If one goes beyond the fully-connected case, a different kind of global ordering arises at high pumping $\mathcal{P}$ which consist in the onset of \emph{phase-waves}, as it is discussed in the introduction to Chap.~\ref{chap:mixedorder}.

In the general case, when the dispersive effects are included, the problem becomes hard to study analytically. However numerical simulations show that the presence of these effects do not change qualitatively the physical scenario of a first-order transition, leading only to a lowering of the critical value of the temperature \eqref{E-TPHOTON} which drives the transition \cite{Gordon03,Antenucci16}.

\section{Multimode Laser Theory in Open and Disordered Media}
The previous description is based on the underlying quantum theory of a multimode laser in an ideally closed cavity, which was first developed in the semiclassical approximation by Lamb in Ref.~\cite{Lamb64} and then generalized to the fully quantum case by Scully and Lamb in Ref.~\cite{Scully67}. In the case of random lasers, a complete quantum theory is still missing given the difficulty of the problem. Besides the disorder of the active medium, which makes the spatial dependence of the electromagnetic field not easy to compute, one more fundamental problem is how to deal with the openness of the system, when the leakages are non-perturbatively relevant. The quantization procedure in this case presents the typical technical issues of quantum systems with dissipation, which are non-Hermitian problems where the spectral theorem for self-adjoint operators does not apply, i.e.~the standard decomposition in a unique complete set of orthogonal eigenvectors corresponding to real eigenvalues is not possible. This problem was already present in quantum optics, even before the theory of Lamb for multimode laser was developed, since when Fox and Li first studied the effect of diffraction losses in a cavity \cite{Fox61}. More recently, several relevant studies have been put forward to overcome the difficulties \cite{Dutra00,Tureci06}, but a part from the exceptional case of a two mode-laser~\cite{Eremeev11}, to the best of our knowledge, the problem remains open. For a comprehensive review on the topic see Ref. \cite{Zaitsev10a}.

Among the approaches that have been proposed, we focus on one based on the standard \emph{system-and-bath} decomposition (see e.g.~Refs.~\cite{Senitzky60,SargentIII78}) which develops the clearest physical intuition and seems to be the most convenient one for the case of random lasers. The experimental observations discussed in the Introduction push towards the development of a theory of the electromagnetic field  which accounts for both a discrete and a continuous part of the spectrum, the former comprised by modes which are confined inside the medium by multiple scattering, the latter by diffusive modes radiating from the medium. The system-and-bath approach developed in Refs. \cite{Viviescas03,Viviescas04} is based on regarding the quantum subsystem composed of electromagnetic cavity modes as embedded in an environment of scattering states into which the states of the system can decay, i.e.~the bath. In the following, we briefly sketch the main features of the approach and report the results. We refer to Ref. \cite{Antenucci16} for a more detailed exposition. 

\subsection{System-and-Bath Decomposition}
The starting point of this approach is the expansion in \emph{modes-of-the-universe} developed in Refs. \cite{Lang73,Glauber91}, where the electromagnetic field quantization is carried out in the presence of a 3-dimensional dielectric medium with spatially dependent permittivity $\epsilon(\bm{r})$ and without specifying boundary conditions. The electromagnetic field can be expressed in terms of its vector potential $\bm{A}(\bm{r},t)$ and of its scalar potential $\Phi(\bm{r},t)$. The Coulomb gauge (transversal gauge), generalized to the case of inhomogeneous media, is defined by the following relations
\begin{eqnarray}
    \Phi(\bm{r},t) &=& 0 \\
    \bm{\nabla} \cdot [\epsilon(\bm{r}) \bm{A}(\bm{r},t)] &=& 0
\end{eqnarray}
and allows one to write the electric and magnetic fields in the form
\begin{eqnarray}
    \bm{E}(\bm{r},t) &=& - \frac{1}{c} \dot{\bm{A}}(\bm{r},t) \\
    \bm{B}(\bm{r},t) &=& \bm{\nabla} \times \bm{A}(\bm{r},t),
\end{eqnarray}
where the dot denotes the time derivative. The Hamiltonian of the system is given by
\begin{equation}
    \mH = \frac{1}{2} \int \de \bm{r} \left\{ \frac{c^2}{\epsilon(\bm{r})}\bm{\Pi}^2(\bm{r},t) + \left[ \bm{\nabla} \times \bm{A}(\bm{r},t)  \right]^2  \right\},
\end{equation}
where $\bm{\Pi} = \epsilon(\bm{r}) \dot{\bm{A}} / c^2$ is the cojugated momentum of the vector potential. The modes-of-the-universe are defined as solutions of the Helmoltz equation
\begin{align} \label{Helm}
        \bm{\nabla} \times [\bm{\nabla} \times \bm{f}_m(\omega, \bm{r})] - \frac{\epsilon(\bm{r}) \omega^2}{c^2}  \bm{f}_m(\omega, \bm{r})  = 0,
\end{align}
where the functions $\bm{f}_m(\omega, \bm{r})$ are defined in all space and satisfy the transversality condition $\bm{\nabla} \cdot [\epsilon(\bm{r}) \bm{f}_m] = 0$. The index $\omega$ is a continuous frequency, but the formalism can be easily adapted to the case of a discrete spectrum by using a discrete index and replacing integrals with sums. The discrete index $m$ specifies the asymptotic boundary conditions far away from the dielectric, including the polarization. We consider asymptotic conditions corresponding to a scattering problem with incoming and outgoing waves. Then $\bm{f}_m(\omega,\bm{r})$ represents
a solution with an incoming wave in channel $m$ and only outgoing waves in all other scattering channels. The definition of the channels depends on the problem at hand: for a dielectric coupled to free space, one may expand the asymptotic solutions in terms of angular momentum states. Then $m$ corresponds to an angular momentum quantum number. On the other hand, for a dielectric connected to external
waveguides, $m$ may represent a transverse mode index \cite{Viviescas03}.

Equation \eqref{Helm} is the classical equation of motion for the field  dynamics in the generalized Coulomb gauge, which can be obtained through a variational principle from the Lagrangian of the electromagnetic field. By defining $\bm{\phi}_m (\omega, \bm{r}) = \sqrt{\epsilon(\bm{r})} \bm{f}_m(\omega, \bm{r})$, the equation can be cast into a well-defined eigenvalue problem for the Hermitian differential operator $\mathcal{L}$: 
\begin{gather}
     \mathcal{L} \bm{\phi}_m (\omega, \bm{r}) = \frac{\omega^2}{c^2} \bm{\phi}_m (\omega, \bm{r}), \label{EigenvalueProb} \\
    \mathcal{L} = \frac{1}{\sqrt{\epsilon(\bm{r})}} \bm{\nabla} \times [\bm{\nabla} \times \frac{1}{\sqrt{\epsilon(\bm{r})}}],
\end{gather}
where the eigenmodes $\bm{\phi}_m (\omega, \bm{r})$ form a complete set in the subspace of $L^2$ functions defined by the transversality condition. The vector potential can be then expressed in terms of the eigenmodes as 
\begin{align}
    \bm{A}(\bm{r},t) = c \sum_m \int \de \omega~q_m(\omega, t)  \frac{\bm{\phi}_m(\omega, \bm{r})}{\sqrt{\epsilon(\bm{r})}},
\end{align}
and a similar expression holds for its conjugated momentum $\bm{\Pi}$ with coefficients $p_m(\omega, t)$. Quantization can be obtained by promoting the coefficients of the expansion to operators and imposing canonical relations on them. 

This normal mode expansion is a consistent field quantization scheme in presence of inhomogeneous media, but does not provide any particular information about the field inside the medium. As showed in Ref.~\cite{Viviescas03}, a separation into cavity (else termed resonator) and radiative (or channel) modes can be obtained by means of a Feshbach projection \cite{Feshbach58}. The eigenmodes of the total system can be projected onto orthogonal subspaces by the operators
\begin{align}
    \mathcal{Q} = \int_{\bm{r} \in V} \de \bm{r} |\bm{r} \rangle \langle \bm{r} | ~~~~~~~~~ \mathcal{P} = \int_{\bm{r} \notin V} \de \bm{r} |\bm{r} \rangle \langle \bm{r} |,
\end{align}
where $V$ is the region of the whole space where the dielectric is present. The eigenmodes $\bm{\phi}_m(\omega, \bm{r})$ can be then written as $|\bm{\phi}\rangle = |\bm{\mu}\rangle + |\bm{\nu}\rangle$, where $|\bm{\mu}\rangle = \mathcal{Q} |\bm{\phi}\rangle$ and $|\bm{\nu}\rangle = \mathcal{P} |\bm{\phi}\rangle$ represent respectively the projections on the cavity and radiative subspaces and $\bm{\phi}(\omega,\bm{r})=\langle \bm{r}|\bm{\phi} \rangle$. Similarly the actual modes-of-the-universe $\bm{f}_m(\omega, \bm{r})$ can be written as $|\bm{f} \rangle = |\bm{u}\rangle + |\bm{v}\rangle$, where $|\bm{u}\rangle$ and $|\bm{v}\rangle$ correspond respectively to $|\bm{\mu}\rangle$ and $|\bm{\nu}\rangle$. The cavity modes vanish outside $V$ and, hence, form a discrete set labeled by a discrete index $\lambda$; vice versa the radiative modes vanish inside $V$ and form a continuum, labeled by a continuous index $\omega$ and a discrete index $m$ specifying boundary conditions at infinity. Each set of modes is a complete and orthonormal set in the subspace of definition, but as whole they can not be considered eigenmodes of the total system.

The eigenvalue problem in Eq.~\eqref{EigenvalueProb} can be rewritten in this formalism and solved with suitable matching conditions at the boundaries. The differential operator $\mathcal{L}$ can be decomposed into resonator $\mathcal{L}_{\mQ\mQ}$, channel $\mathcal{L}_{\mathcal{P}\mathcal{P}}$ and coupling $\mathcal{L}_{\mQ\mathcal{P}},\mathcal{L}_{\mathcal{P}\mQ}$  contributions, in such a way that
\begin{align}
\begin{pmatrix}
\mathcal{L}_{\mQ\mQ} & \mathcal{L}_{\mQ\mathcal{P}} \\
\mathcal{L}_{\mathcal{P}\mQ} & \mathcal{L}_{\mathcal{P}\mathcal{P}}
\end{pmatrix}  \begin{pmatrix}
\bm{\mu}(\bm{r}) \\ \bm{\nu}(\bm{r})
\end{pmatrix}  = \left(\frac{\omega}{c} \right)^2 \begin{pmatrix}
\bm{\mu}(\bm{r}) \\ \bm{\nu}(\bm{r})
\end{pmatrix}
\end{align}
where $\bm{\mu}(\bm{r})=\langle \bm{r} | \bm{\mu} \rangle$ and equivalently for $\bm{\nu}$. The solution yields an exact representation of the eigenstates in terms of cavity and radiative modes
\begin{align}
    |\bm{\phi}_m(\omega)\rangle = \sum_\lambda \alpha_\lambda (\omega) |\bm{\mu}_\lambda \rangle + \sum_m \int \de \omega' \beta_m(\omega,\omega') |\bm{\nu}_m (\omega')\rangle,
\end{align}
where $\bm{\mu}_\lambda$ and $\bm{\nu}_m(\omega)$ are the solutions of the uncoupled problems for $\mathcal{L}_{\mQ\mQ} $ and $\mathcal{L}_{\mathcal{P}\mathcal{P}}$, while the coefficients $\alpha, \beta$ carry the dependence on the coupling operators $\mathcal{L}_{\mQ\mathcal{P}},\mathcal{L}_{\mathcal{P}\mQ}$. The same decomposition holds for the wavefunctions $\bm{f}_m(\bm{r},\omega)$ in terms of their projections $\bm{u}_\lambda$ and $\bm{v}_m(\omega)$.

The vector potential can be expanded in terms of cavity and radiative modes
\begin{align}
    \bm{A}(\bm{r},t) = c\sum_\lambda Q_\lambda(t) \bm{u}_\lambda(\bm{r}) + c \sum_m \int \de \omega~Q_m(\omega,t) \bm{v}_m(\bm{r},\omega),
\end{align}
and similarly for the conjugated momentum $\bm{\Pi}$, with coefficients $P_\lambda(t)$ and $P_m(\omega,t)$ respectively for the discrete and continuoous part of the spectrum. Quantization can be obtained as usual, by promoting the coefficients of the expansion to operators and imposing canonical commutation relations. Eventually, the field Hamiltonian takes the expected system-and-bath form, which in the \emph{rotating-wave} approximation\footnote{The rotating-wave approximation \cite{Bransden84} allows one to neglect fast oscillating terms in the Hamiltonian of an optical system. In the present case we only keep the resonant terms $(a b^\dagger,a^\dagger b)$ in the system-and-bath coupling and neglect the nonresonant ones $(a b,a^\dagger b^\dagger)$, which become relevant only when the frequencies of the modes are spread over a range comparable to their typical frequency. Then, if $\Delta \omega$ is the width of the entire spectrum, the rotating-wave approximation holds as far as $\Delta\omega \ll \omega$, that is, the typical situation for random optically active materials.} reads as
\begin{equation} \label{Syst-Bath-H}
    \begin{split}
        \mH & = \sum_\lambda \hbar \omega_\lambda a_\lambda^\dagger a_\lambda  + \sum_m \int \de \omega \hbar \omega  b_m^\dagger(\omega) b_m(\omega) \\
    & \quad + \hbar \sum_\lambda \sum_m \int \de \omega \left[ W_{\lambda m}(\omega) a_\lambda^\dagger b_m(\omega) + \text{h.c.}  \right],
    \end{split}
\end{equation}
where $a^\dagger_\lambda,a_\lambda$ and $b^\dagger_\lambda,b_\lambda$ are couples of creation and annihilation operators respectively for the cavity and the radiative modes. The first two terms in $\mH$ account for the energy of the resonating system and of the radiative bath separately, while the third one accounts for the interaction energy of the system-and-bath coupling. The procedure followed allows to have explicit expressions for the coupling matrix elements 
\begin{equation}
   W_{\lambda m}(\omega) = \frac{c^2}{2 \hbar \sqrt{\omega_\lambda \omega}} \langle \bm{\mu}_\lambda | \mathcal{L}_{\mathcal{Q}\mathcal{P}} | \bm{\nu}_m(\omega) \rangle.
\end{equation}
Here, however, consistently with the rotating-wave approximation, we consider the matrix elements $W_{\lambda m}$ independent of frequency, at least over a sufficiently large band around the typical mode frequency \cite{Fyodorov97}. This is also compatible with the Markovian limit, which is equivalent to assume a time scale separation so that the typical cavity mode lifetimes are much bigger than the ``bath correlation time'' \cite{Hackenbroich03}.

At this point, it is easy to find coupled dynamical equations for the operators $a_\lambda$ and $b_m(\omega)$ in the Heisenberg representation. From the study of these equations one can find input-output relations based on the scattering matrix formalism, which are useful since the radiative states are the only accessible experimentally. However, here we are only interested in the cavity mode dynamics, which turns out to be given by the following Langevin equation
\begin{align} \label{LangColdCavity}
    \frac{\de a_\lambda}{\de t} = - i \omega_\lambda a_\lambda (t) -  \sum_{\lambda'} \gamma_{\lambda\lambda'} a_{\lambda'}(t) + \eta_{\lambda}(t),
\end{align}
where we have defined the coupling matrix $\gamma_{\lambda\lambda'} = \pi \left[ W W^\dagger \right]_{\lambda\lambda'}$ and the quantum noise operator
\begin{align}
    \eta_{\lambda}(t) = -i \sum_m W_{\lambda m} \int \de \omega e^{i \omega (t-t_0)}  b_m(\omega, t_0).
\end{align}
Therefore, the system-and-bath separation leads to a quantum stochastic dynamical theory in the cavity modes subspace, where the effect of the external bath of radiative modes is included through a noise term. Moreover, an effective linear damping coupling mediated by the radiative modes, i.e. the matrix $\gamma$, acts on the cavity modes. The two main differences with respect to the closed cavity case are represented by: (\emph{i}) the presence of non-diagonal elements $\gamma_{\lambda\lambda'}$ in the interactions;  (\emph{ii}) the fact that the noise is correlated in the mode space, as one can see from the relation
\begin{align}
    \langle \eta^\dagger_\lambda(t_1) \eta_{\lambda'}(t_2) \rangle \propto 2 \gamma_{\lambda \lambda'} \delta(t_1-t_2) \neq \delta_{\lambda \lambda'} \delta(t_1-t_2).
\end{align}

\subsection{Semiclassical Theory of Light-Matter Interaction}
So far, we have managed to deal with the openness of the system. However, in order to complete the theory for active media we have to bring into the game light-matter interactions accounting for the gain. The standard way to go beyond the cold-cavity modes\footnote{By cold-cavity modes we mean solutions of the Helmholtz equation obtained by neglecting scattering and nonlinear effects which could come from the interactions with the active medium. Eq. \eqref{LangColdCavity}, for instance, is written in terms of the cold cavity modes of an open resonator.} is by using the semiclassical Lamb theory and including the gain medium described as a collection of two-level atoms continuously pumped into the excited state. Let us denote by $\rho(\bm{r})$ the atomic density and by $\omega_a$ the atomic transition frequency. If $|g\rangle$ and $|e\rangle$ denote respectively the ground and the exited states and $E_e$ and $E_g$ their energies, then $\omega_a = (E_e - E_g)/\hbar$. Only homogeneous broadening is considered, e.g.~the Doppler effect is neglected in first approximation. The evolution of the atom-field operators can be derived from the Jaynes-Cummings Hamiltonian \cite{Jaynes63,Shore93}, plus the contribution of the damping term accounting for the openness of the system, and can be expressed in the Heisenberg representation by the following set of quantum stochastic nonlinear differential equations \cite{Hackenbroich05}
\begin{subequations}
\begin{gather}
   \dot{a}_\lambda = - i\omega_\lambda a_\lambda - \sum_\mu \gamma_{\lambda\mu} a_\mu + \int \de \bm{r} g_\lambda^\dagger (\bm{r}) \sigma_{-}(\bm{r}) + \eta_\lambda \label{JC1} \\
   \dot{\sigma}_{-}(\bm{r}) = - (\gamma_\perp + i\omega_a)\sigma_{-}(\bm{r}) + 2 \sum_\mu g_\mu(\bm{r}) \sigma_z(\bm{r}) a_\mu + \eta_{-}(\bm{r}) \label{JC2} \\
   \dot{\sigma}_{z}(\bm{r}) = \gamma_\parallel (S \rho(\bm{r}) - \sigma_z(\bm{r})) - \sum_\mu \left(g_\mu^\dagger(\bm{r}) a_\mu^\dagger \sigma_{-}(\bm{r})  + \text{h.c.} \right)+ \eta_{z}(\bm{r}), \label{JC3}
\end{gather} 
\end{subequations}
where $\sigma_{-}^\dagger = |e\rangle \langle g|$ and $\sigma_{-} = |g\rangle \langle e|$ are the atomic raising and lowering operators and $\sigma_z= |e\rangle \langle e| - |g\rangle \langle g|$ is the inversion density operator. The terms $\gamma_\perp$ and $\gamma_\parallel$ are the polarization and population-inversion decay rates, while $S$ is the pump intensity resulting from the interaction between atoms and external baths, which also gives rise to the noise terms $\eta_{-}(\bm{r})$ and $\eta_{z}(\bm{r})$. The noise term $\eta_\lambda$ and the damping matrix $\gamma_{\lambda\mu}$ are the terms previously shown to be induced by the external radiation field. In the electric dipole approximation \cite{Bransden84} the atom-field coupling $g_\lambda(\bm{r})$ are given by
\begin{align}
    g_\lambda(\bm{r}) = \frac{\omega_a }{\sqrt{2 \hbar \epsilon_0 \omega_\lambda}} \bm{p}_{eg} \cdot \bm{\mu}_\lambda(\bm{r}),
\end{align}
where $\bm{p}_{eg}=\langle e | \bm{r} | g \rangle$ is the atomic dipole and $\bm{\mu}_\lambda$ is the complete and orthonormal set of cavity modes previously introduced. 

A full quantum treatment of these equations would require the use of the density matrix formalism to trace over the atomic degrees of freedom, as done in Ref.~\cite{Eremeev11} for the case of a two-mode laser. In general, this is not doable and one resorts to the semiclassical approximation \cite{Hackenbroich05}, where the operators are downgraded to complex numbers corresponding to their expectation values and all the noise sources are neglected (and only later added back). Then, by considering laser media where the characteristic time of atomic pump and loss are much shorter than the lifetimes of the resonator modes, the atomic variables can be adiabatically removed obtaining a set of nonlinear equations for the field modes alone. We will not enter into the details of the procedure, which is carefully described in Ref.~\cite{Antenucci16}, but just sketch the main steps and the final results.

In order to eliminate the atomic variables, we resort to perturbation theory in the mode amplitudes. One can start by neglecting the quadratic term in Eq.~\eqref{JC3}, obtaining the zeroth-order approximation, which replaced in Eq.~\eqref{JC2} gives the first-order approximation, which replaced back in Eq.~\eqref{JC1}
gives the second-order approximation and so on. Once the expressions of $\sigma_{-}$ and $\sigma_{z}$ have been found at a given order of perturbation theory, by replacing them into Eq.~\eqref{JC1} one finally finds the the dynamic equation for the modes alone.

The perturbation series can be resummed obtaining an expression which is valid at all orders in perturbation theory \cite{Zaitsev10b} only in the special case of the \emph{free-running approximation}, for which the lasing modes are considered to oscillate independently from each other. However, this would only be adequate for a theory of random lasing with non-resonant (incoherent) feedback, where the role of interference is neglected, as in the original work by Letokhov \cite{Letokhov68}. As already mentioned in the Introduction, after the observation of structured random laser spectra with sharp peaks (see, e.g., \cite{Cao01,Cao05}), it is generally believed that phases do play an important role in the mode dynamics, determining a coherent lasing action. Therefore, we do not use the free-running approximation and limit ourselves to the third-order theory. In fact, we expect higher orders to become relevant far from the lasing transition and, from the statistical mechanics point of view, not to change universality class of the transition, see, e.g.~Ref.~\cite{Crisanti13}. 

In the third-order theory, the atom-field couplings driving the mode dynamics in the cold-cavity mode basis $\{ a_\lambda\}$ (with Greek letter indices) contain terms of the kind
\begin{gather}
    G_{\lambda_1,\lambda_2}^{(2)} \propto \int \de \bm{r} \rho(\bm{r}) g_{\lambda_1}^*(\bm{r}) g_{\lambda_2}(\bm{r}) \\
    G_{\lambda_1,\lambda_2,\lambda_3,\lambda_4}^{(4)} \propto \int \de \bm{r} \rho(\bm{r}) g_{\lambda_1}^*(\bm{r}) g_{\lambda_2}(\bm{r}) g_{\lambda_3}^*(\bm{r}) g_{\lambda_4}(\bm{r}).
\end{gather}
However, it is convenient to express the mode dynamics in the slow amplitude mode basis, which we have already defined in the previous section, see Eq.~\eqref{slowAmpl}. By denoting with $\{ a_k\}$ (with Latin letters indices) the slow amplitude modes, the following change of variables is performed
\begin{equation}
    a_\lambda = \sum_k A_{\lambda k} a_k,
\end{equation}
which affects all the quantities in the dynamic equation for the modes. The matrix $A$ accounts for the fact that each lasing mode can be thought as a single resonance given by the superposition of many cavity modes, whose fast oscillations can be averaged away. However, the decomposition of a slow amplitude modes in cavity modes is by no means unique: we can use this freedom to choose a basis in which the noise is diagonal, simplifying the stochastic dynamics. Eventually, the resulting equation turns out to be the generalization to random lasers of the SLD Langevin master equation for standard multimode laser Eq.~\eqref{HausME} and reads
\begin{equation} \label{LangevinEq}
\frac{\de a_{k_1}}{\de t} = \sum_{\bm{k} | \text{FMC}(\bm{k})} g_{k_1 k_2}^{(2)} a_{k_2} + \sum_{\bm{k} |\text{FMC}(\bm{k})} g_{k_1 k_2 k_3 k_4}^{(4)} a_{k_2}\overline{a}_{k_3}a_{k_4} + \eta_{k_1}(t),
\end{equation}
where the couplings are
\begin{equation}
    g_{k_1 k_2}^{(2)} = S G_{k_1,k_2}^{(2)} - \tilde{\gamma}_{k_1 k_2} ~~~~~~~~ g_{k_1 k_2 k_3 k_4}^{(4)} = 2 S G_{k_1,k_2,k_3,k_4}^{(4)},
\end{equation}
with
\begin{gather}
    \tilde{\gamma}_{k_1 k_2}= \sum_{\lambda\mu} A_{\lambda k_1}^{-1} \gamma_{\lambda \mu} A_{\mu k_2} \\
    G_{k_1,k_2}^{(2)} \propto \int \de \bm{r} \rho(\bm{r}) g_{k_1}^{L*}(\bm{r}) g_{k_2}^{R}(\bm{r}) \\
    G_{k_1,k_2,k_3,k_4}^{(4)} \propto \int \de \bm{r} \rho(\bm{r}) g_{k_1}^{L*}(\bm{r}) g_{k_2}^{R}(\bm{r}) g_{k_3}^{L*}(\bm{r}) g_{k_4}^{R}(\bm{r}),
\end{gather}
where $g_{k}^{L} = \sum_\mu (A^{-1})^*_{\mu k} g_{\mu}$ and $g_{k}^{R} = \sum_\mu A_{\mu k} g_\mu$, and the proportionality coefficients slightly depend on the frequency \cite{Antenucci16}. Most importantly, in the slow amplitude basis, where by definition $a_k(\omega) \simeq \delta(\omega-\omega_k)$, the relevant terms in the dynamics are selected by the frequency matching condition 
\begin{align} \label{FMC}
\text{FMC}(\bm{k}): | \omega_{k_1} - \omega_{k_2} + \cdots + \omega_{k_{2n-1}} - \omega_{k_{2n}} | \lesssim \gamma ,
\end{align}
which generalizes the selection rule in the case of a comb-like frequency distribution. This can be seen as an adiabatic conservation law coming from averaging over fast mode oscillation.

At this stage, the same techniques developed by the SLD approach can be applied to the Langevin master equation \eqref{LangevinEq}. In order to clearly separate the dissipative contributions from the dispersive ones, we can pass to the real and imaginary parts of the couplings. By defining
\begin{subequations}
\begin{gather}
    G_{k_1 k_2} = \frac{1}{2}\left(g_{k_1 k_2}^{(2)} +  \overline{g}_{k_1 k_2}^{(2)} \right)   ~~~~~~~~~  i D_{k_1 k_2} = \left(g_{k_1 k_2}^{(2)} - \overline{g}_{k_1 k_2}^{(2)} \right) \\
    \Gamma_{k_1 k_2 k_3 k_4} = \frac{1}{2}\left(g_{k_1 k_2 k_3 k_4 }^{(4)} +  \overline{g}_{k_1 k_2 k_3 k_4 }^{(4)} \right)   ~~~~~~~~~  i \Delta_{k_1 k_2 k_3 k_4 }= \left(g_{k_1 k_2 k_3 k_4 }^{(4)} - \overline{g}_{k_1 k_2 k_3 k_4 }^{(4)} \right),
\end{gather}
\end{subequations}
the dynamical equation can be written as
\begin{align}
    \frac{\de a_{k_1}}{\de t} = - \frac{\partial(\mH_{\text{R}} + i \mH_{\text{I}})}{\partial \overline{a}_{k_1}} + \eta_{k_1}(t),
\end{align}
where we have defined
\begin{subequations}
    \begin{gather}
        \mH_{\text{R}} = \sum_{\bm{k} | \text{FMC}(\bm{k})} G_{k_1 k_2} \overline{a}_{k_1} a_{k_2} + \frac{1}{2} \sum_{\bm{k} |\text{FMC}(\bm{k})} \Gamma_{k_1 k_2 k_3 k_4} \overline{a}_{k_1} a_{k_2}\overline{a}_{k_3}a_{k_4} \label{HamiltSemApp} \\
        \mH_{\text{I}} = \sum_{\bm{k} | \text{FMC}(\bm{k})} D_{k_1 k_2}  \overline{a}_{k_1} a_{k_2} +  \frac{1}{2} \sum_{\bm{k} |\text{FMC}(\bm{k})} \Delta_{k_1 k_2 k_3 k_4} \overline{a}_{k_1} a_{k_2}\overline{a}_{k_3}a_{k_4}.
    \end{gather}
\end{subequations}
By considering the purely dissipative limit, i.e. $ \mH_{\text{I}} =0$, and exploiting the fixed-power ensemble defined in \cite{Gordon02}, i.e.~imposing the spherical constraint to model gain saturation, one can prove that the dynamics converges to equilibrium \cite{Antenucci16}. {As for the ordered case of standard multimodal lasers, the more general situation, in which one retains the dispersive part of the dynamics, is not supposed to change the nature of the results that we are going to discuss in the next section.}

\section{The Glassy Random Laser}
In the previous section we have shown that an effective statistical mechanics theory of random lasers can be justified, along the lines of the SLD approach to multimode ordered lasers. The specific features of the mode coupling interaction have been exposed: linear interactions have non diagonal elements accounting for the damping effect due to the openness of the system and a 4-body \emph{disordered} coupling term emerges from the atom-field interaction in the semiclassical approximation. The mode dynamics is described in the slow amplitude basis, where a generalized FMC applies to both the 2-body and the 4-body term of interaction.

However, the model defined by the Hamiltonian in Eq.~\eqref{HamiltSemApp} is still very hard to be addressed. The mean-field fully-connected solution obtained in Refs. \cite{Antenucci15a,Antenucci15b} requires the following additional hypotheses:
\begin{itemize}
    \item \emph{extended modes}: all modes have a spatial wavefunction extended all over the volume $V$, where the dielectric medium is present;
    \item \emph{narrow bandwidth}: the bandwidth of the entire spectrum $\Delta \omega$ is comparable with the typical linewidth of the modes $\gamma$.
\end{itemize}
The extended modes hypothesis guarantees that the only selecting rule in mode coupling is the FMC, while the narrow bandwidth limit ensures that all the modes satisfy the FMC. Hence, the combination of these conditions leads to a model defined on a fully-connected graph of interactions, where each phasor interacts with all the others. Moreover, the mode self-interactions (representing the gain profile) are taken independent from $k$ and set to zero, without loss of generality. 

Another important assumption regards the magnitudes and phases of the couplings, which are related to the spatial overlap among the modes. 
The computation of their values requires a precise knowledge of the spatial structure of the electromagnetic field, which is difficult to access in presence of a disordered medium. Though difficult in practice, it is possible to accomplish the task, and, actually, it has been done in some simple cases \cite{Tureci08,Tureci09,Esterhazy15}. The problem remains however to compute the value of the couplings in the slow amplitude basis, which is used to express the dynamics of the lasing modes. For the construction of the statistical mean-field model, it is then assumed that the couplings are independently drawn from a probability distribution. This is in not true in general, because of the nature of the couplings: for instance, all couplings involving the same mode are correlated. However, these correlations matter only in finite dimensions, while in mean-field theory each coupling coefficient vanishes as $N$ increases and the role of correlation will be quantitatively negligible as far as the system displays enough modes.

By considering all these assumptions together, the mean-field spin-glass model for random lasers is defined by the Hamiltonian
\begin{align} \label{Hamilt2+4}
  \mathcal{H}[\bm{a}] = - \frac{1}{2} \sum_{i,j}^{1,N} J_{ij} \overline{a}_{i}a_{j} - \frac{1}{4!} \sum_{ijkl}^{1,N} J_{ijkl} \overline{a}_i a_j \overline{a}_k a_l,
\end{align}
where the phasors $a_k$ are subjected to the spherical constraint
\begin{align} \label{SpherConstr}
   \sum_{k=1}^N |a_k|^2  = \epsilon N
\end{align}
and the coupling values\footnote{We remind that the couplings are real numbers, since we are considering the purely dissipative limit of the dynamics.} are independently extracted from the Gaussian probability distributions
\begin{align}
    P(J_{i_1,\dots,i_p}) = \frac{1}{\sqrt{2 \pi \sigma_p^2}} \exp\left[- \frac{\left(J_{i_1,\dots,i_p} - \tilde{J}_0^{(p)} \right)^2 }{ 2 \sigma_p^2}  \right].
\end{align}
In order to ensure the extensivety of the Hamiltonian the average $\tilde{J}_0^{(p)}$ and the variance $\sigma_p$ are taken as follows
\begin{align}
    \tilde{J}_0^{(p)} = \frac{J_0^{(p)}}{N^{p-1}} ~~~~~~~~~ \sigma_p = \frac{p! J_p^2}{2 N^{p-1}},
\end{align}
with $J_0^{(p)}$ and $J_p$ independent from $N$. As usual, the variance of the distributions accounts for the strength of the disorder, while their average, by inducing a bias in the extraction of the couplings, acts as an aligning coupling, which tends to induce a long-range ordering in the system at low temperature. To gain a physical intuition of the role played by the free parameters of the model, it is useful to express them in terms of photonic parameters:
\begin{subequations} \label{PHT_par1}
\begin{gather} 
    J_0^{(2)} = (1-\alpha_0) J_0 ~~~~~~~~~ J_0^{(4)} = \alpha_0 J_0, \\
    J_2 = (1-\alpha)J ~~~~~~~~~ J_4 = \alpha J
\end{gather}
\end{subequations}
where $J_0$ and $J$ respectively fix the cumulative strength of the ordered (the coupling average) and disordered (the coupling variance) contributions to the Hamiltonian, while $\alpha_0$ and $\alpha$ fix the strength of nonlinearity in the ordered and disordered parts. Then, we introduce the \emph{degree of disorder} $R_J$ and the pumping rate $\mathcal{P}$ as
\begin{align} \label{PHT_par2}
    R_J = \frac{J}{J_0} ~~~~~~~~~ \mathcal{P} = \epsilon \sqrt{\beta J_0},
\end{align}
where $\beta$ is the inverse of the noise spectral power $T$. The definition of $\mathcal{P}$, like in Eq.~\eqref{E-TPHOTON}, accounts for the equivalence of increasing the optical power per mode $\epsilon$ or decreasing the temperature of the heat bath.

We refer to the model defined by the Hamiltonian \eqref{Hamilt2+4} as spherical (2+4)-phasor model. This is the most general family of mean-field models that has been put forward to study the equilibrium properties of lasing systems, i.e.~the properties of the steady state of a laser, expressed in terms of a thermodynamic equilibrium under the mapping discussed above. Indeed, by changing the values of $J$ and $J_0$ one can tune the degree of disorder and adapt the model to the case of multimode laser with weak disorder or with no disorder at all, and, at the same time depending on $\alpha$ and $\alpha_0$ one can tune the degree of nonlinearity and make the damping effect of the leakages more or less strong. In particular, by choosing $J=0$ and $\alpha_0=1$ one finds back the Hamiltonian \eqref{OrderedHamilt} of the mean-field ordered model defined in \cite{Gat04}.~Therefore, the model results in a comprehensive theory of multimode lasing phenomena.

Before the spherical (2+4)-phasor model was considered, a simpler spin-glass model was proposed in Refs.~\cite{Angelani06a,Angelani06b} which does not take into account the amplitudes of the phasors. The model is a 4-body disordered XY model, defined by the Hamiltonian 
\begin{align} \label{XYHam}
    \mH[\bm{\phi}] = - \sum_{ijkl}J_{ijkl}~\cos(\phi_i -\phi_j +\phi_k - \phi_l),
\end{align}
where $\phi_k$ denotes the phase of the phasor $a_k=A_k e^{i\phi_k}$ and $J_{ijkl}$ are unbiased random couplings. Eq. \eqref{XYHam} can be recovered from the real part of the (2+4)-phasor Hamiltonian in the \emph{strong cavity limit}, for which the damping coupling due to the openness can be neglected, and in the \emph{quenched amplitude approximation}, for which the amplitudes are considered as fixed during the dynamics of the phases and are absorbed in the definition of the couplings. This model was the first mean-field statistical description of random lasers, which goes beyond the free-running approximation, by including the effect of interference. By means of the replica method it was shown for the first time that the competition for amplification in a multimode random optical system can lead to a behavior similar to that of a glass transition. The study was then completed by adding an average to the coupling distribution \cite{Leuzzi09a}, which extends the phase diagram of the model to a globally magnetized phase, and by computing the complexity of the glassy phase \cite{Conti11}.

It is worth noting, that the (2+4)-phasor model we are considering can be regarded as a superposition of the XY (only phases) model of Eq.~\eqref{XYHam}, considered in \cite{Angelani06a}, and the real spherical (2+$p$)-spin (only magnitudes) model considered in \cite{Crisanti04,Crisanti06}, for $p=4$.

\subsection{Quenched Disordered Systems}
In this section we briefly review the replica method, which lies at the heart of the solution of the mean-field model defined by Eq.~\eqref{Hamilt2+4} and of the analytical part of this work. Consider a generic mean-field fully-connected spin-glass model with variables $\bm{\sigma}=\{ \sigma_1,\dots,\sigma_N\}$ and quenched disordered couplings $J$ independently extracted from a probability distribution function $P(J)$. The Hamiltonian $\mH_J[\bm{\sigma}]$ may have pairwise interactions as in the case of the SK model \cite{Sherrington75} or nonlinear interactions as in the case of the $p$-spin model \cite{Gardner85,Crisanti92}. The variables may either take value in a limited, also multi-dimensional domain, in which case the model has $N$ local constraints as in the case of Ising, XY or Heisenberg spins, or be continuous and subject to a global constraint of the kind $||\bm{\sigma}||_\rho = N$, for some choice of the norm. If $\rho=2$, the spherical constraint is recovered. The partition function of the model, which one aims to compute in order to study the equilibrium properties of the system, depends on disorder and is given by 
\begin{align}
   Z_J = \int \mathcal{D}\sigma~e^{-\beta \mH_J[\bm{\sigma}]},
\end{align}
where we have used a shorthand notation for the sum over all the possible configurations of the variables compatible with the constraints. A fundamental quantity is the overlap
\begin{equation}
    q=\frac{1}{N}\bm{\sigma}\cdot\bm{\tau} =\frac{1}{N} \sum_{i=1}^N \sigma_i\tau_i
\end{equation}
among two configurations $\bm{\sigma}$ and $\bm{\tau}$ extracted from the Gibbs measure with the same Hamiltonian $\mH_J$. We call $P_J(q)$ the overlap probability distribution function for a given realization of the quenched disordered couplings $J$.

The meaning of quenched disorder is that the coupling values, once extracted from $P(J)$, remain fixed during the dynamics. Generally, this assumption is justified on the basis of a time scale separation between the dynamics of the system variables and the dynamics of the couplings, which evolve on a much larger time scale. This applies particularly well to the case of random lasers where the light mode amplitudes have a very fast dynamics when compared to changes in the displacements of the particles of the medium, which determine the time evolution of the couplings. The opposite case is the \emph{annealed} one: when variables and couplings evolve on the same time scale, the disorder averages out leaving the system qualitatively equal to its ordered counterpart but for a rescaling of the free parameters. To perform an annealed average of the disorder, one just has to compute
\begin{align}
    \overline{Z_J} = \int \mathcal{D} J P(J) \int \mathcal{D}\sigma~e^{-\beta \mH_J[\bm{\sigma}]},
\end{align}
from which one sees that in this case the disorder is just an additional thermodynamic degree of freedom, being at the same level of the $\bm{\sigma}$. Once the partition function is averaged, no dependence on the $J$'s remain. 

In principle, every macroscopic observable of a quenched disordered system measured at equilibrium depends on the particular realization of the disorder, leading to the idea of dealing with an \emph{ensemble of systems}. However, observables whose fluctuations with respect to the $J$'s decrease as $1/N^{1/2}$ are expected to take the same value in the large-$N$ limit irrespectively of the specific values of the $J$'s. These quantities are called \emph{self-averaging} and in most cases the free energy density is one of such kind. For self-averaging quantities it is sufficient to compute the average value over disorder to make comparisons with the experimental typical values measured on macroscopic samples. On the other hand, because of much stronger fluctuations, non-self-averaging quantities, such as the overlap distribution function $P_J(q)$, do not lose their dependence on the disorder in the thermodynamic limit, so that their averaged value $P(q)=\overline{P_J(q)}$ is generally different from the typical $P_J(q)$.

We are interested in computing the quenched average
\begin{align}
    f = - \lim_{N \rightarrow \infty} \frac{1}{\beta N} \overline{\log Z_J} =  - \lim_{N \rightarrow \infty} \frac{1}{\beta N}  \int \mathcal{D} J P(J) \log \int \mathcal{D}\sigma~e^{-\beta \mH_J[\bm{\sigma}]}.
\end{align}
In this case, the average over the couplings has to be taken \emph{after} the sum over the configurations has been performed at fixed $J$. To avoid the problem of averaging the logarithm of a complicated function one can resort to the replica method, which is based on the following trick
\begin{align}
    \log x = \lim_{n \rightarrow 0 } \frac{x^n - 1}{n},
\end{align}
where $x$ is a generic variable. Once applied to the partition function, the average reduces to 
\begin{align}
    f = - \lim_{N \rightarrow \infty} \frac{1}{\beta N} \overline{\log Z_J} = - \lim_{N \rightarrow \infty} \lim_{n \rightarrow 0} \frac{1}{\beta N} \frac{\overline{Z_J^n} - 1}{n},
\end{align}
where
\begin{align}
    \overline{Z_J^n} = \int \mathcal{D} J P(J) \int \prod_{a=1}^n \mathcal{D}\sigma^a~e^{-\beta \sum_{a=1}^n \mH_J[\bm{\sigma}^a]}.
\end{align}
The replica trick allows us to pass from the average of a function of the random partition function $Z_J$ to the computation of the integer moments of the partition function distribution. This is more than just a simple algebraic trick: the $n$ independent and identical copies of the system are of crucial importance for the study of the equilibrium properties. Once the average over disorder is carried out, a coupling among replicas is found, which naturally leads to the introduction of the global overlap matrices
\begin{equation}
    Q_{ab} = \frac{1}{N}\bm{\sigma}^a \cdot \bm{\sigma}^b = \frac{1}{N} \sum_{i=1}^N \sigma_i^a\sigma_j^b.
\end{equation}
In terms of these quantities (and possibly of other global parameters) the replicated partition function is such that the free energy reads
\begin{align}
    f = - \lim_{N \rightarrow \infty} \lim_{n \rightarrow 0} \frac{1}{\beta N} \frac{\int \mathcal{D}Q~e^{N S(Q)} - 1}{n},
\end{align}
which can be computed with the saddle-point method provided that the order of the limits is exchanged. This is the prescription of the so-called replica method, which leads to
\begin{align}
    f = -\lim_{n \rightarrow 0} \frac{1}{\beta n} S(Q_{\text{SP}}),
\end{align}
where $Q_{\text{SP}}$ is saddle-point value of the matrix $Q$. 

In order to solve the saddle point problem one may restrict the search for $Q_{\text{SP}}$ to a specific matrix space, find self-consistency equations for the parameters and then check the solution \emph{a posteriori} from the behavior of the thermodynamic potentials. The correct solution to this optimization problem is not always given by the intuitive replica symmetric (RS) ansatz, where the overlap matrix is parameterized by only one parameter $q_0$. Usually the RS ansatz describes the high temperature paramagnetic solution of the model, where the system is ergodic and there is only one pure state. If, however, an ergodicity-breaking transition takes place at a certain temperature to a phase where the Gibbs-Boltzmann measure breaks down in many pure states, the solution of the optimization problem can be captured by the more sofisticated Parisi replica symmetry breaking (RSB) scheme \cite{Parisi79a,Parisi80a}, where the the overlap matrix is parameterized by more than one number. The solution has a very deep significance for the physics of complexity, in terms of understanding the structure of the states in the low temperature phase of quenched disordered systems \cite{Parisi83,Mezard87}. Though the mathematical foundations of the replica method have not been laid yet, it has been rigorously proved by Guerra \cite{Guerra03} and Talagrand \cite{Talagrand06} that the Parisi RSB scheme provides the correct solution for the free energy of the SK model. 

Quenched disordered systems can exhibit ergodicity breaking transitions corresponding to different kinds of replica symmetry breaking. Some transitions can be described by a finite number $k$ of steps of replica-symmetry breaking ($k$RSB), where the overlap matrix is parameterized by $k+1$ numbers $q_0,q_1\dots,q_{k}$, whereas others lead to full replica symmetry breaking scheme (FRSB), where the overlap matrix is not parameterized by a discrete set of numbers, but rather by a continuous function $q(x)$ defined on the interval $[0,1]$. Transitions which require a 1RSB ansatz, where the overlap can only take the values $q_0$ and $q_1$, are usually discontinuous, with a jump in the order parameter at the transition point and, at the same time, with a thermodynamic anomaly in the susceptibilities. This kind of phenomenology is known as Random First Order Transition (RFOT) and is the proxy of the glass transition in structural glasses \cite{Kirkpatrick87}. The prototype for the RFOT is the spherical $p$-spin model. FRSB transitions are instead continuous and are the paradigm of the spin-glass transition in the context of magnetic systems, where the ergodicity broken phase space is organized in a hierarchical way. In this case the prototype model is the SK model. Historically, a distinction between 1RSB and FRSB models was made: this distinction has gradually faded over time, as soon as it was realized that more rich and variegated situations exist. In the case of the Ising $p$-spin model, for instance, a ``glass to spin-glass'' transition has been found in Ref.~\cite{Gardner85}, the so-called Gardner transition: by lowering the temperature the system undergoes first a transition from the paramagnetic RS phase to a 1RSB phase and, then, a transition to a FRSB phase. A similar scenario can be found in $p$-spin mixtures with spherical variables \cite{Crisanti04}, where also $k$RSB phases or hybrid $1$-FRSB phases are possible.

\subsection{Replicated Partition Function}
\label{SS-REPLICATED}
In this section we present the solution of the spherical (2+4)-phasor model, by sketching the main steps of the mean-field replica computation and describing the most relevant results. 

The partition function of the model defined by the Hamiltonian \eqref{Hamilt2+4} with the spherical constraint \eqref{SpherConstr} is given by
\begin{align} \label{partFunc-2+4}
    \mZ = \int \prod_{k=1}^N \de a_k \de \overline{a}_k~e^{-\beta \mH[\bm{a}]} \delta\left(\epsilon N - \sum_{k=1}^N |a_k|^2 \right),
\end{align}
i.e.~the sum over all the phasor configurations on the complex hypersphere of radius $\sqrt{\epsilon N}$. In the following, the mode amplitudes will be expressed either in terms of real and imaginary parts or in terms of modulus and phase as 
$$a_k = \sqrt{\epsilon}(\sigma_k + i \tau_k) = A_k e^{i\phi_k}.$$ 
Notice that both the moduli $A_k$ and the phases $\phi_k$ are dynamical variables, i.e.~they actually depend on time. However, under the general assumption that the dynamics of lasing systems is so fast that they can be considered, at least partially (cf.~Sec.~\ref{IFO-theovsexp}), at equilibrium, we are interested in the equilibrium properties of the system, which can be studied through the analysis of the partition function \eqref{partFunc-2+4}. A different choice of the origin of time is not supposed to change the equilibrium properties of the system.

The average over disorder of the replicated partition function naturally leads to the introduction of the following global overlaps matrices\footnote{Actually, the computation also requires the introduction of the overlap matrix
\begin{equation*}
    T_{\alpha\beta} = \frac{1}{\epsilon N} \sum_{k=1}^N \Im\left[a_k^\alpha a_k^\beta \right] = \frac{2}{N}  \sum_{k=1}^N \sigma_k^\alpha \tau_k^\beta
\end{equation*}
which however can be set to zero without loss of generality, as a consequence of the symmetry of the Hamiltonian \eqref{Hamilt2+4} under a global phase rotation $a \rightarrow a e^{i\phi}$ \cite{Angelani06b,Antenucci15b}.}
\begin{gather}
    \mQ_{\alpha \beta} = \frac{1}{\epsilon N} \sum_{k=1}^N \Re\left[a_k^\alpha \overline{a}_k^\beta \right] = \frac{1}{N}\sum_{k=1}^N \left(\sigma_k^\alpha \sigma_k^\beta +  \tau_k^\alpha \tau_k^\beta\right) \\
    \mR_{\alpha \beta} = \frac{1}{\epsilon N} \sum_{k=1}^N \Re\left[a_k^\alpha a_k^\beta \right] = \frac{1}{N}\sum_{k=1}^N \left(\sigma_k^\alpha \sigma_k^\beta -  \tau_k^\alpha \tau_k^\beta\right) 
\end{gather}
and the \emph{coherence} vector
\begin{gather}
    m^\alpha = m_\sigma^\alpha + i m_\tau^\alpha = \frac{1}{N} \sqrt{\frac{2}{\epsilon}}  \sum_{k=1}^N a_k^\alpha \\
    m_\sigma^\alpha = \frac{\sqrt{2}}{N} \sum_{k=1}^N \sigma_k^\alpha ~~~~~~~~~  m_\tau^\alpha = \frac{\sqrt{2}}{N} \sum_{k=1}^N \tau_k^\alpha, \nonumber
\end{gather}
which play the role of the order parameters of the model. In the following we will often refer to the parameter $m$ as magnetization, in analogy with the language of spin-glass models.

It is useful to discuss the physical meaning of the quantities defined above in terms of their connection with the optical properties of the system. The diagonal elements of the overlap matrix $\mQ$ encode the stationarity of the optical intensity, being fixed by the spherical constraint
\begin{equation*}
    \mQ_{\alpha\alpha} = \frac{1}{\epsilon N} \sum_{k=1}^N A_k^2 = 1.
\end{equation*}
The magnetization $m$ and the diagonal part of the overlap matrix $\mR$ are directly connected to the coherence property of the corresponding optical regime
\begin{equation*}
    m = \frac{1}{N} \sqrt{\frac{2}{\epsilon}} \sum_{k=1}^N A_k e^{i \phi_k},  ~~~~~~~~~ R_{\alpha \alpha} = \frac{1}{\epsilon N} \sum_{k=1}^N A_k^2 \cos(2 \phi_k).
\end{equation*}
In the photonic language, a globally magnetized phase corresponds to a regime in which all phasors point in the same direction in the complex plane, i.e.~their phases are all equal. The off-diagonal terms of the overlap matrices can be written in terms of phases and magnitudes of modes in different replicas of the system, as
\begin{gather} \label{Overlphases}
    \mQ_{\alpha \beta} = \frac{1}{\epsilon N} \sum_{k=1}^N A_k^\alpha A_k^\beta \cos(\phi_k^\alpha - \phi_k^\beta) \\
    \mR_{\alpha \beta} = \frac{1}{\epsilon N} \sum_{k=1}^N A_k^\alpha A_k^\beta \cos(\phi_k^\alpha + \phi_k^\beta)
\end{gather}
The presence of more than one value in the off-diagonal part of the overlap matrices, i.e. the breaking of replica symmetry, is as usual interpreted as the existence of a nontrivial structure of thermodynamic states.

Eventually, the averaged replicated partition function reads as
\begin{equation}
    \overline{\mZ^n} = \int \mathcal{D} \Phi \mathcal{D} \hat{\Phi} \exp\left\{-N [\mathcal{B}(\Phi,\hat{\Phi}) - \log \mathscr{Z}_{eff}(\hat{\Phi})]  \right\},
\end{equation}
where the shorthand notations for the set of the order parameters $\Phi=\{\mQ,\mR,m\}$ and for their Lagrange multipliers $\hat{\Phi}=\{\hat{\mQ},\hat{\mR},\hat{m}\}$ have been introduced. The functional $\mathcal{B}$ in the previous expression reads
\begin{equation} \label{FuncB}
\begin{split}
    \mathcal{B}(\Phi,\hat{\Phi}) &= - \frac{\xi_2}{2} \sum_{\alpha\beta}^{1,n} (\mQ_{\alpha\beta}^2 + \mR_{\alpha\beta}^2) 
    - \frac{\xi_4}{4} \sum_{\alpha\beta}^{1,n} (\mQ_{\alpha\beta}^4 + \mR_{\alpha\beta}^4 + 4 \mQ_{\alpha\beta}^2 \mR_{\alpha\beta}^2) \\
    &\quad - b_2 \sum_{\alpha=1}^n [(m_\sigma^\alpha)^2+(m_\tau^\alpha)^2] - b_4 \sum_{\alpha=1}^n [(m_\sigma^\alpha)^2+(m_\tau^\alpha)^2]^2 \\
    &\quad + \sum_{\alpha\beta}^{1,n} (\hat{\mQ}_{\alpha\beta}\mQ_{\alpha\beta} + \hat{\mR}_{\alpha\beta}\mR_{\alpha\beta} ) + \sum_{\alpha=1}^n (\hat{m}_\sigma^\alpha m_\sigma^\alpha + \hat{m}_\tau^\alpha m_\tau^\alpha)
\end{split}
\end{equation}
and the local partition function, which contains the integration over the phasors, is given by
\begin{equation}
\begin{split}
       \mathscr{Z}_{eff}(\hat{\Phi}) &= \int \prod_{\alpha=1}^n \de \sigma^\alpha \de \tau^\alpha \exp\left\{ \sum_{\alpha\beta}^{1,n} \left[\sigma^\alpha (\hat{\mQ}_{\alpha\beta} + \hat{\mR}_{\alpha\beta}) \sigma^\beta + \tau^\alpha (\hat{\mQ}_{\alpha\beta} - \hat{\mR}_{\alpha\beta}) \tau^\beta \right] \right\} \\
       &\quad\quad \times \exp \left\{\sum_{\alpha=1}^n \left[\hat{m}_\sigma^\alpha \sigma^\alpha + \hat{m}_\tau^\alpha \tau^\alpha  \right] \right\}.
\end{split}
\end{equation}
In the previous expressions the following symbols for the external parameters have been introduced for convenience 
\begin{gather}
    b_2 = \frac{\epsilon}{4} \beta J_0^{(2)}  ~~~~~~~~~ b_4 = \frac{\epsilon^2}{96} \beta J_0^{(4)} \\
    \xi_2 = \frac{\epsilon^2}{4} \beta^2 J_2^2 ~~~~~~~~~ \xi_4 = \frac{\epsilon^4}{6} \beta^2 J_4^2,
\end{gather}
where we notice that $b_2$ and $b_4$ vanish if one considers zero-mean probability distributions for the couplings. The explicit expression of these parameters in terms of the photonic quantities $J,J_0,\alpha,\alpha_0,R_J$ and $\mathcal{P}$ is reported in Ref.~\cite{Antenucci16}. Moreover, we notice that the ratios $b_2/b_4$ and $\xi_2/\xi_4$, which are of crucial importance in determining the nature of the phases of the model, do not depend on temperature, but only on the ratios between the free parameters of the coupling distributions, i.~e.~of the linear and non-linear contributions.

The local partition function $\mathscr{Z}_{eff}$ can be computed by performing the multidimensional Gaussian integration in $\sigma$ and $\tau$, while the other Lagrange multipliers can be eliminated by exploiting their saddle-point expressions in terms of overlap matrices and magnetization. After the integration over all the auxiliary variables has been carried out, one is left with
\begin{align} \label{PartFunc2+4}
    \overline{\mZ^n} = \int \prod_{\alpha < \beta}^{1,n} \de \mQ_{\alpha \beta} \prod_{\alpha \leq \beta}^{1,n} \de \mR_{\alpha \beta} \prod_{\alpha=1}^n\left[\de m_\sigma^\alpha \de m_\tau^\alpha \right] ~e^{- N G[\mQ,\mR,m_\sigma,m_\tau]},
\end{align}
where the action functional $G$ reads as
\begin{equation}
\begin{split}
    - G[\mQ,\mR,m_\sigma,m_\tau] &= \frac{1}{2} \sum_{\alpha \beta}^{1,n} g(\mQ_{\alpha \beta}, \mR_{\alpha \beta}) + n k(m_\sigma,m_\tau) + \frac{1}{2} \log \det (\mQ+\mR) \\
   & \quad + \frac{1}{2} \log \det (\mQ-\mR) - \frac{m_\sigma^2}{2} \sum_{ab}^{1,n} (\mQ+\mR)^{-1}_{ab} - \frac{m_\tau^2}{2} \sum_{ab}^{1,n}(\mQ-\mR)^{-1}_{ab},
\end{split}
\end{equation}
and the functions $g$ and $k$ are defined as follows
\begin{gather}
    g(x,y) = \xi_2 (x^2 + y^2) + \frac{\xi_4}{2} (x^4+y^4+4x^2y^2) \\
    k(x,y) = b_2 (x^2 + y^2) + b_4 (x^2 + y^2)^2.
\end{gather}
In the following section we describe the results of the replica computation and present the phase diagram of the model.
 
\subsection{Phase Diagram of the Glassy Laser Transition}
The saddle-point method applied to Eq.~\eqref{PartFunc2+4} leads to a set of stationary equations for the functional $G$, which can be solved with an appropriate ansatz on the structure of the matrices depending on the value of the external parameters. The precise expression of the saddle-point equations in the RS and RSB ansatzes can be found in Ref.~\cite{Antenucci16}, where their solution is discussed in detail for all the various cases. Here, we just aim to describe the phenomenology of the model. The phase diagram obtained by the solution of the saddle-point equations is comprised by four different phases distinguished by the values of the order parameters $\mQ,\mR$ and $m$:
\begin{itemize}
    \item \emph{Paramagnetic phase} (PM): it is the RS solution with all the order parameters equal to zero (with the exception of $\mQ_{\alpha \alpha}=1$); it corresponds to the \emph{Continuous Wave} (CW), where all the modes oscillate incoherently; it is the only phase at high (low) enough temperature (pumping);
    \item \emph{Spin-Glass phase} (SG): it is the RSB phase with vanishing global magnetization $m=0$; it is characterized by the freezing of the modes in configurations where the coherence of oscillations is frustrated by the presence of a nontrivial structure of states; it corresponds to the \emph{Random Laser} (RL); it is the only phase at low enough temperature if $\xi_{2}$ and $\xi_4$ are large enough with respect to $b_2$ and $b_4$ (the degree of disorder $R_J$ is large enough);
    \item \emph{Ferromagnetic phase} (FM): it is the set of all the phases with nonzero magnetization, regardless of possible replica symmetry breaking; all the modes oscillate coherently with the same phase; it corresponds to the \emph{Standard Mode-Locking Laser} (SML); it is the only phase at low enough temperature if $b_2$ and $b_4$ are large enough with respect to $\xi_2$ and $\xi_4$;
    \item \emph{Asymmetric Paramagnetic phase} (APM): it is the RS solution with the order parameters all vanishing except for the diagonal elements of the overlap matrix $\mR$, so there is a partial phase locking, without global magnetization, where the phases take different values but are locked; in the photonic langauge, we refer to this phase as \emph{Phase Locking Wave} (PLW); it is an intermediate phase between the CW and the RL (or SML) phase, which exist only if $\xi_4 \neq 0$. 
\end{itemize}

Both the FM and the SG phases are expected to present different kinds of replica symmetry breaking, depending on the values of the control parameters. In particular, one expects a FRSB structure if the 2-body term in the Hamiltonian is dominating, while a 1RSB one if the 4-body interaction prevails. When the interactions have comparable magnitudes, an intermediate 1-FRSB phase is expected in analogy to the case of real spherical spins \cite{Crisanti04}. The three cases can be distinguished either by the ratio $\xi_2/\xi_4$, or, equivalently, by the photonic parameter $\alpha$, which measures the strength of the nonlinearity in the disordered part of the interactions. On the other hand, the system chooses between the FM and the SG phases depending on the strength of $b_2$ and $b_4$, or equivalently on the value of the photonic parameter $R_J$. 

Before presenting the phase diagram of the model, let us give a general description of all the system phases. Let us first consider the case in which $b_2$ and $b_4$ are low enough. Then, by lowering the temperature from the PM phase, the system may either enter the APM phase or, only in the case when $\xi_4=0$, remain in the RS phase with a non-vanishing value of the overlap\footnote{The fact that for $\xi_4=0$ the solution is always replica symmetric is expected in analogy to the $p=2$ spherical model \cite{Kosterlitz76}. The RS solution with non-vanishing overlap is only marginally stable: the addition of an infinitesimal perturbation to the Hamiltonian (in this case represented by an arbitrary small disordered nonlinearity) causes the solution to become unstable.}. From the APM phase, the system either enters the SG phase through a RFOT if $\xi_2/\xi_4$ is low enough, in which case the structure of the states is of the 1RSB kind, or it undergoes a continuous phase transition towards the SG phase with a FRSB structure in the opposite case when $\xi_2/\xi_4$ is high enough. On the other hand, if $b_2$ and $b_4$ are high enough, by lowering the temperature from both the PM and the APM phase, a transition towards a RS-FM phase is obtained, which can be either continuous (for $b_2/b_4$ high enough) or discontinuous (vice versa). For intermediate values of $b_2$ and $b_4$ the system is in the FM phase with the same kinds of RSB as the SG phase, depending on the $\xi_2/\xi_4$ ratio. 

As already mentioned, non-zero coupling averages yield the alignment of the phasors, acting as an effective field. In Ref.~\cite{Antenucci15b} the model has been mapped into an equivalent model with zero averages and a suitable effective field, which turns out to be related to the magnetization in the following way $h = 2 b_2 m + 4 b_4 m^2$. A non-vanishing value of the field signals that the system is globally magnetized, and, hence, is in the FM (SML) phase. We stress that it is sufficient that $b_2=b_4=0$ for the field to vanish, but it is not necessary: if $b_2$ and $b_4$ are small enough compared to $\xi_2$ and $\xi_4$, then the field vanishes because of $m=0$. Having developed the replica computation including the magnetization has the remarkable advantage of bridging with the ordered case. In this way, the theory describes general multimode laser phenomena, both standard and random and can be adapted to intermediate situations such as weakly disordered systems. However, for the purpose of this work, we are mainly interested in the glassy phase of light: hence, in order to simplify the picture, we present the phase diagram of the model at zero effective field, where no trace of the FM phase is present. The complete phase diagram of the model has an additional axis accounting for positive values of the effective field and can be found in Refs.~\cite{Antenucci15a,Antenucci16}.

\begin{figure}[t]
\centering
\includegraphics[width=0.9\textwidth]{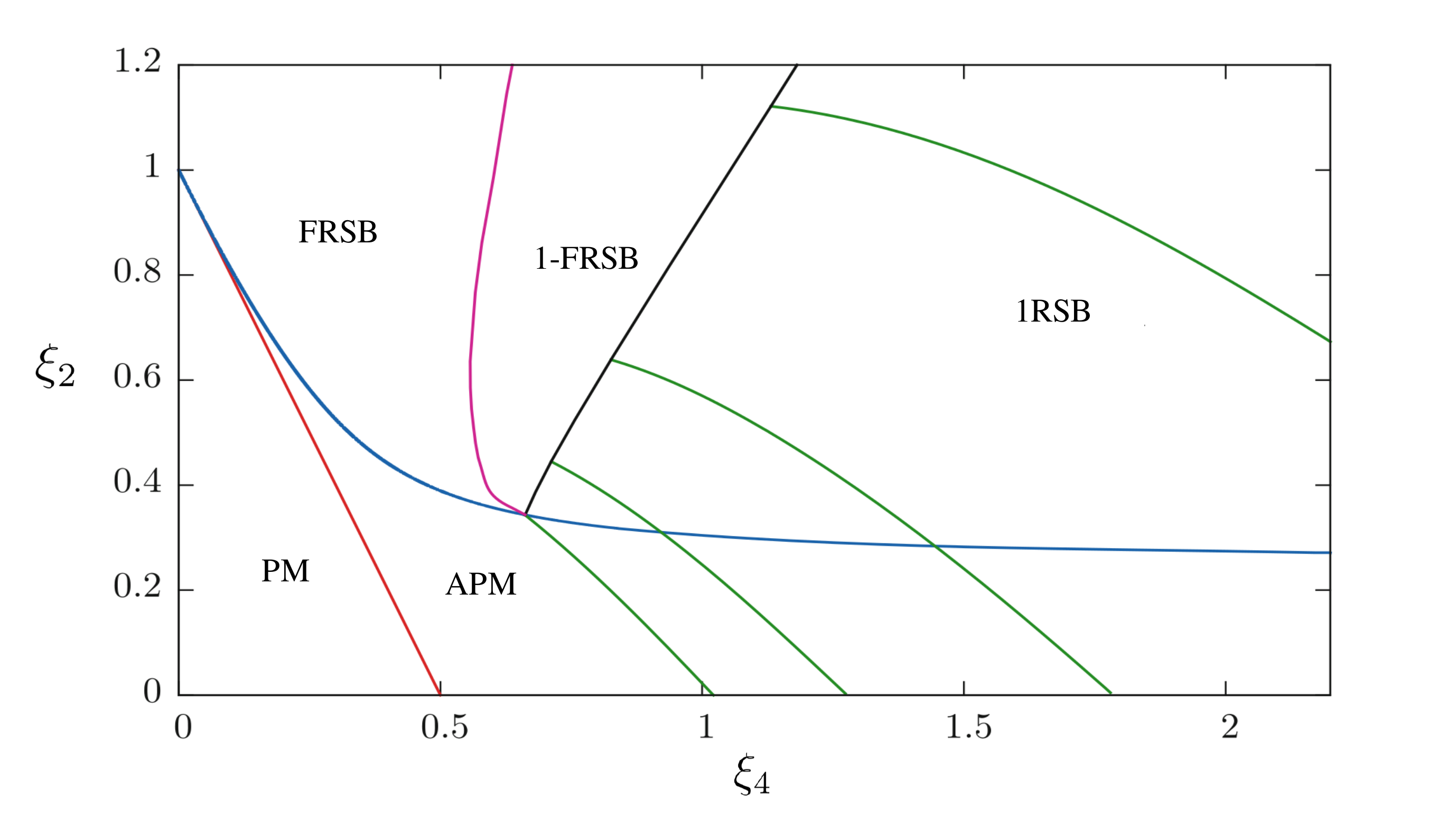}
\caption{Phase diagram of the glassy laser transition (fieldless case). \emph{Red line}: appearence of the APM phase; \emph{blue line}: APM becomes unstable; \emph{green lines}: 1RSB solutions for fixed values of the breaking parameter $x=1,x=0.8,x=0.6,x=0.4$; \emph{black line}: the 1RSB solution becomes unstable. To the left of the black line an intermediate 1-FRSB emerges, which turns into a FRSB phase on the \emph{magenta line}. We notice that the $\xi_4=0$ line is always RS.}
\label{fig:2+4PhDiag1}
\end{figure}

In Fig.~\ref{fig:2+4PhDiag1} the $(\xi_4,\xi_2)$-phase diagram is displayed. The transition lines are obtained through the study of the phase stability, which can be performed with the standard method, by looking at the vanishing of the replicon, i.e.~the highest eigenvalue of the stability matrix \cite{deAlmeida78,DeDominicis06}. Let us briefly describe the results summarized in the phase diagram. Starting from a value of $\xi_2 < 0.3434 \dots$, i.e.~the three-critical point in Fig.~\ref{fig:2+4PhDiag1}, by increasing $\xi_4$ one has the following scenario: below the red line the only stable solution is the PM one; on the red line the PM solution becomes unstable in favor of the emergence of the APM solution, which in turn becomes unstable on the blue line. The stability of both the PM and the APM phases is revealed by the vanishing replicon of the corresponding RS solution ($\lambda_{\text{RS}}^{\text{PM}}=0$ and $\lambda_{\text{RS}}^{\text{APM}}=0$). When the APM solution becomes unstable, a transition towards the 1RSB phase takes place: the first green line corresponds to the static transition at $x=1$, where $x$ is the breaking parameter. At the transition, the usual mixed-order behavior of the RFOT is found: a jump in the order parameters $\mQ$ and $\mR$ is present, but the internal energy remains continuous, a signature of no latent heat exchange. 

The study of the stability of the 1RSB solution reveals that, as anticipated, the 1RSB phase is not stable over the whole region of the parameters where the RS solution is unstable: the replicon of the 1RSB solution vanishes ($\lambda_{1\text{RSB}}=0$) on the black line of Fig.~\ref{fig:2+4PhDiag1}. Ideally, by starting from a value of $\xi_2 > 0.3434 \dots$ in the 1RSB phase and lowering the value of $\xi_4$, the expected ``glass to spin-glass'' transition takes place: first the system enters a mixed 1-FRSB phase and then the FRSB phase emerges (magenta line). 

\subsection{The glassy state of light}
The replica-symmetry broken phase represents the amorphous state of light, in analogy to the low temperature behavior of glass-forming liquids predicted by mean-field theory. All the concepts coming from the mean-field theory of structural glasses are then predicted by this model for optical waves in disordered media. When the glass transition is approached from the RS phase, the system exhibits a critical slowing down and dynamical arrest on the transition line, as could be revealed by study of time correlation functions. The cause of this behavior is, as usual, the breaking of ergodicity in a number $\mathcal{N}$, increasing exponentially with the system size, of degenerate metastable states, which dominate the dynamics, before the static transition is reached. The role of these states can be revealed by the study of the complexity\footnote{Notice that the complexity in disordered systems is the only intrinsically dynamical quantity that can be computed from the statics.}, i.e.~the configurational entropy $\Sigma = N^{-1} \log \mathcal{N}$, which also allows to find the spinodal line of the transition, corresponding to the value of the parameters where the 1RSB states are dynamically accessible. The complexity decreases when passing from the 1RSB to the 1-FRSB phase, until it reaches zero on the magenta line of the continuous transition to the FRSB phase (see Fig.~\ref{fig:2+4PhDiag1}). 

What of this dynamical scenario may be actually observed in real random lasers is not so clear: as already mentioned, the dynamics of light modes is so fast that dynamical phenomena (like aging) connected to the presence of metastable states may be difficult to reveal. However, besides the presence of exponentially many metastable states, the theory predicts a static transition to a ergodicity broken phase with multiple equilibria, which most likely can be put in correspondence with the experimental observations. What can be stated is that the theory predicts that lasing in random media displays a glassy coherent behavior with the following properties: (\emph{i}) the subset of modes which are activated and actually lase is randomly chosen from all the cavity modes and (\emph{ii}) the set of activated modes behave coherently and belong to one out of many possible states. 

\begin{figure}[t]
\centering
\includegraphics[width=0.80\textwidth]{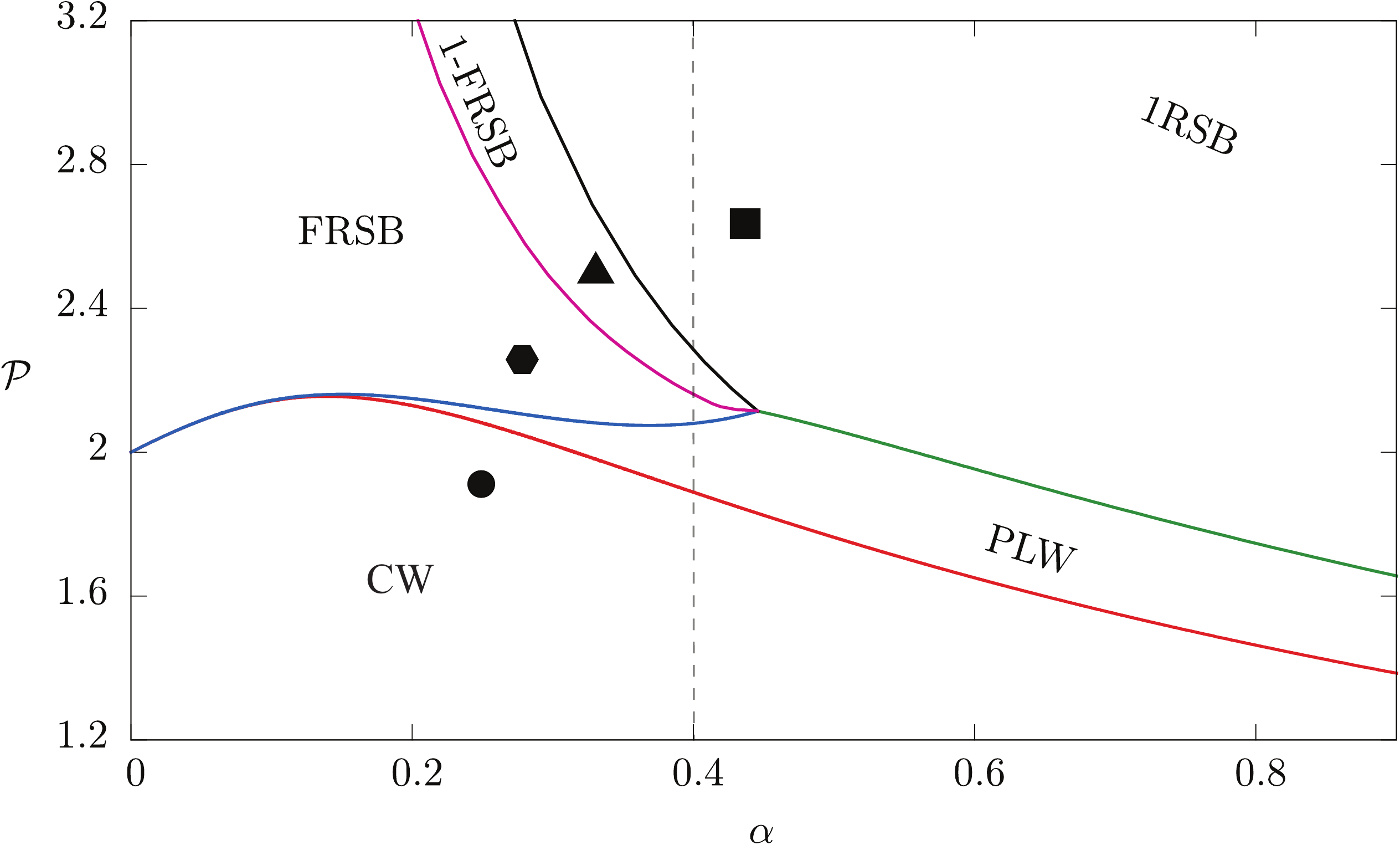}
\caption{Phase diagram of the glassy laser transition (fieldless case) in the photonic parameters, fixing $R_J=\beta J = 1.1$. The region above is the RL phase and it is divided in the three subregions corresponding to the different RSB ansatzes depending on the value of the parameters: FRSB, 1-FRSB and 1RSB. The dashed line at fixed $\alpha=0.4$ is a possible path which goes through all the possible phases of the system. The symbols $\CIRCLE, \hexagonblack, \blacktriangle$ and $\blacksquare$ denote the RS and the three kind of RSB phases and are used for labeling the respective plots of the overlap distribution functions in Figs.~\ref{fig:IFOexp} and \ref{fig:IFOteo}.}
\label{fig:2+4PhDiag2}
\end{figure} 

It is useful to visualize the phase diagram of the glass transition in the photonic parameters $\mathcal{P}$ and $\alpha$ for a fixed value of the degree of disorder $R_J$ (Fig.~\ref{fig:2+4PhDiag2}). Actually, in this case the complete phase diagram has an additional axis for $R_J$: the fieldless case is compatible with values of $R_J > 1$, i.e.~$J>J_0$, where the phenomenology of the model is described in terms of CW, PLW and RL phases\footnote{In order to visualize the RL-SML transition one has to consider a $(R_J,\mathcal{P})$-section of the complete phase diagram at fixed $\alpha$. We recall that it is not necessary that $J_0=0$ (i.e.~$b_2=b_4=0$) to be in the fieldless case: $h$ is zero if $m=0$, which may happen also if $J_0$ is small compared to $J$.}. The diagram in figure is the same diagram presented before, but with respect to $\mathcal{P}$ and $\alpha$. In this case, we gain a clearer physical intuition about the behavior of the model. In particular, by fixing the strength of the nonlinearity (as it is in a real random laser), we can isolate the role of the pumping. If we choose a value of $\alpha$ to the left of the tricritical point in Fig.~\ref{fig:2+4PhDiag2}, for instance $\alpha=0.4$ which corresponds to the dashed vertical line in figure, we see that starting from the CW phase, by increasing the value of the pumping, the laser first enters the PLW phase, where there is partial coherence, and, then, reaches the glassy coherent phase, by going through all the RSB phases: first FRSB, then 1-FRSB and, eventually, 1RSB. On the other hand, if we choose a value of $\alpha$ on the right of the tricritical point, after the intermediate PLW phase, the laser enters directly in the 1RSB phase, by crossing the green line. This last regime will be the working setting of the original part of this Thesis.

\subsection{Intensity Fluctuation Overlap} \label{IFO-theovsexp}
This section is devoted to the introduction of the key observable which allows to connect spin-glass theory to experiments on RLs: the Intensity Fluctuation Overlap (IFO). We refer in particular to the experiments already mentioned in the Introduction, where activated modes are observed to change in spectra acquired at different times from the same sample \cite{Mujumdar07,vanderMolen06,Papadakis07}. From these observations it is not possible to extract the mode phases needed to compute the overlap matrices previously defined, see Eqs.~\eqref{Overlphases}. In particular, the random lasing emission is generally not intense enough to successfully use techniques based on second-harmonic generation to reconstruct the phases of the
modes \cite{Antenucci16}. However, from the statistical mechanics point of view, the phenomenology presented by the experiments strongly suggests that an ergodicity breaking transition controlled by the pumping rate is taking place in real random lasers. To compare the theory with the experiments it would be great to define a quantity, which is experimentally measurable and is related in some way to the order parameters defined in the mean-field analysis.  

In Ref.~\cite{Ghofraniha15}, shot-to-shot intensity fluctuations have been interpreted in terms of an overlap between intensity fluctuations of two real replicas, i.e.~replicas with the same quenched disorder. Provided that the sample is kept in the same experimental conditions for all the data acquisition time, then real replicas can be associated to the different shots, each one thermalized into different equilibrium states characterized by a specific spectral profile of activated modes and sharp peaks. Thermalization is guaranteed by the fact that during a single pulse of the external pumping, several stimulated emission phenomena take place for each mode frequency ensuring a long enough mode dynamics. To be precise, this would only be a partial thermalization, since for a disordered system in the ergodicity broken phase, a complete thermalization would require the system to visit all possible states. Given the development of the theory and the current interpretation of experiments, we cannot say how many equilibrium states a random laser actually visits for each spectral shot, but we can quite safely say that the system has reached the static transition predicted by mean-field theory. This is not only a consequence of fast light mode dynamics, but also of a number of modes which is small compared to usual thermodynamics degrees of freedom (e.g. $\sim 10^{23}$) and of the fact that these modes are highly connected as in the dense interaction network of a mean-field model.

Let us first define the fluctuation of the intensity $I_k^\alpha$ of the resonance at the frequency $\omega_k$ in a single spectrum $\alpha$ with respect to the  spectral intensity  at that frequency averaged over all $N_s$ acquired spectra as
\begin{equation}
    \Delta_k^\alpha = I_k^\alpha - \frac{1}{N_s} \sum_{\gamma =1}^{N_s}I_k^\gamma.
\end{equation}
Each spectrum represents the realization of a replica. The experimental IFO measured between two real replicas can be represented by the following matrix
\begin{equation} \label{ExpIFO}
        \mathcal{C}_{\alpha \beta}^{\text{exp}} = \frac{\sum_k \Delta_k^\alpha\Delta_k^\beta}{\sqrt{\vphantom{\big|} \sum_k (\Delta_k^\alpha)^2}\sqrt{\sum_k (\Delta_k^\beta)^2}} 
\end{equation} 
defined in the interval $[-1,1]$, where $I_k^\alpha$ denotes the intensity of the mode $k$ in the spectrum corresponding to the replica (shot) $\alpha$, with $\alpha= 1,\dots, N_s$. Since thermalization is assumed, the experimental value $I_k^\alpha$ can be thought as the equilibrium average of the intensity, i.e. $I_k^\alpha \equiv \frac{1}{\mathcal{T}} \int_{t_0}^{t_0+\mathcal{T}} \de t |a_k^\alpha(t)|^2 $, where $\mathcal{T}-t_0$ is the time interval corresponding to random laser lifetime, slightly longer than
the pumping pulse. The overlap is defined between intensity fluctuations rather than directly between intensities, in order to exclude the effect of amplified spontaneous emission on the measurements. Fluctuations are taken with respect to the intensity averaged over many different replicas. From the $N_s$ measured spectra one can extract $N_s(N_s-1)/2$ values of the IFO and determining their distribution by building the histogram
$P(\mathcal{C}) = \sum_{\alpha < \beta} \delta(\mathcal{C} - \mathcal{C}_{\alpha \beta})$. 

\begin{figure}[t]
\centering
\includegraphics[width=0.9\textwidth]{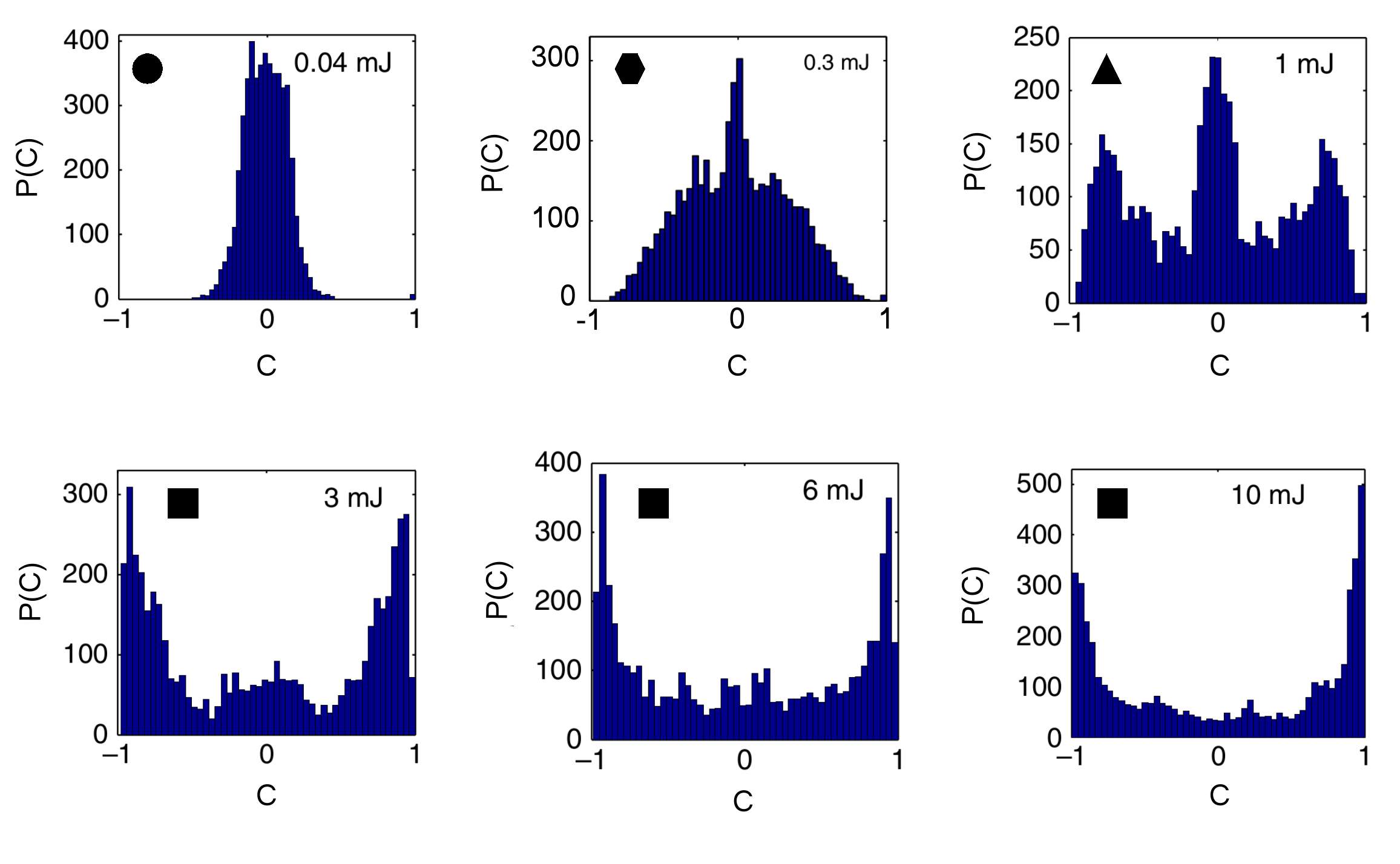}
\caption{Six histograms representing the probability distribution function of the experimental intensity fluctuation overlap \eqref{ExpIFO} built from the data obtained by many single shot spectra collected at six increasing values of the pumping rate (Reprinted from \cite{Ghofraniha15}). Notice that $P(\mathcal{C})$ gradually changes from a Gaussian-like distribution centered in zero for low values of the $\mathcal{P}$ to a bimodal distribution. The symbols represent the region of the phase diagram in Fig.~\ref{fig:2+4PhDiag2} to which these experimental $P(\mathcal{C})$ might correspond. The correspondence is clarified by the analytical study of the IFO distribution function, whose results are presented in Fig.~\ref{fig:IFOteo}.}
\label{fig:IFOexp}
\end{figure}

We report in Fig.~\ref{fig:IFOexp} the results obtained in Ref.~\cite{Ghofraniha15}, concerning the measurement of the IFO distribution. At low pumping rate, $P(\mathcal{C})$ appears as a Gaussian-like distribution centered in $\mathcal{C}=0$. Then, for increasing values of $\mathcal{P}$, the distribution develops a nontrivial structure with three distinguished peaks, one in $\mathcal{C}=0$ and two symmetric side-peaks, and a continuous part between them. Eventually, $P(\mathcal{C})$ reduces to a double-peaked distribution for high values of $\mathcal{P}$: in this last case, as the pumping is varied, $\mathcal{C}$ can in principle take all possible values in the interval $[-1,1]$, while for a given value of $\mathcal{P}$ the position of the peaks is fixed. This kind of behavior resembles the one of the Parisi overlap distribution function in the replica symmetry broken phase of the model. 

\begin{figure}[t]
\centering
\includegraphics[width=0.65\textwidth]{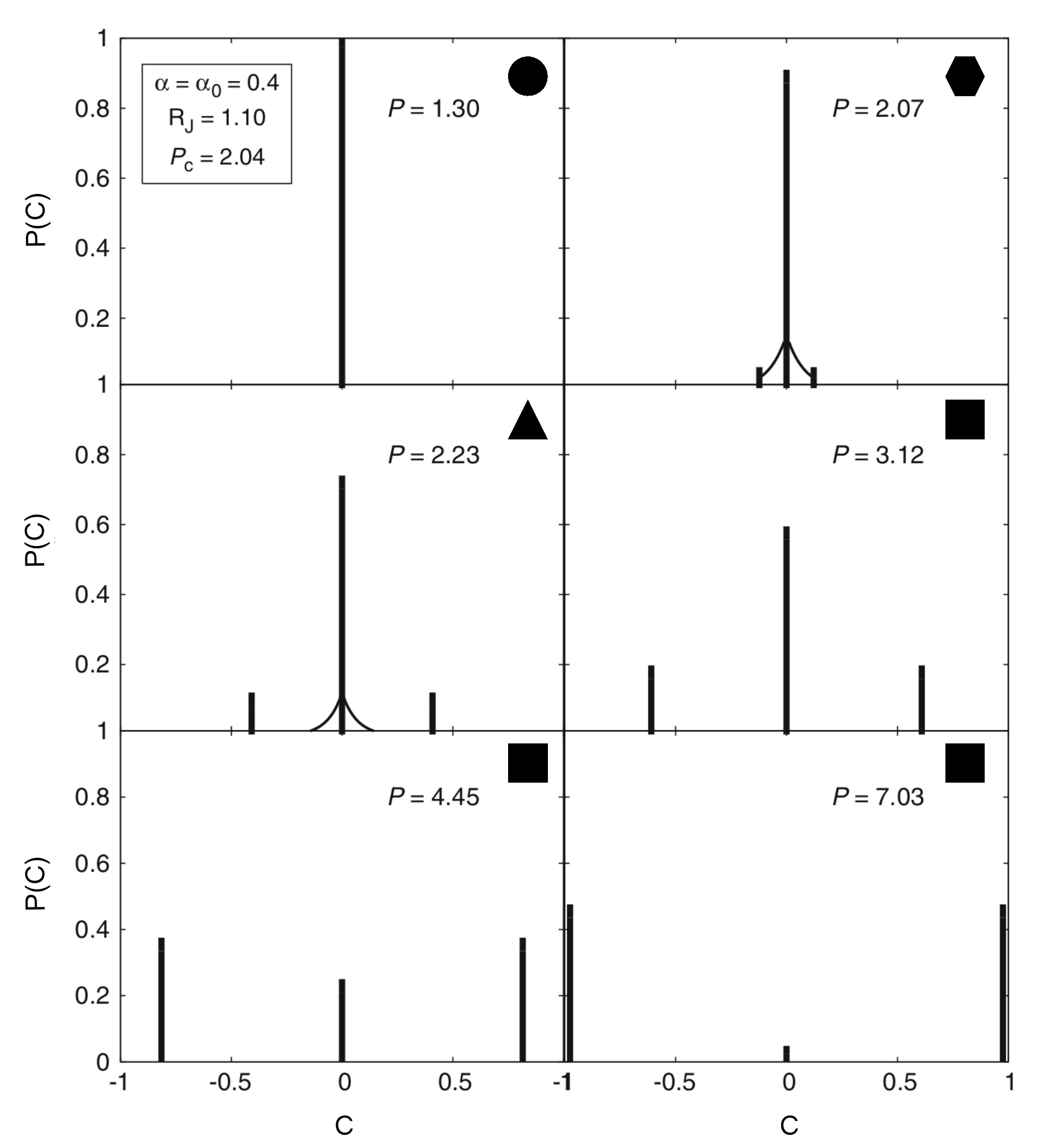}
\caption{Probability distribution of the intensity fluctuation overlap $P(\mathcal{C})$ for competing linear and nonlinear couplings ($\alpha = \alpha_0 = 0.4$), and in the fieldless case $R_J = 1.1$ (reprinted from \cite{Antenucci16}). The pumping rate $\mathcal{P}$ increases from left to right and from top to bottom, along the dashed line in Fig.~\ref{fig:2+4PhDiag2}. At $\mathcal{P}=1.3$ the distribution is a Dirac delta centered in zero. As soon as the pumping rate exceeds the lasing transition threshold, all the RSB regime are displayed: FRSB ($\mathcal{P}=2.07$), 1-FRSB ($\mathcal{P}=2.23$) and 1RSB ($\mathcal{P}=3.12,4.45,7.03$).}
\label{fig:IFOteo}
\end{figure}

In order to build a precise correspondence at least in the mean-field fully-connected model, one has to define a quantity, depending only on intensity fluctuations to be analytically related to the Parisi overlap. In Ref.~\cite{Antenucci15c}, the IFO is expressed (in absolute value) by the following matrix 
\begin{align} \label{IFOtheo}
    \mathcal{C}_{\alpha \beta} = \frac{1}{8 \epsilon^2 N} \sum_{k=1}^N \left[ \langle |a_k^\alpha|^2 |a_k^\beta|^2 \rangle - \langle |a_k^\alpha|^2 \rangle \langle |a_k^\beta|^2 \rangle  \right]
\end{align}
defined in $[0,1]$, where the average is taken with respect to the Gibbs-Boltzmann measure of the spherical (2+4)-phasor model. Clearly this distribution depends on the realization of the disordered couplings $J$ and has to be averaged $P(\mathcal{C}) = \overline{P_J(\mathcal{C})}$. 

It is worth stressing that the definition of emission spectra in statistical mechanical models is only possible in a model which takes into account both phases and intensities, and hence the introduction of the IFO distribution could not be possible within the phase-only approach originally developed in Refs.~\cite{Angelani06a,Angelani06b,Leuzzi09a,Conti11}. While the role of the phases is essential for reproducing the phase transition phenomenology of mode-locking, the role of the intensities is of crucial importance for bridging with the experiments, where we do not have access to the phases.

The crucial result obtained in Ref.~\cite{Antenucci15c} is that the IFO matrix defined in Eq. \eqref{IFOtheo} can be expressed in terms of the overlap matrices $\mQ$ and $\mR$ as
\begin{align}
    &\mathcal{C}_{\alpha\beta} = \mQ_{\alpha\beta}^2 - \frac{m^4}{4} ~~~~~~~~~ a \neq b \\
    &\mathcal{C}_{\alpha\alpha} = \frac{1+\mathcal{R}_{\alpha\alpha}^2}{2} - \frac{m^4}{4},
\end{align}
element by element, whatever the structure of $\mQ$ and $\mathcal{R}$. This result reveals that if a RSB structure is present at the level of the configuration overlap, the same holds also for the IFO: in other terms the structure of the state organization is the same whether we look at the configurations of the modes or at their intensity spectra.

In Fig.~\ref{fig:IFOteo} six different plots of the analytical IFO probability distribution are displayed for increasing values of the pumping rate $\mathcal{P}$ along the dashed line at $\alpha=0.4$ in Fig.~\ref{fig:2+4PhDiag2}, which goes through all kinds of RSB phases. At low pumping rate the distribution is a Dirac delta centered in zero, meaning that no correlations are present among the intensities and the modes are independent and non-interacting (CW phase); increasing the pumping the mode coupling becomes relevant and, accordingly, the overlap distribution function is nontrivial, since the system enters a phase where the modes are highly frustrated by disorder. First, $P(\mathcal{C})$ develops a small continuous part around the central peak in $\mathcal{C}=0$, denoting the typical continuous FRSB shape (second panel); then, besides the continuous part, also symmetric side Dirac deltas emerge, corresponding to the 1-FRSB phase (third panel). Eventually, the analytical $P(\mathcal{C})$ looses the continuous part and becomes a linear combination of Dirac deltas, which is the usual 1RSB structure. The resemblance with the experiments, though only qualitative, is quite remarkable. 

Despite this similarity, it should be noted that the correspondence between theory and experiments is still under construction and many criticisms, both on the theoretical and experimental sides, can be raised with the perspective of improving it. The most obvious criticism is that, for now, no experimental data corresponding to different samples are yet available for averaging over the disorder. Due to the non-self-averageness of overlap probability distribution functions, the average over disorder is essential to observe the typical behavior of these observables and compare it with the theoretical predictions. Furthermore, a major problem lies in the fact that the emission of a random laser occurs in every direction, while the acquisition of spectra is not performed in the whole solid angle (at least not in Ref.~\cite{Ghofraniha15}). This, coupled with spectral resolution issues, casts doubt on whether the detected spectra correspond to all the laser modes oscillating in the sample. Moreover, experimental data inevitably contain a part of dynamical relaxation to equilibrium, which should be taken into account by the theory in order to improve the comparison. Clearly, the analytical results obtained for the IFO distribution in Ref.~\cite{Antenucci15c} and reported in Fig.~\ref{fig:IFOteo} are purely at equilibrium.

An additional issue that has to be addressed on the theoretical side is going beyond the narrow bandwidth limit, which includes the FMC in a trivial way. This is precisely the big goal of this Thesis work. In particular, we aim to develop a numerical tool to simulate the model diluted with the FMC both out-of and at equilibrium. Hopefully, this will also provide useful insight on the diluted model, in view of its analytical solution.

\newpage

\part{Numerical Simulations}

\chapter{Mixed-Order Glass Transition in Random Lasers} \label{chap:mixedorder}

The main goal of this work is to go beyond the fully-connected solution of the spin-glass model for random laser presented in the previous chapter. By this, we mean to release the narrow bandwidth approximation and include in a nontrivial way the mode coupling selection induced by the FMC.
Out of the narrow bandwidth approximation, the interaction network of the glassy random laser can no longer be considered fully-connected, but has to be diluted by removing all the bonds that do not match the condition on the frequencies. The FMC is of key importance for the study of mode-locking and for the reproduction of real random laser spectra: for this reason its inclusion is essential to bridge with the experiments. However, when the solution of the mode-locked model is approached analytically, subtle technical difficulties emerge, which require the development of new techniques with respect to standard mean-field methods for disordered systems. The analytical approach will be developed in the second part of this work. Here, we resort to numerical simulations to get useful insights on the mode-locked model.

A first step towards the inclusion of the FMC has been taken on the ordered version of the model \cite{Antenucci15d,Antenucci15e}. In this case, strong deviations from the fully-connected behavior have been put in evidence. In particular, the mode-locked low temperature (high pumping) phase exhibits lack of global order due to the onset of \emph{phase waves}\footnote{Phase waves is an evocative term reminding of \emph{spin waves} in pairwise spin models with $O(2)$ global symmetry, such as the XY model. In these models, when the symmetry is spontaneously broken, the global magnetization is reduced by the onset of collective excitations analogous to the Goldstone bosons in quantum field theory. In dimensions $d \leq 2$ this phenomenon leads to lack of global magnetic order, as stated by the Mermin-Wagner theorem \cite{Mermin66}; however, in the special case of $d=2$, topological transitions of the Kosterlitz–Thouless type \cite{Kosterlitz73} may be allowed. In the case of multimode lasers, however, we are dealing with dense models, so the comparison can not be pushed too far.} (see also Ref.~\cite{Marruzzo15}) produced by the tendency to align of modes which are close in frequency, due to the FMC. In particular the phases $\phi_k$ of the modes are not all equal as in the fully-connected case, but satisfy a linear relation with their frequencies $\omega_k$, which for a linear comb can be written as $\phi_k = \phi_0 + k \Delta$, where $\Delta$, approximately independent from $k$, is the slope of the phase wave and is a configuration-dependent quantity \cite{Antenucci16}. As a result, the coherency of the laser is not trivial as in the narrow bandwidth limit, where the output consists of a train of almost perfectly delta-like (unchirped) pulses, but there is a phase delay in the emitted pulses, which depends on the slope $\Delta$, and hence on the configuration. Therefore, in the magnetic analogy, when considering the thermal average, the model has vanishing magnetization. This effect has not been found in a model with ordered couplings and a random dilution of the same order of the FMC: in this case, the physical properties of the system are coherent with the fully-connected solution. Similarly, phase waves are not expected in the presence of quenched disordered couplings. 

After that, equilibrium numerical simulations of the spin-glass mode-locked model have been performed in Ref.~\cite{Gradenigo20}, where evidence of a mixed-order phase transition has been found and put in connection with an equipartition-breaking transition at the same critical temperature. The common root of the two transitions can be traced back to the same underlying phenomenon: the breaking of ergodicity. In the present chapter and in the following one, we focus on the mixed-order phase transition, while the analysis of the equipartition-breaking transition will be deepened in Chap. \ref{chap:Condens}. With respect to the mean-field picture described in the previous chapter, we are particularly interested in checking what of the RFOT scenario remains in the diluted mode-locked model, which is much closer to real random lasers than the fully-connected one. 

In the following, first, we present the simulated spin-glass model, which is a slightly simplified version of the mode-locked spherical (2+4)-phasor model. Particular attention is devoted to the role played by the FMC in affecting the topology of the model interaction graph. The numerical technique is explained in detail, by presenting the Exchange Monte Carlo algorithm implemented to shorten
the thermalization time to equilibrium at low temperature. Moreover, the simulated model presents the additional problem of being defined on an interaction graph, which, though diluted, is still very dense. To address this problem, parallel computing on graphic processing units has been adopted. The results pointing towards the presence of a static glass transition are collected and their problematic nature is discussed. In particular, the unexpected scaling of the critical region found in \cite{Gradenigo20} motivates the need for a new campaign of numerical simulations of the model aimed at collecting data less affected by finite-size effects. 

\section{The Mode-Locked 4-Phasor Model}
The simulated model is described by the following Hamiltonian
\begin{align} \label{HamML4}
    \mH[\bm{a}] &= - \sum_{\bm{k} | \text{FMC}(\bm{k})} J_{k_1 k_2 k_3 k_4} \overline{a}_{k_1} a_{k_2} \overline{a}_{k_3} a_{k_4} + \text{c.c.} \nonumber \\
    &= - \sum_{\bm{k} | \text{FMC}(\bm{k})} J_{k_1 k_2 k_3 k_4} A_{k_1} A_{k_2} A_{k_3} A_{k_4} \cos(\phi_{k_1} - \phi_{k_2} + \phi_{k_3} - \phi_{k_4}),
\end{align}
where $\bm{a}=\{a_1,...,a_N\}$ is a $N$-dimensional complex vector of electromagnetic field mode amplitudes. In the second expression, $A_{k}$ and $\phi_k$ represents respectively the modulus and the phase of the mode amplitude $a_k=A_k e^{i\phi_k}$ and a factor 2 has been absorbed in the definition of the random couplings. Configurations are constrained to the complex hypersphere of radius $\sqrt{\epsilon N}$, where $\epsilon = \mathcal{E}/N$ measures the average optical power per mode available in the system. The quenched disordered coupling constants $J_{\bm{k}} = J_{k_1 k_2 k_3 k_4}$ are independently drawn from a zero-mean Gaussian distribution 
\begin{align} \label{Gauss}
    P(J_{\bm{k}}) = \frac{1}{\sqrt{2 \pi \sigma^2}} \exp\left\{ \frac{J_{\bm{k}}^2}{2 \sigma^2}    \right\},
\end{align}
with variance $\sigma^2 = \overline{J_{\bm{k}}^2} = 1/N^{2}$, ensuring the extensivity of the energy. The scaling of the coupling distribution variance takes into account the dilution order of the interaction graph, which is determined by the condition
\begin{align} \label{FMC-4body}
\text{FMC}(\bm{k}): |\omega_{k_1} - \omega_{k_2} + \omega_{k_3} - \omega_{k_4} | \lesssim \gamma,
\end{align}
where $\omega_k$ are the frequencies of the modes and $\gamma$ denotes their typical linewidth. We refer to this simplified version of the general (2+4)-phasor model discussed in the previous chapter as mode-locked (ML) 4-phasor model. With respect to the general model and in terms of the photonic parameters introduced in the previous chapter, see eqs.~\eqref{PHT_par1} and \eqref{PHT_par2}, we are here working in the limits $R_J \rightarrow \infty$ and $\alpha=1$. The motivation for considering this model is that we are mainly interested in the study of the non-linear term of the Hamiltonian defined in Eq.~\eqref{Hamilt2+4}, which is the most relevant one for reproducing the phenomenology of optical waves in disordered media near the lasing transition. Indeed, the behaviour of multimode optical systems in this regime is generally believed to be dominated by non-linear mode interactions, see e.g.~Refs.~\cite{Haus00,Andreasen11}. 

The simplest choice for the frequency distribution is to consider a linear comb as in the case of standard lasers, see Eq.~\eqref{FreqComb}. In this case the FMC \eqref{FMC-4body} can be mapped into a relation among the indices of the interaction graph:
\begin{align} \label{FMC-indices}
    |k_1 - k_2 + k_3 - k_4| = 0.
\end{align}
More realistic dilution rules based on random frequency distributions will be considered in future works in order to improve the modeling of real random lasers. Besides being the simplest possible choice, the frequency comb distribution is compatible with the strong-cavity approximation \cite{Conti11}, which amounts to neglect the off-diagonal elements of the linear interaction term in the Hamiltonian of the 2+4 model. In fact, the 2-body FMC, which generally looks like $|\omega_{k_1}-\omega_{k_2}| \lesssim \gamma$, admits off-diagonal terms only in the case of modes whose frequencies differ less than the threshold fixed by $\gamma$. In principle, modes of this kind exist in random lasers \cite{Andreasen11}. However, in the case of a high-finesse linear comb, these modes are excluded: the condition for mode selection reduces to $|k_1-k_2|=0$, which leaves only the diagonal terms. 

Furthermore, by assuming a flat gain curve, i.e.~by taking $J_{kk}=g$ with constant $g$, the linear part of the interactions becomes an additive constant, which is only responsible for a shift of the energy and can be neglect it. This follows as a consequence of the spherical constraint. The assumption of a flat gain curve is also compatible with the regime we aim to explore through numerical simulations: as shown in~\cite{Antenucci15e} for the case of standard multimode lasers, the inclusion of a more complex gain profile only affects the fluorescence regime, while the transition and the lasing regime are stable under perturbations of the gain.  

The effective distribution of the phasor configurations which will be sampled in numerical simulations is given by
\begin{align} \label{ProbDistr}
    \mathcal{P}[\bm a] \propto e^{-\beta \mathcal{H}[\bm{a}] } \delta\left( \epsilon N - \sum_{k=1}^N |a_k|^2 \right),
\end{align}
where $\beta$ is the inverse of the spectral power of noise $T$. We notice that, by rescaling the variables as $\tilde{a}_k = a_k/\sqrt{\epsilon}$, the new variables are constrained on
a fixed hypersphere at the cost of introducing the effective inverse temperature $\beta_{\rm photonic} = \beta \epsilon^2 = \mathcal{P}^2$, which corresponds to the photonic temperature introduced in Eq.~\eqref{E-TPHOTON}. In these rescaled variables the probability distribution of configurations reads as
\begin{align} \label{ProbDistrP2}
    \mathcal{P}[\tilde{\bm a}] \propto e^{- \mathcal{P}^2 \mathcal{H}[\tilde{\bm{a}}] } \delta\left(N - \sum_{k=1}^N |\tilde{a}_k|^2 \right),
\end{align}
making explicit the role of the pumping rate (i.e. a parameter accounting for both noise and external pumping) as the true control parameter of the system. Now, $\mathcal{P}$ can be tuned in numerical simulations either by varying the effective temperature $T = \beta^{-1}$ and fixing the optical power $\epsilon$, or by working at fixed temperature and varying the value of $\epsilon$. Simulations are performed at $\epsilon=1$ varying the temperature $T$ in order to have a clear correspondence with the literature on glassy systems, but results are often described in terms of pumping rate $\mathcal{P}$. One simply needs to remember that, since $\epsilon=1$, the photonic temperature reduced to the spectral power of noise $T$, and so $\mathcal{P} = 1/\sqrt{T}$.

\subsection{Topological Properties}
In this section, we aim to provide some details about the topology of the interaction network of the ML 4-phasor model. A mode-locked graph can be defined in full generality as a hypergraph whose hyperedges are selected according to the FMC \eqref{FMC}. Equivalently a mode-locked graph can be also defined on a factor graph, with fixed connectivity of the function nodes and connectivity of the variable nodes determined by the FMC. The following treatment is restricted to the case of interest, which is characterized by comb-like frequencies and 4-body interactions, but it can be extended to more general situations. 

An interesting quantity to compute is the total number of hyperedges that are left by the FMC in the interaction graph of the ML 4-phasor model, with respect to the fully-connected case. Let us denote by $N^{(f)}_4$ the number of tetrads in the fully connected graph, which is given by
\begin{align}
    N^{(f)}_4 = \frac{N(N-1)(N-2)(N-3)}{4!} \sim \frac{N^4}{4!},
\end{align}
with $N \gg 1$. We notice that even if the adjacency tensor defined by the FMC with equispaced frequencies is not completely symmetric under permutations of the indices, each term entering the Hamiltonian \eqref{HamML4} has some symmetry. The condition \eqref{FMC-indices} can be satisfied by 24 permutations of the indices, which can be grouped into 3 independent orderings with 8 equivalent permutations each. 

Given a tetrad of indices  $\bm{k} =\{k_1,k_2,k_3,k_4\}$, the 3 non equivalent orderings in Eq. (\ref{FMC-indices}) can be chosen to be
\begin{enumerate}
\item{FMC$_1$. } 
$\mathcal{P}_1 = (k_1,k_2,k_3,k_4)$ identifying the combination $k_1-k_2+k_3-k_4=0$ and all indices permutations;

\item{FMC$_2$. } 
$\mathcal{P}_2=(k_1,k_3,k_2,k_4)$  identifying $k_1-k_3+k_2-k_4=0$ and all indices permutations;

\item{FMC$_3$. } 
$\mathcal{P}_3 = (k_2,k_1,k_3,k_4)$  identifying $k_2-k_1+k_3-k_4=0$ and all indices permutations;
\end{enumerate}
All the permutations inside each of the 3 groups correspond to terms inside the Hamiltonian \eqref{HamML4} which have the same value of the energy. Consider for example the first ordering:
\begin{align} \label{FMC_1}
    \text{FMC}_1: \omega_{k_1} + \omega_{k_3}= \omega_{k_2} + \omega_{k_4} \rightarrow k_1 + k_3 = k_2 + k_4,
\end{align}
where we have used Eq.~\eqref{FreqComb}. Following Ref.~\cite{Marruzzo18}, we consider uniformly distributed indices, i.e. $P(k)=1/N$ for $k \in [1,N]$. In this case, the probability for the sum of two indices $k_{ij}^+=k_i+k_j$ to take the value $k^+ \in [2,2N]$ can be easily determined:
\begin{align}
    P_+(k^+) = \left\{ \begin{array}{rcl}
    \begin{aligned}
     & \frac{k^+ - 1}{N^2} & \mbox{if} & ~~~k^+\in [2,N+1] \\ 
     & \frac{2N - (k^+ -1)}{N^2} & \mbox{if} &~~~k^+\in [N+2,2N].
     \end{aligned}
\end{array}\right.
\end{align}
We can now evaluate the probability that the quadruplet $\bm{k}$ satisfies the condition $\mbox{FMC}_1$ as the probability that the left and right hand side of Eq.~\eqref{FMC_1} take the same value $k^+$:
\begin{align}
    P(\mbox{FMC}_1) &= \sum_{k^+=2}^{2N} P_+(k^+)^2 \nonumber \\
    &= \frac{1+2N^2}{3N^3} \sim \frac{2}{3N}, \nonumber
\end{align}
where the last relation holds in the large-$N$ limit. The same occurs for the other independent orderings, say $\mbox{FMC}_{2,3}$. This leads to the removal of the factor $1/3$ from the probability that the FMC is satisfied by any ordering. Eventually, the number of couplings in the interaction network of the ML 4-phasor model is
\begin{align} \label{N4true}
    N_4^* = \frac{2}{N}\left[1 + \mathcal{O}\left( \frac{1}{N}\right)  \right] N_4^{(f)}.
\end{align}
Hence, the FMC tends to cut $\mathcal{O}(N)$ couplings from the fully-connected graph, reducing the total number of couplings in the system to $\mathcal{O}(N^3)$. This prevision has been verified numerically with great accuracy. It is worth stressing that, though diluted, the graph is still dense.

The FMC condition is a deterministic selection rule, which induces non-trivial correlations among the interacting modes. To gain an insight into the kind of correlations, we consider an analogy with random networks, following Ref.~\cite{Antenucci15e}. A way to build random but correlated networks is to introduce a distance among the nodes: in the case of a random graph, e.g.~Erd\"os-Rényi graph, the distance can be chosen as the absolute value of the difference of the node indices $d_{k_1 k_2} = |k_1 - k_2|$. Then, one can select bonds accordingly to a probability that depends on that distance. In the case of the ML 4-phasor model, which has factor nodes of connectivity 4, one needs a 4-indexed metrics in order to select the interacting quadruplets. This metrics can be taken as $d_{\bm{k}}= |k_1 -k_2 +k_3 -k_4|$: therefore, the FMC with equispaced frequencies is equivalent to
including only quadruplets presenting the minimum value, $d_{\bm{k}}=0$. In this way, the mode frequencies are not degrees of freedom, but coordinate driving correlations playing the role of a distance on a graph. It should be stressed again that in the present case the mode coupling is deterministic and not random: no probability is associated with the distance. As a result, modes that are in the center of the spectrum are preferred for combinatorial reasons. Indeed, the central modes have a higher probability of having close frequencies in the sense of the distance $d_{\bm{k}}$. This is the reason for the narrowing of the intensity spectrum observed through the lasing transition, see e.g.~Fig.~\ref{fig:spettriIntro}, when the external pumping is increased (or equivalently the temperature is reduced).

\subsection{Generation of a Mode-Locked Graph} \label{Graph-gen}
Let us here describe in detail how the FMC is implemented in our code in order to build the interaction network of the ML 4-phasor model. 

First, a virtual complete graph with $N_4^{(f)}=\binom{N}{4} \sim \mathcal{O}(N^4)$ interactions is generated with ordered quadruplets of indices $k_1<k_2<k_3<k_4$. Then the FMC is applied to the complete graph. Notice that for each ordered quadruplet of indices, the FMC can be satisfied only by the permutation class $\mathcal{P}_3(k_1,k_2,k_3,k_4)$. Each time a quadruplet of indices matches the previous condition, the corresponding interaction is added to the real graph and a random value extracted from the Gaussian distribution Eq.~\eqref{Gauss} is assigned to it. This procedure is repeated by randomly picking a quadruplet from the complete graph until a preassigned number $N_4 \sim \mathcal{O}(N^3)$ of interactions for the ML graph is reached. 

In order to be able to perform a neat finite-size scaling analysis, the number $N_4$ is chosen to be the largest power of $2$ below the total number of couplings satisfying the FMC, which is given by the quantity $N_4^*$ computed in Eq.~\eqref{N4true}. In practice, the number $N_4$ is chosen first and then the corresponding size $N$ to be simulated is selected in order to minimize the difference $\Delta N = N_4-N_4^*$. 

We notice that this way of building the mode-locked interaction network introduces an artificial source of disorder in the model, besides the original one. Each one of the $N_{\text{s}}$ simulated disordered samples is characterized by a realization of the couplings $\{J_{\bm{k}}\}$ that differs from the others both in the quadruplet network and in the numerical values. However, the fluctuations of the observables with respect to the randomness of the quadruplet topology turn out to be much smaller than the fluctuations due to the numerical values of the couplings, already for the smallest simulated sizes. In fact, as it has been observed also for the ordered mode-locked graph \cite{Antenucci16}, when compared on the log-scale the energy fluctuations occurring during the equilibrium dynamics (i.e.~the ensemble fluctuations) are at least two orders of magnitude larger than the graph-to-graph fluctuations, which are therefore negligible for all practical purposes. 

\section{Numerical Analysis}
This section is devoted to present the details of the Monte Carlo algorithm implemented for the simulation of the model. 

\subsection{Exchange Monte Carlo Algorithm}
The numerical simulations of the ML 4-phasor model have been performed by means of an Exchange Monte Carlo algorithm parallelized on GPU's \footnote{Graphic Processing Units. The code, written in CUDA, has been running on three types of GPU:
Nvidia GTX680 (1536 cores), Nvidia Tesla K20 (2496 cores) and Nvidia Tesla V100 (5120 cores).} in order to sample the equilibrium probability distribution Eq.~\eqref{ProbDistr}. In this section, we briefly describe the most salient features of the method, by following Ref.~\cite{Newman99}; then, we provide a few details on our implementation for the simulation of the ML 4-phasor model.

The Exchange Monte Carlo method, else called Parallel Tempering (PT), was introduced by Hukushima and Nemoto in Ref.~\cite{Hukushima96} as a variation of simulated tempering~\cite{Marinari92}. In fact, PT is the most simple and general form of simulated tempering, which has been proposed as a finite-temperature generalization of the famous simulated annealing~\cite{KirkpatrickS83}. All these algorithms have been developed in order to cope with complex optimization problems, characterized by the presence of many minima of the cost function, where usual Monte Carlo algorithms are not feasible. In particular, PT has been widely used for equilibrium simulations of finite-dimensional spin glasses, see e.g.~Refs.~\cite{Ballesteros00,Hasenbuch08,Papakonstantinou14}, which are known to be  ``hardly-relaxing'' systems. Indeed, for glassy systems standard iterative algorithms tend to get stuck in small regions of the state space from which they cannot escape, facing the so called critical slowing down.

One explanation for this phenomenon is that each state in the Markov chain of a Monte Carlo algorithm is chosen from the previous one and is in some sense close to it. Therefore, starting from a certain initial configuration, there are states that can be reached with a small number of moves, while there are other states, farther in the configuration space, which can only be reached in a large number of moves. The state space of glassy models contains many stable and metastable states, as it can be revealed by several techniques \footnote{Metastable states can be revealed by directly studying the model dynamics, typically in the Martin-Siggia-Rose functional integral formalism (developed by De Dominicis and Janssen for disorderd models, see e.g.~Ref.~\cite{DeDominicis78}), but also by the study of the complexity or by an analysis based on the Thouless-Anderson-Palmer (TAP) equations \cite{Thouless77}. The prototype case is that of the spherical p-spin model, for which the dynamical equations have been closed and solved in Ref.~\cite{Crisanti93,Cugliandolo93} and the TAP approach has been developed in Refs.~\cite{Kurchan93,Crisanti95}. Moreover, the structure of the metastable states has been carefully analyzed by means of the Franz-Parisi potential in Refs.~\cite{Franz95b, Cavagna97}, while their basins of attraction have been studied in Ref.~\cite{Barrat98}.}. It turns out that the metastable states have a relatively low energy with respect to the states by which they are surrounded. If a simulated system is initialized in a configuration close to a metastable state (or directly in a stable state), to escape its basin of attraction, the algorithm must pass through one of the surrounding states with higher energy, an occurrence that has an exponentially low probability, since configurations are sampled with Boltzmann weights, which depend on the energy difference. 

It has to be pointed out that, of course, multiple states are also present in much simpler systems, such as the standard Ising model below the critical temperature. In this case there are only two low temperature states in which the Gibbs measure breaks down when the system size is sent to infinity: a positively magnetized state and a negatively magnetized one. In fact, to pass from one state to the other, the system configuration has to jump an energy barrier whose height scales exponentially with the size of the system. This event has a probability that is exponentially small, similarly to the corresponding case in spin-glass models. However, in this case the low temperature states are symmetric under spin reversal transformation and one gets the same information on the measured properties of the system, no matter whether the initial configuration is chosen close to a state or to the other. Conversely, in glassy models the states are usually not related by any symmetry, so it happens that for different simulations the algorithm gets stuck in different basins each time depending on the initial condition, giving completely different answers for the observables. This can be regarded as a finite-size evidence of ergodicity breaking.

To avoid this situation, PT has been developed based on the idea that system thermalization may be facilitated by a reversible Markovian dynamics of configurations among heat baths at close temperatures. In particular, configurations belonging to copies of the system at higher temperature may help the copies at lower temperature to jump out of the local minima of the rugged free energy landscape. While the dynamics is carried out in parallel for all the heat baths simulated, once after a fixed number of steps an exchange of configurations between baths at neighboring temperatures is proposed. We refer to this kind of move as swap, to distinguish it from the usual Monte Carlo step. A swap is proposed sequentially for all pairs of neighboring inverse temperatures $\beta_i$ and $\beta_{i+1}$, with the following acceptance probability implementing detailed balance with the equilibrium Boltzmann distribution for each thermal bath:
\begin{align} \label{PT-accProb}
p_{\text{swap}} = \min\ [1 \ ,\ e^{(\beta_i - \beta_{i+1})( \mathcal{H}[\bm{a}_i] -  \mathcal{H}[\bm{a}_{i+1}])}].
\end{align}
Thus, each state in each of the simulations is sampled with exactly its Boltzmann weight, so that in PT simulations measurements can be performed in the same way as in a usual Monte Carlo simulation. We will come back on the measurement process in the following. 

It should be noted that a key role is played by the time after which a move is proposed. This time has to be large enough in order to avoid exchanges among configurations that are very similar to those just exchanged, and, on the other hand, not too large, otherwise thermalization will require a very long time. In one word, one needs the time between two subsequent swaps to be the smallest possible in order to make the most of a PT algorithm.

\subsection{The choice of temperatures}
The reason why the PT method overcomes energy barriers is strictly related to the choice of the simulation temperatures. To give a more intuitive understanding of the situation, we go through the following argument, taken from Ref.~\cite{Newman99}. 

Let us focus of two copies of the system at temperatures $T_1<T_2$, one below and the other above the glass transition of the system. Suppose that the two copies start from configurations which belong to the same energy basin. Since the high temperature simulation do not show ergodicity breaking, it will freely explore the phase space on a time scale similar to that of a simulation of a normal, non-glassy system. On the contrary, the low temperature copy of the system will remain stuck in the initial energy basin. If one attempts to swap the states of the two simulations, Eq.~\eqref{PT-accProb} says that, unless the swap does not increase the energy of the low temperature simulation by a great deal, then it is unlikely to be accepted. However, from time to time, it will happen that the system at $T_2$ finds its way into a region of low energy, that is another basin with respect to the initial one, where the simulation at $T_1$ is stuck. When this happens a swap will quite likely be accepted. Thus the low temperature copy is transported in one move to another energy basin, and the high temperature one finds itself back in the basin that it started in. By repeating the process over a long time, the low temperature simulation is moved repeatedly to new energy basins. Thus, the PT algorithm effectively overcomes the problem of barrier crossing, which makes simulation of glassy systems so hard, and allows us to sample a significant fraction of the state space, while still sampling with the correct Boltzmann weights for a temperature below the glass transition \cite{Newman99}.

In view of this argument, before running the simulations one must have an approximate knowledge of the critical temperature of the model, in order to establish properly the temperature interval, i.e. define a $\beta_\text{min}$ and a $\beta_{\max}$ such that the critical temperature falls inside the interval. Moreover, one has to bear in mind that temperatures should be close enough, so that the typical configuration domains at nearby temperatures overlap. If this does not occur, the energy distributions at some nearby heat-baths might display no sensitive overlap, thus yielding an extremely low probability of a swap between them. If a critical point is there, this is likely to occur when one heat-bath is at a temperature above the critical one and the other one at a temperature below it. If this is the case there will be a drastic drop in the exchange frequencies (swapping rate) between these two temperatures, above and below the phase transition, making the algorithm extremely inefficient.

One essential criterion to decide if the algorithm is working efficiently, both regarding on how often we propose a swap move and the choice of temperatures, is to compute the swapping rate between adjacent temperatures, that is the fraction of accepted swaps. The algorithm works efficiently only if, for all couples of temperatures is not too small. An optimal value lies between 0.6 and 0.8 and this interval has been taken as a reference in this work.

\subsection{Details of the Algorithm}
In what follows, the specific PT algorithm designed for the simulation of the ML 4-phasor model is described. The first step is to build the mode-locked graph of interaction is built according to the procedure described before. Then a PT dynamics is run for copies of the system with the same quenched disorder at different temperatures. Each of the system copies follows its own dynamics, except when a swap is accepted among neighboring heat baths. In the following, we focus on the main features of the algorithm.

\subsubsection{Local update}
The algorithm uses local Metropolis updates for the dynamics of each PT copy of the system. Then, a configuration update has to be proposed  with the requirement to keep
\begin{align*}
    \sum_{k=1}^N | a_k |^2 = \text{const}.
\end{align*}
In order fulfill the constraint, each update of the configurations is carried out by choosing at random two variables $a_i = A_i e^{i\phi_i}$ and $a_j=A_je^{i\phi_j}$ and extracting three random numbers \cite{Antenucci16, Gradenigo20}: $x,y \in [0,2\pi]$ and $z \in [0, \pi/2]$. The first two correspond to the new (attempted) phases $x=\phi_i'$ and $y=\phi_j'$, the third one mixes the intensities of the modes selected by preserving the spherical constraint
\begin{align*}
    A_j' = \sqrt{A_i^2 + A_j^2} \cos z ~~~~ A_j' = \sqrt{A_i^2 + A_j^2} \sin z
\end{align*}
Then the attempted update is accepted according to the usual Metropolis formula, in order to implement detail balance.

\subsubsection{Parallel computation}
It is worth noticing that for the case of the ML 4-phasor model the updates have to performed sequentially, rather than in parallel. In fact, the parallel Monte Carlo algorithm of a system of interacting variables needs a sparse network of interaction, such as nearest neighbors, e.g., to
be implemented, in order to split the system in smaller non-interacting sub-systems, that can be updated in parallel. This procedure clearly speeds up the computation of the update. In our system, however, this is not possible due to the density of the mode-locked interaction network, in which each variable participate in $O(N^2)$ interacting quadruplets. 

However another kind of code optimization can be implemented for the ML 4-phasor model, by exploiting the computing capability of GPU's: the parallelization of the energy computation in the local Metroplis update. In order to accept or reject the update of two spins $a_i$ and $a_j$, one has to compute the energy difference between the attempted configuration and the current one. This operation has a computational complexity which scales like the number of quadruplets involved in the computation, i.e. $N_4^{(i,j)} = O(N^2)$:
\begin{align*}
    \Delta E = \sum_{k=1}^{N_4^{(i,j)}} \Delta E_k,
\end{align*}
where $\Delta E_k$ denotes the energy difference between each quadruplet. The computation of each $\Delta E_k$ is realized in parallel on a distinct kernel on GPU.

Last, but not least, also the PT dynamics at different temperatures has been parallelized on GPU's. The two kinds of parallelization considered together reduce the execution time of the entire simulation by a factor of 8 \cite{Gradenigo20}.

\subsection{Observables of Interest}
Before listing the observables considered, let us clarify the measurement procedure. In order to properly estimate statistical errors, time correlations have been taken into account. A correlation time $\tau_{\rm corr}$ can be identified as the maximum among all the correlation times of each thermal bath dynamics. Consequently, the observables can be measured every $\tau_{\rm corr}$ Monte Carlo steps. If $N_{\rm MCS}$ is the total amount of Monte Carlo steps of the simulation, for each disordered sample the number of thermalized, uncorrelated configurations is given by
\begin{align} \label{ConfigTot}
    \mathcal N \equiv \frac{N_{\rm MCS}-\tau_{\rm eq}}{\tau_{\rm corr}},
\end{align}
where $\tau_{\text{eq}}$ denotes the thermalization time of the replica with the lowest temperature. We will come back in a while on how the equilibrium time is defined. Then, for a given observable $O$ function of the configurations $\bm{a}$, the ensemble average is estimated by the following time average 
\begin{align} \label{timeAver}
    \langle O[\bm{a}] \rangle = \frac{1}{ \mathcal N}\sum_{t=\tau_{\rm eq}/\tau_{\rm corr}}^{N_{\rm MCS}/\tau_{\rm corr}} O[\bm{a}_t]. 
\end{align}
On top of that, the average over the quenched randomness of the couplings is defined as follows. For each $\{J_{\bm{k}}\}$ realization we have a thermal average $\langle O[\bm{A}] \rangle_J$.  Averaging over the random samples yields the least fluctuating finite-$N$ proxy for the average in the thermodynamic limit:
\begin{align}
    \overline{O[\bm{a}]} = \frac{1}{N_{\text{s}}} \sum_{j=1}^{N_{\text{s}}}  \langle O[\bm{a}] \rangle_J^{(j)}.
\end{align}

The observables that will be considered in the present chapter and in the next one are the following:
\begin{itemize}
    \item \emph{intensity spectrum}: normalized to the square root of the temperature, in order to connect with the physical intensities
    
    \begin{align} \label{Spectrum}
        I_k = \frac{A_k^2}{\sqrt{T}}, 
    \end{align}
   
    \item \emph{specific heat}: measured as the equilibrium energy fluctuations as
    
    \begin{align} \label{SpecificHeat}
        c_{V_N} = \frac{1}{N} \frac{\overline{\langle \mH^2 \rangle - \langle \mH \rangle^2}}{T^2}, 
    \end{align}
    
    \item \emph{Parisi overlap distribution}, $P(q)$, where for the ML 4-phasor model the overlap among configurations (see Eq.~\eqref{Overlphases}) is given by
    
    \begin{align} \label{OverlapOB}
        q_{\alpha \beta} &= \frac{1}{N} \mbox{Re } \sum_{k=1}^N \overline{a}_k^\alpha a_k^\beta \nonumber \\
        &= \frac{1}{N} \sum_{k=1}^N A_k^\alpha A_k^\beta \cos(\phi_k^\alpha - \phi_k^\beta),
    \end{align}
    
    \item \emph{plaquette overlap distribution}, $P(\mathcal{Q})$, where the overlap among the plaquettes of two replicas is defined as \cite{Gradenigo20}
    
    \begin{align} \label{PlaqOB}
        \mathcal{Q}_{\alpha \beta} = \frac{1}{N_4} \sum_{\bm{k}} \mathcal{E}_{\bm{k}}^\alpha \mathcal{E}_{\bm{k}}^\beta
    \end{align}
    with the plaquette overlap given by
    \begin{equation}
        \mathcal{E}_{\bm{k}}^\alpha = a_{k_1} \overline{a}_{k_2} a_{k_3} \overline{a}_{k_4} + \text{c.c.}
    \end{equation}
    and $N_4$ the number of quadruples satisfying the FMC, as discussed in Sec. \ref{Graph-gen}.
    
    \item \emph{intensity fluctuations overlap distribution}, $P(\mathcal{C})$, where the intensity fluctuation overlap (IFO) among two replicas of the system (see Eq.~\eqref{IFOtheo}) is given by 
    
    \begin{align} \label{IfoOB}
     \mathcal{C}_{\alpha \beta} = \frac{1}{N} \sum_{k=1}^N \Delta_k^\alpha \Delta_k^\beta,    
    \end{align}
    where the intensity fluctuations are defined as
    \begin{align}
        \Delta_k^\alpha = \frac{I_k^\alpha - \langle I_k^\alpha \rangle}{2 \sqrt{2}\epsilon}.
    \end{align}
\end{itemize}

\subsubsection{Thermalization}
In order to guarantee that the data used to compute the displayed observables are taken from correctly equilibrated samples thermalization can be tested in several ways. 

First, one can look at energy relaxation on sequential time windows whose length is each time twice the length of the previous one. For each simulated heat bath dynamics, a minimal requirement is that the time average of the energy $\langle \mH \rangle$ takes the same value at least on the last and second but last windows. A similar test is performed on the specific heat, by computing energy fluctuations over the last and the second-last “logarithmic” window and checking that the values obtained for each temperature match inside their statistical errors. 

However, one can not rely solely on the trend of the energy and of its fluctuations over time to assess thermalization. In fact, as mentioned before, due to the presence of energy barriers that scale exponentially with the system size, one may mistake a single local minimum for a good equilibrium state. Therefore, a further and stronger requirement for thermalization is the symmetry of the Parisi overlap distribution $P_J(q)$ for each disordered sample, which can be tested by checking that its skweness is approximately zero in the low temperature phase. 

Once dynamical thermalization to equilibrium has been assessed and a thermalization time $\tau_{\rm eq}$ identified, the time average Eq. \eqref{timeAver} coincides with the canonical ensemble average. The number of Monte Carlo steps necessary to reach thermalization for each simulated size are reported in Tables \ref{tab1} and \ref{tab2}, for the simulations performed to obtain the results of Ref.~\cite{Niedda22a}.

\section{Evidence of a Random First Order Transition}
In this section we present the results of numerical simulations, where evidence of a glass transition in the ML 4-phasor model has been first reported \cite{Gradenigo20}. Here, we are just interested in discussing the physical picture drawn from these simulations: more technical details on measurement and data analysis will be provided in the next chapter, where the results of new simulations are presented.

As already mentioned in the previous chapter, a Random First-Order Transition (RFOT), the paradigm of the glass transition \cite{Kirkpatrick87,Lubchenko07,Leuzzi08}, is a mixed-order phase transition, characterized by the divergence of the thermodynamic susceptibilities and, at the same time, by the discontinuity of the order parameter at the static transition point. The former is the signature of a critical phenomenon, which is determined by a continuous second-order phase transition; the latter is, instead, a feature which is typical of first-order transitions, where the new dominant thermodynamic state is already present before the transition, differently from the continuous case, where it arises at the transition. 

The observables which help to investigate the presence of a RFOT\footnote{Remember that it is the static transition to be relevant for experiments on random lasers, and not the dynamic one, as it usually is in structural glasses. This is why our attention is devoted to the simulation of the equilibrium properties of the model at the static glass transition.} in the ML 4-phasor model are the specific heat \eqref{SpecificHeat} and the overlap distribution function \eqref{OverlapOB}. A singularity in the specific heat puts in evidence the second-order nature of the transition, whereas a jump in the overlap probability distribution $P(q)$ is a signature of its first-order nature. In models with continuous variables, the $P(q)$ is expected to be a distribution with a single peak in $q=0$ in the high temperature phase and to develop side peaks, as well, in the low temperature glassy phase. At finite $N$, of course, exact Dirac delta peaks in the $P(q)$ appear as smoothed functions of $q$, due to finite-size effects. One does not have to confuse this finite-size behavior of the $P(q)$ in a RFOT, with the behavior of the $P(q)$ in the spin-glass transition of the SK model~\cite{Sherrington75}, where the overlap distribution is expected to take a non-trivial shape, different from a bimodal one, even in the $N\rightarrow \infty$ limit.

\begin{figure}[t]
\centering
\includegraphics[width=\textwidth]{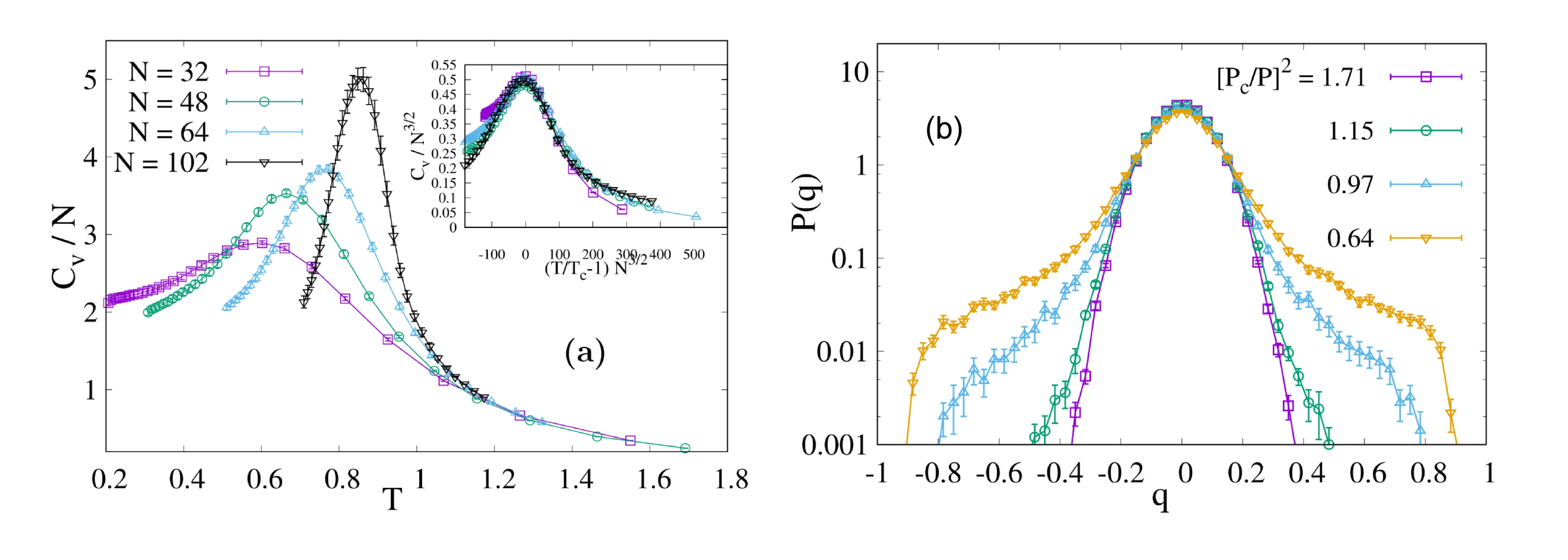}
\caption{(a) Specific heat $c_{V_N} = \overline{\langle \mH^2 \rangle - \langle \mH \rangle^2}/(N T^2)$ as a function of $T$; Different curves represent different system size. (Inset) Specific heat as a function of $\tau N^{3/2}$ where $\tau=(T/T_c(N) - 1 )$: curve collapse in the scaling region. The four sizes are $N = 32, 48, 64, 96$. (b) Configuration overlap probability distribution for $N = 64$ at temperatures $T / T_c = 1.71, 1.15, 0.97, 0.64$. Reprinted from \cite{Gradenigo20}.}
\label{fig:cV_Grade20}
\end{figure}

In Fig. \ref{fig:cV_Grade20}, we display the behaviour of the specific heat and of the configuration overlap distribution function. The specific heat diverges as the size increases: in the inset panel data are collapsed in the critical region with an exponent $3/2$, which turns out not to be compatible with a mean-field theory of second-order phase transitions. The reason why this comes about will be clarified in the next chapter: there, we will see how this unexpected exponent $3/2$ turns out to be a preasymptotic finite size effect. 

The configuration overlap distribution function in Fig. \ref{fig:cV_Grade20} turns out to be Gaussian in the low-$\mathcal{P}$ phase. Then, for $\mathcal{P} > \mathcal{P}_c$, the distribution shows a clear deviation from Gaussianity, but only ``shoulders'' are displayed at the simulated sizes, rather than proper side peaks. 

These results can be compared with those obtained through numerical simulations of a 4-phasor model with random dilution of the same order of that induced by the FMC, see Ref.~\cite{Gradenigo20}. The comparison reveals two important differences: first, the scaling of the critical region in the case of random dilution yields an exponent of $1/2$ which perfectly matches the expectations of standard $\phi^4$ mean-field theory; secondly, in the case of random dilution the overlap distribution function exhibits clear secondary peaks at a finite distance from the origin in the high-$\mathcal{P}$ phase, signaling a glassy RSB phase. It is evident, then, that the finite-size effects are stronger in the mode-locked model than in the randomly diluted one. This is also quite intuitive: at the small simulated sizes the correlations induced by a deterministic selection rule are not negligible. In fact, the diluted mode-locked graph of interactions, though comprised by an extensive number of couplings less than the fully-connected graph, is still dense, suggesting compatibility with mean-field theory. However, up to the precision of the study reported here, the question whether the ML 4-phasor model is a mean-field theory or not remains open and needs a more refined analysis to be answered.

\begin{figure}[t]
\centering
\includegraphics[width=\textwidth]{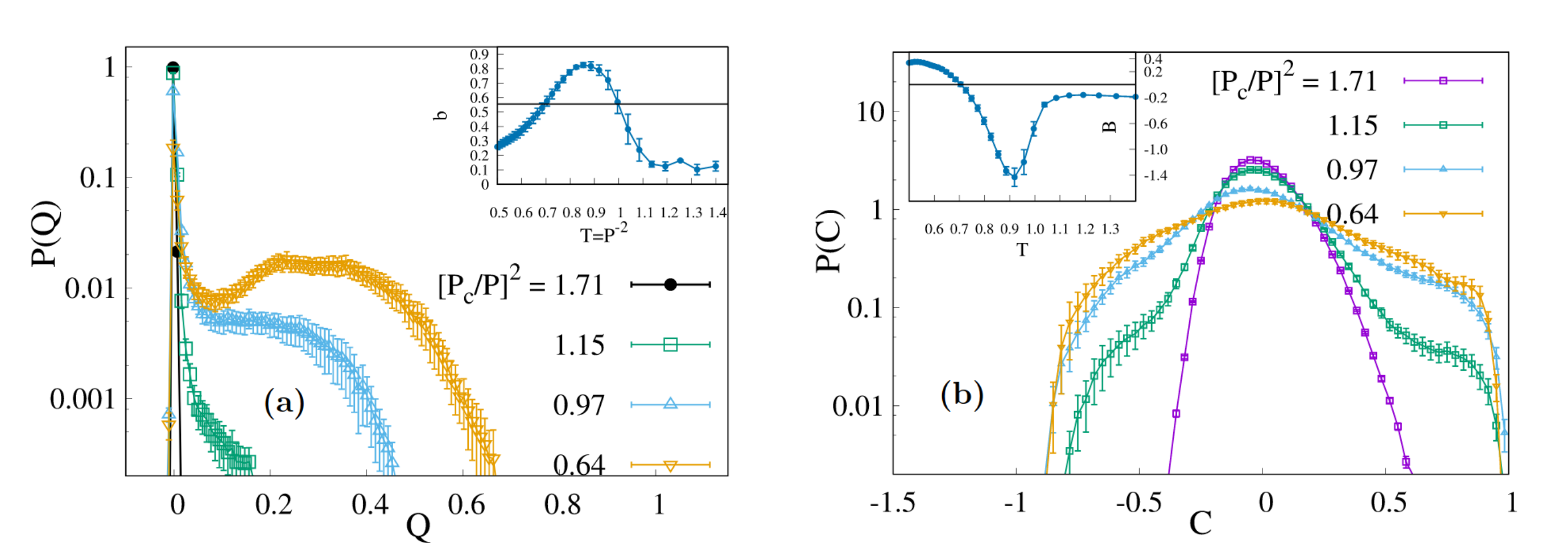}
\caption{(a) Plaquette overlap distribution $P(\mQ)$ for $N = 64$, computed over $2^{14}$ quadruplets and plotted for four different values of the pumping rates: $(\mathcal{P}_c/ \mathcal{P} )^2 = 1.71, 1.15, 0.97, 0.64$ (Inset) Multimodality parameter $b$ measured for $P(\mQ)$ as a function of $T=\mathcal{P}^{-2}$: values above the threshold $b^* = 5/9$ (full black line) indicate a bimodal distribution. (b) Intensity fluctuation overlap (IFO) probability distribution $P(\mathcal{C})$ for system size $N = 64$, plotted for the same values of the pumping rate as panel (a). (Inset) Binder parameter $\mathcal{B}$ measured for $P(\mathcal{C})$ as a function of $T$ ; the behavior is typical of first-order transitions, with the transition at the minimum of $\mathcal{B}$. Reprinted from \cite{Gradenigo20}.}
\label{fig:PQ_PC_Grade20}
\end{figure}

Regarding the first-order nature of the transition a stronger indication comes from the study of the plaquette overlap distribution \eqref{PlaqOB}. At variance with the configurational overlap, which is computed over $N$ variables, the plaquette overlap is computed over $\mathcal{O}(N^3)$ quadruplets, hence it is less plagued by finite-size effects. In Fig.~\ref{fig:PQ_PC_Grade20} we display both the plaquette overlap and the IFO \eqref{IfoOB} probability distribution functions. In the low-$\mathcal{P}$ phase the plaquette overlap has a very peaked distribution in $\mQ \simeq 0$, while for $\mathcal{P} \approx \mathcal{P}_c$ there is clear evidence of a secondary peak at $\mQ > 0$. A similar behavior is shown by the IFO distribution function: in this case, although at first sight there is no clear evidence of secondary peaks at high pumping rates, for the same reason why this happens in the $P(q)$, non-Gaussian tails appear in the vicinity of the transition.

To deepen the analysis one can study the moments of these distributions. In particular, it is useful to consider the third and fourth moments, which are related to the symmetry of the distribution and to the swelling of its tails. By denoting with $q$ a general overlap (be it the Parisi, the plaquette or the intensity fluctuation overlap) the skewness and the kurtosis of its distribution function are given by 
\begin{equation}
    \gamma = \frac{\overline{\Delta q^3}}{(\overline{\Delta q^2})^{3/2}} ~~~~~~~~  k = \frac{\overline{\Delta q^4}}{(\overline{\Delta q^2})^2}.
\end{equation}
In the high temperature phase, where the order parameter is zero, the distribution at finite size will be a Gaussian centered in zero, with $k = 3$. At the phase transition and in the low temperature
phase the distribution will be different from a Gaussian. A very useful quantity to measure the deviation from Gaussianity is the Binder parameter \cite{Binder81,Binder84}, which is defined as
\begin{equation}
    \mathcal{B} = \frac{1}{2} \left(3 - k \right).
\end{equation}
If we are dealing with a first-order phase transition, then the Binder parameter displays a nonmonotonic reversed bell behavior with a maximal deviation from Gaussianity in the coexistence
region, where the distribution is bimodal. This is precisely what is observed for the IFO distribution function and has been reported in the inset of panel (b) in Fig.~\ref{fig:PQ_PC_Grade20}, where $\mathcal{B}$ is plotted as a function of the effective temperature $T=\mathcal{P}^{-2}$. 

As a term of comparison, we refer to Ref.~\cite{Picco01}, where the Binder parameter is computed in numerical simulations of both the SK model and the Ising $p$-spin model (with $p=3$). In the former case, characterized by a continuous transition to a FRSB phase, the Binder parameter is zero at high temperature and then increases monotonically. Moreover, the Binder parameter for different sizes exhibits a crossing of the curves for different sizes, from which the critical temperature of the model can be estimated (cf. also Ref.~\cite{Bhatt88} for the case of the Edwards-Anderson model in two or three dimensions). On the other hand, in the case of the Ising $p$-spin model, characterized by a discontinuous transition to a 1RSB phase (regarding the distribution of the order parameter) the Binder parameter exhibits the reverse bell behavior that has been here reported in the case of the IFO probability distribution function.

For the study of the plaquette overlap distribution, one can not use the Binder parameter as a good indicator because, although clearly bimodal at the transition, the distribution is not Gaussian far from the transition. However, in order to study the bimodal nature of the distribution at the transition, one can introduce another parameter, the so-called \emph{bimodality} parameter, which is defined as
\begin{equation}
    b = \frac{\gamma^2 + 1}{k + \frac{3(n-1)^2}{(n-2)(n-3)}}
\end{equation}
where $n$ is the number of data composing the histogram of the probability distribution. In the inset of panel (b) in Fig.~\ref{fig:PQ_PC_Grade20} the behavior of $b$ as a function of temperature is reported: the region where the parameter $b$ signals a bimodal distribution of the overlap is precisely the interval of pumping rates around $\mathcal{P}_c$.

\newpage

\chapter{Universality Class of the Transition} \label{chap:Univ}
In this chapter, the study of the static glass transition in the ML 4-phasor model is improved with respect to the previous analysis (Chap.~\ref{chap:mixedorder}). The main goal is to determine the universality class of the model, through a refined finite-size scaling analysis of the transition. In particular, we aim to assess whether the unexpected scaling exponent of the critical region found in Ref.~\cite{Gradenigo20} is a genuine non-mean-field feature of the mode-locked diluted model or if it is only a preasymptotic effect, due to the very strong finite-size effects, which derive from the difficulty to simulate large enough sizes in dense disordered models.

First, a preliminary argument for the scaling of the critical region in mean-field theory is presented in a very general way, which does not require any specific knowledge about the glass transition, besides its second-order nature concerning the divergence of the susceptibilities at the critical temperature. The outcome of the analysis is that the exponent for the scaling of the critical region in a generic mean-field model must fall in a compact interval and its specific value is related to the order of the leading nonlinear term of interaction. The result obtained is coherent with the scaling exponent of the Random Energy Model (REM), the simplest model exhibiting a glass transition \cite{Derrida80,Gross84}.

In the second part of the chapter, results obtained from new numerical simulations are presented \cite{Niedda22a}. Besides performing simulations of the original ML 4-phasor model increasing the number of simulated sizes, temperatures and samples, a slightly different version of the model is introduced, in order to reduce the specific kind of finite-size effects induced by the FMC. In particular, the bond filtering action of the FMC is considered on a frequency space with periodic boundary conditions, leading to simulations of the bulk spectrum of the original model. The study of the critical region in both cases leads to a reduction of the value of the exponent for the scaling of the critical region, which turns out to fall into the interval defined by the mean-field argument, inside the experimental uncertainty. Finally, the study of the glass transition is completed by analyzing data in order to obtain the overlap distribution functions introduced in the previous chapter.

\section{A Mean-Field Argument for the Scaling of the Critical Region}
Second-order phase transitions are characterized by scale invariance at criticality: in finite-dimensional systems (e.g.~lattices in $d$ dimensions), fluctuations extend over regions of all possible sizes and a characteristic scale of length no longer exists. As a consequence, the behavior of physical quantities near the critical point is described with respect to some control parameter, e.g.~temperature, by power laws, whose exponents are universal, in the sense that they depend only on very general properties of the system (such as the dimensionality of space $d$, the dimensionality of the order parameter and the symmetries of the Hamiltonian), but not on the details of the microscopic interactions \cite{ZinnJustin02}. Out of the mean-field approximation, the critical exponents can be computed with renormalization group techniques, such as the $\epsilon$-expansion \cite{Parisi88}, or can be extrapolated from the study of the system at finite size $L$ (i.e. the linear size of a lattice), through the finite-size scaling analysis \cite{Fisher72}. 

However, here we are interested in the case of infinite-dimensional models defined on graphs not embeddable into any space with finite dimensions $d$, for which mean-field theory is exact. Even in this case, the scaling regime amplitude has a dependence on the system size $N$ (i.e. the number of nodes in the graph of interactions), governed by an exponent, which we will denote as $\nu_{\rm eff}$, through the following relation
\begin{align}
    |T-T_c| \sim 1/N^{1/\nu_{\rm eff}},
\end{align}
where $T$ is an effective control parameter for the transition - from now on we will refere to it as temperature - and $T_c$ denotes the critical point. We will make sense of $\nu_{\rm eff}$ in terms of standard critical exponents in the following. The prototype of a mean-field theory of continuous phase transitions is represented by the Landau effective potential\footnote{Let us stress that $\phi$ is a global quantity, not a local magnetization field $\phi(x)$ as in the Landau-Ginzburg $\lambda \phi^4$ theory. The potential $V(\phi)$ in Eq.~\eqref{potLandau} is the result of the Landau approximation of the $\lambda \phi^4$ field theory, which consist in taking $\phi(x)=\phi$ for all points in space, hence neglecting the Laplacian term: this is nothing but a mean-field approximation.} 
\begin{align} \label{potLandau}
    V(\phi) = \frac{\tau}{2}\phi^2 + \frac{g}{4!}\phi^4,
\end{align}
where $\phi$ is the global order parameter of the transition, $\tau = T/T_c - 1$ is the reduced temperature and $g$ is the coupling constant. The probability distribution of the order parameter $\phi$ is 
\begin{align}
    p(\phi) = \frac{e^{-N V(\phi)}}{Z},
\end{align}
where the partition function $Z$ is given by 
\begin{align}
    Z = \int \de \phi~e^{-N V(\phi)}.
\end{align}
In a second-order transition, the relevant quantities that have to be considered are the fluctuations of the order parameter around the minimum of the effective potential
\begin{align} 
\delta \phi^2 = \langle \phi^2 \rangle - \langle \phi \rangle^2,
\end{align}
where the brackets denote the average with respect to the distribution $p(\phi)$. The critical behaviour of the susceptibilities, which include the specific heat, is related to the fluctuations of the field near the critical point. In order to estimate the value of the exponent $\nu_{\rm eff}$ we aim to match the fluctuations above and below $T_c$. 

When $\tau \gtrsim 0$ the effective potential is well approximated by $V(\phi) \approx \frac{1}{2} \tau \phi^2$ for values of the order parameter close enough to the minimum $\phi^* = 0$. The probability distribution of $\phi$ is a zero-mean Gaussian distribution and its normalization is given by 
\begin{align}
Z \approx \int \de \phi~e^{- \frac{N\tau}{2}  \phi^2} \sim \frac{1}{\sqrt{N\tau}}.
\end{align}
In this regime, the fluctuations of the order parameter centred around the minimum $\phi^* = 0$ are given by the variance of the Gaussian distribution: \begin{equation}
\delta\phi^2_{ _{T>T_c}} \sim 1/N\tau.    
\end{equation}
We notice that the previous equation is the usual scaling law for the linear susceptibility above the critical point in the standard mean-field $\phi^4$ theory:
\begin{equation}
    \chi \sim \frac{1}{N} \tau^{-\gamma},
\end{equation}
with $\gamma = 1$, see, e.~g.~Ref.~\cite{Cardy96}.

On the other hand, when $\tau \lesssim 0$ the quartic term of the potential becomes relevant and can not be neglected. In this regime, the fluctuations of the order parameter are centered around one of the two symmetric minima of the potential, namely $\phi_\pm = \pm \phi^*$, depending on the initial conditions. Since we are interested in matching the fluctuations above and below the critical temperature, we assume the temperature to be sufficiently close to $T_c$ in order for the amplitude of the fluctuations to be of the order of the distance from the origin:
\begin{equation}
    \delta\phi^2_{ _{T<T_c}} \sim (\phi^*)^2.
\end{equation}
Clearly, this is no longer valid well below $T_c$, where the curvature of the potential has to be considered. The minima $\phi_\pm$ can be easily determined according to the saddle-point approximation of the partition function
\begin{align}
Z = \int  \de \phi~e^{ - N V(\phi)} \approx e^{ - N V(\phi^*)},
\end{align}
where $\phi^*$ such that $\frac{\de V(\phi)}{\de \phi}\Big|_{\phi^*} = 0$ is
\begin{equation} \label{ordparamscal}
    \phi^* = \sqrt{-\frac{6 \tau }{g}}.
\end{equation}
Hence, $\delta\phi^2_{ _{T<T_c}} \sim -\tau/g$. Therefore, we have an estimate of the fluctuations on the two sides of the critical point, respectively
\begin{align}
\delta\phi^2_{ _{T>T_c}} &\sim \frac{1}{N\tau} \label{susceptscal} \\
\delta\phi^2_{ _{T<T_c}} &\sim -\tau/g.     
\end{align}
By matching the previous expressions we find
\begin{align} \label{matchingArg}
\delta\phi^2_{ _{T>T_c}} \sim \delta\phi^2_{ _{T<T_c}} ~~~\Longrightarrow~~~|\tau| \sim \frac{1}{N^{1/2}},
\end{align}
yielding the estimate $\nu_{\rm eff} = 2$. 

Notice, that Eq.~\eqref{ordparamscal} corresponds to the usual scaling law for the order parameter below $T_c$ \cite{Cardy96}
\begin{equation}
    \phi \sim - \tau^{\beta}
\end{equation}
with exponent $\beta = 1/2$ for the mean-field $\phi^4$ theory. In terms of the exponents $\beta$ and $\gamma$ defined by the usual scaling laws, Eq.~\eqref{matchingArg} can be written as
\begin{equation*}
    \tau^{2 \beta} \sim \frac{1}{N} \tau^{-\gamma} ~~~\Longrightarrow~~~|\tau| \sim \frac{1}{N^{\frac{1}{2\beta +\gamma}}}.
\end{equation*}
This allow us to identify the scaling exponent $\nu_{\rm eff}$ for the critical region of infinite-dimensional models with the expression
\begin{equation}
    \nu_{\rm eff} = 2 \beta + \gamma.
\end{equation}
Therefore, $\nu_{\rm eff}$ can be taken just as a shorthand symbol for the expression $2 \beta + \gamma$.

This argument can be straightforwardly extended to more general mean-field potentials, in order to obtain a range of values for the exponent $\nu_{\rm eff}$. Let us consider the family of potentials 
\begin{align}
V(\phi) = \frac{1}{2} \tau\phi^2 + \frac{g}{(2n)!} \phi^{2n},
\end{align}
where the choice of an even lowest order non-linearity is still compatible with the phenomenology of a second-order phase transition. The fluctuations of the order parameter above the critical temperature are the same as in the case $n=2$, which means that the exponent $\gamma=1$ for every $n$. Below the critical temperature, by using the saddle-point method we find
\begin{align}
\phi^* = \left[-\frac{(2n-1)!~\tau }{g} \right]^{\frac{1}{2(n-1)}}.
\end{align}
Incidentally, this means that the general expression of the exponent $\beta$ for the scaling of the order parameter in a $\phi^{2n}$ mean-field theory is $\beta = 1/2(n-1)$. Therefore, the amplitude of the fluctuations above and below $T_c$ are given by
\begin{align}
\delta\phi^2_{_{T>T_c}} &\sim \frac{1}{N\tau} \\
\delta\phi^2_{_{T<T_c}} &\sim (-\tau/g)^{\frac{1}{(n-1)}},
\end{align}
and their matching leads to
\begin{align}
\delta\phi^2_{ _{T>T_c}} \sim \delta\phi^2_{ _{T<T_c}} ~~~\Longrightarrow~~~|\tau| \sim \frac{1}{N^{\frac{n-1}{n}}}.
\end{align}
The range of values that $\nu_{\rm eff} = 2\beta + \gamma$ can take in a mean-field model can be found  by taking $n=2$ and $n \rightarrow \infty$ in the previous expression. Thus, we have
\begin{align} \label{EXPIntervals}
 1 < \nu_{\rm eff} \leq 2 ~~~ \Longleftrightarrow ~~~ \frac{1}{2} \leq \frac{1}{\nu_{\rm eff}} <  1 .
\end{align}
With respect to this argument, it is clear that the value $1/\nu_{\text{eff}} = 3/2$ found in Ref.~\cite{Gradenigo20}, is out of the interval found for mean-field values of the exponent $\nu_{\text{eff}}$.

It is worth noting that the validity of this argument is restricted to the large-$N$ limit, where the saddle-point approximation holds. This is something that we have to keep in mind, when comparing results of numerical simulations at finite $N$ with the estimate on the scaling exponents obtained from the previous argument.

\subsection{The Random-Energy Model}
In order to compare the previous argument with a well-known model we have performed a numerical analysis of the Random-Energy Model (REM), which is the reference mean-field model for disordered systems with non-linear interactions. Let us briefly review the main features of the model, following Ref.~\cite{Mezard16}, before presenting our results.

The REM is the simplest statistical mechanics model of a disordered system exhibiting a phase transition \cite{Gross84}. It was originally introduced by Derrida in Refs.~\cite{Derrida80,Derrida81} as an exactly solvable model arising from the limit $p\rightarrow\infty$ of the p-spin model. As a result of this procedure, the model does not take into account any specific interaction among the variables: the energy is a random process rather than a deterministic function. For this reason, the REM has the remarkable advantage that results obtained through the replica method can be compared with formal mathematical approaches, see e.g.~Refs.~\cite{Gardner89,Bovier06,Ogure09,Guerra13,Dotsenko11}. 

Given $M=2^N$ energy levels (possibly corresponding to the configurations of $N$ Ising spin variables) the corresponding energies $\{ E_\nu \}_{\nu \in \{1,\dots,M\}}$ are taken as independent random Gaussian variables\footnote{A generalized version of the model, where correlations among the energy levels are introduced in a hierarchical way, has been formulated by Derrida in Ref.~\cite{Derrida85b} and exactly solved by Derrida and Gardner in Ref.~\cite{Derrida86}. Here, we are just interested in the original version of the model, since the generalized version is in the same universality class.} extracted from the distribution function
\begin{align} \label{REM-Gaussian}
    p(E) = \frac{1}{\sqrt{\pi N J^2}} \exp \left( -\frac{E^2}{N J^2} \right),
\end{align}
where the scaling of the variance with $N$ ensures the extensivity of the thermodynamic potentials and $J$ is a parameter. An instance of the quenched disorder corresponds to an extraction of the $M$ energy levels. A Boltzmann weight $p_\nu = \exp(-\beta E_\nu) / \mZ$ is then assigned to each configuration, where $\mZ$ is the partition function of the model
\begin{align} \label{REM:part_func}
   \mZ = \sum_{\nu=1}^{M} e^{-\beta E_\nu}.
\end{align}

We have developed a simple enumeration algorithm to study the REM, which works as follows: for each disorder sample of a given system size $N$ the energy levels $\{E_\nu\}$ are generated, by independently extracting a set of $2^N$ random numbers from the Gaussian distribution Eq.~\eqref{REM-Gaussian} with $J=1$. A set of equispaced temperatures $T$ is generated in the interval $[T_{\text{min}}, T_{\text{max}}]$. The internal energy of the model is computed as a function of temperature by evaluating the following thermal average
\begin{align}
    \langle E \rangle = \frac{\sum_\nu E_\nu~e^{-\beta E_\nu}}{\sum_\nu e^{-\beta E_\nu}},
\end{align}
for each of the $\beta=1/T$ values extracted before. The specific heat is computed from the fluctuations of the internal energy as 
\begin{align}
    c_{V_{N,i}}(T) = \frac{1}{N} \frac{ \langle E^2 \rangle -  \langle E \rangle^2}{T^2},
\end{align}
where the index $i$ accounts for the sample. The procedure is repeated for several independent extractions of the random energies $\{E_\nu\}$. Eventually, when a sufficiently large number of samples $N_{\text{sam}}$ is collected for a certain simulated size, the disorder average of the specific heat is computed by averaging over the samples:
\begin{align}
     c_{V_{N}}(T) = \frac{1}{N_{\text{sam}}} \sum_{i=1}^{N_\text{sam}} c_{V_{N,i}}(T).
\end{align}
The number of samples $N_{\text{sam}}$ is chosen in such a way that the estimated value of the specific heat remains stable upon fluctuations of the samples included in the average. 

\subsection{Finite-Size Scaling Analysis}
Let us consider momentarily a model defined on a $d$-dimensional lattice of linear size $L$. 

We consider an observable $Y_L$ that in the thermodynamic limit scales like 
$$Y_\infty(T) \approx A t^{-\psi}$$ 
near the critical point, where $t$ is the modulus of the reduced temperature, $A$ is some constant and $\psi$ some critical exponent. The fundamental assumption of the Finite-Size Scaling (FSS) Ansatz~\cite{Fisher72,Cardy88} is that the behaviour of $Y_L$ near the critical temperature is controlled by the ratio $\xi_\infty/L$. The parameter $\xi_\infty$ denotes the correlation length of the infinite-size system that scales as
\begin{align} \label{FSS:xi}
    \xi_\infty(T) \approx \xi_0 t^{-\nu},
\end{align}
where $\xi_0$ is a constant. The previous equation is the definition of the critical exponent $\nu$ governing the scaling of the correlation length. The scaling hypothesis for $Y_L$ can be then written as
\begin{align} \label{FSS:ScalHp1}
Y_L(T) = L^\omega f_Y(\xi_\infty/L),
\end{align}
where $\omega$ is the critical exponent for the scaling of the peak of the observable and $f_Y$ is a dimensionless function that depends on the observable $Y$. The function $f_Y$ is such that in the limit $L \rightarrow \infty $ one recovers the scaling law $Y_\infty(T) \approx A t^{-\psi}$, therefore $\omega = \psi/\nu$~\cite{Fisher72}. Moreover, by exploiting Eq.~\eqref{FSS:xi} the scaling relation~\eqref{FSS:ScalHp1} can be rewritten as
\begin{align} \label{FSS:ScalHp2}
Y_L(T) = L^{\psi/\nu} \hat{f}_Y(L^{1/\nu} t_L)
\end{align}
where $t_L = |T/T_c(L) - 1|$, being $T_c(L)$ the finite-size critical temperature, and $\hat{f}_Y$ is another scaling function that depend on $Y$. 

It is important to notice that, since the lattice size $L$ is linked to the number of degrees of freedom $N$ by $L=N^{1/d}$, the scaling relation for the observable $Y$ can be written as
\begin{align}
Y_N(T) = N^{\psi/ \nu d} \hat{f}_Y(N^{1/\nu d} t_N)
\end{align}
i.e. in terms of $N$, rather than $L$. By considering the well-known hyperscaling relation $\nu d = 2\beta + \gamma$ \cite{Cardy96}, we find
\begin{align}
Y_N(T) = N^{\psi/(2\beta + \gamma)} \hat{f}_Y(N^{1/(2\beta + \gamma)} t_N).
\end{align}
This is the only possible scaling relation that one can use when studying an infinite-dimensional model, where there is no lattice size $L$, but the only finite-size parameter is $N$.

In the case of the specific heat, the previous finite-size scaling law takes the following form (see e.g.~\cite{Itzykson89})
\begin{align} \label{ScalHpCv}
c_{V_N}(T) = N^{\alpha/\nu_{\rm eff}} \hat{f}_{C_{V_N}}(N^{1/\nu_{\rm eff}} t_N),
\end{align}
where $\alpha$ denotes the critical exponent of the specific heat peak divergence, and we have used fact that $\nu_{\rm eff} = 2\beta + \gamma$. Since the dimensionless function $\hat{f}$ is scaling invariant, if one uses the correct values of the exponents $\alpha$ and $\nu_{\rm eff}$, the curves $c_{V_N}(T)/N^{\alpha/\nu_{\rm eff}}$ for different values of $N$ should collapse on the same curve.

\begin{figure}[t]
\centering
\includegraphics[width=0.55\textwidth]{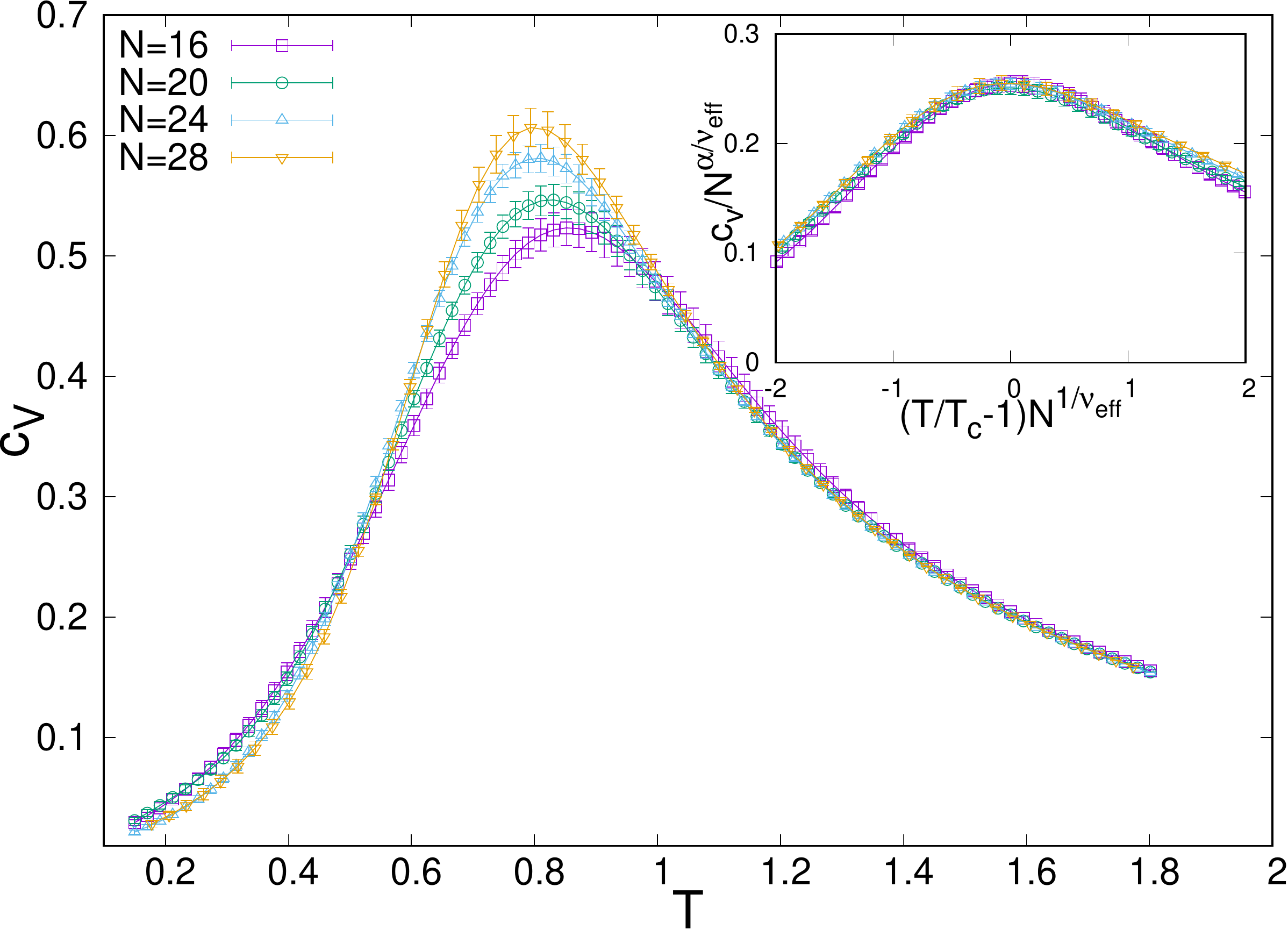}
\caption{Specific heat of the Random Energy Model. Different colors represent different simulated sizes. The simulated sizes are $N=16,20,24,28$. (Inset) Specific heat scaled by $N^{\alpha/\nu_{\rm eff}}$ as a function of $\tau N^{1/\nu_{\rm eff}}$. The data have been collapsed with exponents $\alpha=0.52 \pm 0.07$ and $\nu_{\rm eff}=1.94 \pm 0.22$.}
\label{fig:cV_REM}
\end{figure}

In Fig.~\ref{fig:cV_REM} we show the finite-size behaviour of the specific heat around the critical temperature. In the main panel, the specific heat is plotted as a function of temperature, for different system sizes. The simulated sizes are $N=16,20,24,28$, and for each size $N_{\text{sam}}=100$. In the inset data belonging to different sizes are collapsed near the critical temperature with exponents $\alpha = 0.52 \pm 0.07$ and $\nu_{\rm eff} = 2\beta + \gamma = 1.94 \pm 0.22$. As expected, due to the fact that the scaling hypothesis holds near the critical point, the collapse of the data succeeds around the peaks of the specific-heat curves. Moving away from the critical temperature the curves begin to separate one from another. In the inset we show only the interesting part of the collapse. At variance with the result found in Ref. \cite{Gradenigo20} for the ML 4-phasor, the result $1/\nu_{\rm eff} = 1/2$ for the scaling exponent of the critical region is in perfect agreement with the mean-field argument we have presented at the beginning of the chapter, in the case of a Landau potential. Moreover, we notice that the same behavior of REM specific heat divergence, has been found in a 4-phasor model with a random diluted topology \cite{Gradenigo20}. Both these model belong to the same universality class, the one of the standard $\phi^4$ Landau theory.


\section{Simulation with Periodic Boundary Conditions in the Frequency Space}
There are many sources of finite-size effects when simulating a dense model, most of which can be coped with by using powerful algorithms, such as the PT Monte Carlo algorithm. However, the ML 4-phasor model has another specific source of finite-size effects induced by the FMC. As discussed in the previous chapter, because of this condition, modes near the boundaries of the spectrum ($k\gtrsim 1$, $k\lesssim N$) interact much less than modes whose frequency lays in the middle of the spectrum ($k\sim N/2$). Though their dynamic evolution is taken into account in the simulations, the edge modes are less and less important as the external pumping increases. To circumvent this problem we introduce a slightly different model network, imposing periodic boundary conditions on the frequencies \cite{Niedda22a}.

\begin{figure}[t]
   \begin{minipage}{0.48\textwidth}
     \centering
     \includegraphics[width=0.95\textwidth]{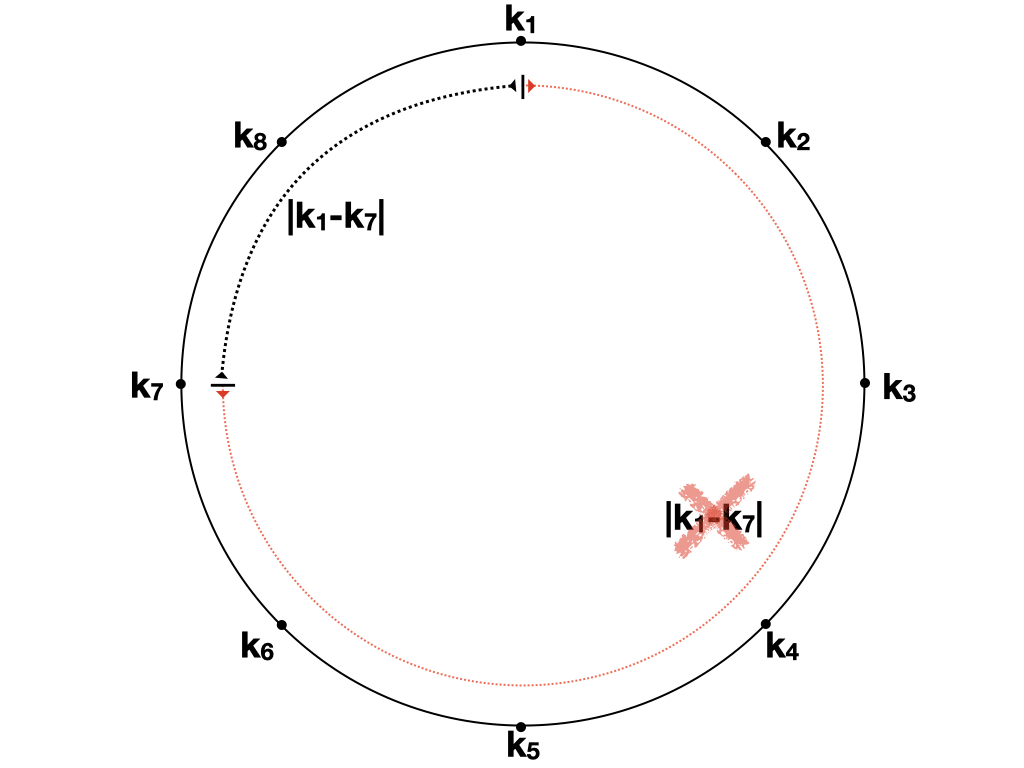}
    \caption{Periodic boundary conditions on the mode frequency indexes for the frequency matching condition. The modes can be thought as distributed on a ring, instead of a linear chain.}
    \label{fig:pbc}
   \end{minipage}\hfill
   \begin{minipage}{0.48\textwidth}
     \centering
     \includegraphics[width=\textwidth]{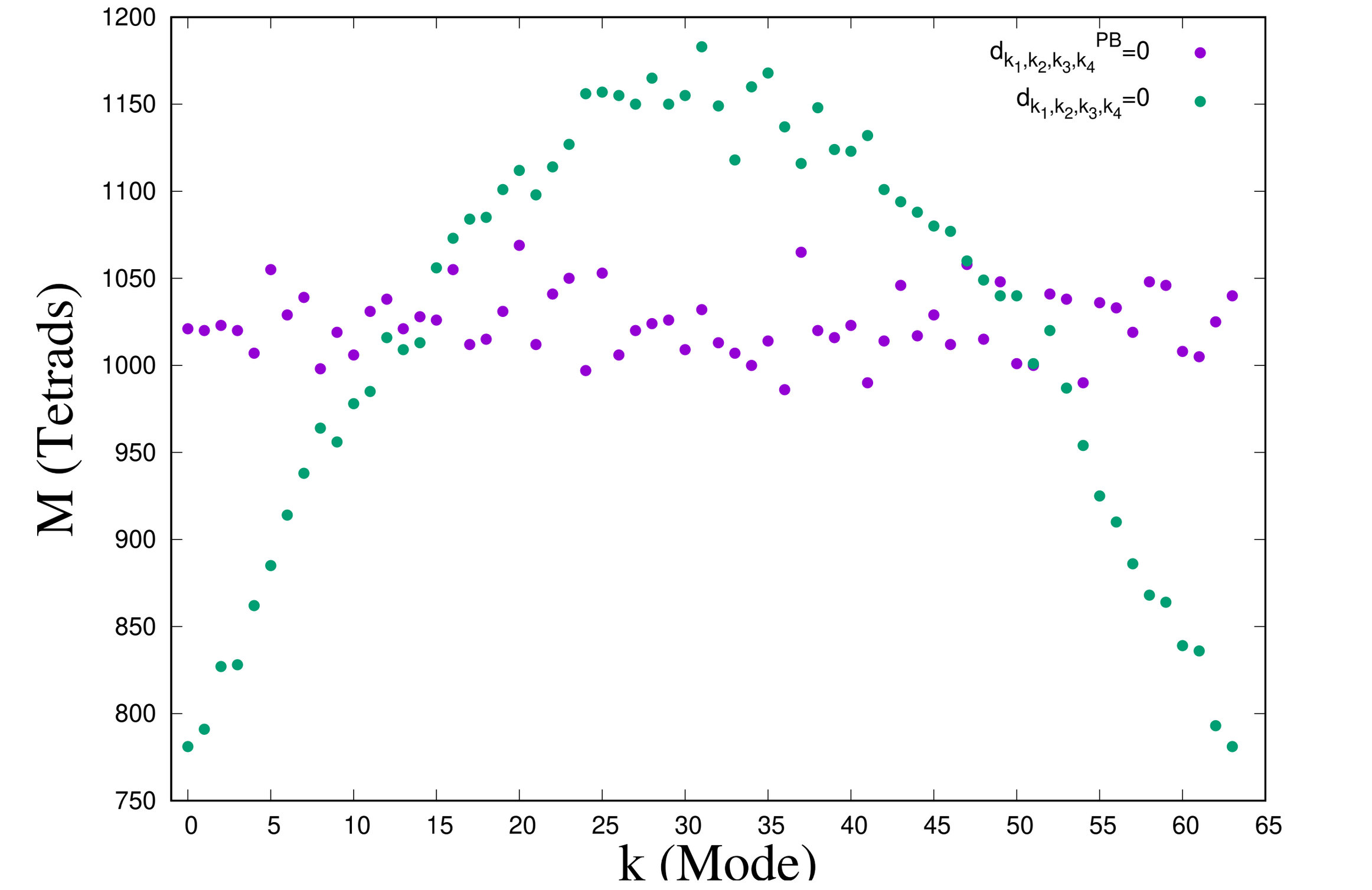}
    \caption{Number of tetrads $M$ to which each mode $k$ of the size $N=64$ participate. Green: standard FMC; purple: FMC with periodic boundary conditions. Notice the absence of a central peak in the purple distribution. }
    \label{fig:pbcvsfbc}
   \end{minipage}
\end{figure}

In the perspective in which the FMC can be thought as a distance on the graph, we can introduce a distance with periodic boundary conditions. This has the effect of eliminating band-edge modes, or, equivalently, it is like considering only modes at the center of the spectrum, as if pertaining to a larger system. The periodic boundary conditions on the frequencies are obtained in practice by representing the frequency indices as variables on a ring, see Fig.~\ref{fig:pbc}, and taking their distance as the smallest one between any two of them
\begin{align} \label{E-FPBC}
    |k_i-k_j| = \left\{ \begin{array}{c c} 
    |k_i-k_j| & \mbox{if }~~~|k_i-k_j|\leq \frac{N}{2}
    \\ 
    &
    \\
    \frac{N}{2}-|k_i-k_j|\  \mbox{mod}\left(\frac{N}{2}\right)
    & \mbox{if }~~~|k_i-k_j| > \frac{N}{2}
    \end{array}\right .
\end{align}
The generalization to the case of quadruplets is straightforward and will be denoted as $d_{\bm{k}}^{\text{PB}}$. In Fig.~\ref{fig:pbcvsfbc} we plot the distribution of tetrads per mode $k$. Data in green pertain to the original mode-locked network built by selecting modes according to $d_{\bm{k}}=0$; data in purple pertain to the mode-locked network with periodic boundary conditions on the frequencies, where modes are selected according to $d_{\bm{k}}^{\text{PBC}}=0$. The distribution is approximately flat in the case of modes selected by the FMC with periodic boundary conditions, whereas it is peaked on the center of the spectrum in the original case.

From now on, we will refer to the version of the model with periodic boundary conditions on the frequencies as PBC, whereas the original one, with free boundary conditions will be termed FBC.

An important remark is that, for a certain number of modes $N$, the total number of quadruplets $N_4^*$ satisfying the FMC with periodic boundary conditions is slightly greater than the corresponding number in the case of free boundary conditions, since all the modes participate in approximately the same number of interactions. However, the order of the dilution with respect to the fully-connected graph is the same in the two cases.

It is appropriate to anticipate that the fact that with periodic boundary conditions the central modes are no more preferred by the interactions clearly affects the intensity spectrum of the model, by eliminating the global narrowing at high pumping, see e.g.~Fig.~\ref{fig:spectra}. Actually, one can think that there is no more distinction among central modes and edge modes, though the interacting quadruplets are still selected through a deterministic condition. We are loosing the ability of qualitatively reproducing the central narrowing of random laser spectra, in favor of a significant reduction of finite-size effects. 

\begin{table}[t]
\centering
\begin{tabular}{lrrrrrrr}
\hline \hline
$N$ & $N_4$ & $T_{\text{min}}$ & $T_{\text{max}}$ & $N_{\text{PT}}$  & $N_{\text{MCS}}$ & $N_{\text{rep}}$ & $N_{\text{s}}$ \\
\hline 
18 & $2^{8}$  & 0.35 & 1.6 & 50 & $2^{19}$ & 4 & 400 \\
32 & $2^{11}$  & 0.45 & 1.6 & 46 & $2^{19}$ & 4 & 350 \\
48 & $2^{13}$  & 0.5 & 1.6 & 44 & $2^{20}$ & 4 & 300 \\
62 & $2^{14}$  & 0.55 & 1.6 & 42 & $2^{20}$ & 4 & 250 \\
62 & $2^{14}$  & 0.3 & 1.6 & 52 & $2^{20}$ & 4 & 100 \\
96 & $2^{16}$  & 0.65 & 1.6 & 38 & $2^{20}$ & 2 & 100 \\
120 & $2^{17}$  & 0.7 & 1.6 & 36 & $2^{20}$ & 2 & 75 \\
\hline \hline
\end{tabular}
\caption{Details for the simulations of the ML 4-phasor with free boundary conditions on the frequencies. Notice that for the size $N=62$ a second series of simulation has been performed with lower $T_{\text{min}}$ and $T_{\text{max}}$, in order to better explore the low temperature phase.}
\label{tab1}
\end{table}
\begin{table}[h]
\centering
\begin{tabular}{lrrrrrrr}
\hline \hline
$N$ & $N_4$ & $T_{\text{min}}$ & $T_{\text{max}}$ & $N_{\text{PT}}$  & $N_{\text{MCS}}$ & $N_{\text{rep}}$ & $N_{\text{s}}$ \\
\hline 
18 & $2^{9}$  & 0.05 & 1.2 & 46 & $2^{19}$ & 2 & 200 \\
28 & $2^{11}$  & 0.1 & 1.2 & 44 & $2^{19}$ & 2 & 200 \\
42 & $2^{13}$  & 0.2 & 1.2 & 40 & $2^{20}$ & 2 & 150 \\
54 & $2^{14}$  & 0.25 & 1.25 & 40 & $2^{20}$ & 4 & 100 \\
66 & $2^{15}$  & 0.25 & 1.25 & 40 & $2^{20}$ & 2 & 100 \\
82 & $2^{16}$  & 0.3 & 1.3 & 40 & $2^{20}$ & 2 & 100 \\
104 & $2^{17}$ & 0.35 & 1.3 & 38 & $2^{21}$ & 2 & 80 \\
\hline \hline
\end{tabular}
\caption{Details for the simulations of the ML 4-phasor model with periodic boundary conditions on the frequencies.}
\label{tab2}
\end{table}

\subsection{Details of the Simulation}
We have performed numerical simulations of the ML 4-phasor model, both with free and with periodic boundary conditions on the frequencies. 

First, the number of interacting quadruplets $N_4$ is chosen as a power of $2$. Then, the corresponding size $N$ is selected in such a way that the difference between $N_4$ and the true number of tetrads $N_4^*$ is minimum, as explained in the previous chapter.  Since $N_4^*$ is greater with PBC rather than with FBC, we managed to perform simulations in the PBC case with networks which are larger from the point of view of the number of couplings, though with smaller sizes. In both cases, however, the simulated network with the highest number of quadruplets has $N_4=2^{17}$.

For each size $N$ of the simulated systems, we have run PT simulations with $N_{\text{PT}}$ replicas of the system at temperatures $T_i \in [T_{\rm min}, T_{\rm max}]$, with $i=1,\dots,N_{\rm PT}$. On top of that, we have performed simulations on $N_{\mbox{rep}}$ identical replicas of the PT simulations in order to compute overlap distributions. The temperature interval has been chosen properly in order to have a sufficient number of replicas above the size-dependent critical temperature $T_c(N)$ and ensure faster thermalization of the replicas in the low temperature glassy phase. We have chosen equispaced temperatures with spacing $\Delta T = 0.025$ and a number of Monte Carlo steps $N_{\text{swap}}=64$ after which a swap of configurations between adjacent heat baths is proposed. Both these choices have been checked to be compatible with high acceptance ratios for the swaps for the whole duration of the simulation. Each copy of the system at each temperature share the same realization of quenched disordered couplings $\{J_{\bm{k}}\}$. The number of simulated disordered samples has been chosen in relation with the size of the system. The values of the simulation parameters are reported in Table \ref{tab1} for the model with FBC and Table \ref{tab2} for the model with PBC.

\section{Numerical Results}
We devote this section to the results of the Monte Carlo analysis of the ML 4-phasor model Eq.~\eqref{HamML4} with FBC and PBC on the frequencies.

\subsection{Spectra}
\begin{figure}[h]
\centering
\includegraphics[width=\textwidth]{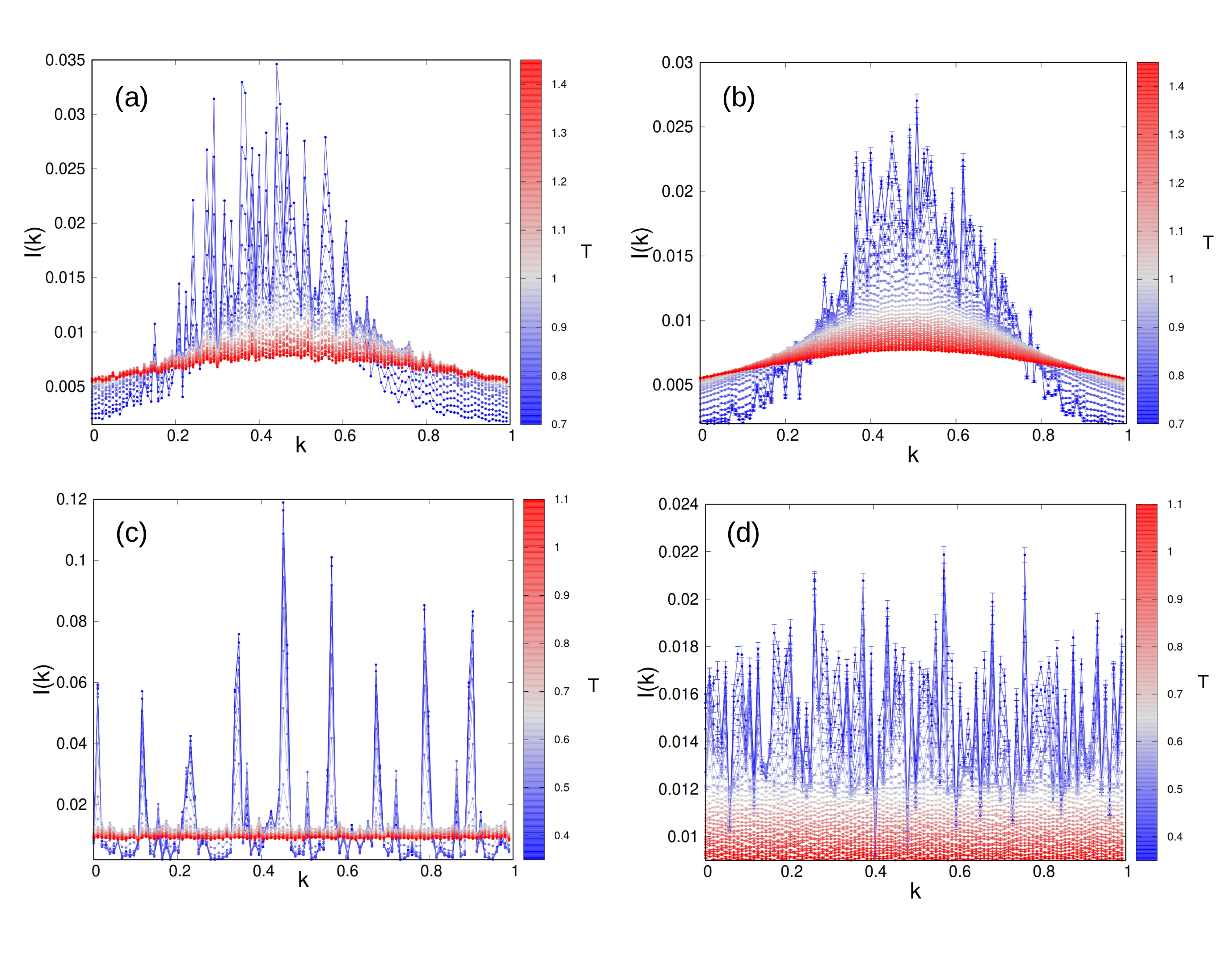}
\caption{(a,c) Intensity spectrum for a single realization of quenched disorder of the ML $4$-phasor model with free (periodic) boundary conditions on the frequencies and $N=120$ ($N=104$) modes. Color map: temperature $T\in [0.7,1.45]$ ($T\in  [0.35,1.1]$). (b,d) Intensity spectrum with free (periodic) boundary conditions on the frequencies and $N=120$ ($N=104$) averaged over $N_{\text{s}}=75$ ($N_{\text{s}}=80$) instances of quenched disorder. All spectra are normalized to their area and modes $k$ are divided by $N$.}
\label{fig:spectra}
\end{figure}
The first observable we display is the intensity spectrum Eq.~\eqref{Spectrum}. Let us, first,
briefly comment on the relationship between the physical intensities $I_k$ and the complex amplitude variables of the simulated model \eqref{HamML4}.  In 
real experiments the heat bath temperature $T$ is typically kept fixed
(there are exceptions like, e.g., in Ref.~\cite{Wiersma01}) and the overall
system energy ${\mathcal E=\epsilon N}$ is varied by tuning the pumping
power. As already discussed (Chap.~\ref{chap:mixedorder}), in our simulations, $\epsilon$ is fixed and kept equal to one in the
spherical constraint, $\sum_k |a_k|^2 = \sum_k A_k^2 = N$, whereas $T$
is varied. Therefore, according to $\mathcal{P}=\epsilon/\sqrt{T}$, a change in
the pumping rate $\mathcal P$ because of a shift in the energy
$\epsilon$ pumped into the system corresponds to a shift of
$1/\sqrt{T}$. If we rescale the intensity of the mode $k$ as in Eq.~\eqref{Spectrum}, i.e.~$I_k = A_k^2/\sqrt{T} $, we have $\sum_k I_k = N/\sqrt{T} = N \epsilon$, as in
Eq.~\eqref{SpherConstr}.

In Fig.~\ref{fig:spectra}, we show the emission spectra at equilibrium for the ML $4$-phasor model with FBC and PBC, both for a single instance of disorder (panels (a,c)) and averaged over
roughly a hundred instances of disorder (panels (b,d)). The most relevant difference between the model with FBC and PBC is the complete absence of global narrowing in the spectrum of the PBC case,
which corresponds to the absence of band-edge modes: all modes
interact with identical probability with the rest of the system. 

On the other hand in the FBC case, one can observe the typical central narrowing occurring in random lasers \cite{Cao99,Cao00,Wiersma08,Ghofraniha15} as the pumping energy increases. This phenomenon becomes particularly evident in the averaged spectrum (panel (b)), which is smoother than the single sample one. Finally, we notice that the averaged spectrum in the PBC case looks like the central part of the FBC spectrum. 

One of the most relevant
features of all the intensity spectra shown in
Fig.~\ref{fig:spectra} is that they become more and more
structured and heterogeneous upon decreasing the temperature. The pattern of the peaks is disordered and strongly depends on the random sample and on the single dynamic history. A first analysis of this phenomenon in terms of intensity equipartition breaking among the different modes has been performed
in~\cite{Gradenigo20}, and a deepening of the collective inhomogeneous behavior of the modes will be presented in Chap.~\ref{chap:Condens}.

\subsection{Specific Heat}
\begin{figure}[t]
   \begin{minipage}{0.48\textwidth}
     \centering
     \includegraphics[width=\linewidth]{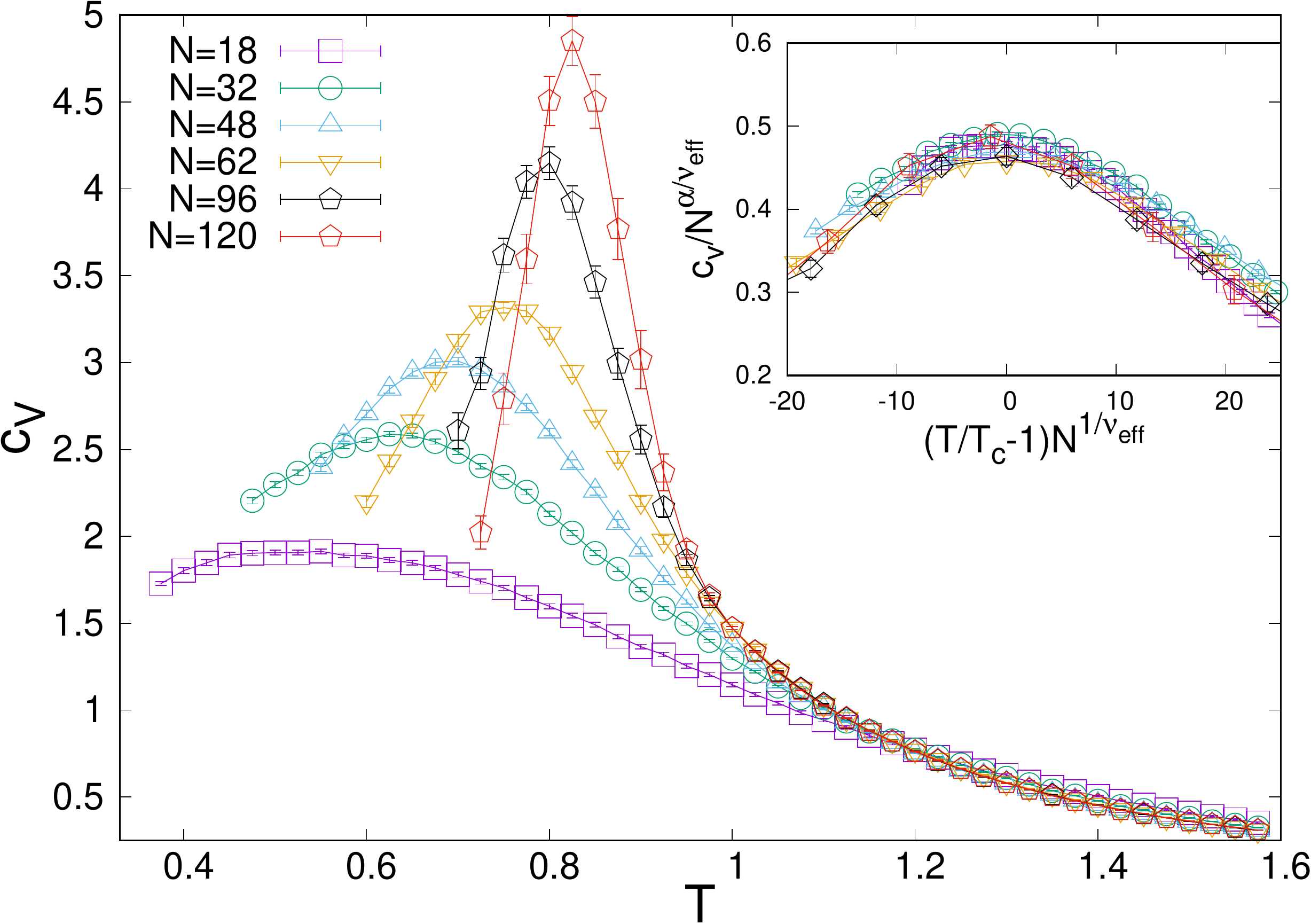}
     \caption{Specific heat $c_{V_N}$ for the ML 4-phasor model with free boundary conditions on the frequencies as a function of $T$. Different curves represent different simulated sizes of the system. The simulated sizes are $N=18,32,48,62,96,120$. (Inset) Specific heat scaled by $N^{\alpha/\nu_{\rm eff}}$ as a function of $\tau N^{1/\nu_{\rm eff}}$, with $\alpha=0.48$, $\nu_{\rm eff}=0.9$.} \label{fig:cV_FBC}
   \end{minipage}\hfill
   \begin{minipage}{0.48\textwidth}
     \centering
     \includegraphics[width=\linewidth]{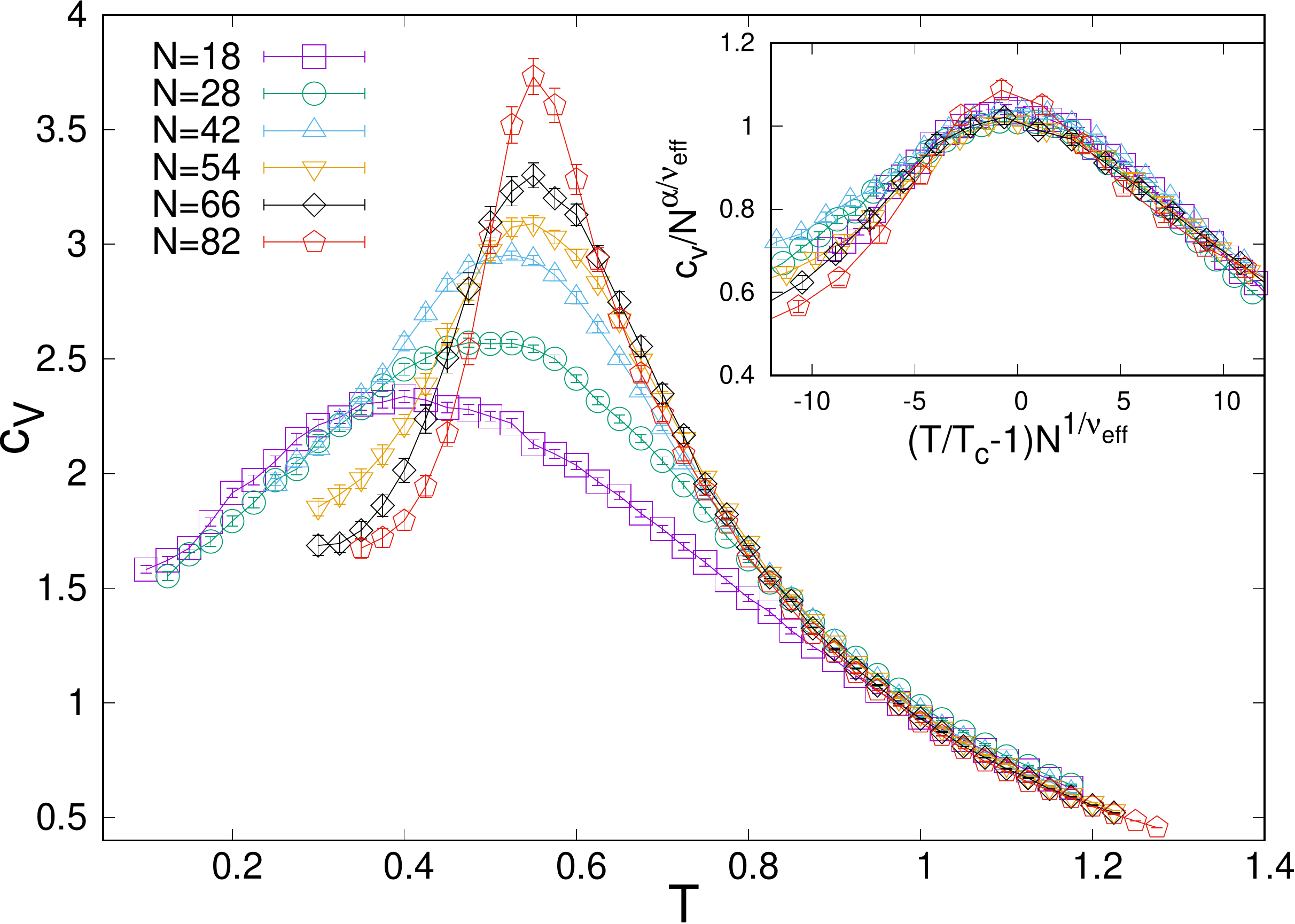}
     \caption{Specific heat $c_{V_N}$ for the ML $4$-phasor model with periodic boundary conditions on the frequencies as a function of $T$. Different curves represent different simulated sizes of the system. The simulated sizes are $N=18,28,42,54,66,104$. (Inset) Specific heat scaled by $N^{\alpha/\nu_{\rm eff}}$ as a function of $\tau N^{1/\nu_{\rm eff}}$, with $\alpha=0.27$, $\nu_{\rm eff}=1.16$.} \label{fig:cV_PBC}
   \end{minipage}
\end{figure}

In Figs.~\ref{fig:cV_FBC} and~\ref{fig:cV_PBC}  we show the specific heat respectively for the ML 4-phasor model with FBC and PBC. In the main panel data are plotted with respect to temperature: each point corresponds to the equilibrium fluctuations of energy within a certain heat bath averaged over the disordered samples. In the inset, data are collapsed according to the scaling hypothesis previously discussed. The finite-size scaling analysis of the specific heat has been performed in a more refined way with respect to the simple case of the REM. In order to get the two exponents $\alpha$ and $\nu_{\rm eff}=2\beta + \gamma$ of Eq.~\eqref{ScalHpCv} from our numerical data we follow the method proposed in Refs.~\cite{Baity13,Leuzzi15}. 

First, the size-dependent critical temperatures $T_c(N)$ are more precisely assessed by fitting the points around the peak of each curve in the main panels of Figs.~\ref{fig:cV_FBC} and~\ref{fig:cV_PBC} with a quadratic function of the temperature $f_N(T)=a_N+b_NT+c_NT^2$. The critical temperatures are identified with the maximum of each of the fitting functions $T_c(N)=-b_N/(2c_N)$, with a statistical error estimated accordingly. The results of this procedure are reported in Table \ref{tab3}. The critical temperature $T_c(\infty)$ of the models can be extrapolated from the fit of the finite-size critical temperatures with the following function: $T_c(N) = T_c(\infty) + a N^{-b}$, where the exponent $b$ gives a first rough estimate of the critical exponent $1/\nu_{\rm eff}$. The results of the fit are:
\begin{align}
\mbox{FBC:}~~& T_c(\infty) = 0.86 \pm 0.03, \quad b = 1.6 \pm 0.5 ,
\\
\mbox{PBC:}~~& T_c(\infty) = 0.61 \pm 0.03, \quad b = 0.98 \pm 0.3.
\end{align}
As one can see, already from this rough estimate the model with PBC has an exponent $1/\nu_{\rm eff}$ falling inside the interval derived for generic mean-field models \eqref{matchingArg} at the beginning of this chapter. In comparison to the estimate provided from FBC data, indeed, there appears to be a drastic reduction of finite-size effects.

\begin{table}[t]
\centering
\begin{tabular}{c| c c c | c c c }
\hline \hline
        & \multicolumn{3}{c|}{FBC}      & \multicolumn{3}{c}{PBC} \\
\hline
$N_{4}$ & $N$ & $T_c$ & $\Delta T_{c}$  & $N$  & $T_c$ & $\Delta T_{c}$ \\   \hline
$2^8$   & 18  & 0.55  & 0.04            &   -  &  -    & -      \\
$2^9$   &  -  &  -    &   -             &  18  & 0.42  & 0.02  \\
$2^{11}$  & 32  & 0.63  & 0.025           &  28  & 0.49  & 0.02  \\
$2^{13}$  & 48  & 0.69  & 0.02           &  42  & 0.52  & 0.02  \\
$2^{14}$   & 62  & 0.75  & 0.03           &  54  & 0.55  & 0.03  \\
$2^{15}$  &  -  &  -    &  -              &  66  & 0.56  & 0.04   \\
$2^{16}$  & 96  & 0.8   & 0.07            &  82  & 0.56  & 0.05   \\
$2^{17}$   & 120 & 0.83  & 0.09            &  -  &  -     &  -     \\
\hline \hline
\end{tabular}
\caption{Values of the critical temperatures for each simulated size of the ML 4-phasor model with fixed and periodic boundary conditions on the frequencies.}
\label{tab3}
\end{table}

Then we take the following Ansatz on the form of the scaling function $\hat{f}$ in Eq.~\eqref{ScalHpCv}
\begin{align}
\hat{f}(x) = A + C x^2,
\end{align}
where $x=N^{1/\nu_{\rm eff}} t_N$, with $t_N$ computed by using the $T_c(N)$ reported in  Table~\ref{tab3}. In the previous Ansatz we have not included the linear term, since the points are translated in order for the peak of each curve to be in the origin and we expect the linear term not to matter. With this Ansatz the scaling hypothesis for the specific heat Eq.~\eqref{ScalHpCv} reads as
\begin{align}
c_{V_N}(T) = \tilde{A}_N + \tilde{C}_N t_N^2,
\end{align}
where $\tilde{X}_N = X_N N^{\frac{\alpha + x}{\nu_{\rm eff}}}$, with $X_N = \{A_N,C_N\}$ and $x=\{0,2\}$. For each size we select a temperature interval centered in $T_c(N)$ corresponding to the points plotted in the insets of Figs.~\ref{fig:cV_FBC} and~\ref{fig:cV_PBC}. We fit the points in the selected interval with the previous function and determine the values of the coefficients. We notice that the behaviour of the logarithm of the coefficients absolute value, i.e.
\begin{align}
    \log|\tilde{X}_N| = \log |X_N| + \frac{\alpha + x}{\nu_{\rm eff}} \log N
\end{align}
is linear in $\ln N$; hence, the estimates of $\alpha$ and $\nu_{\rm eff}$ can be obtained by linear interpolation.

\begin{figure}[t]
\centering
\includegraphics[width=0.8\textwidth]{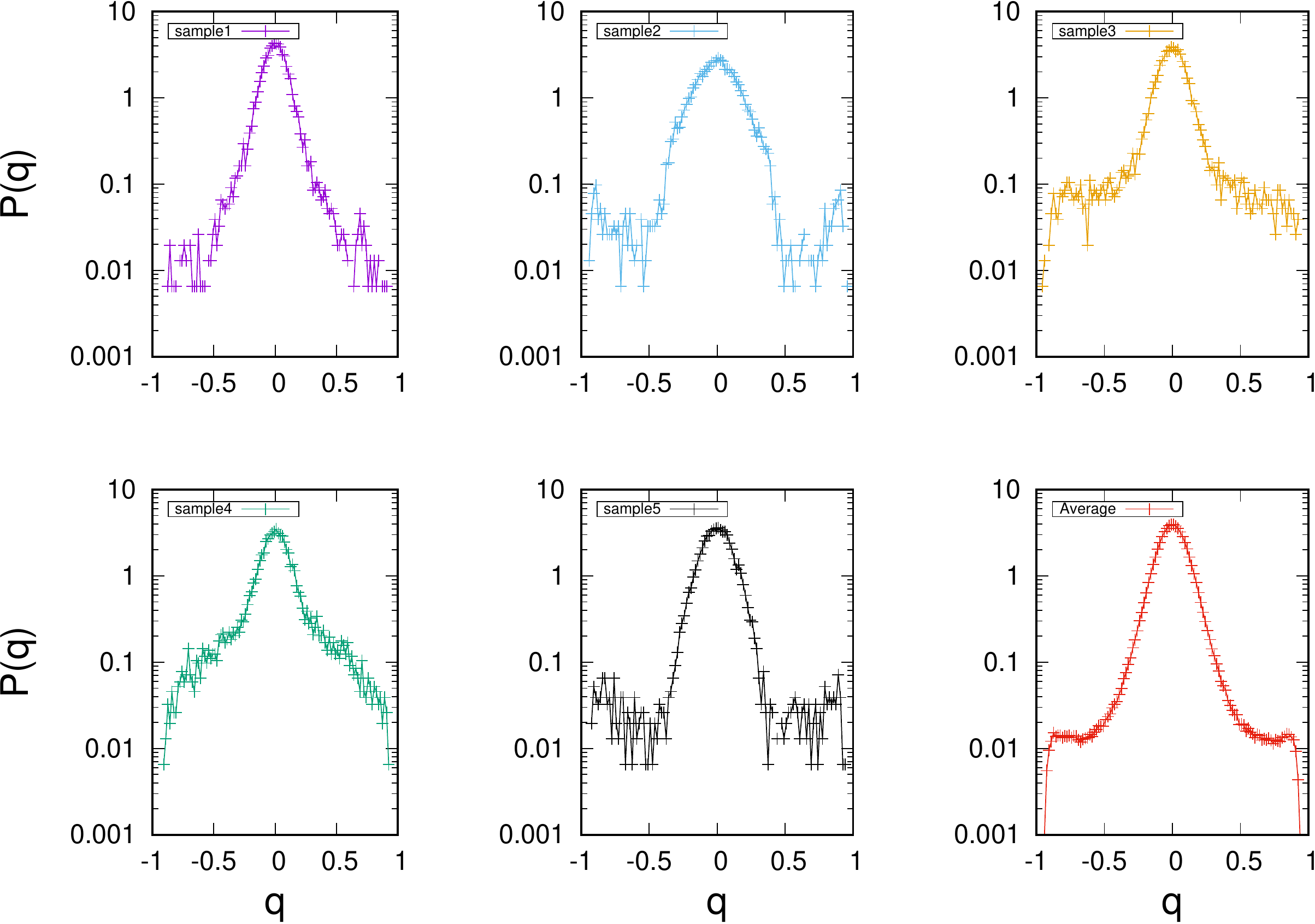}
\caption{Overlap distributions for five instances of disorder and for the average over all instances at $T= 0.25 \simeq 0.45~T_c$. Simulation size $N=54$ of the ML 4-phasor model with periodic boundary conditions on the frequencies.}
\label{fig:pq_singsam}
\end{figure}

For the systems with FBC, this finite-size analysis provides the following values for the critical exponents  
\begin{align}
\mbox{FBC:}~~ \alpha=0.48 \pm 0.05 , \quad  1/\nu_{\rm eff}=1.1 \pm 0.16.
\end{align}
With respect to the estimate $1/\nu_{\rm eff}\simeq 3/2$ found in~\cite{Gradenigo20}, a much larger statistics allows to find an estimate of $1/\nu_{\rm eff}$ closer to, and compatible with, the mean-field threshold, suggesting that deviations from mean-field theory might be due to pre-asymptotic effects in $N$. The confirmation that this is, indeed, the origin of the anomalous value previously found for $1/\nu_{\rm eff}$ comes from the analysis with PBC. In this case, the critical exponents turn out to be
\begin{align}
\mbox{PBC:}~~ \alpha=0.27 \pm 0.05 , \quad  1/\nu_{\rm eff}=0.86 \pm 0.14.
\end{align}
With PBC we find an estimate of  $1/\nu_{\rm eff}$  well below the threshold for a mean-field universality class. Therefore, up to the precision of our analysis, despite being possibly still of a different universality class with respect to the REM, for which $1/\nu_{\rm eff} =1/2$, we can assess the mean-field nature of the glass transition in the ML 4-phasor model.

\subsection{Overlap Distribution Functions}

In this section we complete the study of the glass transition in the ML 4-phasor model, by presenting results of our numerical simulations about the first-order nature of the transition. 

\subsubsection{Parisi overlap}
\begin{figure}[t]
   \begin{minipage}{0.48\textwidth}
     \centering
     \includegraphics[width=\linewidth]{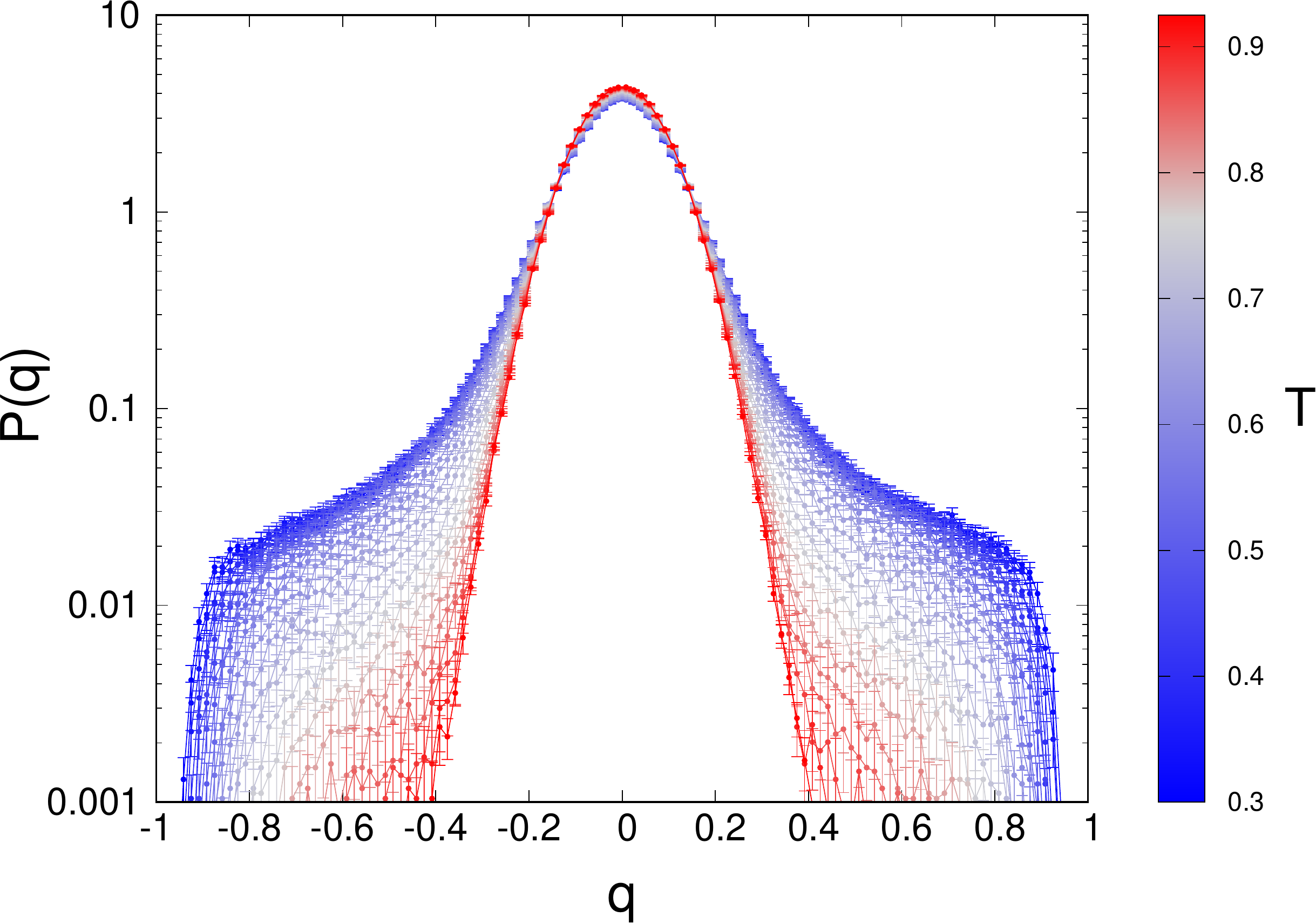}
     \caption{Parisi overlap distribution for the size $N=62$ of the ML 4-phasor model with free boundary conditions on the frequencies. The distribution is averaged over $N_{\text{s}} = 100$ instances of disorder. Color map: same as Fig.~\ref{fig:spectra}. The blue curve corresponding to the lowest temperature is at $T\simeq 0.4 T_c$, with $T_c=0.86(3)$.} \label{fig:pq_FBC}
   \end{minipage}\hfill
   \begin{minipage}{0.48\textwidth}
     \centering
     \includegraphics[width=\linewidth]{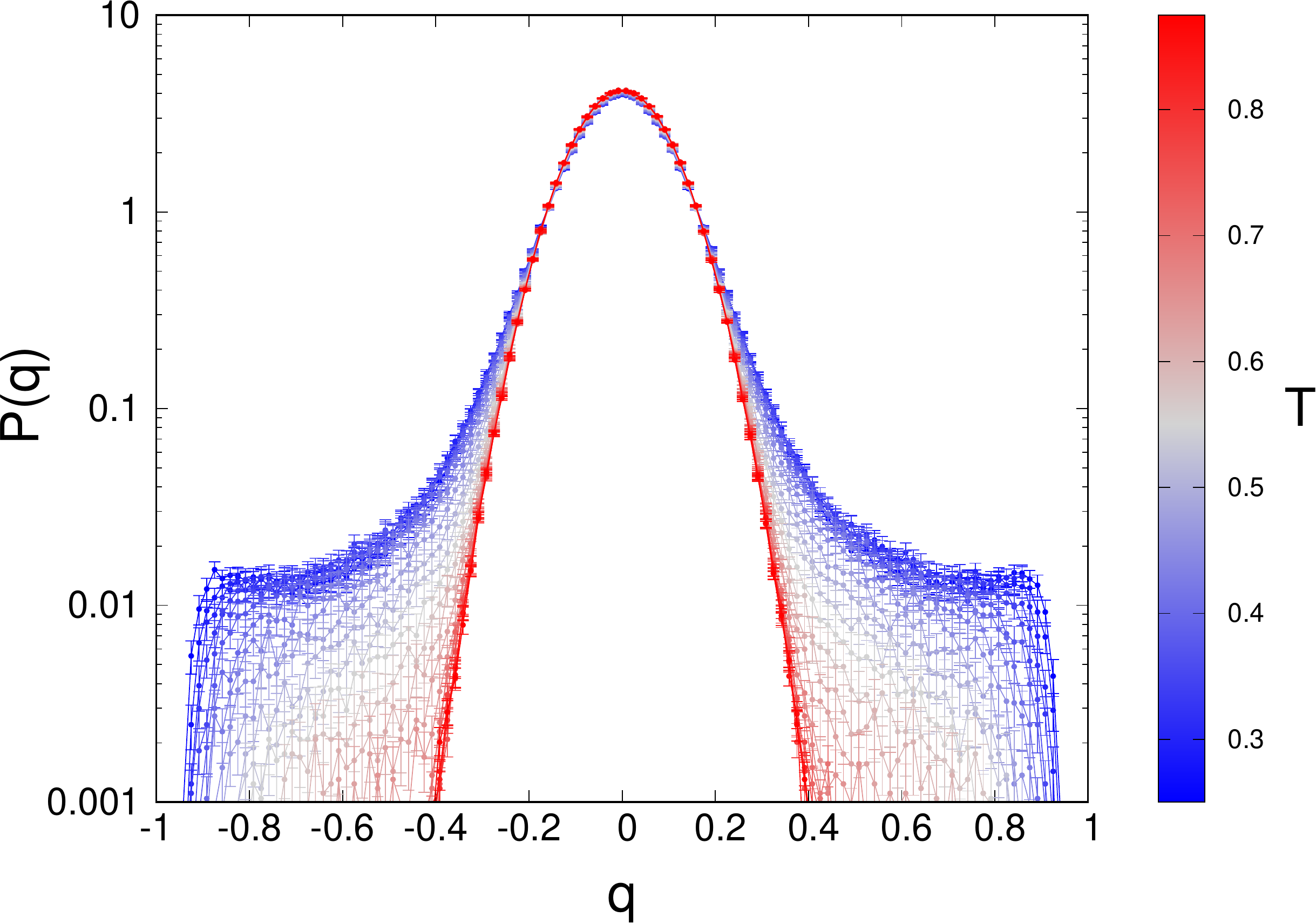}
     \caption{Parisi overlap distribution for the size $N=54$ of the ML 4-phasor model with with periodic boundary conditions on the frequencies. The distribution is averaged over $N_{\text{s}} = 100$ instances of disorder. Color map: same as Fig.~\ref{fig:spectra}. The blue curve corresponding to the lowest temperature is at $T\simeq 0.45 T_c$, with $T_c=0.61(3)$.} \label{fig:pq_PBC}
   \end{minipage}
\end{figure}
Let us first discuss the case of the Parisi overlap distribution \eqref{OverlapOB}. The protocol used in numerical simulations to measure overlaps corresponds to the definition of replicas as independent copies of the system with the same quenched disorder. For each sample, i.e.~each realization of disorder, we run dynamics independently for $N_\text{rep}$ replicas of the system, starting from randomly chosen initial phasor configurations. In this way, replicas explore different regions of the same phase space, passing through configurations typically belonging to separated equilibrium states, if there are many of them, and, sometimes, to the same state.

To study the  behavior of the $P_J(q)$ we choose $N_\text{rep}=4$, so that at any measurement time six values of the overlap are available $q_{\alpha\beta}=\{q_{01}, q_{02}, q_{03}, q_{12}, q_{13}, q_{23}\}$. Hence, passing from two replicas to four increases the statistics by a factor six. In order to accumulate statistics, we measure the value of $q_{\alpha\beta}$ using $\mathcal{N}$ equilibrium, time uncorrelated, configurations of replicas at the same iteration of the simulated dynamics, see Eq.~\eqref{ConfigTot}. Hence, for each disordered sample the $P_J(q)$ histograms are built with $\mathcal{N} \times  N_\text{rep}(N_\text{rep}-1)/2 $ values of the overlap. 

The overlap distribution functions $P_J(q)$ are computed as the normalized histograms of the overlaps for each one of the samples. This has been done for each simulated size of the ML 4-phasor model with both FBC and PBC. In Fig.~\ref{fig:pq_singsam} we present the overlap distributions for five samples at the temperature $T=0.25\simeq0.45T_c$ of the size $N=54$ of the ML 4-phasor model with PBC, together with the overlap distribution averaged over  $100$ samples. 
Given the fluctuations of $P_J(q)$ among the different samples, it is clear that the only physical quantity to be considered in order to assess the glass transition is the averaged $P(q)\equiv {\overline{P_J(q)}}$.

This is particularly important in the case of the overlap distribution function, since, it is not a self-averaging quantity~\cite{Mezard87}, i.e., the average $P(q)$ cannot be reached simply by increasing the size of the system over which a single sample $P_J(q)$ is built, as, for instance in the case of the free energy or of the specific heat, but only by averaging over disorder.

In Fig.~\ref{fig:pq_FBC} and Fig.~\ref{fig:pq_PBC} the average overlap distribution function of the ML 4-phasor model with FBC and PBC are, respectively, reported for the whole simulated temperature range in systems whose size correspond to $N_4=2^{14}$, i.e.~respectively $N=62$ and $N=54$ spins.
The reduction of the finite-size effects obtained by using periodic boundary conditions in the choice of interacting modes leads to display $P(q)$ with more distinct secondary peaks in the case of the ML 4-phasor model with PBC. 

\begin{figure}[t]
     \centering
     \includegraphics[width=0.65\linewidth]{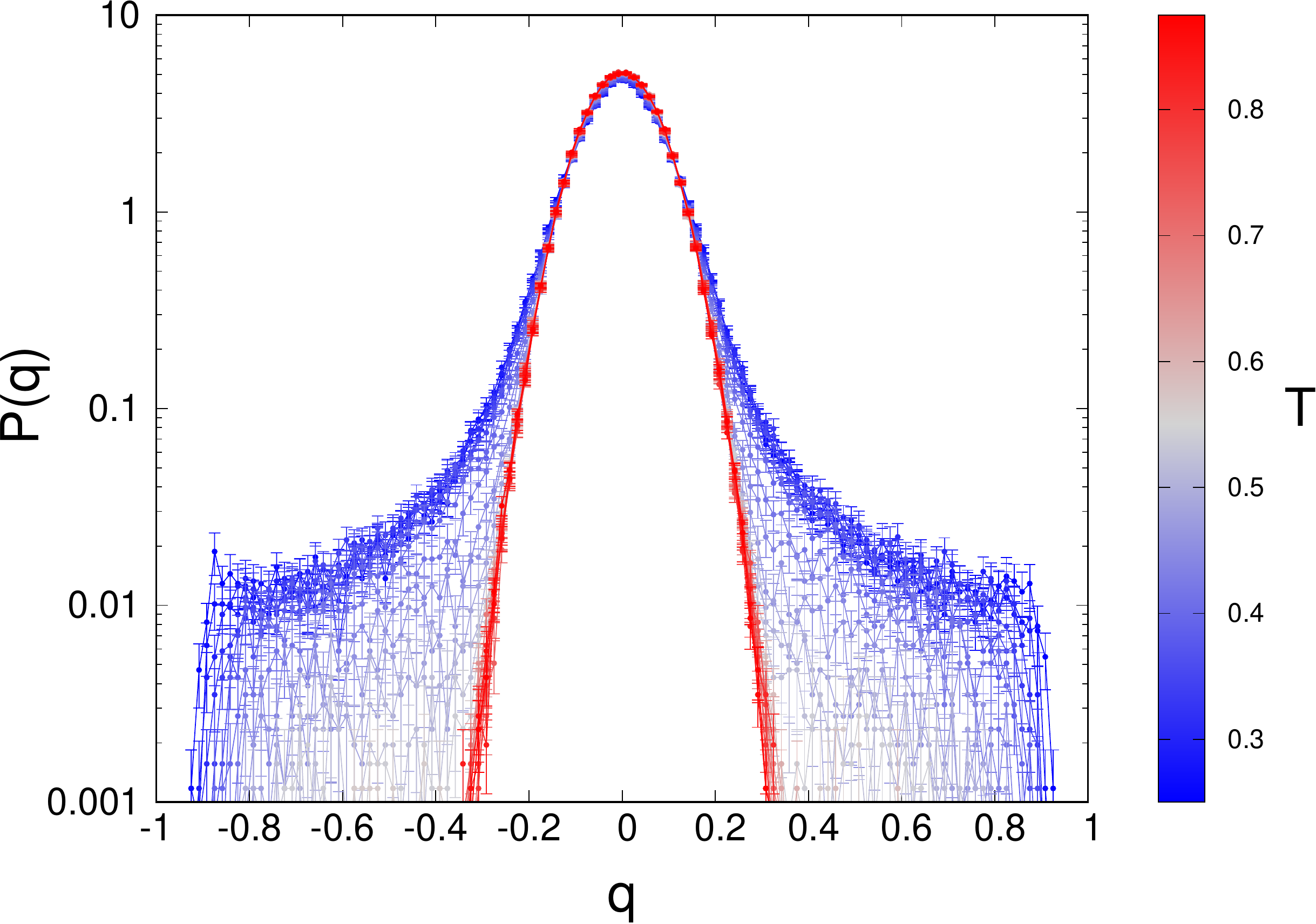}
     \caption{Parisi overlap distribution for the size $N=54$ of the ML 4-phasor model with with periodic boundary conditions on the frequencies. The distribution is averaged over $N_{\text{s}} = 100$ instances of disorder. Color map: same as Fig.~\ref{fig:spectra}.} \label{fig:pq82_PBC}
\end{figure}

We have showed the overlap distribution functions for the $N=62$ (for the model with FBC) and $N=54$ (for the model with PBC), because these are the largest sizes for which we performed simulations with $N_{\text{rep}}=4$. However, it can be useful to show also the behavior of the $P(q)$ for a higher size. In Fig.~\ref{fig:pq82_PBC} we show the overlap distribution function for the size $N=82$ of the model with periodic boundary conditions. One can see that the side peaks of the distribution at low temperatures are not appreciably more pronounced than than those of the distribution for $N=54$. However, a reduction of the finite size effects with respect to $N=54$ can be appreciated at high temperature, where the distribution is still a Gaussian, but it slightly narrows around the peak in $q=0$. We expect that in the large $N$ limit the overlap distribution function at high temperature becomes a delta function peaked in $q=0$.

\subsubsection{Plaquette overlap and IFO}
The study of the plaquette overlap and IFO distribution functions has been performed both for the model with FBC and PBC. In the following we only display results for the PBC case, which is less affected by finite-size effects. 

\begin{figure}[t]
     \centering
     \includegraphics[width=0.65\linewidth]{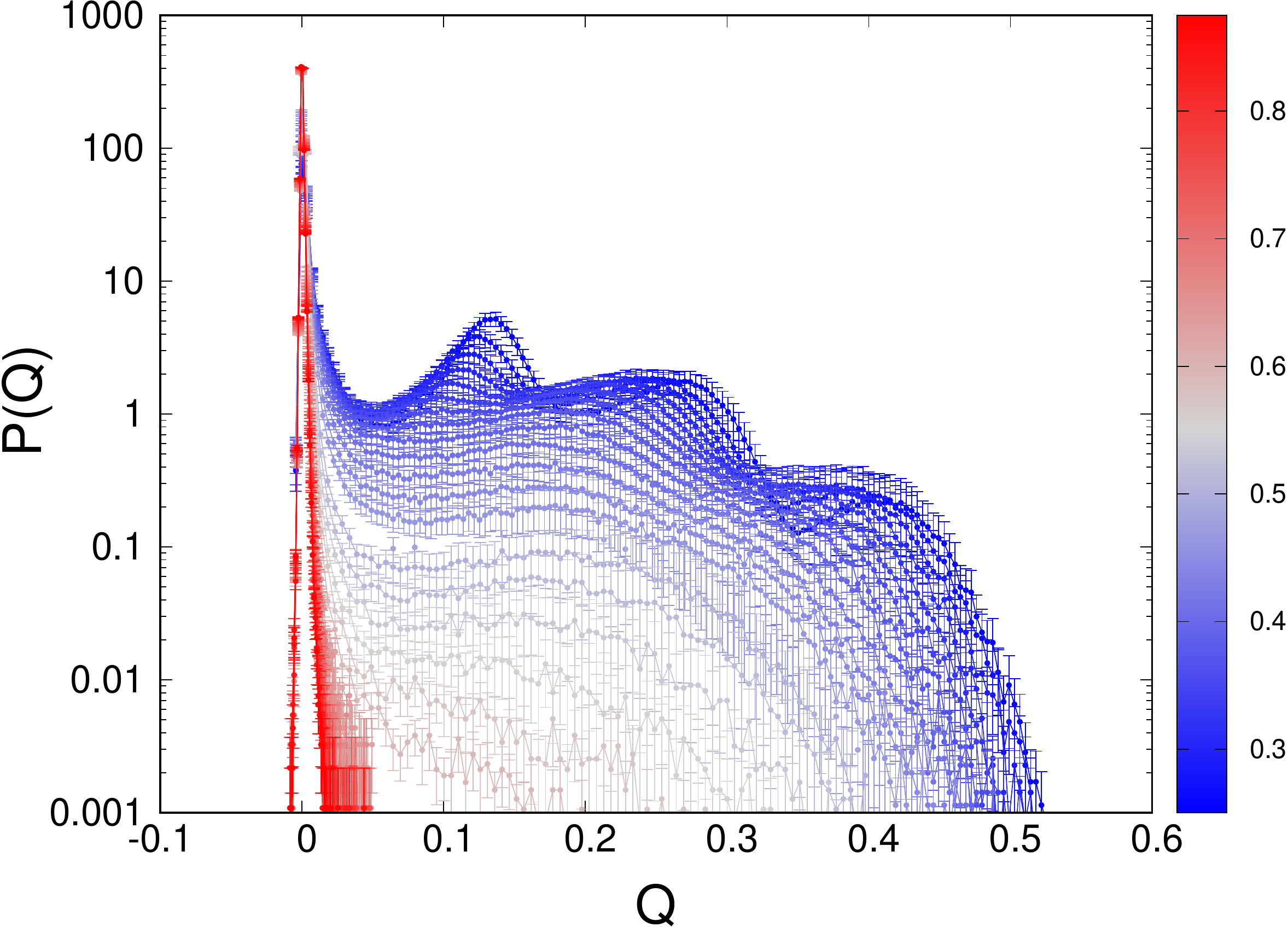}
     \caption{Plaquette overlap distribution for the size $N=54$ of the ML 4-phasor model with periodic boundary conditions on the frequencies. The distribution is averaged over $N_{\text{s}} = 100$ instances of disorder. Color map: same as Fig.~\ref{fig:spectra}. The blue curve corresponding to the lowest temperature is at $T\simeq 0.45 T_c$, with $T_c=0.61(3)$.} \label{fig:pPLAQ_PBC}
\end{figure}

In Fig.~\ref{fig:pPLAQ_PBC} the plaquette overlap distribution is plotted for all the simulated temperatures and for the system size $N=54$. For each of the $\mathcal{N}$ uncorrelated configurations at equilibrium, the plaquette overlaps are computed over $N_4 \sim O(N^3)$ quadruplets, leading to a reduction of the statistical error on the overlap values. Then, for each sample, the distribution is computed with $\mathcal{N} \times  N_\text{rep}(N_\text{rep}-1)/2 $ values of the plaquette overlap. Data in figure are averaged over $N_{\rm s} = 100$ disordered samples. 

\begin{figure}[t]
    \centering
    \includegraphics[width=0.8\linewidth]{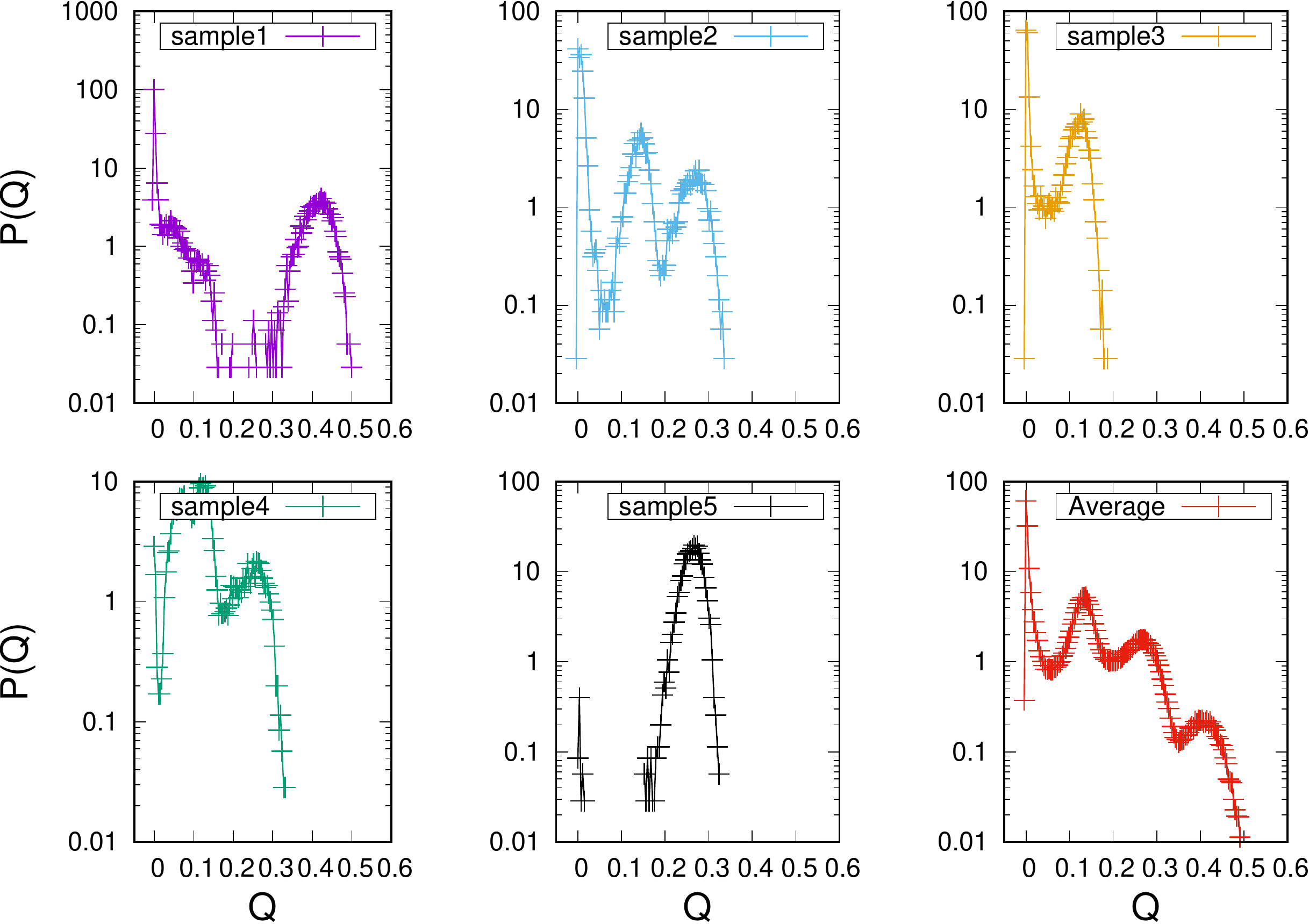}
    \caption{Plaquette overlap distributions for five instances of disorder and for the average over all instances at $T= 0.25 \simeq 0.45~T_c$. Simulation size $N=54$ of the ML 4-phasor model with periodic boundary conditions on the frequencies. As for the Parisi overlap distribution, each sample exhibits a particular distribution shape, with one or two peaks appearing at different positions: the relevant quantity in the thermodynamic limit is the averaged distribution in the sixth panel.}
    \label{fig:pPLAQ_singsam}
\end{figure}

In the low temperature region, the distribution clearly develops a nontrivial shape, with a heavy tail for values of $\mQ > 0$. The presence of the three visible peaks in the distribution tail at the lowest temperatures is a consequence of the single sample behavior of the $P(\mQ)$, which we present in Fig.~\ref{fig:pPLAQ_singsam} for the simulated temperature $T=0.25$. As one can see, the shape of the distribution dramatically changes from sample to sample: in particular, for some instances of disorder a single peak, for others two peaks are displayed in the $P(\mQ)$, at sample dependent positions. The effect of taking the average over all the simulated samples is reported in the sixth panel in Fig.~\ref{fig:pPLAQ_singsam}, which corresponds to the lowest $T$ curve in Fig.~\ref{fig:pPLAQ_PBC}. We notice that, even with $N_s = 100$ disordered samples, the average is far from a smooth function, displaying the strong lack of self-averaging of the plaquette distribution.

\begin{figure}[ht]
    \centering
    \includegraphics[width=0.8\linewidth]{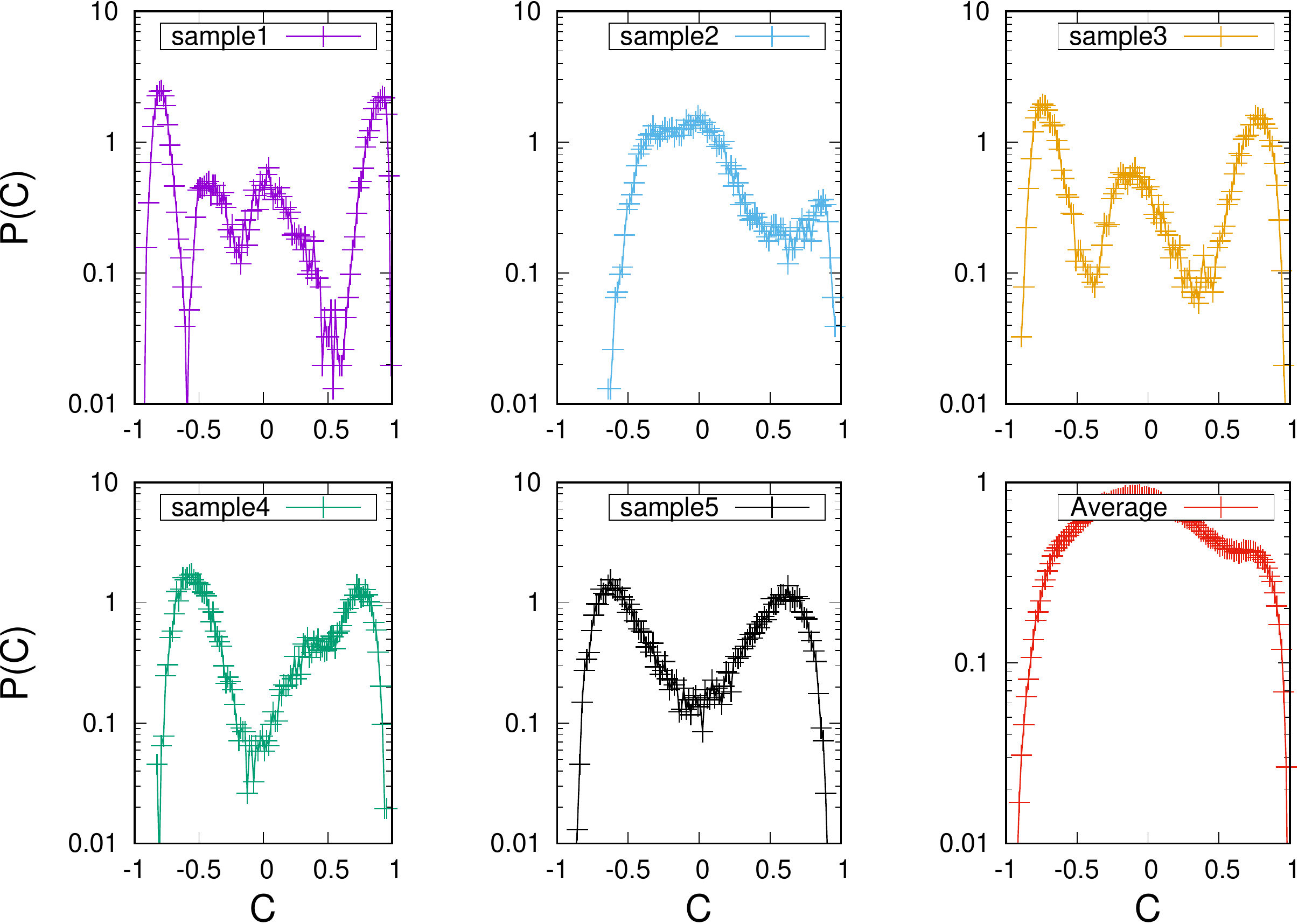}
    \caption{IFO distributions for five instances of disorder and for the average over all instances at $T= 0.25 \simeq 0.45~T_c$. Simulation size $N=54$ of the ML 4-phasor model with periodic boundary conditions on the frequencies.}
    \label{fig:pIFO_singsam}
\end{figure}

In Fig.~\ref{fig:pIFO_singsam} the averaged IFO distribution is displayed for the system size $N=54$, at all simulated temperatures. Even with the significant reduction of finite-size effects obtained through the PBC, the distribution obtained with our data at equilibrium does not show clear side peaks in the low temperature region. Actually, side peaks are more evident very close to the transition temperature (gray curves in figure), rather than for the lowest simulated temperatures. Once again, we get a clear understanding of the averaged distribution by looking at single instances of disorder. In Fig.~\ref{fig:pIFO_singsam}, one can see that, while many samples have a distribution which is still peaked in $\mathcal{C}=0$, a signature that they may not have entered in the glassy phase yet, in many others the $P(\mathcal{C})$ has developed nice side-peaks at the temperature $T=0.25$. However, when taking the average, the combined effect of samples of the first kind and of the variable position of the peaks in samples of the second kind, leads to the smoothed and almost flat shape of the IFO distribution at low temperature.

\begin{figure}[t]
   \centering
     \includegraphics[width=0.65\linewidth]{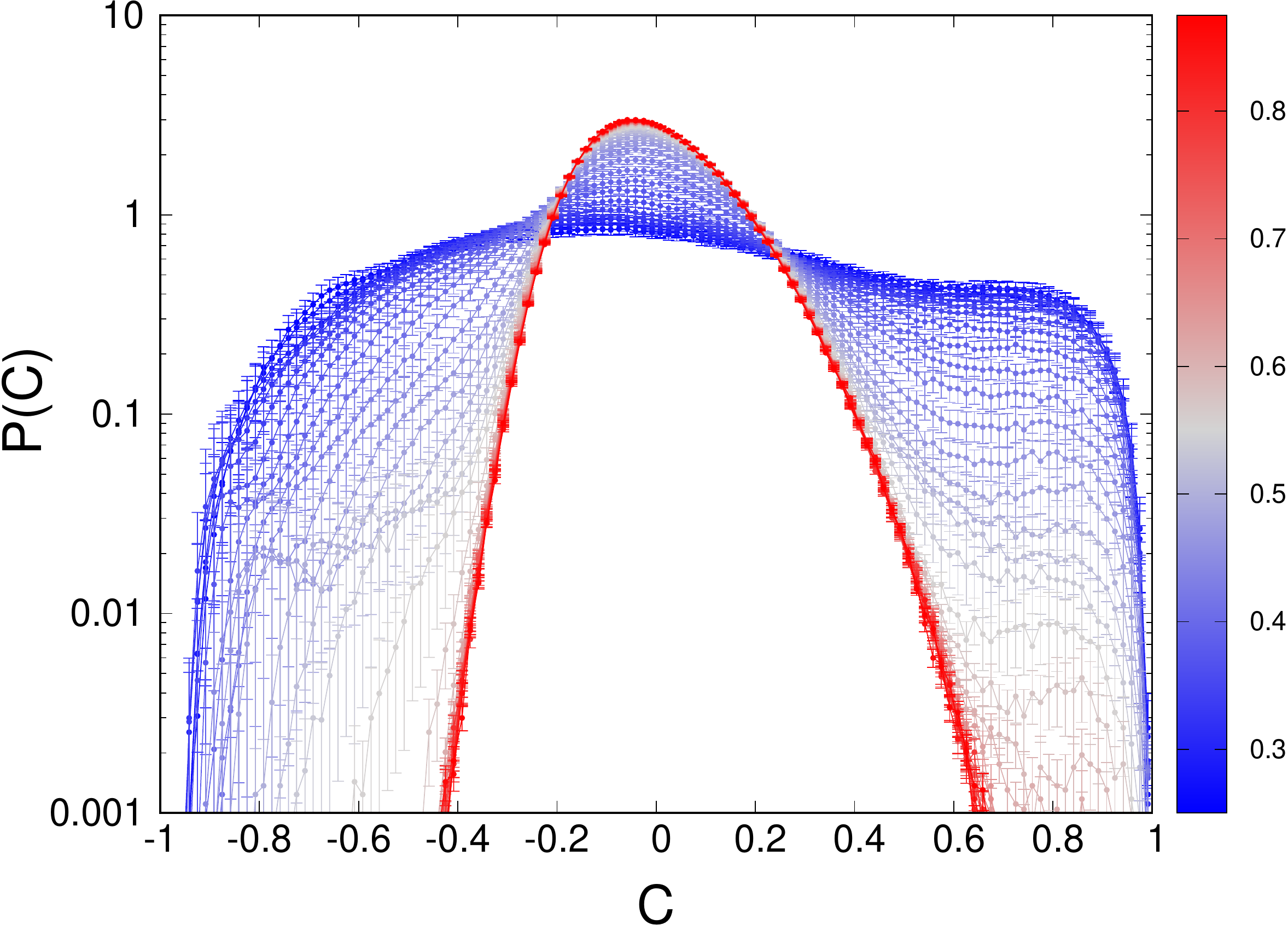}
     \caption{IFO distribution for the size $N=54$ of the ML 4-phasor model with with periodic boundary conditions on the frequencies. The distribution is averaged over $N_{\text{s}} = 100$ instances of disorder. Color map: same as Fig.~\ref{fig:spectra}. The blue curve corresponding to the lowest temperature is at $T\simeq 0.45 T_c$, with $T_c=0.61(3)$.} \label{fig:pIFO_PBC}

\end{figure}

\newpage

\chapter{Breaking of Equipartition and Pseudo-Localization at the Transition} \label{chap:Condens}
The present chapter is devoted to the study of a particular phenomenon taking place in the ML 4-phasor model, which has been mentioned previously in this work, when describing the intensity emission spectra of random lasers: intensity localization of light, else termed \emph{power condensation} \cite{Antenucci15c,Antenucci15d}, and its relationship to the high pumping replica symmetry breaking (RSB) random lasing phase. 

Localization is a widespread phenomenon in physics, which has always been related to the breaking of ergodicity. In the pioneering work of Anderson on semiconductors, the spatial localization of the electron wavefunction induced by a large degree of disorder was identified as the underlying mechanism for a semiconductor/insulator transition \cite{Anderson58}. The absence of diffusion in the insulating-localized phase was interpreted as a clear manifestation of dynamical ergodicity breaking. Besides the seminal work of Anderson, localization was related to ergodicity breaking also in classical systems, since the famous numerical study of Fermi-Pasta-Ulam-Tsingou (FPUT) on the anharmonic chain \cite{Fermi54}. In this case, localization was observed in the Fourier space of the chain modes. Starting from an atypical initial condition, e.g.~only the lowest harmonic excited, one would expect that a slight anharmonicity is sufficient to cause the system to relax on a state where the energy is equally divided among all the modes\footnote{If one initializes a harmonic chain on a eigenmode of the Hamiltonian, the time evolution will leave the system on that eigenmode, resulting in a breaking of ergoditicy. Nonlinearity is believed to facilitate the recovery of ergodicity, because it introduces a coupling which makes the eigenmodes of the unperturbed Hamiltonian more connected.}. Contrary to expectations, the system showed a recurrent dynamics for all the duration of the experiment, with no sign of relaxation to equipartition. Therefore, localization was coming along with dynamical ergodicity breaking also in
the case of energy localization in the Fourier power spectrum.

Ergodicity breaking was understood as a purely dynamical phenomenon until the '80s, when the theory of replica symmetry breaking was established as a new thermodynamic paradigm and statistical ensembles formalism for ergodicity breaking transitions in complex disordered systems.~Quite interestingly, while localization phenomena in quantum many-body systems have been widely investigated during the last decades~\cite{Altshuler97,Basko06,Nandkishore15,Ros15,Vidmar16,Abanin17,Alet18}, they have seldom been probed in disordered systems, apart from the attempts of Refs.~\cite{Mossi18,Gradenigo20}. The lack of a broad analysis of localization phenomena in the context of disordered glassy systems is mainly due to the nature of the variables which are customary for these systems: Ising, XY or Heisenberg spins, all locally bounded, $|\vec s|=1$. In those models, such as the spherical $p$-spin model, where variables are used with continuous locally unbounded magnitude as a proxy for magnetic spins in spin-glasses or density fluctuations in structural glasses, in order to be able to perform analytical computations, the interaction network is usually fully connected, thus hindering any sort of magnitude localization.~Indeed, this kind of mean-field representation on a complete graph, together with a local potential (soft spins) or a global constraint (spherical spins) guarantees magnitude equipartition.

Therefore a careful investigation of how magnitude localization coexists with replica symmetry breaking in disordered systems is a gap that needs to be filled. Once again, the spin-glass theory of random lasers represents a very fortunate research field for this kind of analysis. The fundamental variables, i.e.~the amplitudes of the light modes, are naturally continuous and locally unbounded; the global constraint which they are subjected to is a quite natural requirement for the stationary regime of a lasing system (see Chap.~\ref{chap:IntroCap}); the presence of dilution is a direct consequence of the specific selection rule in light mode coupling, i.e.~the FMC. Hence, these systems have all the ingredients necessary to exhibit a power condensation transition. 

As a disclaimer, it is worth stressing that this is not the generalization to light waves of the spatial wavefunction localization occurring in Anderson theory, that is known to be inhibited in 3D random lasers because of the vectorial nature of light waves \cite{Skipetrov14,Sperling16}. Rather, it is a condensation of the overall magnitude on a few variables in a set of locally unbounded variables subjected to a global constraint, i.e.~a global conservation law. This kind of condensation has already been observed and studied analytically in non-interacting systems \cite{Majumdar05,Evans06,Trombettoni01}. When the value of this globally conserved quantity exceeds a given (non-universal) threshold the system undergoes a transition where a macroscopic fraction of the conserved quantity concentrates on a finite portion of the system. Given the clarification, from now on we will use the terms  `localization' and  `condensation' interchangeably to refer to the phenomenon of our interest.

This chapter is organized as follows: first, a few details are provided regarding the condensation transition in non-interacting systems, pointing out the emergence of a pseudo-localized phase, which has been discovered in Refs.~\cite{Gradenigo21a,Gradenigo21b}; then we clarify to what extent, in the presence of an interaction network, the order of dilution is key to understand possible regimes of equipartition or localization of a globally conserved quantity; finally, results from the numerical simulations of the ML 4-phasor model are presented, revealing the presence of a \emph{hybrid} phase analogous to the pseudo-localized phase of non-interacting systems, where the intensity of light modes is neither equipartitioned among all modes nor really localized on a few of them.

\section{Pseudo-Localization in Non-Interacting Systems}
Condensation of a global quantity on a finite number of degrees of freedom has been found and very precisely described in the framework of large deviation calculations and ensemble inequivalence in the case of mass-transport models~\cite{Majumdar05,Evans06} or for bosonic condensates in optical lattices described by the Discrete Non-Linear Schr\"odinger Equation (DNLSE)~\cite{Trombettoni01}. The fundamental ingredient involved in the localization phenomenology for non-interacting systems is the
presence of a global constraint. 

Let us consider, for instance, the DNLSE. We focus on the infinite temperature limit in which the hopping term, i.e.~the kinetic two-body term in the Hamilotnian of the DNLSE, can be neglected. In this case, the partition function of the model reads
\begin{align}
  \Omega(\mu,E) = \int \prod_{i=1}^N \de\psi_i\de\bar \psi_i \ e^{-\mu \sum_{i=1}^N |\psi_i|^2}~\delta\left(E - \sum_{i=1}^N |\psi_i|^4\right),
  \label{eq:semican-bec}
\end{align}  
where $\mu$ is the chemical potential. Clearly the quantity
\begin{align}
  A = \sum_{i=1}^N |\psi_i|^2,
  \label{eq:spheric-DNLS}
\end{align}
represents the mass of the condensate, so that the partition function in Eq.~\eqref{eq:semican-bec} corresponds to an ensemble where exact conservation of energy is enforced by means of a Dirac delta function,
whereas mass is conserved only on average by means of a field $\mu$. In the following, we will refer to the former way of imposing a global conservation law as ``hard'' constraint and to the latter as ``soft'' constraint.

In the case of the DNLSE, as in all
cases where it takes place, the physical quantity that localizes is the
one controlled by the hard constraint, hence in this case it is the
energy. It is only thanks to the \textit{global} action of the
constraint on the total energy that configurations with a strongly
heterogeneous distribution of energy on lattice sites are allowed. The analytical calculations
of Refs.~\cite{Gradenigo21a,Gradenigo21b} show that as soon as energy is constrained
above a certain critical value, $E>E_c$, these localized
configurations dominate the partition function.~It can be shown
analytically, but it can be easily guessed by looking at
Eq.~\eqref{eq:semican-bec}, that localization cannot take place for a
quantity controlled ``on average''. It is not possible to have
strongly inhomogeneous fluctuations and/or localization of something
which is controlled \textit{homogeously} by means of a Lagrange
multiplier like the chemical potential in Eq.~\eqref{eq:semican-bec}. This is precisely the same mechanism  characterizing the Bose-Einsten condensation,
which is a form of localization in Fourier space: the
condensed phase cannot be reached by controlling density with a chemical potential; density 
must be tuned directly, for instance, by decreasing the volume for a given
number of bosons~\cite{Huang87}. The fact that the Bose-Einstein condensation cannot be implemented by tuning the chemical potential of a
reservoir in contact with the system is  analogous to the
fact that energy localization cannot be
achieved  by tuning the temperature of a
thermostat, i.e.~by studying the partition function where 
conservation of energy is imposed ``on average'', as $\exp(-\beta
\mathcal{H})$,  rather than exactly, as $\delta(E-\mathcal{H})$. In both the above
examples localization entails lack of statistical ensemble
equivalence: for the energy localization in the DNLSE it is the lack of
equivalence between fixed temperature (canonical) and fixed energy (microcanonical) ensembles, for
the Bose-Einstein condensation it is the lack of equivalence between fixed
chemical potential (grancanonical) and fixed density (canonical) ensembles.

One important feature that has been predicted for the DNLSE~\cite{Gradenigo21a,Gradenigo21b} is the existence of a pseudo-localized phase similar to what we are going to show in numerical simulations of the ML 4-phasor model. The peculiarity of the localization phenomenon in the case of the DNLSE is that it takes place in two steps. First, by increasing the energy, i.e.~the  quantity which controls localization, one first encounters a second-order transition at a value of the energy --  $E_{\rm th}$ --  where the equivalence of ensembles breaks down and temperature becomes negative. Then, at a larger value of energy, $E_c > E_{th}$, there is a first-order transition to a localized phase. For $E\in[E_{\rm th},E_c]$ the system finds itself in a pseudo-localized phase, where a thermodynamic anomaly is indicated by the lack of ensemble equivalence and the presence of negative temperature, but localization has not been really achieved yet.

The main difference between the ML $4$-phasor model of the glassy random laser and DNLSE, is that in the random laser case the joint distribution of the variables over which the global constraint is imposed is not factorized. Therefore, the analytical results discussed in~\cite{Gradenigo21a,Gradenigo21b} cannot be straightforwardly extended to this case. In the following, first we try to derive some understanding of localization in generic interacting systems; then, we resort to numerical simulations in order to probe the localization transition in the ML 4-phasor model.


\section{Scaling Argument for the Occurrence of Intensity Localization in Interacting Systems} \label{SS-SCALING-LOC}
In this section we go through a scaling argument on generic $p$-spin interacting systems (both with ordered and disordered interactions) with continuous variables and a global constraint, of which the ML 4-phasor model with the spherical constraint \eqref{SpherConstr} is a special case. 

Let us consider the $p$-spin Hamiltonian
\begin{equation}
    \mathcal H[\sigma] = -\sum_{k_1\ldots k_p}^{\# N^A} J_{k_1\ldots k_p} \ \sigma_{k_1}\ldots \sigma_{k_p}, 
    \label{eq:Hp}
\end{equation}
whose $N$ continuous spin variables $\sigma$ are subjected to a generic $\rho$-metrical constraint (e.g.,~$\rho=2$ is the case of the spherical constraint)
\begin{equation}
    \sum_{i=1}^N \sigma_i^{\rho} =N,
    \label{eq:SpherCon}
\end{equation}
where $N^A$ on top of the sum, with $A\in [1,p]$,
denotes the scaling with the size of the number of $p$-uples contributing to the energy and the spin indices  $k_{i}$ run from $1$ to $N$. If $A=p$ we have a fully-connected interaction graph, i.e.~each spin contributes in $\mathcal(N^{p-1})$ $p$-uples. At the other extreme, if $A=1$ the graph is {\em sparse}, i.e.~each spin only interacts in a finite  number of $p$-uples, not growing with the size of the system. All dilutions in between will be considered hereafter. 

For simplicity, we will take the variables as real valued here, yet  keeping the word {\em intensity} for the spin magnitude $|\sigma|$. We notice that the glassy random laser is a model belonging to this family, with $p=4$ but with complex spins. Though complex variables yield new physical features (see Chap.~\ref{chap:IntroCap}), we stress that these are not relevant for what concerns the influence of connectivity on the possible onset of intensity localization. 

\subsection{Ordered Couplings}
First we consider the ordered case $J_{k_1\ldots k_p}=J$. The typical ground state spin configurations in the canonical ensemble, where equipartition is expected to hold (in the average, not strictly), are those minimizing Eq.~(\ref{eq:Hp}). The energy is extensive
$$E=\mathcal H[\sigma_{\rm gs}]=\mathcal O(N)$$
provided that the coupling constant scales as \begin{equation}
    J\propto \frac{1}{N^{A-1}}.
    \label{eq:JNscale}
\end{equation}

In the occurrence of intensity localization, that is, if only a few modes take the overall intensity, equal to $N$ according to  Eq.~(\ref{eq:SpherCon}), whereas all the other are zero, we are interested in the energy contribution of a localized spin configuration.
First of all, let us notice that in order to have a non-zero contribution the intensity at least  $p$ coupled spins must localize. If we represent by $\square$ such a localizing $p$-uple, the intensity localized configuration is
\begin{equation} \label{eq:localization}
    \{\sigma_{\rm loc}\}: \qquad \sigma_{k \in \square} \propto N^{1/\rho}, \qquad  \sigma_{k \notin \square} =0.
\end{equation}

According to Eqs. (\ref{eq:Hp}), (\ref{eq:JNscale}) the energy of such a configuration of spins scales with $N$ like 
$$E_{\rm loc}=\mathcal H[\sigma_{\rm loc}] = 
\mathcal O\left(\frac{N^{p/\rho}}{N^{A-1}}\right).$$

To figure out whether intensity condensation might occur and dominate, one eventually has to compare the scaling behaviors of the energies of an equipartitioned and a localized configuration:
$$\mathcal O(N) \quad \mbox{vs}\quad  \mathcal O(N^{\frac{p}{\rho}+1-A}).$$ Hence, we notice that the kind of global constraint imposed is also key to understand whether a localization transition may take place. The following cases occur for $\rho=2$, depending on the interaction connectivity scaling $N^A$:
\begin{enumerate}
    \item $A>p/2$. Any possible intensity localized configuration of spins would yield subextensive contributions to the energy. The equipartition regime is, therefore, dominant. The case $A=p$ is the fully connected interaction graph.
    
    \item $A=p/2$. Both kinds of spin configurations yield an $\mathcal O(N)$ contribution to the energy. In this case a pseudo-localized phase might occur. 
    
    \item $A<p/2$. Intensity localization provides the most prominent contribution to the energy, that is, $\mathcal O(N^{>1})$. The case $A=1$ is the sparse case. 
\end{enumerate}

\begin{figure}[t!]
\centering
\includegraphics[width=0.7\textwidth]{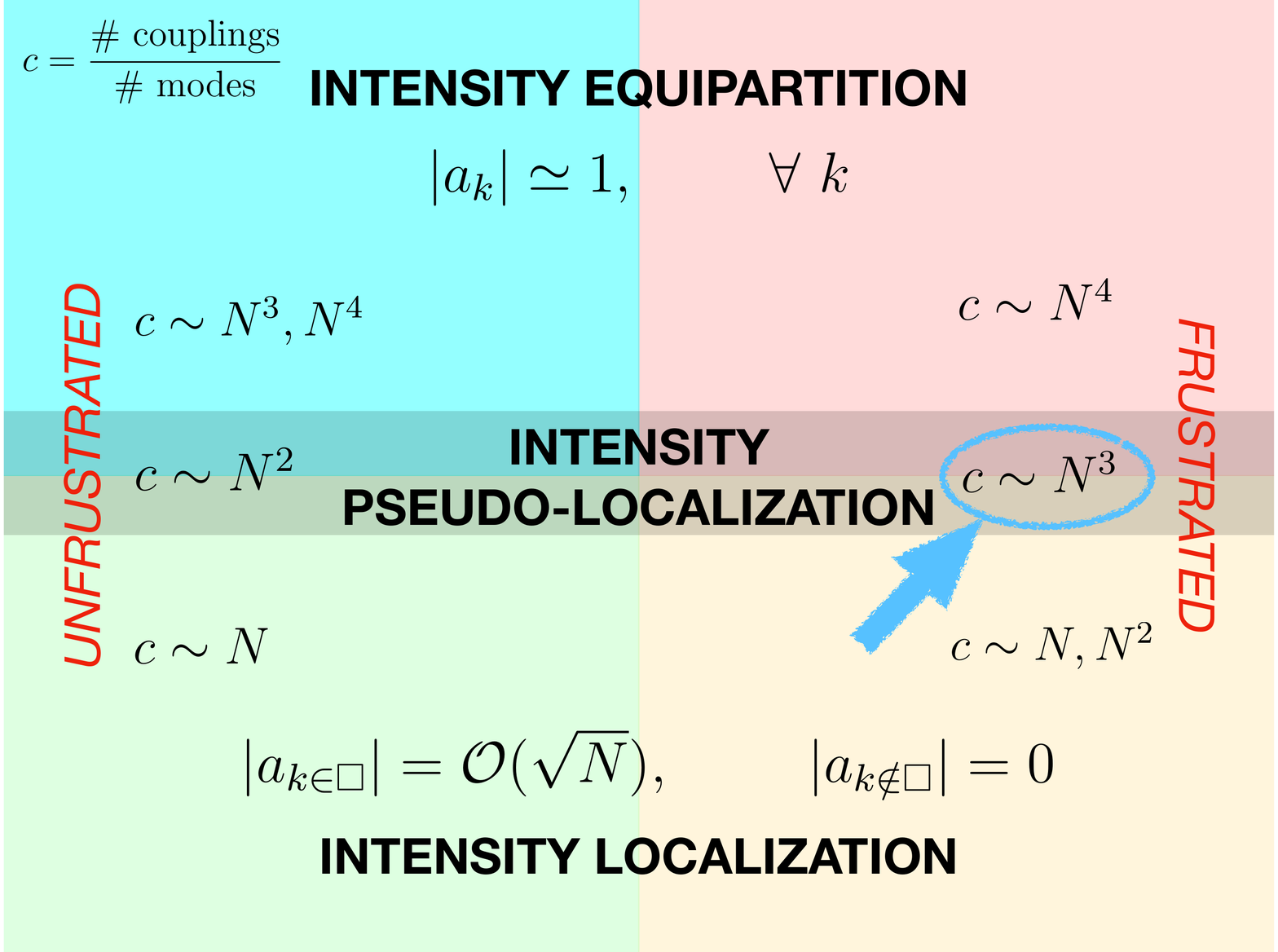}
\caption{Schematic representation of the possible regimes of low temperature intensity equipartition, intensity localization or pseudo-localization with respect to the coupling dilution in the $4$-spin spherical model, both for the ordered unfrustrated version (left) and the quenched disordered, frustrated one (right). The case of the mode-locked random laser whose connectivity $c \sim N^3$ is pointed by the arrow. Here, we have used the phasor notation $a$, instead of $\sigma$, used in the text for real variables.}
\label{fig:dilution}
\end{figure}

\subsection{Quenched Disordered Couplings}
If the interaction couplings $J_{k_1\ldots k_p}$ are quenched disordered, independently distributed and with zero mean ${\overline{J_{k_1\ldots k_p}}}=0$, the typical ground state of the Hamiltonian (\ref{eq:Hp}) is extensive -- $E=\mathcal H [\sigma_{\rm gs}]=\mathcal O(N)$ --  if the variance of the distribution of the couplings scales like
\begin{equation}
    \label{eq:JNdisscale}
{\overline{J^2_{k_1\ldots k_p}}}\propto \frac{1}{N^{A-1}}.
\end{equation}
If the total intensity of the system is localized in a single interacting $p$-uple, as in (\ref{eq:localization}), Eqs.~(\ref{eq:Hp}), (\ref{eq:JNdisscale}) imply that the energy scales with the size like
$$E_{\rm loc}=\mathcal H[\sigma_{\rm loc}] = 
\mathcal O\left(\frac{N^{p/\rho}}{N^{(A-1)/2}}\right).$$
Comparing  the equipartitioned contribution $E=\mathcal O(N)$ and the localized contribution $E_{\rm loc}$
we find for $\rho=2$ the following three regimes as the exponent $A$ varies:

\begin{enumerate}
    \item $A=p$. Localized energy contibutions are subextensive. The equipartition regime is dominant. This  is the fully connected interaction graph case.
    
    \item $A=p-1$. Both kinds of spin configurations yield an $\mathcal O(N)$ contribution to the energy. This is the case, e.g., of the mode-locked glassy random laser, where $p=4$, $A=3$.
    In this case one might conjecture the occurrence of a pseudo-localized phase. 
    
    \item $A<p-1$. Intensity localization provides superextensive contributions to the energy.  

\end{enumerate}

In figure (\ref{fig:dilution}) we summarize the scaling argument predictions on a pictorial diagram with the known cases of equipartition, localization and pseudo-localization for the $4$-spin model.

\section{Evidence of Pseudo-Localization in the ML 4-Phasor Model}
Let us turn back to our system. Besides the presence of interactions, there is another important difference between the glassy random laser and the case of the DNLSE. In the partition function of the ML 4-phasor model, which reads as
    \begin{equation}
    \label{eq:ZRL}
    Z_N(\beta,\mathcal E) = \int \prod_{k=1}^N \de a_k \de \bar a_k \ e^{- \beta \mathcal H[\bm a]}\delta\left( \mathcal E- \sum_{k=1}^N |a_k|^2 \right),
\end{equation}
with $\mH$ given by Eq.~\eqref{HamML4}, the conservation of energy is realized on average, by imposing homogeneously a temperature for all interacting quadruplets, while the conservation of the total intensity is realized exactly, by means of the hard global constraint (\ref{SpherConstr}). It is then possible to guess that in the ML $4$-phasor model a localization transition might occur at the level of intensity, rather than energy. If that were the case, though, the algorithms used in our numerical simulations would not be suitable anymore. An intensity localized system, indeed, would display a small, fixed number of variables whose energy fluctuations are no longer dependent on temperature.

In a way analogous to the condensation in the DNLSE, intensity localization is achieved by tuning the physical quantity $\mathcal E = \epsilon N$ controlled by the overall hard constraint (\ref{SpherConstr}). More precisely, when a certain threshold is overcome in the controlling parameter of the constraint, $\mathcal E>\mathcal E_c$, configurations for which a finite amount of the overall intensity is stored in $\mathcal{O}(1)$ quadruplets might become thermodynamically dominant. However, as already discussed in Chap.~\ref{chap:mixedorder}, numerical simulations are performed by keeping the optical power fixed ($\epsilon=1$) and varying the temperature $T$ or the pumping rate $\mathcal{P}$. We notice that this is equivalent to sample configurations from the equilibrium distribution
\begin{equation}
    P[\bm{\hat{a}}] \propto e^{-\mH[\bm{\hat{a}}]} \delta \left( \mathcal{P} N - \sum_{k=1}^N |\hat{a}_k|^2 \right)
\end{equation}
as one can see by performing the change of variables $\hat{a}_k = a_k \beta^{-1/4}$. A decrease (increase) in $T$ (in $\mathcal{P}$) can be read off equivalently as an increase of the spherical constraint value. Therefore, the occurrence of a localization transition will be revealed in terms of a critical temperature $T_c$ or equivalently a critical pumping rate $\mathcal{P}_c$. 

Qualitative information about the presence of a localization transition can be already traced in the behaviour of the emission spectra, when the temperature is lowered (or equivalently the pumping rate is increased), see Fig.~\ref{fig:spettriIntro}. It can be very clearly seen that, as the pumping is increased the overall intensity is heterogeneously distributed among the modes. This might hint that a localization phenomenon in intensity occurs, but it is not enough to establish it. 

In the following sections the analysis is refined by introducing and studying suitable observables for localization and equipartition breaking transitions. Data are referred to the simulations of the ML 4-phasor model with PBC, whose details are reported in Table \ref{tab2}.

\subsection{Participation Ratio}
Whether the system truly localizes or not can be ascertained only from the study of the participation
ratio, i.e.~the localization order parameter, which for our system is defined as
\begin{align} 
  Y_2 = \left\langle \frac{\sum_{k=1}^N {I}^2_k
  }{\left(\sum_{k=1}^N {I}_k\right)^2} \right\rangle =\frac{1}{N^2} \left\langle
  \sum_{k=1}^N {I}^2_k
  \right\rangle, 
    \label{Y2amp}
\end{align}
where ${I}_k = |a_k|^2$ and we have used the fact that  
\begin{align}
\sum_{k=1}^N {I}_k = N
\end{align}
because of equation \eqref{SpherConstr}, with $\epsilon=1$. In this case, it is irrelevant to normalize the intensities to the square root of the temperature (as done for the spectra in Eq.~\eqref{Spectrum}), since this factor cancels out between numerator and denominator. As in the previous chapters, $\langle \cdot \rangle$ denotes the thermal average, i.e. the average computed over all the uncorrelated equilibrium configurations, which have been sampled.

The dependence on the number of degrees of freedom of $Y_2$ can be
easily rationalized in two extreme situations: equipartition and
localization of the mode intensity. Let us consider localization first: in this case a
finite fraction of the whole intensity is taken by  a
finite number of modes that does not increase with  $N$. That is, in the localized phase, a few modes $k$ have intensity
\begin{align}
{I}_k \propto N,
\end{align}
whereas all the others have $ I_{\neq k}= 0$. 
Then, we have 
$$\sum_{k=1}^N I^2_k\simeq \sum_{k=1}^{\# {\rm loc\  modes}} \hspace{-.3cm} I^2_k\propto N^2.$$
This implies that in a localized phase the participation ratio $Y_2$  in
the limit $N\rightarrow \infty$ is a constant that does not depend on
$N$:
\begin{align}
\text{localization}~\Longleftrightarrow~  \lim_{N\to\infty }Y_2 =\mbox{const}.
\end{align}
On the contrary, in the equipartition phase of nearly homogeneous spectral intensities, any of the $N$ modes has intensity
${I}_k=\mathcal{O}(1)$, so that in the thermodynamic limit  
\begin{figure}[t!]
\centering
\includegraphics[width=0.65\textwidth]{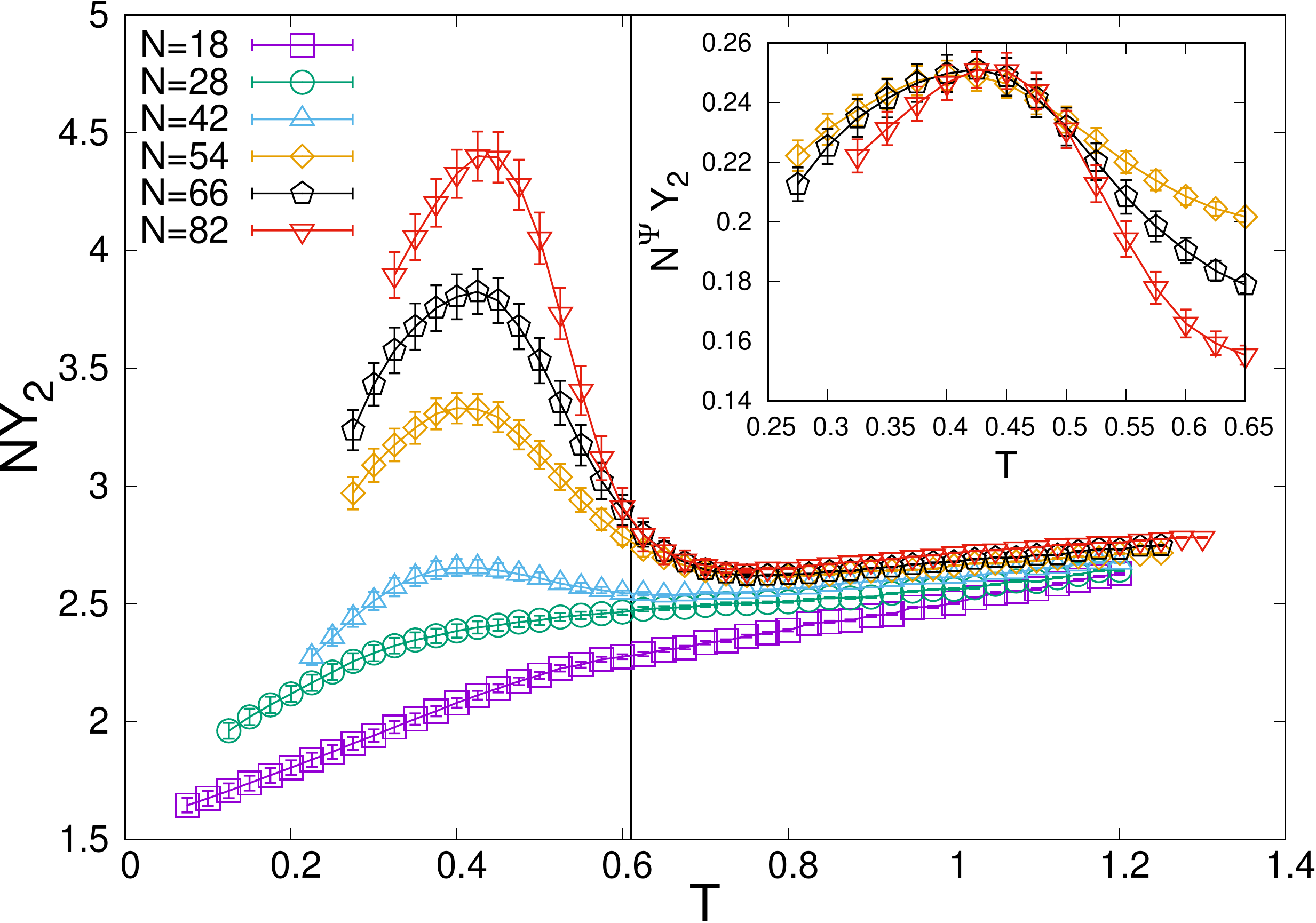}
\caption{Main: Participation ratio $NY_2$ of the mode intensities versus $T$ at different systems sizes $N$. The vertical line is the asymptotic value of the critical temperature for the system $T_c\simeq 0.61$ (from \cite{Niedda22a}, see also Table \ref{tab3}). Inset: scaling of the participation ratio near the peak for the three largest sizes, as $N^\Psi
  Y_2$ versus $T$, with $\Psi=0.35>0$. The peak height scaling is less than $N$, thus Eq. (\ref{Y2amp}) tends to  zero as $N\to\infty$.}
\label{fig:Y2-ampiezze}
\end{figure}
\begin{align}
\text{equipartition}~\Longleftrightarrow~ Y_2 \sim \frac{1}{N}.
\end{align}

Now we are ready to display,  in
Fig.~\ref{fig:Y2-ampiezze}, the first important quantitative information
obtained from the study of the equilibrium distribution of the
intensity among modes. In the
figure we have plotted for convenience the average over quenched disorder of $ N Y_2(T)$, which we
expect of $ \mathcal{O}(1)$ in the equipartition phase and of
$\mathcal{O}(N)$ in a possible localized phase.
We observe that in the high temperature phase $NY_2\sim \mbox{const}$ and, therefore,  the system is in the 
equipartition regime. Below the critical point, indicated by the vertical line at $T_c=0.61$, which is the glass transition temperature (see \cite{Niedda22a}), we find, instead, an anomaly. Despite the
main panel of Fig.~\ref{fig:Y2-ampiezze} gives a clear indication that
below the glass temperature  $NY_2(T)$ grows with the system size, the collapse of
data in the inset  shows that the scaling of the growth is definitely less than $N$. Therefore,  the regime is not localized, though  it is not equipartitioned, either. 
 In fact, for $T\lesssim T_c$ it occurs to be  $N Y_2(T)\sim N^{1-\Psi}$
with a value close to $\Psi \simeq 1/3$. This means that, since
\begin{equation*}
    Y_2 = \frac{1}{N^2} \sum_{k=1}^N \langle |a_k|^4 \rangle \sim N^{-\Psi},
\end{equation*} 
those modes $k$ on which the intensity is mostly concentrated scale like
\begin{equation}
    |a_k|^4 \sim N^{2-\Psi} \qquad \rightarrow \qquad |a_k|^2 \sim N^{1-\Psi/2}.
\end{equation}

The difference from the localization scaling $N$ cannot be accounted for as a finite size effect, as we might hypothesize in the estimate of critical exponents of a second-order phase
transition. These effects are usually due to  the cutting
of long-wavelength fluctuations in a finite size
simulation lattice. Localization is, instead, controlled by a first-order
mechanism where a finite fraction of the whole localizing
quantity concentrates on a few variables in such a way that
$Y_2$  is strictly independent from $N$.
In the glassy phase we, thus, have a
phase that might show some signature of incipient localization but it
is certainly not localized in intensity. Intensity equipartition is broken but no finite group of modes (i.e.~independent of $N$)  takes all the intensity of the system. 

\subsection{Spectral Entropy}
What
can play the same role played by temperature in DNLSE to help us recognizing that
we are in a non-trivial phase, rather than in a localized one?
A possible answer
is to look for an indicator of equipartition.  In Ref.~\cite{Gradenigo20} both the spectral entropy and the effective number of degrees of freedom were considered for the ML 4-phasor model. 
The spectral entropy is defined as 
\begin{align}
S_{\text{sp}} = - \sum_{k=1}^N \hat{\mathcal{I}}_k \ln(\hat{\mathcal{I}}_k),
\end{align}
where $\hat{\mathcal{I}}_k$ is the thermodynamic averaged intensity of the
mode $k$ normalized to the total intensity of the spectrum
\begin{align}
\hat{\mathcal{I}}_k = \frac{\langle {I}_k \rangle}{\sum_{k=1}^N \langle {I}_k \rangle} = \frac{\langle |a_k|^2 \rangle}{N\epsilon}.
\end{align}
\begin{figure}[t]
\centering
\includegraphics[width=0.65\textwidth]{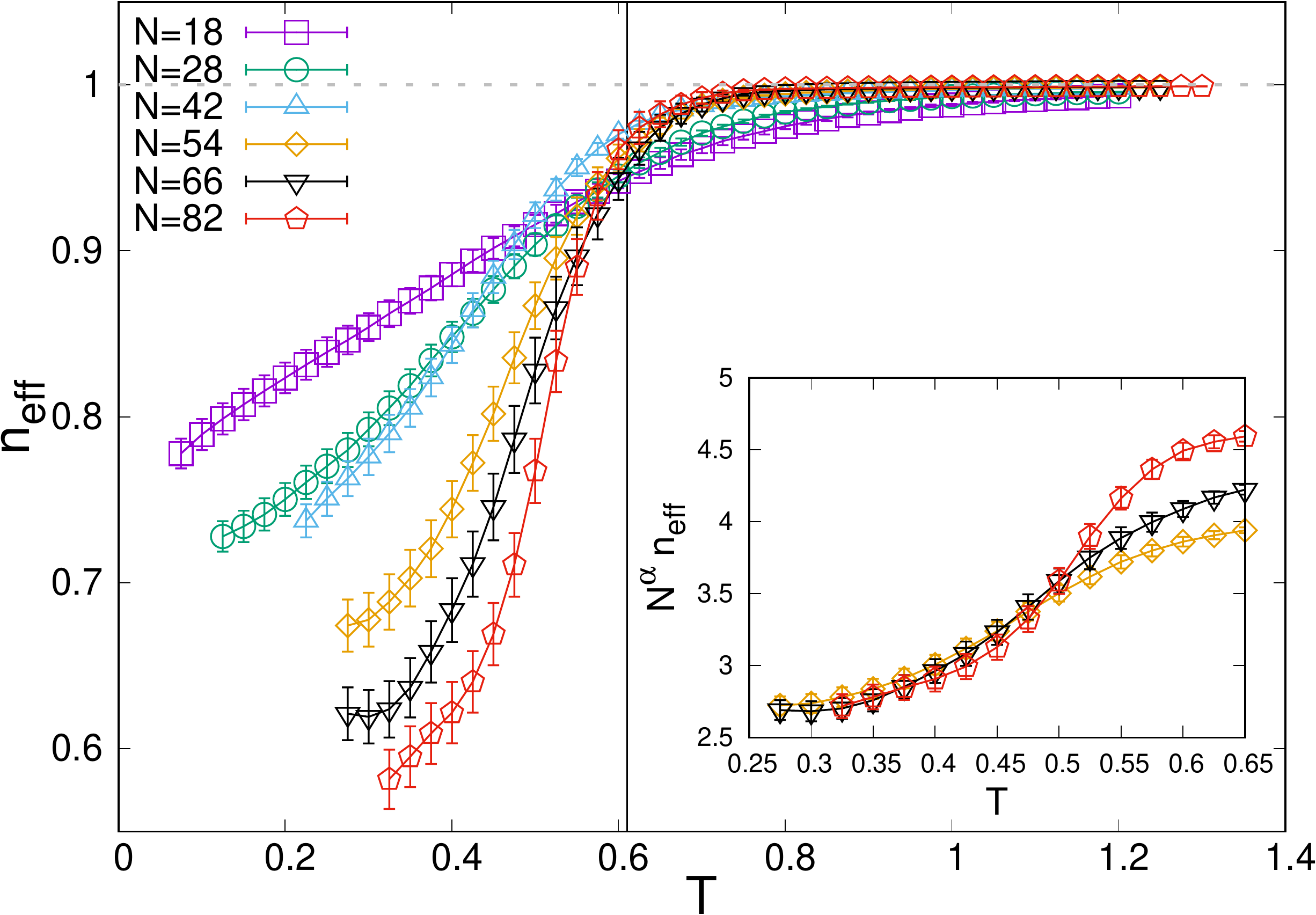}
\caption{Main: Effective number of degrees of freedom $n_{\text{eff}}$ versus $T$ for the mode
  intensities $|a_k|^2$ for different systems sizes. Inset: $N^\alpha  Y_2$ versus $T$, $\alpha=0.33$.}
\label{fig:Neff-ampiezze}
\end{figure}
The effective number of degrees
of freedom, which is a function of the spectral entropy and is more easily understandable, is defined as
$$
n_{\text{eff}} = \frac{e^{S_{\text{sp}}}}{N} .$$
The behaviour of $n_{\text{eff}}$, averaged over quenched disorder, is reported in
Fig.~\ref{fig:Neff-ampiezze} and shows the clear signature of a phase transition, where equipartition breaks down, at the same critical temperature where the glass transition takes place. The transition is first-order, as shown by the analysis of the Binder and bimodality parameters of the probability distribution of $n_{\text{eff}}$. This study has been performed on the first simulations of the ML 4-phasor model in Ref.~\cite{Gradenigo20}.

We notice that in the high temperature phase all the curves
perfectly approach one, while they decrease to a size-dependent
quantity for low temperatures: the larger the size, the steeper
${n}_{\text{eff}}$ decreases. In the inset, we plot the rescaled
effective number of degrees of freedom for the three  largest sizes. In the low temperature phase they collapse on each other with an exponent\footnote{In order to avoid any possible source of confusion, we stress that this exponent has nothing to do with the exponent $\alpha$ of the specific heat peak scaling, obtained in the previous chapter.}
$\alpha=1/3$. Thus, in the thermodynamic limit, this quantity tends
to $0$ in the low temperature phase, marking the breaking of
equipartition.

The transition taking place at $T_c$, besides being a static glass transition as it has been discovered in Refs.~\cite{Gradenigo20,Niedda22a} and discussed in the previous two chapters, can be 
characterized as the transition to a phase with a thermodynamic
anomaly which consist in the breaking of equipartition, in an analogous way
 for which we have a breaking of ensemble equivalence in a non-interacting system such as 
the DNLSE. 

\subsection{Amplitude Marginal Distribution}
\begin{figure}[ht!]
\centering
\includegraphics[width=0.65\columnwidth]{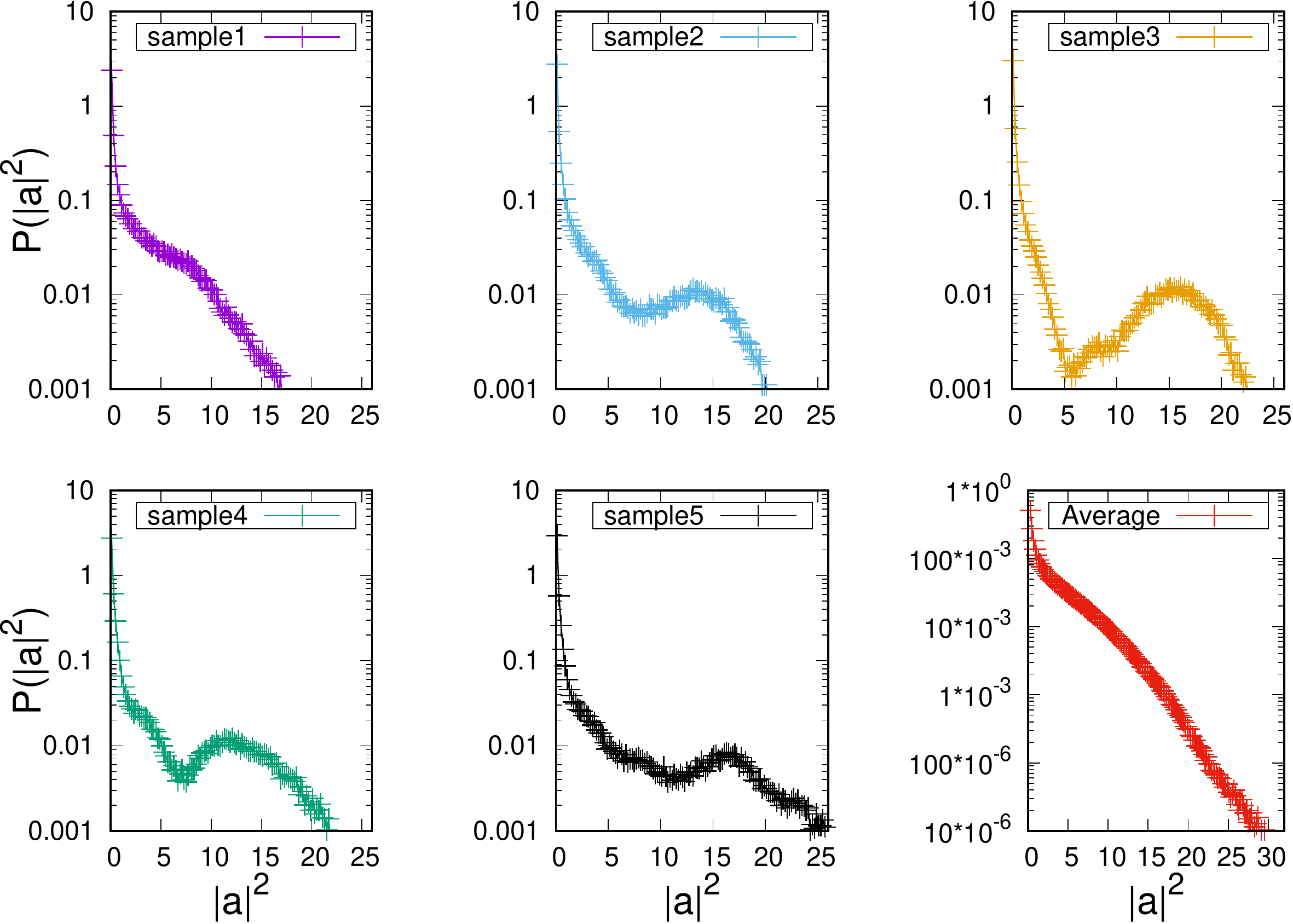}
\caption{Marginal intensity distributions for single instances of the quenched disorder. Data belong to the simulated size $N=82$ at the temperature $T=0.3$. The sixth panel contains the average over all the disordered samples, which corresponds to the lowest temperature intensity distribution in Fig.~\ref{fig:marginal-1}.}
\label{fig:marginal-2}
\end{figure}

\begin{figure}[ht!]
\centering
\includegraphics[width=0.65\columnwidth]{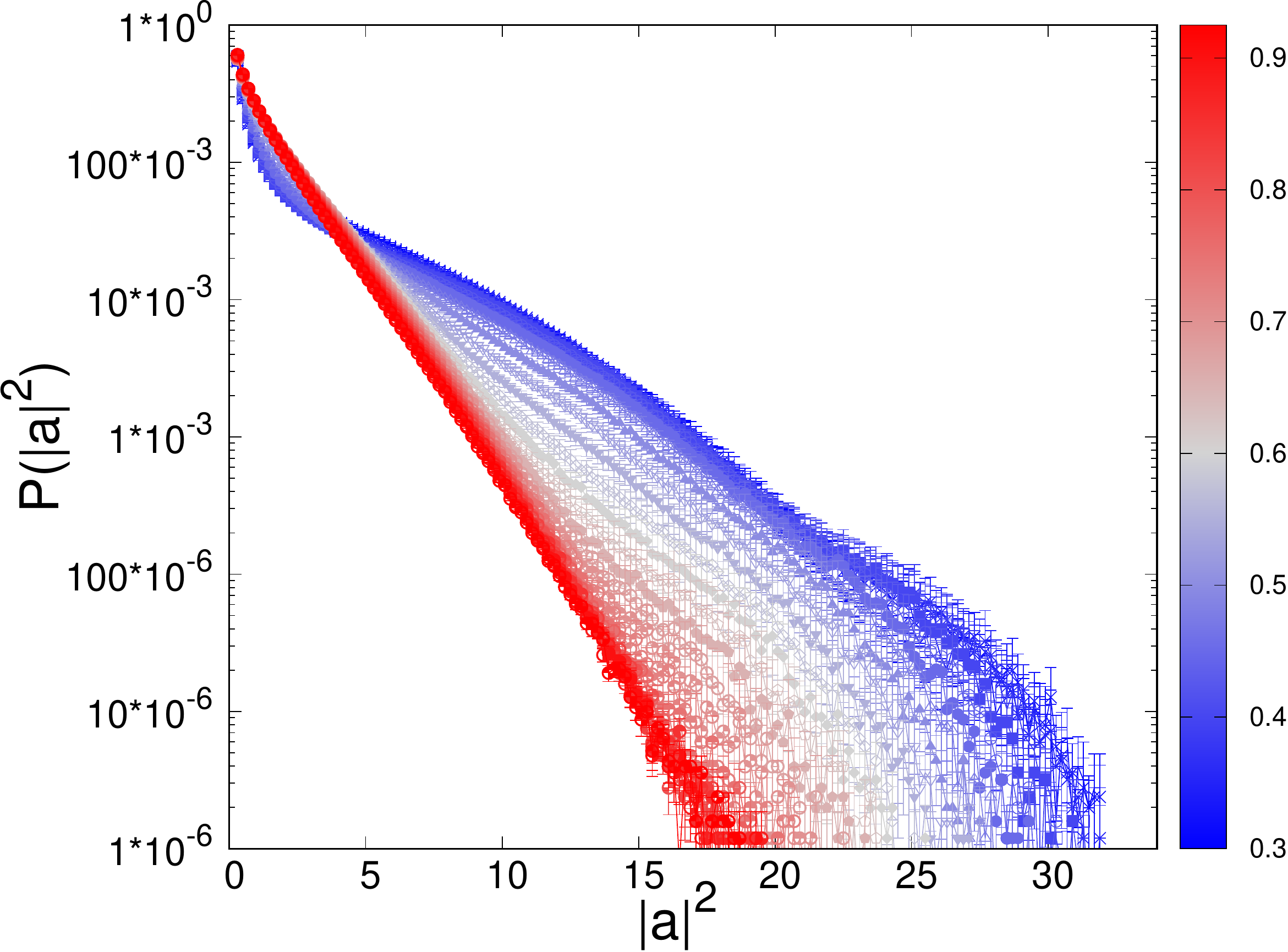}
\caption{Marginal distribution of spectral intensity averaged over quenched disorder for the system size $N=82$ at all the simulated temperatures. Color map: as in Fig.~\ref{fig:spectra}. At high temperature the distribution is exponentially decaying as revealed by the linear trend in semilogarithmic scale (red curves). By lowering the temperature deviation from exponential decay can be observed (blue curves) in relation to the onset of the pseudo-localized phase.}
\label{fig:marginal-1}
\end{figure}

The fact that this thermodynamic anomalous phase with lack of
equipartition is, indeed, a phase with incipient localization is
signaled, as in the case of DNLSE (see  \cite{Gradenigo21b}), by the
non-monotonic shape of the spectral intensity distribution $P(I_k)$. 
In Fig.~\ref{fig:marginal-2} we display $P(I_k)$ for single instances of 
quenched disorder for the size $N=82$ at the lowest simulated
temperature $T=0.3$. We notice that some of the samples (such as sample 2,3 and 4) 
exhibit a clear peak in the tale of the distribution corresponding to accumulation of the intensity on a single mode. The position of the peak is sample dependent. On the other hand, many samples behave as sample 1, exhibiting only a deviation from monotonicity in the marginal intensity distribution. Other samples behave in an intermediate way, such as sample 5: the peak is developing, but still barely visible. The effect of averaging over disorder is reported in the sixth panel. Finally, Figs.~\ref{fig:marginal-1} shows the behavior of the averaged $P(I_k)$ for the size $N=82$ when lowering the temperature.

Notice that the variance of the marginal distribution of the intensities is related to the participation ratio \eqref{Y2amp}. This can be seen in the following way: by denoting the sample average over $P(I_k)$ as $\langle \cdot \rangle_I$, which accounts both for the thermal average and for the average over the modes, the variance of $I_k$ is given by
\begin{align}
    \sigma_{I}^2 &= \langle (I_k-\langle I_k \rangle_I)^2 \rangle_I = \langle I_k^2 \rangle_I - \langle I_k \rangle_I^2 \nonumber \\
    &= \frac{1}{N} \left\langle \sum_{k=1}^N I_k^2 \right\rangle - \frac{1}{N^2} \left\langle \sum_{k=1}^N I_k \right\rangle^2 \nonumber \\
    &= \frac{1}{N} \left\langle \sum_{k=1}^N I_k^2 \right\rangle - 1,
\end{align}
where the spherical constraint has been used. Therefore, we have
\begin{equation}
    \frac{1}{N} \sum_{k=1}^N I_k^2 = \sigma^2 + 1.
\end{equation}
Since $Y_2$ has the expression in Eq.~\eqref{Y2amp}, we find that $NY_2 = 1+\sigma_I^2$.

\newpage

\part{Analytical Approach}

\chapter{The Merit-Factor Problem} \label{chap:MF}
In the first part of this work, we have presented results obtained through Monte Carlo numerical simulations of the ML 4-phasor model for optical waves in random media.~In particular, we have shown that, notwithstanding the deterministic dilution of the interaction network induced by the FMC, the model is still compatible with mean-field theory, though, up to the accuracy of our analysis, it may not be in the same universality class as the REM. However, as already mentioned, the analytical solution of the model is not achievable through standard mean-field techniques for disordered systems. Consider, for instance, the replica method: due to the dilution, the heterogeneities induced by the quenched disorder do not disappear as in fully connected mean-field models after averaging over the couplings and a nasty dependence on the site indices remains in the computation of the free energy, which impairs the introduction of the usual global order parameter of the glass transition, i.e.~the configuration overlap between replicas. Much of this discussion will be made more explicit in Chap.~\ref{chap:ML}. Furthermore, implementing the model on a sparse graph and adopting the cavity, or belief propagation, methods \cite{Mezard01,Franz01} is not possible in this case, because the variables are not locally constrained and intensity localization would occur because the interaction network is too diluted, as thoroughly illustrated in Sec. \ref{SS-SCALING-LOC}.

In order to develop the analytical technique to address the solution of the ML 4-phasor model, in the present chapter we temporarily turn to a different problem, which, in fact, has some striking formal resemblance to our model. The Merit Factor problem (MF), see Ref.~\cite{Jedwab05} for a survey paper on the topic, is a long standing problem in digital sequence design, with applications in many communication engineering problems, such as synchronization, pulse formation and especially radar \cite{Bohemer67,Turyn68}. The problem lies in finding Low Autocorrelation Binary Sequences (LABS), according to some suitable measure \cite{Golay77}. The merit factor was first introduced by Golay \cite{Golay82,Golay83} as an important measure of the kind, which is maximized by the LABS. Though an upper bound has been conjectured \cite{Golay82}, the problem of finding the merit factor highest value has resisted decades of attempts by mathematicians and it is still an open issue \cite{Jedwab05,Packebusch16}. Interestingly, the MF problem is not only related to the LABS: determining the best asymptotic merit factor is also an unsolved problem in complex analysis, which was proposed by Littlewood \cite{Littlewood61,Littlewood66} even before Golay's definition and until the early 00's was studied along independent lines.

From the point of view of theoretical physics, a major contribution to this line of research was given by Bernasconi in Ref.~\cite{Bernasconi87}, where the problem of finding sequences which maximize the merit factor has been reformulated in statistical mechanics terms as the problem of determining low-energy configurations of a specific spin model with long range 4-spin interactions. The Bernasconi model represents the formal connection between the MF problem and the problem of finding the solution of the glassy random laser, in which we are primarily interested. The non linear 4-body nature of the couplings is the first feature shared between the two systems, though in the case of the Bernasconi model the couplings are antiferromagnetic, rather than extracted from a zero mean, symmetric probability distribution. Moreover, the most important resemblance between the models is that the interaction graph of the Bernasconi model has the same structure determined by the FMC on the mode-locked graph, a quite fortunate occurrence, given the very different fields from which the two models originated. To be more specific, if one starts with a 4-spin antiferromagnet defined on the fully connected graph, with tetrads of spins denoted by $\bm{k} = \{k_1,k_2,k_3,k_4\}$ in order to obtain the Bernasconi model, one should dilute the interaction network with the rule $k_1-k_2+k_3-k_4=0$, which is precisely the FMC in the case of a linear comb (cf.~Eq.~\eqref{FMC-indices}). Our perspective is to take advantage of these similarities between the two models, in order to develop and test in a simpler environment the analytical methods, with which we aim to address the solution of the ML 4-phasor model.

After Bernasconi's reformulation, the merit factor problem has captured the attention of physicists working in the field of spin glasses and disordered systems, see Refs.~\cite{Marinari94a,Marinari94b,Bouchaud94,Migliorini94,Krauth95,Parisi95}. The model proposed by Bernasconi belongs to a class of models which exhibit frustration and glassy features without structural disorder - besides the reference just cited, see also the interesting case studied in Ref.~\cite{Franz95a}. Indeed, the finite-size analysis performed in Ref.~\cite{Bernasconi87} with the simulated annealing procedure provides results which are compatible with the properties of systems characterized by complex energy landscapes, leading to conjecture an ergodicity-breaking phase transition at finite temperature. The outcome of the numerical analysis performed in Ref.~\cite{Marinari94a} on a slightly modified version of the Bernasconi model points in the same direction, an evidence which has led the authors to develop a technique which is based on the introduction of random unitary matrices in the model and allows to perform a replica computation. It is precisely this technique that we aim to carefully review and master in order to apply it to the case of the ML 4-phasor model. 

In the first part of this chapter, after introducing the model and presenting a high temperature approximation, which was already discussed in Refs.~\cite{Golay82,Bernasconi87,Marinari94a}, we perform new numerical analyses of the model, broadening the results of Ref.~\cite{Marinari94a}. In the second part of the chapter, we complete the replica analysis of the model, along the lines of Ref.~\cite{Marinari94a}, and draw some tentative conclusion on the low temperature nature of the model.

\section{The Bernasconi model}
Consider a binary sequence of length $N$ denoted by
$\bm{s} = \{s_1,\dots,s_N\}$, where the variables are Ising spins $s_i = \pm 1$. The autocorrelation of the sequence at distance $k$ is given by the scalar product of the sequence with itself shifted by $k$. Two kinds of autocorrelations can be defined, leading to two versions of the model: aperiodic correlations
\begin{align}
    R_k = \sum_{j=1}^{N-k} s_j s_{j+k},
\end{align}
where the summation has to be stopped at $N-k$, and periodic correlations
\begin{align}
    R_k = \sum_{j=1}^{N} s_j s_{(j+k-1)(\text{mod}N)+1},
\end{align}
where the summation contains $N$ terms at distance $k$ and $(j+k-1)(\text{mod}N)+1$ is a formal way to implement periodic boundary conditions. If we think of the sequence as of a chain, then the model with aperiodic correlations is defined on an open chain, whereas the model with periodic correlations is defined on a closed chain. Historically, the problem of determining the LABS was defined for aperiodic correlations. The quality of a LABS can be measured either by minimizing the quantity $\max \{ |R_k|, k \neq 0 \}$, or by maximizing the merit factor \cite{Golay82,Golay83}, which is defined as 
\begin{align*}
    F = \frac{N^2}{2 \sum_{k=1}^{N-1}R_k^2}.
\end{align*}
In the Bernasconi model \cite{Bernasconi87}, one considers the equivalent problem of minimizing a cost function defined as the inverse of the merit factor
\begin{align}
    \mH = \frac{1}{N-1}\frac{N^2}{2 F} = \frac{1}{N-1}\sum_{k=1}^{N-1} R_k^2
\end{align}
which can be considered to represent the energy function of a one-dimensional spin system with long-range 4-body interactions. In this formulation the MF problem turns into an optimization problem with which we are more familiar in statistical mechanics.

Following Ref.~\cite{Marinari94a}, in the rest of this work we will be only concerned with the periodic model, due to some particular features which allow a deeper investigation and a
generalization to the ML random laser models. We, thus, consider the Hamiltonian
\begin{align} \label{MF:HamOriginal}
    \mH = \frac{1}{N-1} \sum_{k=1}^{N-1} R_k^2 = \frac{1}{N-1} \sum_{k=1}^{N-1} \sum_{i=1}^N\sum_{j=1}^N s_i s_{i+k} s_j s_{j+k},
\end{align}
where we implicitly assume periodic boundary conditions. Incidentally, we notice that this Hamiltonian is equivalent in the large-$N$ limit to the Hamiltonian obtained from a fully-connected 4-spin antiferromagnet, 
\begin{align} \label{MF:HamAFM}
    \mH =\frac{1}{N-1}  \sum_{i_1,i_2,i_3,i_4|\star}^{1,N}  s_{i_1} s_{i_2} s_{i_3} s_{i_4} , \qquad  \star: i_2-i_1 = i_4-i_3 \in [1,N-1]
\end{align}
which is diluted with the selection rule $i_1-i_2+i_4-i_3 = 0$. For this reason, we recognize in this problem the same topology as the mode-locked graph.


\subsection{The Golay-Bernasconi Approximation}
In this section, we discuss the Golay-Bernasconi (GB) approximation \cite{Golay82,Bernasconi87} of the periodic model, in which the correlations $R_k$ are assumed to be Gaussian distributed independent random variables with variance $N$, i.e.~extracted from $ p(R) = e^{-R^2/2N}/{\sqrt{2 \pi N}} $. We notice a certain resemblance between this approximation and the way the REM \cite{Derrida80} is built: here the correlations, rather than the energy levels, are taken as independent random variables, but the spirit is the same. However, while the REM is defined at all temperatures and exhibits a non-trivial low temperature behavior, in the GB approximation a negative entropy is found at finite temperature, which is not acceptable with discrete variables. Hence, we are dealing with a high temperature approximation, which breaks down when the entropy becomes negative.

Notice that the periodic correlations satisfy $R_k = R_{N-k}$. This is very easy to see as follows: by implying the operation of $\text{mod} N$ whenever the index of summation yields a value greater than $N$ in the subscript of the spins, we have
\begin{align*}
    R_{N-k} &= \sum_{i=1}^N s_i s_{i+N-k} = \sum_{i=1}^N s_{i+k} s_{i+N} \\
    &= \sum_{i=1}^N s_{i+k} s_i = R_k,
\end{align*}
where we have used $i \rightarrow i+k$ and  $(i+N) (\text{mod} N) = i$. Therefore, we can rewrite the Hamiltonian of the periodic model by taking into account only one half of the contributions and multiplying by a factor two. If $N$ is odd, we can write
\begin{align} \label{MF:HamiltPeriod}
    \mH = \frac{2}{N-1} \sum_{k=1}^{\frac{N-1}{2}} R_k^2,
\end{align}
whereas if $N$ is even
\begin{align*}
     \mH = \frac{2}{N-1} \sum_{k=1}^{\frac{N-2}{2}} R_k^2 + \frac{1}{N-1} R_{N/2}^2,
\end{align*}
as can be checked with simple renaming of the summation indices. However, since the difference between the two cases is completely irrelevant in the large-$N$ limit, in the following we use the expression of the Hamiltonian for $N$ odd, which is easier to handle.


The solution of the model in the GB approximation is immediate. The partition function can be computed as follows
\begin{align*}
    \mZ &= \sum_{\bm{s}} e^{-\beta \mH[\bm{s}]} =  \sum_{\bm{s}} \exp \left[-\frac{2 \beta}{N-1} \sum_{k=1}^{\frac{N-1}{2}} R_k^2(\bm{s})\right] \\
    &= \sum_{\bm{s}} \int \prod_{k=1}^{\frac{N-1}{2}} \left[\de R_k~\delta \left( R_k - \sum_{j=1}^N s_j s_{j+k} \right) \right] \exp \left[-\frac{2 \beta}{N-1} \sum_{k=1}^{\frac{N-1}{2}} R_k^2 \right] \\
    &= \int \prod_{k=1}^{\frac{N-1}{2}} \left[\de R_k~e^{-\frac{2 \beta}{N-1} R_k^2} \right]  \prod_{k=1}^{\frac{N-1}{2}} \left[ \sum_{\bm{s}}  ~\delta \left( R_k - \sum_{j=1}^N s_j s_{j+k} \right) \right] ,
\end{align*}
where we have changed variables using the relation 
\begin{align*}
    1 = \prod_{k=1}^{\frac{N-1}{2}} \int \de R_k~\delta \left( R_k - \sum_{j=1}^N s_j s_{j+k} \right).
\end{align*}
The quantity $\sum_{\bm{s}} \delta \left(R_k - \sum_{j=1}^N s_j s_{j+k} \right)$ accounts for how many times a certain value $R_k$ is found over the configurations: it is therefore a measure of the entropy of the variable $R_k$. Given the statistical independence of the correlations, this quantity is just $2^N$ times the probability of $R_k$ and therefore we have that
\begin{align*}
    \mZ &= 2^N \prod_{k=1}^{\frac{N-1}{2}} \int \frac{\de R_k} {\sqrt{2 \pi N}} \exp \left[-\left( \frac{1}{2N} + \frac{2 \beta}{N - 1} \right) R_k^2\right] \nonumber \\
    &= 2^N \left(\frac{N-1}{N-1+4\beta N}\right)^{\frac{N-1}{4}},
\end{align*}
which in the large-$N$ limit eventually yields
\begin{align}
    \mZ = \exp \left[ N  \left(\log 2 -\frac{1}{4} \log(1 + 4\beta) \right) \right].
\end{align}

From the partition function we can deduce the behavior of the thermodynamic observables of the model. The expression of the free energy, of the energy and of the entropy densities in the GB approximation are reported below
\begin{eqnarray}\nonumber
    f(\beta)& =&  -\frac{1}{\beta} \log \mZ = -\frac{1}{\beta} \log 2 + \frac{1}{4 \beta} \log(1 + 4\beta) \\
    \nonumber
    s(\beta) &=& \beta^2 \frac{\partial f}{\partial \beta} = \log 2 - \frac{1}{4} \log (1 + 4\beta) + \frac{\beta}{1 + 4\beta} \\
    \nonumber
    u(\beta)  &=& \frac{\partial (\beta f)}{\partial \beta}  = \frac{1}{1+4\beta}.
\end{eqnarray}
It is clear, then, that the approximation breaks down at low temperature, since the entropy becomes negative for $\beta > 10.3702$, i.e. $T < 0.0964$, and this is not possible for a model with discrete variables.

\begin{figure}
    \centering
    \includegraphics[width=.65\linewidth]{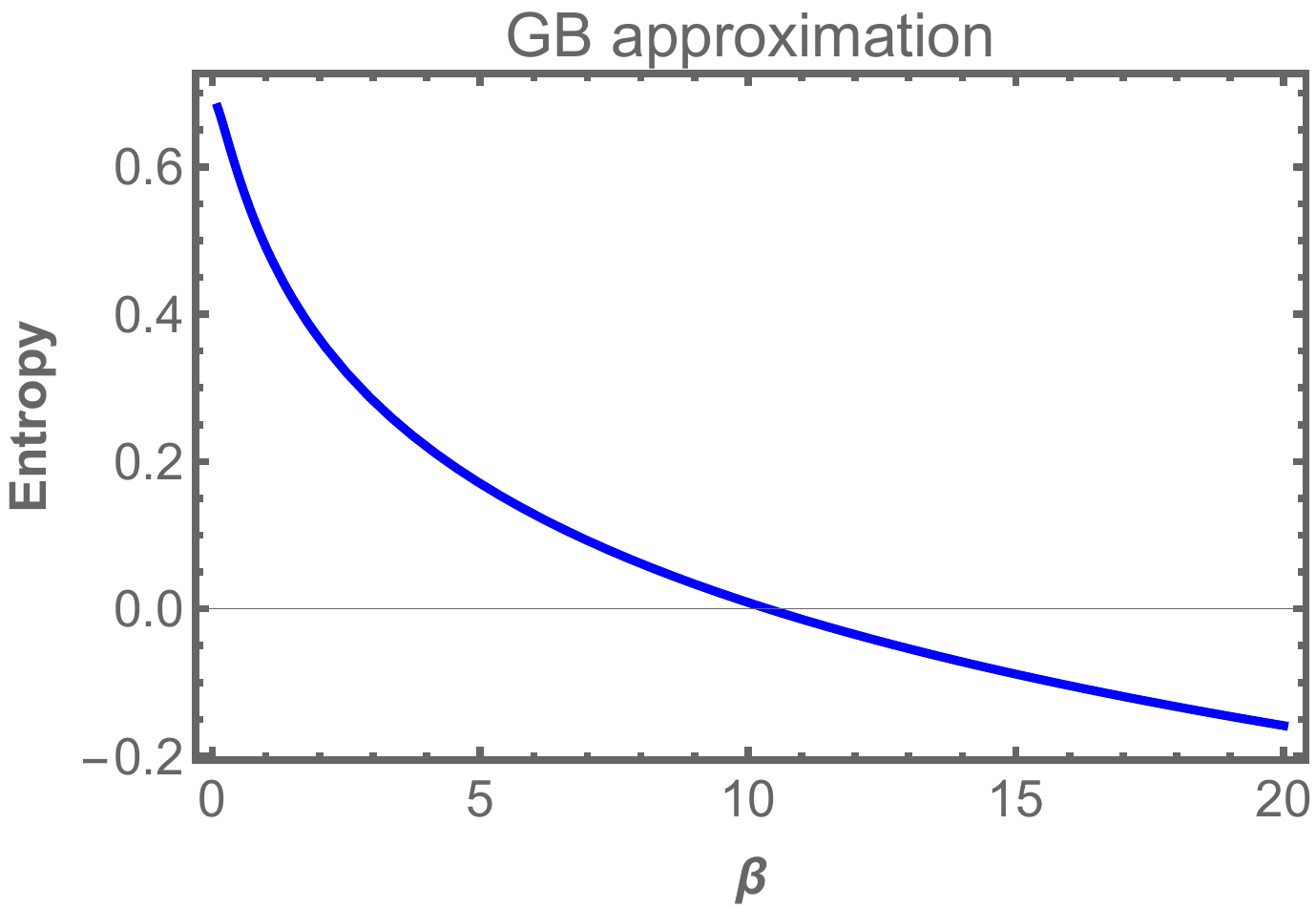}
    \caption{Entropy of the Bernasconi model in the GB approximation as a function of $\beta$. The change of sign at $\beta > 10.3702$ is a signature of the breakdown of the approximation.}
    \label{fig:entropiaGB}
\end{figure}

As shown in Ref.~\cite{Marinari94a}, the GB approximation can be recovered in the high tempererature regime of a disordered model with a Hamiltonian that looks like Eq.~\eqref{MF:HamiltPeriod}, but where the variables $R_k$ are given by
\begin{align} \label{CorrSelIndDis}
    R_k = \sum_{ij}^N J^{(k)}_{ij} s_i s_j,
\end{align}
where $J_{ij}^{(k)}$ are random connectivity matrices, independent for different $k$'s, whose values are extracted from some probability distribution.~The same kind of strategy was developed also in Ref.~\cite{Bouchaud94} for the aperiodic model. The replica analysis of the disordered model reveals a phenomenology which is compatible with the REM, with a stable 1RSB solution and zero entropy at the transition point.~The physical interpretation behind this scenario might be that of  ``self-induced'' disorder~\cite{Bouchaud94}. However, following Ref.~\cite{Marinari94a}, we notice that this result is not sufficient to draw definitive conclusions about the low temperature behavior of the original model without quenched disorder. The disordered model defined by Eqs.~\eqref{MF:HamiltPeriod} and \eqref{CorrSelIndDis} has to be regarded just as a test model, which is only capable of reproducing a high temperature approximation to the deterministic model. It would be too simplistic (if not wrong) to think that the glassy phase of this model provides an explanation of the low temperature complex behavior of the Bernasconi model. 




\section{Numerical Study}
\begin{figure}[t]
    \centering
    \includegraphics[width=\textwidth]{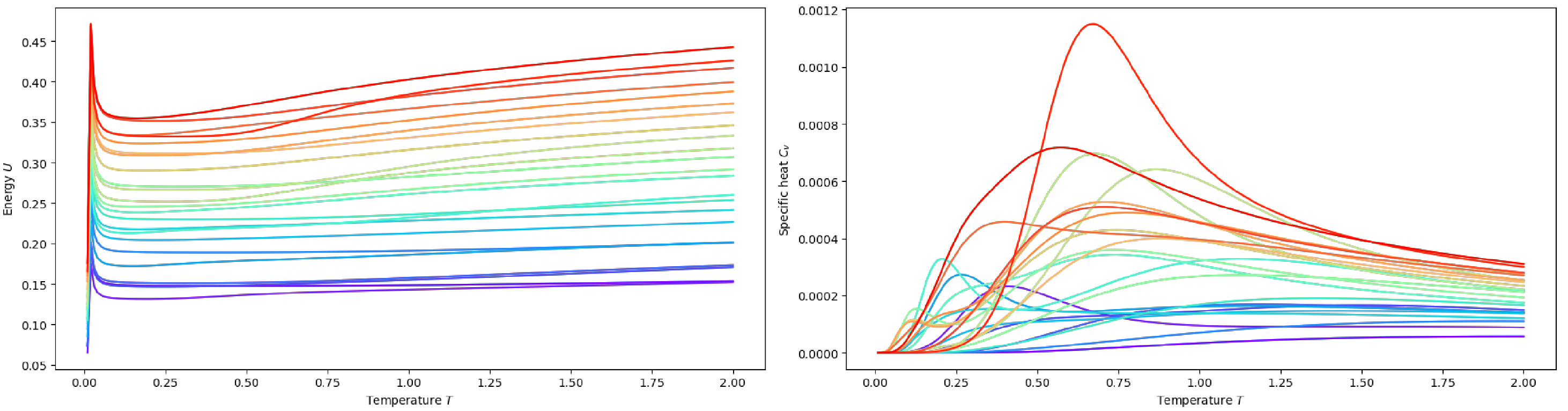}
    \caption{Energy and specific heat as a function of temperature for many different sizes of the Bernasconi model. Different colors correspond to different sizes: from $N=10$ (purple) to $N=31$ (red). The crossing of the energies corresponding to different sizes, as well as the unclear scaling of the specific heat peak position, reveal a nontrivial behavior of the system in the thermodynamic limit.}
    \label{fig:MFnum1}
\end{figure}

This section is devoted to present some numerical results obtained by studying the Hamiltonian \eqref{MF:HamOriginal} with periodic correlations. The ground state of the model is not known in general and is still the object of research: the effort involves both number theory approaches \cite{Schroeder09} and extensive searches \cite{Mertens96,Packebusch16}. We have no claim here to compete with the most recent achievements reached in the field, but rather to replicate and extend the finite-temperature study proposed in Ref.~\cite{Marinari94a}, in order to improve our knowledge of the model phenomenology.

Although no systematic procedure to construct ground state configurations for a generic size is known, \emph{ad hoc} constructions based on number theory exist for some specific values of $N$. One of these constructions works for prime numbers of the kind $N=4n+3$, with $n \in \mathbb{N}$. In this case configurations with the lowest energy are given by the Legendre sequences 
\begin{equation}
    \sigma_k = k^{(N-1)/2}\text{mod} N,
\end{equation}
which gives $\sigma_k = \pm 1$ for all $k$, but $k=N$ where $\sigma_N=0$. This is a consequence of a theorem by Fermat \cite{Schroeder09}, which states that unless $k$ is a multiple of $N$, then $k^{N-1}\mod N =1$, so that in this case $k^{(N-1)/2}\mod N =\pm 1$. To obtain a legal binary sequence, then, one has to see what happens by replacing the last bit, which now is zero, with $\pm 1$. This operation leads to an increase of the energy by a finite amount, with respect to the value computed before changing the last bit, apart from the lucky case $N=4n+3$, which leaves the energy untouched. The degeneracy of the ground state (and of the other energy levels as well) is related to the symmetries of the Hamiltonian\footnote{The Hamiltonian of the Bernasconi model is invariant under translation, i.e.~the shift of all the spins of a given number of positions, and under parity, i.e.~the flipping of all the spins. Actually, the reflection of a configuration is another symmetry of the Hamiltonian, but it can be obtained as a combination of translation and parity, hence not contributing to the total degeneracy.}. Other ground states can be
constructed from linear shift register sequences based on primitive polynomials over Galois
fields. This construction requires $N=2^p - 1$ with $p$ prime, see \cite{Marinari94a}.

In the case where no such constructions exist, one may resort to brute force algorithms: for a given size $N$, one lists all the $2^N$ possible sequences and computes the corresponding values of the energy. Then, the ground state configurations can be found by sorting the obtained values. The computation of the energy through the long-range Hamiltonian \eqref{MF:HamOriginal} is very demanding, since it requires $O(N^3)$ operations. Even considering the degeneracy of the energy levels, in order to exclude from the list of configurations those connected by the symmetries of the Hamiltonian, one cannot reach very high values of $N$ in reasonable times. To the best of our knowledge, the largest size studied with this method is $N=66$ in Ref.~\cite{Packebusch16}, where although parallel computing is also exploited, it took 55 days of machine time to obtain the results. Alternatively, one can resort to Monte Carlo optimizations of the Hamiltonian to find approximately optimal sequences for arbitrarily large values of $N$.

\begin{figure}[t]
    \centering
    \includegraphics[width=1.015\textwidth]{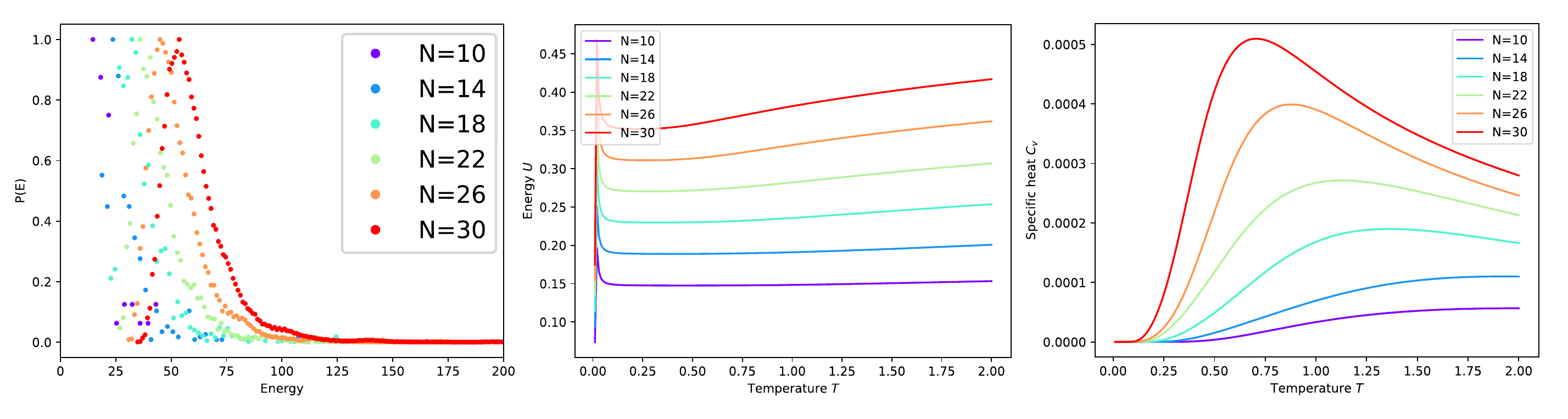}
    \caption{From left to right: density of states, energy and specific heat for the sizes in the legend.}
    \label{fig:MFnum2}
\end{figure}

\begin{figure}[t]
    \centering
    \includegraphics[width=\textwidth]{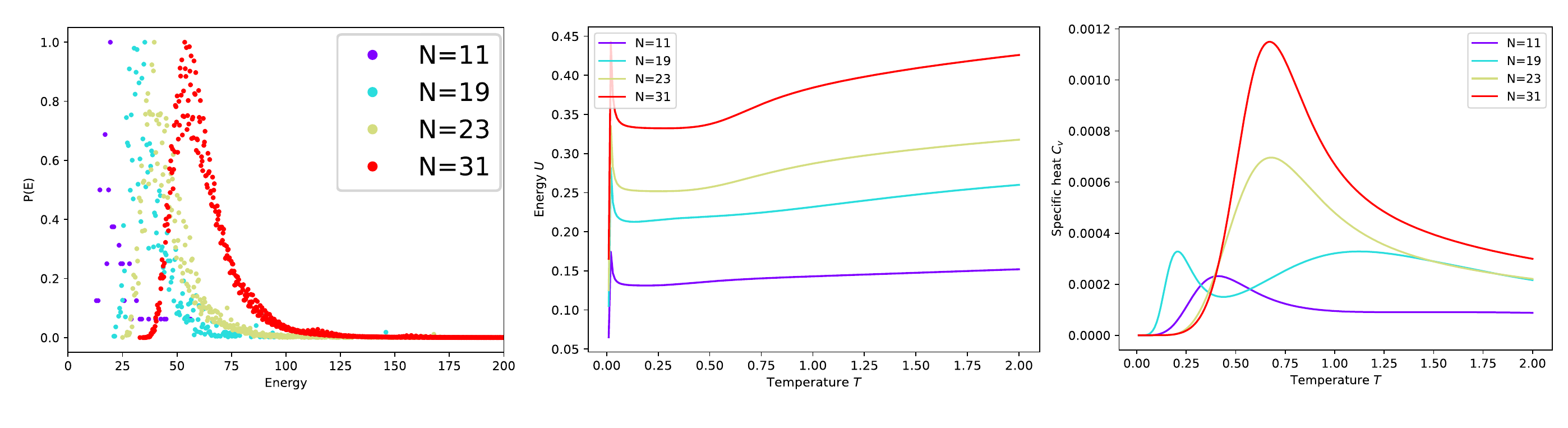}
    \caption{From left to right: density of states, energy and specific heat for the sizes in the legend. These sizes correspond to the ``good primes'' for which the ground state is given by the Legendre sequence $\sigma_k = k^{(N-1)/2}\text{mod} N $.}
    \label{fig:MFnum3}
\end{figure}

In this work, we used a simple exhaustive search algorithm and studied sizes up to $N=32$. We are interested in all energy levels, not only the ground state, since as soon as the temperature is added to the system, there is a finite probability of finding the system in a level with higher energy than the lowest one. The finite-temperature behavior of the model can be deduced from the density of states expressed as a function of the energy $E$. We, then, build the histograms
\begin{equation}
    p_N(E) = \sum_{\bm{s}} \delta\left(E - \mH_N[\bm{s}] \right),
\end{equation}
in terms of which the canonical partition function of the model can be written as
\begin{equation}
    \mZ_N(\beta) = \sum_{\bm{s}} e^{-\beta \mH_N[\bm{s}]} = \int \de E~p_N(E) e^{-\beta E},
\end{equation}
where the subscript $N$ is just a reminder of the finite size nature of the quantities here studied. From the partition function, we get the free energy of the model, from which, in turn, we derive the specific heat. 

In Fig.~\ref{fig:MFnum1} we display the energy and the specific heat for most of the sizes studied.
Fluctuations from one volume size $N$ to a similar one are large and macroscopic.
Such fluctuations forbid any simple extrapolation to the limit $N \rightarrow \infty$ from the sizes analyzed. 
They decrease however for increasing $N$. The pronounced peak in the specific heat 
strongly suggests that in the infinite volume limit the system undergoes a phase transition.
However, the position in temperature of the specific heat peak changes a lot from size to size 
and seems to decrease towards a small value of $T$, even if in a irregular pattern, which makes it difficult to extrapolate an estimate of the critical temperature of the model. This becomes clearer if one selects
specific sequences of sizes: for instance in Figs.~\ref{fig:MFnum2} we plot the thermodynamic observables for even sizes increasing with a $\Delta N = 4$, and in Fig.~\ref{fig:MFnum3} for the sequence of ``good primes'', whose ground state can be constructed analytically. 

\begin{figure}[t]
    \centering
    \includegraphics[width=\textwidth]{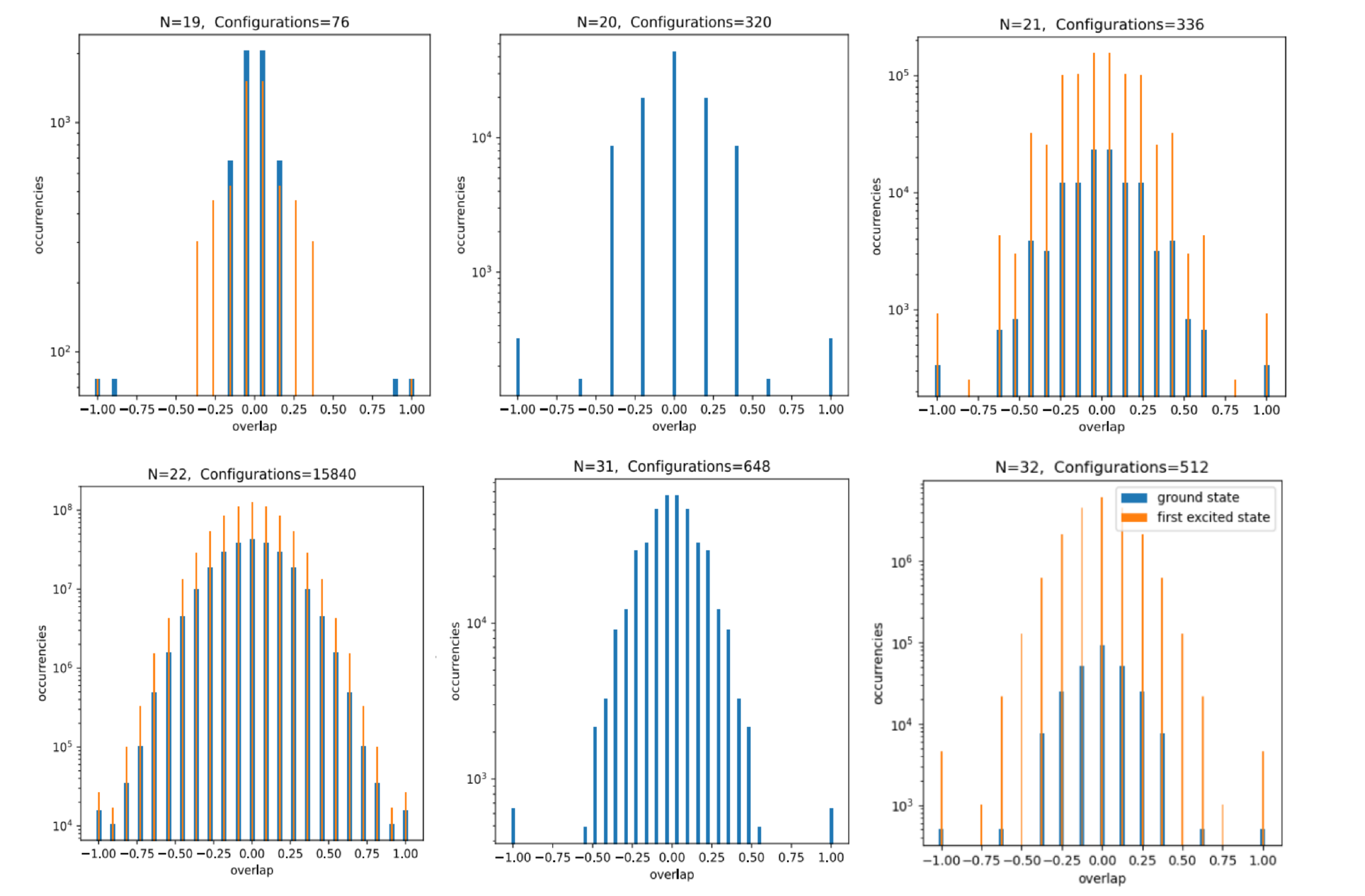}
    \caption{Histograms built with the frequencies of the overlap values $q = \sum_{i=1}^N \sigma_i \tau_i / N$, for all the pairs of configurations $\sigma$ and $\tau$ of the ground (blue) and first excited (orange) states for the system sizes $N=19,20,21,22,31,32$.} 
    \label{fig:MFnum4}
\end{figure}

Finally, we have computed the overlaps between all the configurations belonging to the ground state and to the first excited state. In Fig.~\ref{fig:MFnum4} we display the histograms built with the overlap frequencies for some of the studied sizes. The fact that the overlap can take many values is evidence of a nontrivial structure of the ground state: most of the configurations minimizing the energy are not correlated. This study, repeated for the first excited state, basically retraces the results obtained for the ground state, meaning that a small thermal excitation of the system does not change dramatically how the configurations are organized.

Taken together, these results constitute the phenomenology of a phase transition to a complex low temperature phase, which is well captured by the theory of replica symmetry breaking in spin-glass models. However, given the small sizes considered, the output of this study cannot be taken as a proof that a phase transition occurs in the thermodynamic limit.

\section{The Random-Unitary Model}
The main reason why we focus on the periodic model is that it is suitable for an analysis in Fourier space \cite{Marinari94a}. Let us first introduce the Fourier-space version of the model and then discuss a first tool of investigation. The discrete Fourier transformation (DFT) of the spin variables and its inverse are defined as follows
\begin{align} \label{DFT}
    B_p = \frac{1}{\sqrt{N}} \sum_{j=1}^N e^{i \frac{2 \pi p}{N}j} s_j ~~~~~~~ s_j = \frac{1}{\sqrt{N}} \sum_{p=1}^N e^{-i \frac{2 \pi j}{N}p} B_p,
\end{align}
where the symmetric convention for the normalization has been adopted. Then, the correlations can be easily expressed in terms of the Fourier variables as
\begin{align}
    R_k = \sum_{p=1}^N \sum_{q=1}^N e^{-i \frac{2 \pi k}{N} q} B_p B_q \delta_{p,-q} = \sum_{p=1}^N e^{i \frac{2 \pi k}{N} p} |B_p|^2,
\end{align}
where the definition of the Kronecker delta as the inverse DFT of 1 has been used, together with the fact that $B_p = \overline{B}_{-p}$, which holds since the original spin variables are real. Hence, the periodic Hamiltonian \eqref{MF:HamiltPeriod} can be rewritten as 
\begin{align} \label{HamiltDFT}
    \mH &= \sum_{p=1}^{N/2} |B_p|^4,
\end{align}
where an irrelevant additive constant has been neglected. Moreover, both the two previous expressions are correct in the large-$N$ limit, while at finite $N$ one should take into account the precise definition of the periodic correlations.

A useful tool to yield physical insight about the model, which has been implemented in Ref.~\cite{Marinari94a}, is the high-temperature expansion. In principle this technique can be applied in a very straightforward way; in the present case, however, the non-locality of the interaction term causes complications, such as the fact that the expansion coefficients do not behave well in the large-$N$ limit. It turns out that the procedure simplifies in Fourier space, allowing for the computation of the thermodynamic observables at the first orders in $\beta$, which better represent the high temperature regime of the Bernasconi model, with respect to the GB approximation. This is expected, since the high temperature expansion does not require any \emph{ad hoc} approximations, but to take carefully the limit $\beta \rightarrow \infty$. 

At this stage, one would like to define a model based on the Hamiltonian \eqref{HamiltDFT} and capable of: (\textit{i}) resumming the high temperature expansion and (\textit{ii})  showing (hopefully) a non-trivial low temperature phase that reproduces the complex phenomenology observed in numerical simulations. This has led the authors of Ref.~\cite{Marinari94a} to introduce a model based on random unitary matrices, which we will refer to as Random Unitary model and to which the rest of the chapter is dedicated.

\subsection{Definition of the Model} \label{defRUM}
The core idea of the Random Unitary model is to substitute the standard Fourier transformation of the spin variables with a generic unitary transformation. In fact, the DFT matrix, which can be denoted as $U_{pj} = (u^{pj}/\sqrt{N})_{p,j=1,\dots,N}$ with $u=e^{-2i\pi/N}$, is just a particular choice of matrix belonging to the unitary group, and one may be interested in studying a more general case. In the language of Lagrangian mechanics, a random unitary transformation brings us from spin variables to a set of random generalized coordinates in the configuration space: in order to visualize this operation, one can think of a rotation with an arbitrary angle. In fact, the introduction of random rotations defines a very similar class of models, which were extensively studied in Refs.~\cite{Marinari94b,Parisi95}, namely the Random Orthogonal models. In the language of disordered systems, the unitary group plays the role of quenched disorder, of which the DFT matrix is a particular instance. 

When passing from the DFT matrix to a generic unitary matrix, one has to pay attention to a subtle aspect of the procedure. In fact, if one starts with real variables, i.e.~the spins $s_i$, the Fourier transformed variables $B_p$ of Eq.~\eqref{DFT} satisfy the property $B_p=\overline{B}_{-p}$, as a consequence of the particular expression of the Fourier transformation. In general, this is not true for a unitary matrix\footnote{The only property which defines a matrix representing an element of the unitary group is that $U^\dagger=U^{-1}$, so that $U U^\dagger=U^{\dagger}U=1$. This leads to the fact that a unitary matrix can always be written in an exponential form, such as $U=e^{ih}$ where $h$ is a generic Hermitian matrix, i.e. $h=h^\dagger$. However, it is not necessary that the elements of $h$ satisfy the relation $h_{pj}=\overline{h}_{j(-p)}$ which implies the property of complex conjugation $B_p=\overline{B}_{-p}$ in the special case of a DFT matrix}. However, it turns out that this property of complex conjugation is crucial if one aims to reproduce the results of the high temperature expansion of the original model through a model defined with generic unitary matrices \cite{Marinari94a}. In other terms, if one just substitutes to Eq.~\eqref{DFT}, the following relation 
\begin{equation*}
    B_p=\sum_{pj}^N U_{pj} s_j
\end{equation*}
where $U_{pj}$ are the elements of a generic unitary matrix, one finds different results from the high temperature expansion already at the first order. 

One way to solve the problem is to introduce a model based on a double orthogonal transformation with Hamiltonian
\begin{equation} \label{doubleROM}
    \mH = \sum_{p=1}^{N/2}|A_{2p-1} + i A_{2p}|^4,
\end{equation}
where the variables $A$ are related to the variables $B$ as $B_p=A_{2p}+iA_{2p+1}$ and are defined in terms of the spin variables as
\begin{equation}
    A_p = \sum_{j=1}^N O_{pj}s_j,
\end{equation}
with $O_{pj}$ orthogonal matrices, over which we aim to integrate. We call this model double Random Orthogonal model. However, by following Ref.~\cite{Marinari94a} we define $N/2$ complex spin variables in the following way $\tau_j = s_{2j-1} + i s_{2j}$ and apply a random unitary transformation to $\tau_j$. Notice that this is just a redistribution of the degrees of freedom of the theory, not changing the physical content of the model. Clearly, in this case, the unitary transformation has to be represented by a $N/2\times N/2$ matrix. With these prescriptions, the Hamiltonian of the Random Unitary model is given by
\begin{align} \label{MF:RandUnitModelH}
    \mH[\bm{\tau}] = \sum_{p=1}^{N/2} |C_p(\bm{\tau})|^4,
\end{align}
where the dynamic variables of the model are the unitary-transformed of the complex spins $\tau_j$: 
\begin{align}
    C_p = \sum_{j=1}^{N/2} U_{pj} \tau_j.
\end{align}

\subsection{Replicated Partition Function}
In this section we develop the replica computation for the model defined by the Hamiltonian~\eqref{MF:RandUnitModelH}, where the dynamic variables $C_p=\sum_j U_{pj} \tau_j$ are unitary-transformed of the complex spins $\tau_j = s_{2j-1} + i s_{2j}$ and $U_{pj}$ are the elements of a random matrix belonging to the unitary group. The matrices $U$ play the same role in the computation as the quenched couplings in usual spin-glass models. The partition function of the model for a specific transform $U$ is given by
\begin{align}
    \mZ_U = \sum_{\bm{\tau}} e^{-\beta \mH[\bm{\tau}]} = \sum_{\bm{\tau}} \exp\left[ - \beta \sum_{p=1}^{N/2} |C_p(\bm{\tau})|^4 \right],
\end{align}
where $\sum_{\bm{\tau}} = \prod_{j=1}^{N/2}\sum_{\{\tau_j\}} $ and the sum runs over the four possible values of the complex numbers $\tau_j$. First of all, let us change variables in the computation of the partition function, by exploiting the relation
\begin{align*}
    1&=\prod_{p=1}^{N/2}\int \de C_p \de \overline{C}_p \delta\left(C_p-\sum_{j=1}^{N/2} U_{pj}\tau_j \right) \delta\left(\overline{C}_p-\sum_{j=1}^{N/2} \overline{U}_{pj}\overline{\tau}_j \right) \\
    &= \prod_{p=1}^{N/2}\int \de C_p \de \overline{C}_p \de \lambda_p \de \overline{\lambda}_p \exp \left[ i \overline{\lambda}_p \left(C_p-\sum_{j=1}^{N/2} U_{pj}\tau_j \right) - i \lambda_p  \left(\overline{C}_p-\sum_{j=1}^{N/2} \overline{U}_{pj}\overline{\tau}_j \right) \right],
\end{align*}
where we used the Fourier integral representation of the delta functions and neglected irrelevant constant factors. We have
\begin{equation} \label{MF:PartFunc1}
\begin{split}
    \mZ_U &= \int \prod_{p=1}^{N/2} \left[\de C_p \de \overline{C}_p \de \lambda_p \de \overline{\lambda}_p \right] \exp\left[- \beta \sum_{p=1}^{N/2} |C_p|^4 + \sum_{p=1}^{N/2} \left(i C_p \overline{\lambda}_p - i \overline{C}_p \lambda_p \right) \right]  \\
    & \times  \sum_{\bm{\tau}} \exp\left[ \sum_{p=1}^{N/2} \sum_{j=1}^{N/2}\left(-i U_{pj} \tau_j \overline{\lambda}_p + i \overline{U}_{pj} \overline{\tau}_j \lambda_p \right) \right].
\end{split}
\end{equation}

The free energy of the model depends on the choice of the matrices $U$. Since we are interested in the typical behaviour of the system with respect to this source of randomness, the average of the free energy over the unitary group has to be computed. At this level, we expect the system to exhibit a low temperature glassy phase, which can only be revealed by computing a quenched average. This can be done through the replica method. Therefore our goal is to find the following quantity
\begin{align} \label{freeEn1}
f(\beta)= \lim_{N\rightarrow\infty} - \frac{1}{\beta N} \overline{ \log \mZ_U} = \lim_{n \rightarrow 0} \lim_{N\rightarrow\infty} - \frac{1}{\beta nN} \log \overline{\mZ_U^n},
\end{align}
where $\overline{(\cdots)}$ denotes the integration over the unitary group and, as usual, the two limits have been exchanged, in order to compute first the saddle point of the averaged replicated partition function. As a consequence, in the following we will use several times the fact that we are taking the large-$N$ limit at finite $n$.

Considering Eq.~\eqref{MF:PartFunc1}, the $n$-th power of the partition function reads as
\begin{equation} \label{replPartFunc}
\begin{split}
    \mZ_U^n &= \int \prod_{a=1}^n \prod_{p=1}^{N/2} \left[\de C_p^a \de \overline{C}_p^a \de \lambda_p^a \de \overline{\lambda}_p^a \right] \exp\left[-\beta \sum_{a=1}^n\sum_{p=1}^{N/2} |C_p^a|^4 + \sum_{a=1}^n\sum_{p=1}^{N/2} \left(i C_p^a \overline{\lambda}_p^a - i \overline{C}_p^a \lambda_p^a \right) \right]  \\
    & \times  \prod_{a=1}^n\left[\sum_{\bm{\tau}^a}\right] \exp\left[ \sum_{a=1}^n\sum_{p=1}^{N/2} \sum_{j=1}^{N/2}\left(-i U_{pj} \tau_j^a \overline{\lambda}_p^a + i \overline{U}_{pj} \overline{\tau}_j^a \lambda_p^a \right) \right].
\end{split}
\end{equation}
By defining the auxiliary variables $\Omega_{pj} = i \sum_a^n \overline{\tau}_j^a \lambda_p^a $, the disorder dependent term of the replicated partition function can be compactly written as 
\begin{align} \label{DisDepPart}
    \exp\left[ \sum_{a=1}^n\sum_{p=1}^{N/2} \sum_{j=1}^{N/2}\left(-i U_{pj} \tau_j^a \overline{\lambda}_p^a + i \overline{U}_{pj} \overline{\tau}_j^a \lambda_p^a \right) \right] = \exp\left[\Tr(U \Omega^\dagger + U^\dagger \Omega) \right].
\end{align}
In order to average the replicated partition function, we have to compute an integral over the Haar measure of the unitary group. This problem was first encountered in the large-$N$ limit of lattice gauge theories in Ref.~\cite{Brezin80a,Brezin80b}, where the authors considered an approach which is analogous to standard mean-field theory for magnetic systems and leads to the computation of partition functions of the kind
\begin{align} \label{ExtFieldProblem}
    Z = \int \de U \de U^\dagger \exp N \left[\Tr(U A^\dagger + U^\dagger A) \right],
\end{align}
where $A$ is an arbitrary matrix source. This is exactly what we aim to compute, when averaging the right hand side Eq.~\eqref{DisDepPart} over disorder, if we replace the generic matrix $A$ with $\Omega/N$. The integral was solved in full generality by Brezin and Gross and the result is reported in Eq.~(33) of Ref.~\cite{Brezin80b}. However, as noted in~\cite{Marinari94a}, in the present case at finite non-zero $n$, only terms containing a single trace operator survive in the large $N$ limit. Hence, the cited result reduces to 
\begin{align} \label{MF:AverDis}
   \overline{ \exp\left[\Tr(U \Omega^\dagger + U^\dagger \Omega) \right]} = \exp\left[\frac{N}{2} \Tr \mG\left( \frac{\Omega^\dagger\Omega}{N^2} \right) \right],
\end{align}
where $\mG$ is a function of the eigenvalues of a matrix defined as
\begin{align} \label{MF:G}
    \mG(z) = -\log\left(1 + \sqrt{1+z}\right) + \sqrt{1+z}.
\end{align}
This is a main technical part of the computation: in Appendix~\ref{app:UG} we provide the demonstration of this result in a particularly simple case, where the integral can be directly performed with the saddle point method. 

As is usual in standard replica computations, the average over quenched disorder leads to the coupling of the originally independent copies of the system. The coupling among replicas suggests what are the global order parameters of the theory. In the present case, let us define the following overlap matrices
\begin{align} \label{Overlaps}
    \mQ_{ab} = \frac{1}{N} \sum_{j=1}^{N/2} \tau_j^a \overline{\tau}_j^b ~~~~~~ \Lambda_{ab}=\frac{1}{N}\sum_{p=1}^{N/2}\lambda_p^a\overline{\lambda}_p^b.
\end{align}
The right hand side of Eq.~\eqref{MF:AverDis} can be expressed in terms of the overlaps as follows
\begin{align} \label{TrGoverl}
    \exp\left[\frac{N}{2} \Tr \mG\left( \frac{\Omega^\dagger\Omega}{N^2} \right) \right] = \exp\left[\frac{N}{2} \Tr \mG\left(\mQ\Lambda \right) \right].
\end{align}
The previous relation can be shown for the expansion of of the function $\mG$, i.e.~for any integer power $K$ of its argument:
\begin{align*}
    \Tr \left( \frac{\Omega^\dagger \Omega}{N^2} \right)^K &=  \frac{1}{N^{2K}} \sum_{j_1=1}^{N/2} (\Omega^\dagger \Omega)^K_{j_1j_1}  \\
    &= \frac{1}{N^{2K}} \sum_{j_1,\dots,j_K}^{1,N/2} \sum_{p_1,\dots,p_K}^{1,N/2} \overline{\Omega}_{p_1 j_1}\Omega_{p_1 j_2} \cdots \overline{\Omega}_{p_K j_K} \Omega_{p_K j_1}  \\
    &= \frac{1}{N^{2K}} \sum_{a_1,\dots, a_K}^{1,n} \sum_{b_1,\dots, b_K}^{1,n} \sum_{j_1,\dots,j_K}^{1,N/2} \tau_{j_1}^{a_1} \overline{\tau}_{j_1}^{b_K} \cdots \tau_{j_K}^{a_K} \overline{\tau}_{j_K}^{b_{K-1}} \sum_{p_1,\dots,p_K}^{1,N/2} \overline{\lambda}_{p_1}^{a_1} \lambda_{p_1}^{b_1} \cdots \overline{\lambda}_{p_K}^{a_K} \lambda_{p_K}^{b_K} \\
    &= \sum_{a_1,\dots, a_K}^{1,n} \sum_{b_1,\dots, b_K}^{1,n}  \mQ_{a_1b_K} \overline{\Lambda}_{a_K b_K} \mQ_{a_K b_{K-1}} \cdots \mQ_{a_2 b_{1}}  \overline{\Lambda}_{a_1b_1} \\
    &= \Tr  \left(\mQ \Lambda\right)^K,
\end{align*}
where we have used the fact that $\Lambda_{ab} = \overline{\Lambda}_{ba}$.

In principle, both the matrices defined in Eq.~\eqref{Overlaps} are Hermitian; however, since the overlaps of the original spin variables $\sum_{j=1}^N s_j^{(a)}s_j^{(b)}$ are symmetric quantities, we can take both $\mQ$ and $\Lambda$ as real valued. The symmetry $\mQ_{ab}=\mQ_{ba}$ (and analogously $\Lambda_{ab}=\Lambda_{ba}$) is indeed preserved by any replica symmetry breaking ansatz\footnote{In a replica symmetry breaking ansatz it is not the symmetry of the matrices for index exchange to be broken, but the symmetry under the $n$-dimensional permutation group of replicas.}; so even if we carry on the computation for complex order parameters, we will end up with real valued matrices at the level of the saddle point equations. In Eq. \eqref{replPartFunc}, we change variables to the overlap matrix $\mQ$ through the following relations
\begin{align*}
  1 &= \prod_{a < b}^{1,n} \int \de \mQ_{ab}~\delta\left(N \mQ_{ab}-\sum_{j=1}^{N/2} \tau_j^a \overline{\tau}_j^b\right) \\
  & = \int \prod_{a < b}^{1,n} \de \mathcal{Q}_{ab} \int_{-i\infty}^{+i\infty} \prod_{a < b}^{1,n} \frac{N \de \mathcal{R}_{ab}}{2 \pi i}e^{\frac{1}{2}\sum_{a \neq b}^n \mathcal{R}_{ab} \left( N\mathcal{Q}_{ab}-\sum_r^{N/2}\tau_j^a\overline{\tau}_j^b \right)}
\end{align*}
for the off-diagonal terms and
\begin{equation*}
    1 = \prod_{a=1}^n \int \de \mathcal{Q}_{aa} \delta(N\mathcal{Q}_{aa}-N) = \int \prod_{a=1}^n \de \mathcal{Q}_{aa} \int_{-i\infty}^{+i\infty} \prod_{a=1}^n\frac{N \de \mathcal{R}_{aa}}{4 \pi i} e^{\frac{1}{2}\sum_a^n \mathcal{R}_{aa}(N\mathcal{Q}_{aa}-N)}
\end{equation*}
for the diagonal terms. Similarly, we use following relations for the overlap matrix $\Lambda$  
\begin{align*}
    1 &= \prod_{a < b}^{1,n} \int \de \Lambda_{ab}\delta\left(N\Lambda_{ab}-\sum_{p=1}^{N/2}\lambda_p^a\overline{\lambda}_p^b\right) \\
    & = \int \prod_{a < b}^{1,n} \de \Lambda_{ab} \int_{-i\infty}^{+i\infty} \prod_{a < b}^{1,n}    \frac{N \de \mathcal{M}_{ab}}{2 \pi i} e^{\frac{1}{2}\sum_{a \neq b}^n \mathcal{M}_{ab} \left( N\Lambda_{ab}-\sum_p^{N/2}\lambda_p^a\overline{\lambda}_p^b\right)}
\end{align*}
and
\begin{align*}
1 = \prod_{a=1}^n \int \de \Lambda_{aa} \delta\left(N\Lambda_{aa}-\sum_{p=1}^{N/2}|\lambda_p^a|^2\right) = &\int \prod_{a=1}^n \de \Lambda_{aa} \int_{-i\infty}^{+i\infty} \prod_{a=1}^n\frac{N \de \mathcal{M}_{aa}}{4 \pi i} \\
&\times e^{\frac{1}{2}\sum_a^n \mathcal{M}_{aa}(N\Lambda_{aa}-\sum_p^{N/2}|\lambda_p^a|^2)}.
\end{align*}
All the delta functions have been represented in terms of their Laplace transformations through the integration over the Lagrange multipliers $\mathcal{R}$ and $\mathcal{M}$ respectively for $\mathcal{Q}$ and $\Lambda$. The factors $1/2$ in front of the summations over the off-diagonal terms follow from the symmetry of the overlap matrices. By considering the previous relations all together and neglecting all constant prefactors and subleading terms, we have
\begin{align*}
    1 &= \int \prod_{a \leq b}^{1,n}  \de \mQ_{ab} \de  \Lambda_{ab} \int_{-i\infty}^{+i\infty} \prod_{a \leq b}^{1,n} \de \mathcal{R}_{ab} \de \mathcal{M}_{ab} \\ &\times \exp\left[\frac{N}{2}\Tr(\mathcal{R}\mathcal{Q})+ \frac{N}{2} \Tr(\mathcal{M}\Lambda)-\frac{1}{2}\sum_{j=1}^{N/2}\sum_{ab}^n \tau_j^a\mathcal{R}_{ab}\overline{\tau}_j^b - \frac{1}{2}\sum_{p=1}^{N/2}\sum_{ab}^n \lambda_p^a\mathcal{M}_{ab}\overline{\lambda}_p^b  \right]
\end{align*}
Therefore, by looking back at Eq.~\eqref{replPartFunc} and using the previous relation together with the result in Eq.~\eqref{TrGoverl}, the averaged replicated partition function reads as
\begin{equation}
\begin{split}
    \overline{\mZ^n_U} &= \int \mathcal{D}C \mathcal{D}\overline{C} \mathcal{D}\lambda \mathcal{D}\overline{\lambda}D\mathcal{Q}D\mathcal{R}D\Lambda D\mathcal{M} \exp \frac{N}{2} \left[\Tr(\mathcal{R}\mathcal{Q})+ \Tr(\mathcal{M}\Lambda) +  \Tr \mG\left(\mQ \Lambda \right)  \right] \\
    & \times \exp\left[-\beta \sum_{a=1}^n\sum_{p=1}^{N/2} |C_p^a|^4 + \sum_{a=1}^n\sum_{p=1}^{N/2} \left(i C_p^a \overline{\lambda}_p^a - i \overline{C}_p^a \lambda_p^a \right) - \frac{1}{2}\sum_{p=1}^{N/2}\sum_{ab}^n \lambda_p^a\mathcal{M}_{ab}\overline{\lambda}_p^b \right] \\
    & \times  \prod_{a=1}^n\left[\sum_{\bm{\tau}^a}\right] \exp\left[ -\frac{1}{2} \sum_{ab}^n \sum_{j=1}^{N/2} \tau_j^a\mathcal{R}_{ab}\overline{\tau}_j^b \right],
\end{split}
\end{equation}
where in order to shorten our notation the integration measures in the global $X=\{\mathcal{Q},\Lambda,\mR,\mM\}$ and local variables $x=\{C,\lambda\}$ have been written respectively as
\begin{equation*}
    DX=\prod_{a \leq b}^{1,n} \de X_{ab} ~~~~~~~ \mathcal{D} x= \prod_{j=1}^{N/2} \prod_{a=1}^n \de x_j^a.
\end{equation*}

The expression of the partition function can be simplified by performing the complex Gaussian integral in the Lagrange multipliers $\bm{\lambda}$. We notice that all the terms involving the variables $\bm{\lambda}$ are diagonal with respect to the local unitary-transformed space index $p$. Hence, the dependence on $\bm{\lambda}$ can be factorized in $N/2$ Gaussian integrals. Each of these integrals can be compactly written in a vector formalism for the replica indices and has the following solution 
\begin{align*}
    \int \de \bm{\lambda}_p \de \overline{\bm{\lambda}}_p \exp\left[ - \overline{\bm{\lambda}}_p \frac{\mathcal{M}}{2}\bm{\lambda}_p+i \overline{\bm{\lambda}}_p \cdot\bm{C}_p-i \overline{\bm{C}}_p \cdot\bm{\lambda}_p \right] &= (2\pi)^n \det(2 \mM^{-1}) \exp\left[2 \overline{\bm{C}}_p \mM^{-1} \bm{C}_p \right] \\
    & = \exp \left[ - \log \det \mM + 2 \overline{\bm{C}}_p \mM^{-1} \bm{C}_p \right].
\end{align*}
To obtain this result one has to complete the square with the change of variables $\bm{\lambda}_p \rightarrow \bm{\lambda}_p + 2i \mM^{-1} \bm{C}_p$ and $\overline{\bm{\lambda}}_p \rightarrow \overline{\bm{\lambda}}_p - 2i  \overline{\bm{C}}_p \mM^{-1}$. Moreover, in the last expression we neglected an overall constant additive terms $(4\pi)^n$. Since we have $N/2$ of these contributions, the partition function reads as
\begin{equation}
\begin{split}
    \overline{\mZ^n_U} &= \int D\mathcal{Q}D\mathcal{R}D\Lambda D\mathcal{M} \exp \frac{N}{2} \left[\Tr(\mathcal{R}\mathcal{Q})+ \Tr(\mathcal{M}\Lambda) +  \Tr \mG\left(\mQ\Lambda\right) - \Tr \log  \mM  \right] \\
    & \times \int \mathcal{D}C \mathcal{D}\overline{C} \exp\left[-\beta \sum_{a=1}^n\sum_{p=1}^{N/2} |C_p^a|^4 + 2 \sum_{ab}^n\sum_{p=1}^{N/2} C_p^a \mM^{-1}_{ab} \overline{C}_p^b  \right] \\
    & \times  \prod_{a=1}^n\left[\sum_{\bm{\tau}^a}\right] \exp\left[ -\frac{1}{2} \sum_{ab}^n \sum_{j=1}^{N/2} \tau_j^a\mathcal{R}_{ab}\overline{\tau}_j^b \right],
\end{split}
\end{equation}
where we have used the relation $\log\det \mM = \Tr\log \mM$. The first line of the previous equation contains entropic contributions in terms of the global order parameters and their Lagrange multipliers, whereas the second and third line correspond to the local contributions obtained by tracing respectively over the unitary-transformed variables $\bm{C}$ and the complex spins $\bm{\tau}$. At this point of the computation the dependence on the local indices both of the direct and of the unitary-transformed space can be factorized in $N/2$ equivalent contributions. By defining the following local free energies 
\begin{gather} \label{localFreeEn}
    f_C(\mM) = \log \int \prod_{a=1}^{n} \left[ \de C_a \de \overline{C}_a \right] \exp\left[-\beta \sum_{a=1}^n |C^a|^4 + 2 \sum_{ab}^nC^a \mathcal{M}_{ab}^{-1}\overline{C}^b  \right], \\
    f_\tau(\mR)= \log \prod_{a=1}^n \left[\sum_{ \{\tau^a \}}\right] \exp\left[ -\frac{1}{2}\sum_{ab}^n\tau^a\mathcal{R}_{ab}\overline{\tau}^b \right],
\end{gather}
the replicated partition function averaged over disorder eventually reads as
\begin{equation} \label{replPartFunc2}
    \overline{\mZ^n_U} = \int D\mathcal{Q} D\mathcal{R} D\Lambda D\mathcal{M} \exp \left[N S(\mQ,\mR,\Lambda,\mM)\right]
\end{equation}
where, after singling out the overall factor $N$, we have defined the action density
\begin{equation} \label{action}
\begin{split}
    S(\mQ,\mR,\Lambda,\mM) &= \frac{1}{2} \{ f_\tau(\mathcal{R}) + f_C(\mathcal{M}) + \Tr(\mathcal{R}\mathcal{Q})+ \Tr(\mathcal{M}\Lambda) \\
    &\quad +  \Tr \mG\left(\mQ \Lambda \right) - \Tr \log  \mM \}.
\end{split}
\end{equation}
The subscripts $\tau$ and $C$ in the notation for the local free energies are just reminders of the variables over which the trace is taken.

\subsection{Reduced Theory}
The free energy Eq.~\eqref{freeEn1} is determined by the stationary point of the action function derived in the previous section, which has to be computed through the saddle point method in the large-$N$ limit and evaluated in the limit $n \rightarrow 0$:
\begin{align}
f(\beta)= -\frac{1}{\beta} \lim_{n \rightarrow 0} \frac{S_{\text{sp}}}{n},
\end{align}
where $S_{\text{sp}}$ is a shorthand notation for the action computed in the solution of the saddle point equations. In order to solve the optimization problem, it is convenient to eliminate some of the variables that have been introduced along the computation. From Eqs.~\eqref{replPartFunc2} and~\eqref{action} it is easy to derive the saddle point equations, by using well-known matrix identities \cite{Petersen12}. The full set of saddle point equations for the action $S$ reads as
\begin{subequations}
\begin{gather}
\frac{\partial S}{\partial\mathcal{R}_{ab}}= \frac{\partial f_\tau (\mathcal{R})}{\partial\mathcal{R}_{ab}}+\mathcal{Q}_{ab} =0 \label{SPE1} \\
\frac{\partial S}{\partial\Lambda_{ab}}= \mM_{ab}+[\mathcal{Q}\mathcal{G}'(\mathcal{Q}\Lambda)]_{ab} =0 \label{SPE2} \\
\frac{\partial S}{\partial\mathcal{Q}_{ab}}= \mathcal{R}_{ab}+[\Lambda\mathcal{G}'(\mathcal{Q}\Lambda)]_{ab} =0  \label{SPE3} \\
\frac{\partial S}{\partial \mM_{ab}}=\frac{\partial f_C(\mM)}{\partial \mM_{ab}} -\mM^{-1}_{ab}+\Lambda_{ab} =0, \label{SPE4}
\end{gather}
\end{subequations}
where $\mathcal{G}'$ formally denotes the derivative of the function $\mathcal{G}$ with respect to its argument. In the following, we adopt the matrix formalism to shorten our notation. 

The main idea to derive the reduced theory is to manipulate the saddle point equations in order to eliminate the matrix $\Lambda$ from the theory. A key ingredient in this procedure is the fact that the function $\mathcal{G}$ satisfies the following relation
\begin{align} \label{G'rel}
(\mathcal{G}'(z))^2=z^{-1}\left(\frac{1}{4}- \mathcal{G}'(z)\right),
\end{align}
which can be checked \emph{a posteriori} by direct substitution of the expression of $\mG$, Eq.~\eqref{MF:G}. In fact, the previous relation can be seen as the ordinary differential equation that defines the function $\mG$: this equation was derived in Ref.~\cite{Brezin80b} to find the solution of the integration over the unitary group and has been reported in Eq.~\eqref{ODEforG}. In the following, we will assume that all the matrices commute: this can be justified in view of the fact that the saddle point value of the action with respect to variations of these matrices will be computed by restricting the optimization on the subspaces of RS or RSB (in the Parisi scheme) matrices which are all subspaces of commuting matrices. From the saddle point equations Eqs.~\eqref{SPE2} and~\eqref{SPE3} one finds
\begin{gather*}
\mathcal{G}'(\mathcal{Q}\Lambda)=-\mathcal{Q}^{-1}\mM \\
\mathcal{G}'(\mathcal{Q}\Lambda)=-\Lambda^{-1}\mathcal{R},
\end{gather*}
from which, by subtracting them, one finds an expression of $\Lambda$ in terms of the other matrices
\begin{align} \label{Lambda}
\Lambda=\mathcal{Q}\mathcal{R}\mM^{-1} .
\end{align}
By plugging the $\Lambda$-independent expression of $\mathcal{G}'$ into Eq.~\eqref{G'rel} and using Eq.~\eqref{Lambda} one gets after some algebra the relation
\begin{align*}
\mathcal{R}\mathcal{Q} -\frac{1}{4}\mathcal{Q}\mM^{-1}= I,
\end{align*}
which, with the change of variables $\mathcal{M} \rightarrow \mM^{-1}/4$, yields the relation
\begin{equation} \label{SPEnew}
(\mathcal{R}-\mathcal{M})\mathcal{Q}= I.
\end{equation}
It is worth stressing that the change of variables performed does not modify the integration measure of the partition function up to a subleading term in the large-$N$ limit. Eq.~\eqref{SPEnew} is an algebraic relation that connects the saddle point value of the overlap matrix to those of the Lagrange multipliers: hence, it can be viewed as a constraint to a new (reduced) system of saddle point equations. The saddle point equations Eq.~\eqref{SPE2} and~\eqref{SPE3} can be then substituted by the constraint Eq.~\eqref{SPEnew}. Finally, by modifying the saddle point equation~\eqref{SPE4} in order to eliminate $\Lambda$ and substituting $\mM$ with $\mM^{-1}/4$, the reduced set of saddle point equations reads as
\begin{subequations}
\begin{gather} 
\frac{\partial f_\tau(\mathcal{R})}{\partial\mathcal{R}_{ab}}+\mathcal{Q}_{ab} =0 \label{RedSPE1} \\
-\frac{\partial f_C(\mathcal{M})}{\partial \mathcal{M}_{ab}} + \mathcal{Q}_{ab} =0 \label{RedSPE2} \\
(\mathcal{R}-\mathcal{M})\mathcal{Q} = I.  \label{AlgConstr}
\end{gather}
\end{subequations}

The reduced theory can be now induced from the previous set of saddle point equations, which in fact can be derived by applying the saddle point method to the following partition function
\begin{align}
\mathcal{Z} = \int D\mathcal{R}D\mathcal{M}  \exp\left[N A(\mathcal{R},\mathcal{M}) \right],
\end{align}
where
\begin{align} \label{Action2}
A(\mathcal{R},\mathcal{M})= \frac{1}{2} \left[f_\tau(\mathcal{R})+f_C(\mathcal{M}) + \Tr\log(\mathcal{R}-\mathcal{M})\right]
\end{align}
and
\begin{gather}
    f_C(\mathcal{M}) = \log \int \prod_{a=1}^{n} \left[\de C_a \de \overline{C}_a \right] \exp\left[-\beta \sum_{a=1}^n |C^a|^4 + \frac{1}{2} \sum_{ab}^nC^a \mathcal{M}_{ab}\overline{C}^b  \right] \label{localFreeEnC} \\
    f_\tau(\mathcal{R})= \log \prod_{a=1}^n \left[\sum_{ \{\tau^a \}}\right] \exp\left[ -\frac{1}{2}\sum_{ab}^n\tau^a\mathcal{R}_{ab}\overline{\tau}^b \right]. \label{localFreeEntau}
\end{gather}
In the redefinition of the first local free energy a numerical factor coming from the change of variables $\mathcal{M} \rightarrow \mM^{-1}/4$ has been absorbed in the integration variables $\bm{C},\overline{\bm{C}}$ and the temperature has been rescaled accordingly. On the other hand, the second local free energy has remained untouched and has been reported here only for the sake of completeness. The saddle point equations~\eqref{RedSPE1} and \eqref{RedSPE2} have a very intuitive physical meaning: they fix the value of the overlap matrix to the thermal average of the product of two different replicas. This can be seen by computing explicitly the derivatives of the local free energies Eqs.~\eqref{localFreeEnC} and \eqref{localFreeEntau}, which yield 
\begin{gather}
\langle \tau^a\overline{\tau}^b \rangle_\tau = 2 \mathcal{Q}_{ab} \\
\langle C^a\overline{C}^b \rangle_C = 2 \mathcal{Q}_{ab},
\end{gather}
where the averages induced by the two free energies are defined as
\begin{align*}
\langle (\cdots) \rangle_\tau = \frac{\prod_{a=1}^n \left[\sum_{ \{\tau^a \}}\right]e^{-\frac{1}{2}\sum_{ab}^n\tau^a\mathcal{R}_{ab}\overline{\tau}^b} (\cdots)}{\prod_{a=1}^n \left[\sum_{ \{\tau^a \}}\right] e^{-\frac{1}{2}\sum_{ab}^n\tau^a\mathcal{R}_{ab}\overline{\tau}^b}}
\end{align*}
and 
\begin{align*}
\langle (\cdots) \rangle_C = \frac{\int \prod_{a=1}^{n} \left[\de C_a \de \overline{C}_a\right] e^{-\beta \sum_a^n |C^a|^4 +\frac{1}{2}\sum_{ab}^nC^a\mathcal{M}_{ab}\overline{C}^b} (\cdots)}{\int \prod_{a=1}^{n} \left[\de C_a \de \overline{C}_a\right] e^{-\beta \sum_a^n |C^a|^4 +\frac{1}{2}\sum_{ab}^nC^a\mathcal{M}_{ab}\overline{C}^b}}.
\end{align*}
Finally, the thermodynamic free energy of the system is given by 
\begin{align} \label{freeEn2}
    f(\beta)= -\frac{1}{\beta} \lim_{n \rightarrow 0} \frac{A_{\text{sp}}}{n},
\end{align}
where $A_{\text{sp}}$ is the action computed in the solution of the reduced saddle point equations \eqref{RedSPE1} and \eqref{RedSPE2} with the algebraic constraint \eqref{AlgConstr}. 

There is an important remark that has to be made regarding this procedure. The reduced theory is not completely equivalent to the original one: the action defined in Eq.~\eqref{Action2} has not been derived through manipulations of the original one Eq.~\eqref{action}, it has rather been guessed from the saddle point equations. For this reason we have chosen a different notation for the two quantities. The identities exploited in order to define the new action function are satisfied only at the saddle point, which however yield the correct free energy in the large $N$ limit. Thus, the reduced theory is expected to correctly reproduce the thermodynamics of the original theory.

\section{Discussion of the results}
\begin{figure}[t]
     \centering
     \includegraphics[width=.9\linewidth]{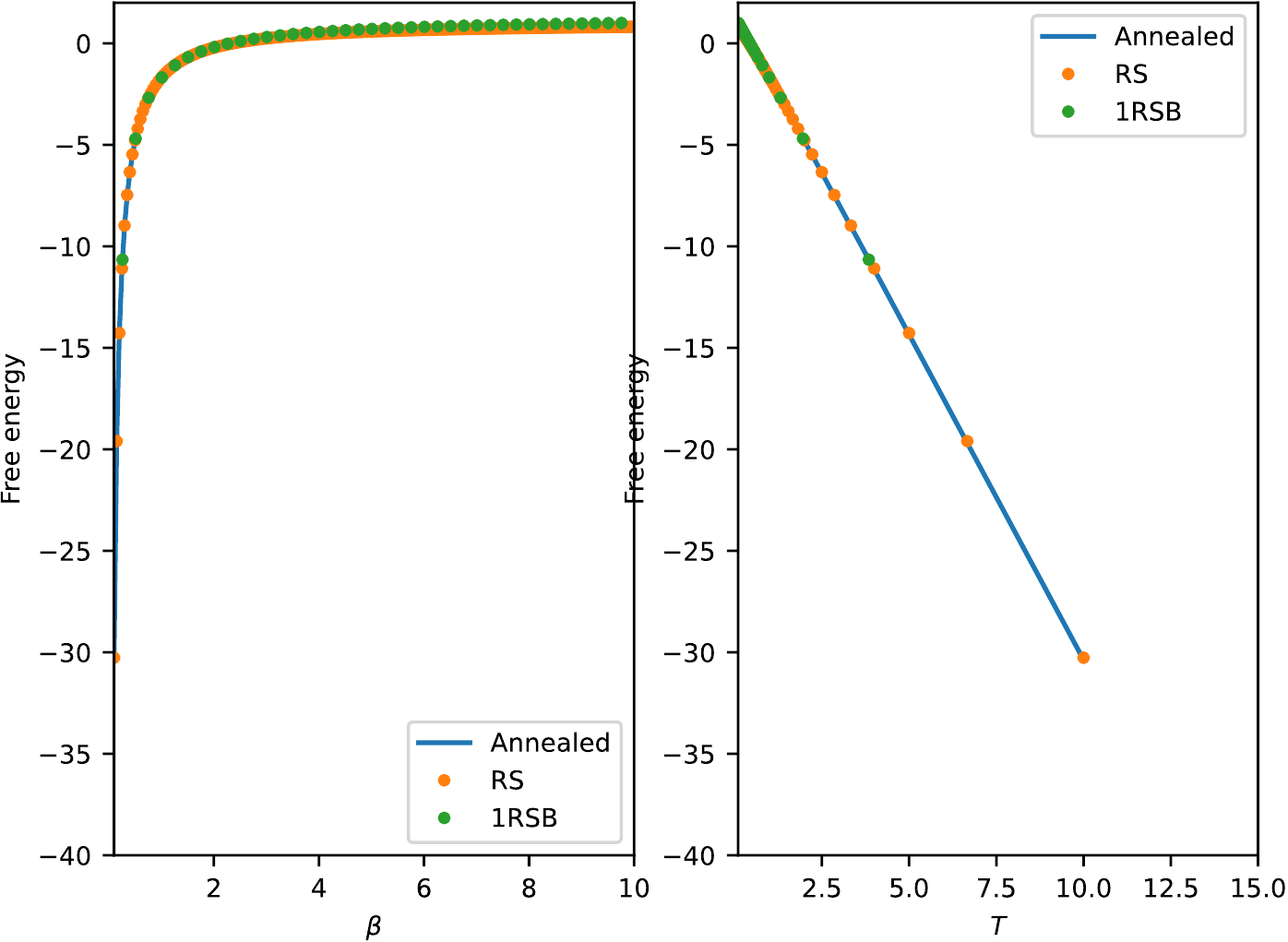}
     \caption{Free energy of the Random Unitary Model for the Merit Factor problem. Left panel: data are plotted as a function of the inverse temperature $\beta$; Right panel: data are plotted as a function of temperature $T$. Blue: free energy of the annealed solution; Orange: Replica Symmetric free energy; Green: one step Replica Symmetry Breaking free energy. From all the computations, the same result is obtained, corresponding to the paramagnetic state of the system.} \label{fig:MFsolutions}
\end{figure}

Before entering into the details of the analysis, we anticipate the discussion of the results in this section for the convenience of the reader. We have performed three kinds of computations: the annealed, the Replica Symmetric (RS) and the one step Replica Symmetry Breaking (1RSB) one. The annealed limit is obtained by considering $n=1$ in the previous equations and yields the paramagnetic solution of the model. The RS and 1RSB computations are based on different parametrizations of the replica matrices: if a phase transition from the paramagnetic phase to another phase occurs, we expect some of the off-diagonal parameters of these matrices to become non-vanishing and starting increasing by further cooling the system. However, contrary to expectations, from the numerical study of the model, neither the RS nor the 1RSB computations give a different solution from the paramagnetic one at finite temperature, up to the accuracy of our analysis. In Fig.~\ref{fig:MFsolutions} we display the free energy of the model both as a function of temperature and of its inverse, obtained from the solution of the saddle-point equations pertaining to each one of the three computations: data fall on the same curve, corresponding to the paramagnetic state.

The fact that the RS solution does not differ from the annealed one reminds of the situation found in the spherical $p$-spin model, where no trace of the glass transition is found at the RS level, i.e.~the off-diagonal element $q_0$ of the overlap matrix is zero at every temperature. Conversely, in the SK model one can observe a phase transition already within the RS ansatz, since a non-vanishing value of $q_0$ can be found at finite temperature. However, this solution becomes unstable on the de-Almeida-Thouless line \cite{deAlmeida78} and, also, leads to a low temperature negative entropy. In the spherical $p$-spin model the RS solution describing the paramagnetic state is always stable, a typical feature of first-order phase transitions, and a negative entropy is not physically inconsistent with continuous variables; nevertheless, the 1RSB solution, with the overlap value of the diagonal blocks $q_1 \neq 0$, gives a higher free energy, which in the replica method means that the 1RSB ansatz yields the thermodynamically dominant phase. If our analysis is correct, the lack of a similar scenario in the Random Unitary Model for the Merit Factor problem leads to the following possible situations: (\emph{i}) the glass transition occurs at zero temperature with a 1RSB ansatz; (\emph{ii}) the transition occurs at finite temperature, but with a different kind of replica symmetry breaking ansatz from the 1RSB one; (\emph{iii}) there is no transition at all in the model with random unitary matrices.

In order to test the first hypothesis, we have to compute the zero-temperature limit of the 1RSB saddle-point equations, solve them and check whether the asymptotic value of the free energy is greater than that of the paramagnetic solution. This is still work in progress, see Sec.~\ref{sec:0temp}. However, it would be very surprising to discover that in this model the addition of an infinitesimal thermal noise would destroy the alleged zero-temperature transition. A useful indication regarding the second hypothesis may come from the study of the stability of the RS solution: proving that the RS replicon is always positive definite at finite temperature can be considered as an evidence in favor of the $p$-spin model scenario. 

In fact, as a result of our analysis, we are led to reconsider the mapping from the original Bernasconi model, with Hamiltonian \eqref{MF:HamiltPeriod} and whose phebomenology seems compatible with a glass transition, to the model with random unitary matrices, which may be not under control. Therefore, the third scenario may be the most reasonable one:  the correct mapping to a disordered model should involve two random orthogonal transformation (see Eq. \eqref{doubleROM}), one for the real and one for the imaginary part of the complex spins $\tau$, instead of a single unitary transform. In order to test this hypothesis, we are performing analytical and numerical studies of the original models, of the model with random orthogonal matrices and of the random unitary model.

\section{Annealed Limit}
In this section we focus on the annealed limit, which yields a great simplification of the theory from the mathematical point of view, since it amounts to consider numbers instead of matrices. The two local free energies boil down to
\begin{gather}
f_\tau (\mR) = \log\sum_{\{\tau\}}e^{-\frac{1}{2}|\tau|^2 \mathcal{R}} = \log4 - \mathcal{R} \\
f_C(\mathcal{M}) =\log \int \de C \de \overline{C} e^{-\beta |C|^4 + \frac{\mathcal{M}}{2}|C|^2}. \label{fcANN}
\end{gather}
In the first expression the term $\log4$ is in place of the usual $\log 2$ for binary variables, since the sum over the configurations of $\tau$ runs over $4$ possible values. However, this fact is compensated by the factor $1/2$ in the definition of the action, which takes into account the correct number of degrees of freedom. Hence, the action reads
\begin{equation} \label{AnnealedAction}
A_{\text{ann}}=\log2 - \frac{\mR}{2} + \frac{1}{2}\log \int \de C \de \overline{C} e^{-\beta |C|^4 + \frac{\mathcal{M}}{2}|C|^2} + \frac{1}{2} \log(\mR- \mathcal{M})
\end{equation}
and the only value of the overlap $q$ is connected to $\mathcal{R}$ and $\mathcal{M}$ through the simple relation
\begin{equation} \label{qRMAnn}
q=\frac{1}{\mR- \mathcal{M}},
\end{equation}
which is the scalar version of the algebraic constraint Eq.~\eqref{AlgConstr}. The saddle point equations can be derived straightforwardly. The first one is simply $\mR-\mM =1$, which, by using Eq.~\eqref{qRMAnn}, gives $q=1$ consistently with the expectation for the annealed case. The other equation, which determines the value of $\mathcal{M}$, is 
\begin{equation} \label{AnnSPeq}
\langle |C|^2 \rangle_C = 2,
\end{equation}
where the average is performed with the probability measure induced by the local free energy $f_C$ Eq.~\eqref{fcANN}.

The annealed limit shows that, from the technical point of view, the theory is not trivial, since even in this simple case we cannot obtain an analytical solution. This is mainly due to the quartic measure, which characterizes the free energy integrated in the unitary-transformed variables. The integrals appearing in the action and in the equation for $\mM$ can be simplified with some change of variables and cast into truncated Gaussian integrals, yielding error functions which have to be computed numerically. By passing to polar coordinates in the complex plane we have
\begin{equation*}
\int \de C \de \overline{C} e^{-\beta |C|^4 + \frac{\mathcal{M}}{2}|C|^2} = 4 \pi \int_0^\infty \de r r e^{-\beta r^4 + \frac{\mathcal{M}}{2} r^2} = 4 \pi\frac{e^{\frac{\mathcal{M}^2}{16\beta}}\sqrt{\pi}}{4\sqrt{\beta}}\left(1+\erf\left(\frac{\mathcal{M}}{4\sqrt{\beta}}\right)\right)
\end{equation*}
and 
\begin{align*}
\int \de C \de \overline{C} e^{-\beta |C|^4 + \frac{\mathcal{M}}{2}|C|^2}|C|^2 &= 4 \pi \int_0^\infty \de r r^3 e^{-\beta r^4 + \frac{\mathcal{M}}{2} r^2}  \\
&=4 \pi \frac{4\sqrt{\beta} + e^{\frac{\mathcal{M}^2}{16\beta}}\sqrt{\pi}\mathcal{M}\left(1+\erf\left(\frac{\mathcal{M}}{4\sqrt{\beta}}\right)\right) }{16 \beta^{\frac{3}{2}}},
\end{align*}
where the final result can be obtained by changing variables to $u=r^2$. Eventually, Eq.~\eqref{AnnSPeq} becomes
\begin{equation} \label{AnnSPeq2}
\frac{\mathcal{M}}{4\beta}+\frac{e^{-\frac{\mathcal{M}^2}{16\beta}}}{\sqrt{\pi \beta}\left(1+\erf\left(\frac{\mathcal{M}}{4\sqrt{\beta}}\right)\right)} = 2,
\end{equation}
and can be solved numerically. The free energy is given by
\begin{equation}
f_{\text{ann}}(\beta)= -\frac{1}{\beta} A_{\text{ann}}
\end{equation}
where $A$ is computed over the solutions of the saddle point equations. Due to Eq.~\eqref{qRMAnn} and to the fact that $q=1$, the logarithm in $A$ vanishes and $\mR$ can be expressed in terms of $\mathcal{M}$. Hence, we have
\begin{equation}
f_{\text{ann}}(\beta)= -\frac{1}{\beta}\left\{\log 2 - \frac{\mathcal{M}+1}{2} + \frac{1}{2} \log \left( \pi^{3/2} \frac{e^{\frac{\mathcal{M}^2}{16\beta}}}{\sqrt{\beta}} \right) + \frac{1}{2} \log \left(1+\erf\left(\frac{\mathcal{M}}{4\sqrt{\beta}}\right)\right) \right\},
\end{equation}
where $\mathcal{M}=\mathcal{M}(\beta)$ is given by the solution of Eq.~\eqref{AnnSPeq2} for each value of $\beta$.

\section{Replica Symmetric Ansatz}
In this section we perform a replica symmetric (RS) ansatz for the solution of the saddle point equations. We consider the following parametrizations for the global order parameters
\begin{subequations} \label{RSansatz}
    \begin{gather}
    \mathcal{Q}_{ab} = \delta_{ab} + q_0(1-\delta_{ab}) \\
    \mathcal{R}_{ab} = \mR_D \delta_{ab} + \mR_0(1-\delta_{ab}) \\
    \mathcal{M}_{ab} = \mM_D \delta_{ab} + \mM_0(1-\delta_{ab}),
    \end{gather}
\end{subequations}
where the diagonal elements of $\mathcal{Q}$ are fixed to one due to the fact that $|\tau^a|^2=2$. The action in Eq.~\eqref{Action2} has to be expressed in terms of the parameters of the RS matrices. In order to lighten the exposition, here we simply limit ourselves to reporting the results and we refer to Appendix \ref{app:RSMF} for the details of the computations. 

To shorten the notation we introduce the function
\begin{align} \label{gbeta0}
    g_{\beta,0}(C|x,y,z) = e^{-\beta |C|^4 + \frac{1}{2} (x-y)|C|^2 + \sqrt{y} \Re(C \overline{z})},
\end{align}
where $x,y \in \mathbb{R}$ and $z,C \in \mathbb{C}$. By integrating $g_{\beta, 0}$ with respect to $C$ we have the function
\begin{align} \label{Ibeta0}
    I_{\beta,0}(x,y,z) = \int \de C \de \overline{C} e^{-\beta |C|^4 + \frac{1}{2}(x-y)|C|^2 + \sqrt{y} \Re(C \overline{z})},
\end{align}
that is a local partition function (related to the local free energy \eqref{localFreeEnC}), which plays exactly the same role played by the $\cosh$ function, which will appear in the other local free energy \eqref{localFreeEntau}, after tracing over the discrete spins. However, in this case, we are not able to reduce it further due to the quartic term in the exponential. The function
\begin{equation} \label{P_g_C}
    \mathcal{P}_{\beta,0}(C|x,y,z) = \frac{g_{\beta,0}(C|x,y,z)}{I_{\beta,0}(x,y,z)}
\end{equation}
defines a probability measure for the unitary transforms $C$ of the complex spin variables $\tau$.

Eventually, the RS action reads
\begin{equation} \label{RSaction}
    \lim_{n \rightarrow 0} \frac{2}{n} A_{\text{RS}} = f_\tau(\mR_0) + f_C(\mM_D,\mM_0) + s_0(\mR_D, \mR_0, \mM_D, \mM_0)
\end{equation}
where the $\mathcal{O}(n)$ expressions of the local free energies are
\begin{gather}
    f_\tau(\mR_D, \mR_0) =  \log 4 - (\mathcal{R}_D - \mathcal{R}_0) + 2 \int\frac{\de h}{\sqrt{2\pi}} e^{-h^2/2} \log \cosh (\sqrt{-\mathcal{R}_0}h) \\
    f_C(\mM_D,\mM_0) = \int \frac{\de z \de \overline{z}}{4\pi} e^{-|z|^2/2} \log I_{\beta,0}(\mM_D, \mM_0, z),
\end{gather}
and the $\mathcal{O}(n)$ expression of the entropic term has been stored into the function 
\begin{equation} \label{ENTRO-MERIT-RS}
s_0(\mR_D, \mR_0, \mM_D, \mM_0) = \log(\mathcal{R}_D -\mathcal{M}_D- \mathcal{R}_0 + \mathcal{M}_0 ) + \frac{\mathcal{R}_0 - \mathcal{M}_0}{\mathcal{R}_D -\mathcal{M}_D- \mathcal{R}_0 + \mathcal{M}_0}.
\end{equation}
Plugging Eq.~\eqref{RSaction} into Eq.~\eqref{freeEn2} yields the RS free energy of the model.

We now have to find self-consistency equations for $\mM_D,\mM_0, \mR_D,\mR_0$, whose solution will give the dynamics of the parameters with respect to temperature variations, which is necessary to evaluate the free energy at each value of $\beta$. These equations can be derived by imposing the vanishing of the RS action gradient components, as shown in detail in Appendix \ref{app:RSMF}. Eventually, the RS self-consistency equations are given by
\begin{subequations}
    \begin{gather}
    q_0 = \int\frac{\de h}{\sqrt{2\pi} }e^{-h^2/2} \tanh^2(\sqrt{-\mathcal{R}_0}h) \label{RS1} \\
    q_0 = \frac{1}{2} \int \frac{\de z \de \overline{z}}{4\pi} e^{-|z|^2/2}  |\langle C \rangle_{0}|^2  \label{RS2} \\
    1 = \frac{1}{2} \int \frac{\de z \de \overline{z}}{4\pi} e^{-|z|^2/2}  \langle |C^2| \rangle_{0}  \label{RS3}
\end{gather}
\end{subequations}
where  
\begin{equation} \label{RSaver}
\begin{split}
        \langle (\cdots) \rangle_{0} &= \int \de C~\de \overline{C}~ \mathcal{P}_{\beta,0}(C | \mM_D,\mM_0,z) (\cdots) \\
    &= \frac{\int \de C~\de \overline{C}~g_{\beta,0}(C | \mM_D,\mM_0, z) (\cdots) }{I_{\beta,0}(\mM_D,\mM_0,z)},
\end{split}
\end{equation}
is the average defined over the probability measure $\mathcal{P}_{\beta,0}(C|\mM_D,\mM_0,z)$, see Eq.~\eqref{P_g_C},  and, therefore, is a function of the complex Gaussian variable $z$ and the saddle point parameters $\mM_{D,0}$. The system of RS equations is completed by the relations
\begin{subequations} \label{AlgEqs}
\begin{gather}
    \mathcal{R}_D-\mathcal{M}_D -(\mathcal{R}_0-\mathcal{M}_0)q_0=1  \label{AlgEq1}   \\
    (\mathcal{R}_0-\mathcal{M}_0)(1-2q_0)+ (\mathcal{R}_D-\mathcal{M}_D)q_0=0.  \label{AlgEq2}
\end{gather}
\end{subequations}
which follow from the RS expression of the algebraic relation \eqref{AlgConstr}. We notice that with the above saddle point equations the entropy \eqref{ENTRO-MERIT-RS} can be rewritten as
$$s_0(q_0)=-\ln(1-q_0)-\frac{q_0}{1-q_0},$$
which clearly vanishes for $q_0=0$. One important remark: if one sets $\mR_0 = \mM_0 = q_0 = 0$ in the RS action \eqref{RSaction}, one recovers the annealed action \eqref{AnnealedAction}.

\subsubsection{Further manipulations}
It is convenient, especially in view of the numerical analysis of the previous set of equations, to write every complex variable in terms of its real and imaginary parts. To have an even lighter notation, let us denote the integral over a real Gaussian variable $x$ as
\begin{align*}
    \mathscr{D} x = \frac{dx }{\sqrt{2\pi}}e^{-x^2/2}.
\end{align*}
We use the following notation: $z=\rho+i \sigma$ and $C=a+ib$. Therefore, we have
\begin{align*}
    \int \frac{\de z \de \overline{z}}{4\pi} e^{-|z|^2/2} \langle (\cdots) \rangle_{0} = \int   \mathscr{D} \rho \mathscr{D} \sigma  \langle (\cdots) \rangle_{0}
\end{align*}
where a factor $1/2$ is cancelled by the modulus of the Jacobian of the transformation $z,\overline{z}\rightarrow \sigma,\rho$. The average induced by the measure $\mathcal{P}_{\beta,0}(a,b|\mM_D,\mM_0,z)$ now reads
\begin{equation}
 \langle (\cdots) \rangle_{0} = \int \de a\  \de b~(\cdots)~\mathcal P(a,b|\mathcal M_D,\mathcal M_0,z)  
 = \frac{\int \de a \ \de b ~(\cdots)~ g_{\beta,0}(a,b|\mathcal M_D,\mathcal M_0,\rho, \sigma)  }{I_{\beta,0}(\mathcal M_D,\mathcal M_0,\rho, \sigma)}
\end{equation}
where in this case the Jacobians of the transformations $C,\overline{C} \rightarrow a,b$ cancel out between numerator and denominator and 
\begin{gather*}
   g_{\beta,0}( a, b|x,y,\rho, \sigma,) = \exp\left[-\beta (a^2+b^2)^2 +\frac{1}{2}(x-y)(a^2+b^2)+\sqrt{y}(a\sigma + b\rho) \right] \\
   I_{\beta,0}(x,y,\rho,\sigma) = \int \de a~\de b~ g_{\beta,0}(a, b|x,y,\rho, \sigma).
\end{gather*}
In this notation, we have
\begin{equation*}
\begin{split}
\int \frac{\de z \de \overline{z}}{4\pi} e^{-|z|^2/2} \langle |C|^2 \rangle_{0} &= \int \mathscr{D} \rho \mathscr{D} \sigma  ~\langle a^2 \rangle_{0} + \int \mathscr{D} \rho \mathscr{D} \sigma~\langle b^2 \rangle_{0} \\
&= 2 \int \mathscr{D} \rho \mathscr{D} \sigma ~\langle a^2 \rangle_{0},
\end{split}
\end{equation*}
where the last identity holds because the two integrals are equal under the simultaneous changes of variables $\rho \leftrightarrow \sigma$ and $a \leftrightarrow b$. Similarly, for the other expectation value in the self-consistency equations, we have
\begin{equation*}
\begin{split}
\int \frac{\de z \de \overline{z}}{4\pi} e^{-|z|^2/2} | \langle C \rangle_{0}|^2  &= \int \frac{\de z \de \overline{z}}{4\pi} e^{-|z|^2/2} (\Re \langle C \rangle_{0})^2 + \int \frac{\de z \de \overline{z}}{4\pi} e^{-|z|^2/2}(\Im \langle C \rangle_{0})^2 \\
&= \int    \mathscr{D} \rho \mathscr{D} \sigma ~\langle a \rangle_{0}^2 + \int   \mathscr{D} \rho \mathscr{D} \sigma ~\langle b \rangle_{0}^2 \\
&=2 \int    \mathscr{D} \rho \mathscr{D} \sigma ~\langle a \rangle_{0}^2,
\end{split}
\end{equation*}
where the last identity is again due to the changes of variables $\rho \leftrightarrow \sigma$ and $a \leftrightarrow b$. 

Given the previous results, we can finally rewrite the set of self-consistency equations for the RS parameters as follows
\begin{subequations} \label{RSsystem}
\begin{gather}
     q_0= \int \mathscr{D}h \tanh^2(\sqrt{-\mathcal{R}_0}h) \label{RSequation1}  \\
      q_0= \int  \mathscr{D} \rho \mathscr{D} \sigma ~\langle a \rangle_{0}^2 \\
       1= \int \mathscr{D} \rho \mathscr{D} \sigma  ~\langle a^2 \rangle_{0} \label{RSequation3}
\end{gather}
\end{subequations}
plus the algebraic constraints Eqs.~\eqref{AlgEqs}. Finally, we report the expression of the RS free energy, which has to be computed at each temperature on the solutions of the self-consistency equations
\begin{equation} \label{freeEnRS}
\begin{split}
f_{\text{RS}}(\beta) &= -\frac{1}{\beta} \Bigg(\log2 - \frac{\mathcal{R}_D - \mathcal{R}_0}{2} + \int \mathscr{D} h \log \cosh (\sqrt{-\mathcal{R}_0}h) \\
&\quad+ \frac{1}{2} \int  \mathscr{D} \rho \mathscr{D} \sigma \log \left[2 I_{\beta,0}(\mM_D,\mM_0,\rho,\sigma)\right] + \frac{1}{2} s_0(\mR_D,\mR_0,\mM_D,\mM_0)  \Bigg) .
\end{split}
\end{equation}
where we recall that $s_0$ contains the expression of the entropic term Eq.~\eqref{ENTRO-MERIT-RS}. The factor 2 multiplying $I_{\beta,0}$ comes from the change of variables $C,\overline{C} \rightarrow a,b$.

\section{One Step of Replica Symmetry Breaking}
The first step of replica symmetry breaking (1RSB), introduced in Ref.~\cite{Parisi79a} for the SK model, is based on a more sophisticated choice of matrices than the more intuitive RS ansatz, in order solve the saddle point optimization problem. When replica symmetry is broken, the group $S_n$ of replica permutations is no more a symmetry group for the theory. However, the theory may still be invariant under a subgroup of replica permutations. In the 1RSB case the symmetry group left is $(S_m)^{\otimes \frac{n}{m}} \otimes S_{\frac{n}{m}}$, for some integer value of $m$, where $(S_m)^{\otimes \frac{n}{m}}$ is the direct product of the permutation group of $m$ objects with itself for $n/m$ times \cite{Mezard87}. This ansatz amounts to consider matrices of the kind
\begin{equation*}
\begin{pmatrix}
  \begin{matrix}
  \mA_D & \mA_1  \\
    \mA_1 & \mA_D  \\
  \end{matrix}
  & \rvline & \mA_0 \\
\hline
  \mA_0 & \rvline &
  \begin{matrix}
   \mA_D & \mA_1  \\
    \mA_1 & \mA_D \\
\end{matrix}
\end{pmatrix}    
\end{equation*}
which are characterized by $n/m$ diagonal blocks of dimension $m \times m$. With this choice, it is clear that $S_m$ corresponds to permutations of replicas inside a block and $S_{\frac{n}{m}}$ corresponds to permutations of the blocks. From the physical point of view, the presence of two values $\mA_0$ and $\mA_1$ signals the organization of states in clusters with one value of the overlap between configurations belonging to different states ($\mA_0$) and only one intrastate overlap value ($\mA_1$). The parameter $m$, which is known as \emph{breaking parameter}, is connected to the probability for the overlap to take one of the two allowed values.

This replica symmetry breaking scheme can be iterated hierarchically, by taking smaller sub-blocks inside each diagonal blocks and so on \emph{ad infinitum}. This procedure with infinite steps of breaking has proved to give the correct free energy for the SK model \cite{Parisi80a,Parisi79b}. However, in analogy with the REM and with the spherical $p$-spin model, we expect the 1RSB ansatz to be the right one to capture the low temperature properties of the model \cite{Gross84}.

An important remark, which will be particularly helpful for the algebra of this kind of matrices, is that 1RSB matrices can be decomposed as follows
\begin{equation} \label{1RSBdecomp}
\begin{pmatrix}
  \mA_0  & \rvline & \mA_0 \\
\hline
  \mA_0  & \rvline & \mA_0 
\end{pmatrix} +
\begin{pmatrix}
   \mA_1 - \mA_0  & \rvline & 0 \\
\hline
   0  & \rvline & \mA_1 - \mA_0
\end{pmatrix}    
+ \text{diag}(\mA_D - \mA_1),
\end{equation}
where the first one is a matrix of all elements equal to $\mA_0$, the second one is a block matrix with $n/m$ diagonal $m\times m$ blocks of all elements equal to $\mA_1-\mA_0$ and the third one is a diagonal matrix with all diagonal elements equal to $\mA_D-\mA_1$. This decomposition provides a nice visualization of a 1RSB matrix. The first useful result, which can be better understood from the decomposition, is the following: a term in the action containing replicated variables $x_a$ - either discrete or continuous - which are coupled through a 1RSB matrix can be written as
\begin{equation*}
\sum_{ab}^nx_a \mA_{ab}x_b = \mA_0 \left(\sum_{a=1}^n x_a
\right)^2 +(\mA_1-\mA_0)\sum_{k=1}^{n/m}\left(\sum_{\substack{ a \in \\ \text{Block}(k)}}^{1,m} x_a\right)^2 +(\mA_D-\mA_1)\sum_{a=1}^n x_a^2,
\end{equation*}
where it is clear that each term corresponds to one of the matrices in the expression \eqref{1RSBdecomp}. Other properties of 1RSB matrices will be discussed when necessary. In the following, we put $\mA = \mR - \mM$.

The precise computation of the action \eqref{Action2} in the 1RSB ansatz is reported in Appendix \ref{app:1RSBMF}. Let us here report and discuss the result. In analogy with the RS case, see Eq.~\eqref{P_g_C}, we define the probability density for the variables $C$ as
\begin{equation}
\label{P_g_C_1}
     \mathcal P_{\beta,1}(C|x,y,t,z,w)\equiv\frac{ g_{\beta,1}(C|x,y,t,z,w)}{I_{\beta,1}(x,y,t,z,w)},
\end{equation}
where 
\begin{align} \label{g1beta}
    g_{\beta,1}(C|x,y,t, z,w) = e^{-\beta|C|^4 + \frac{1}{2}(x-t)|C|^2 + \sqrt{y} \Re(C\overline{z})+\sqrt{t - y} \Re(C\overline{w})},
\end{align}
with $x,y,t \in \mathbb{R}$ and $z,w,C \in \mathbb{C}$
and 
\begin{align}\label{Ibeta1}
    I_{\beta,1}(x,y,t,z,w) = \int \de C \de \overline{C} \  g_{\beta,1}(C|x,y,t, z,w) .
\end{align}
Notice that, by taking $y=t$, the definitions of $g_{\beta,1}$ and $I_{\beta,1}$ are equivalent to those of $g_{\beta,0}$ $I_{\beta,0}$, see Eqs.~\eqref{gbeta0} and~\eqref{Ibeta0}: this is exactly what one expects, since in this case we are taking $\mM_0=\mM_1$, i.e.~a RS matrix. Moreover, in this case we also define the following function, which will appear in the expression of the free energy $f_\tau$ 
\begin{align} \label{xifunc}
   \Xi(\mR_0, \mR_1, h, u) = \sqrt{-\mR_0}~h + \sqrt{\mR_0-\mR_1}~u.
\end{align}
Notice, that by taking $\mR_0=\mR_1$ this function reduces to its RS form, which is simply $\Xi = \sqrt{-\mR_0} h$. 

With respect to these quantities, the 1RSB action reads as
\begin{equation}  \label{1RSBaction}
    \begin{split}
        \lim_{n\rightarrow0} \frac{2}{n} A_{1\text{RSB}} =  f_\tau(\mR_D,\mR_0,\mR_1,m)+ f_C (\mM_D,\mM_0,\mM_1,m) +  s_1(\mA_D,\mA_0,\mA_1,m)
    \end{split}
\end{equation}
where the local free energies have the $\mathcal{O}(n)$ expressions
\begin{gather}
    f_\tau(\mR_D,\mR_0,\mR_1,m) = \log4 - (\mathcal{R}_D-\mathcal{R}_1) + \frac{2}{m} \int \mathscr{D}h \log  \int \mathscr{D}u \cosh^m \Xi(\mR_0, \mR_1, h, u) \label{freeEn1RSBa}\\
     f_C (\mM_D,\mM_0,\mM_1,m) = \frac{1}{m}\int \mathscr{D}[z\overline{z}] \log \int \mathscr{D}[w\overline{w}]~I^m_{\beta,1}(\mM_D,\mM_0,\mM_1 | z,w) \label{freeEn1RSBb} \\
     \mathscr{D}[z\overline{z}] = \frac{\de z \de \overline{z}}{4\pi} e^{-|z|^2/2}
\end{gather}
and the entropic term 
\begin{equation} \label{entr1rsb}
    \begin{split}
       s_1(\mA_D,\mA_0,\mA_1,m) &=  \frac{m-1}{m}\log(\mA_D-\mA_1)  +\frac{1}{m}\log[\mA_D+(m-1)\mA_1 - m \mA_0] \\ 
   &\quad +\frac{\mA_0}{\mA_D+(m-1)\mA_1 - m \mA_0}.  
    \end{split}
\end{equation}

The procedure employed to compute the self-consistency equations for the 1RSB parameters goes along the same line as in the RS case, but the computations are heavier. This is due to the fact that a 1RSB matrix has two more parameters than a RS matrix, leading to an additional level of Gaussian integration in the local free energies. When performing the derivatives, this will lead to nested averages defined over probability measures which have more involved expressions compared to the RS case. In Appendix \ref{app:1RSBMF}, we present in some detail the computation of the derivatives, adopting notations that are compact enough to allow us to write down the main equations, but not so much to obscure their meaning. Here, we just state the result. 

In analogy to the RS case let us define the following average over the measure $\mathcal{P}_{\beta,1}$, see Eq.~\eqref{P_g_C_1},
\begin{equation} \label{AVERAGE_MF_1RSB}
\begin{split}
        \langle (\cdots) \rangle_{1} &= \int \de C~\de \overline{C}~ \mathcal{P}_{\beta,1}(C | \mM_D,\mM_0,\mM_1, z,w ) (\cdots) \\
    &= \frac{\int \de C~\de \overline{C}~g_{\beta,1}(C | \mM_D,\mM_0,\mM_1, z,w) (\cdots) }{I_{\beta,1}(\mM_D,\mM_0,\mM_1, z,w)},
\end{split}
\end{equation}
which is a function of the complex Gaussian variables $z,w$ and the saddle point parameters $\mM_{D,0,1}$. Then, the 1RSB stationary point is given by the solution of the following set of equations
\begin{subequations}
    \begin{gather}
    q_0 = \int \mathscr{D} h \left( \frac{ \int \mathscr{D} u~\cosh^m \Xi \tanh \Xi }{\int \mathscr{D}u \cosh^m \Xi}\right)^2 \\
    q_1 = \int \mathscr{D} h \frac{\int \mathscr{D}u~\cosh^m\Xi \tanh^2\Xi}{\int \mathscr{D}u \cosh^m\Xi} \\
    1 = \frac{1}{2} \int \mathscr{D}[z\overline{z}] \frac{\int \mathscr{D}[w \overline{w}] I_{\beta,1}^m \langle |C|^2 \rangle_{1} }{\int \mathscr{D}[w \overline{w}] I_{\beta,1}^m }  \\
    q_0 = \frac{1}{2} \int \mathscr{D}[z\overline{z}]   \left| \frac{\int \mathscr{D}[w \overline{w}] I_{\beta,1}^m \langle C \rangle_{1} }{\int \mathscr{D}[w \overline{w}] I_{\beta,1}^m  } \right|^2  \\
    q_1 = \frac{1}{2} \int \mathscr{D}[z\overline{z}]  \frac{\int \mathscr{D}[w \overline{w}] I_{\beta,1}^m |\langle C \rangle_{1}|^2 }{\int \mathscr{D}[w \overline{w}] I_{\beta,1}^m  }.
\end{gather}
\end{subequations}
The system is completed by the relations coming from the 1RSB expression of the algebraic constraint \eqref{AlgConstr}, which in the limit $n\rightarrow 0$, reads as
\begin{subequations} \label{Alg1RSB}
\begin{gather}
     \mathcal{A}_D +(m-1)\mathcal{A}_1 q_1 - m\mathcal{A}_0 q_0 =1  \label{Alg1RSBa} \\
      \mathcal{A}_1 + \mathcal{A}_D q_1 + (m-2)\mathcal{A}_1 q_1 - m\mathcal{A}_0 q_0 =0  \label{Alg1RSBb} \\
      \mathcal{A}_D q_0 + (m-1)\mathcal{A}_1 q_0 + \mathcal{A}_0 + (m-1) \mathcal{A}_0 q_1 - 2m \mathcal{A}_0 q_0 = 0. \label{Alg1RSBc}
\end{gather}
\end{subequations}

\subsubsection{The derivative with respect to $\bm{m}$}
In this subsection we compute the derivative of the action with respect to the last 1RSB parameter, the breaking parameter $m$. In fact, this parameter was originally an integer number, such that $m<n$, which denoted the dimension of the diagonal blocks of a 1RSB matrix; when the limit $n \rightarrow 0$ is taken, the more intuitive thing to do would be to send $m$ to zero as well, keeping fixed the ratio $n/m$. However, the prescription of the replica method is that, in order to obtain a well-defined probability distribution function of the overlap, $m$ has to be promoted to a real number in the interval $[0,1]$ and as a result the functions depending on $m$ are analytically continued with respect to $m$ in that interval \cite{Mezard87}. Therefore, $m$ has to be regarded by all means as a variational parameter with respect to which a saddle point self-consistency equation has to be computed. 

As for the other parameters, we compute the derivatives of the action \eqref{1RSBaction} with respect to $m$. The derivatives of the free energies \eqref{freeEn1RSBa} and \eqref{freeEn1RSBb} are 
\begin{align*}
     \frac{\partial  f_\tau}{\partial m}  = -\frac{1}{m^2} \int \mathscr{D} h \log \int \mathscr{D} u~\cosh^m \Xi + \frac{1}{m} \int \mathscr{D} h \frac{\int \mathscr{D}u \cosh^m\Xi \log \cosh \Xi}{\int \mathscr{D}u \cosh^m\Xi}
\end{align*}
and
\begin{align*}
    \frac{\partial f_C}{\partial m}  = -\frac{1}{m^2} \int \mathscr{D}[z\overline{z}] \log \int \mathscr{D}[w \overline{w}] I_{\beta,1}^m  + \frac{1}{m} \int \mathscr{D}[z\overline{z}] \frac{\int \mathscr{D}[w \overline{w}] I_{\beta,1}^m \log I_{\beta,1}}{\int \mathscr{D}[w \overline{w}] I_{\beta,1}^m}.
\end{align*}
The derivative of the entropic term reads as
\begin{align*}
    \frac{\partial s_1}{\partial m} &= - \frac{\mathcal{A}_0(\mathcal{A}_1-\mathcal{A}_0)}{(\mathcal{A}_D+(m-1)\mathcal{A}_1-m\mathcal{A}_0)^2} - \frac{1}{m^2}\log[\mathcal{A}_D+(m-1)\mathcal{A}_1-m\mathcal{A}_0] \\
    &+ \frac{1}{m} \frac{\mathcal{A}_1-\mathcal{A}_0}{\mathcal{A}_D+(m-1)\mathcal{A}_1-m\mathcal{A}_0} + \frac{1}{m^2} \log(\mathcal{A}_D-\mathcal{A}_1).
\end{align*}
Then the equation for $m$, which is too cumbersome to be written here, is given by the sum of the previous three derivatives set equal to zero.

\subsubsection{Further manipulations}
With the same procedure followed in the RS case, we pass to the real and imaginary parts of all the complex variables, with the notations: $C=a+ib$, $z=\rho + i\sigma$ and $w = u + i v$. Each of the square moduli in the equations obtained from the derivatives of $f_C$, gives two contributions, which can be proved to be equal under proper changes of variables. Eventually, we obtain the following set of equations
\begin{subequations} \label{1RSBequations}
\begin{gather}
   q_0 = \int \mathscr{D} h \left( \frac{ \int \mathscr{D} u~\cosh^m \Xi \tanh \Xi }{\int \mathscr{D}u \cosh^m \Xi}\right)^2 \label{1RSBequation1} \\
   q_1 = \int \mathscr{D} h \frac{\int \mathscr{D}u~\cosh^m\Xi \tanh^2\Xi}{\int \mathscr{D}u \cosh^m\Xi} \\
   1 = \int \mathscr{D}\sigma \mathscr{D}\rho \frac{\int \mathscr{D}u \mathscr{D}v  I_{\beta,1}^m \langle a^2 \rangle_{1} }{\int \mathscr{D}u \mathscr{D}v I_{\beta,1}^m } \\
   q_0 = \int \mathscr{D}\sigma \mathscr{D}\rho   \left( \frac{\int \mathscr{D}u \mathscr{D}v I_{\beta,1}^m \langle a \rangle_{1} }{\int \mathscr{D}u \mathscr{D}v I_{\beta,1}^m } \right)^2 \\
   q_1 = \int \mathscr{D}\sigma \mathscr{D}\rho  \frac{\int \mathscr{D}u \mathscr{D}v I_{\beta,1}^m \langle a \rangle_{1}^2 }{\int \mathscr{D}u \mathscr{D}v I_{\beta,1}^m },  \label{1RSBequation5}
\end{gather}
\end{subequations}
where the average induced by the measure $\mathcal{P}_{\beta,1}(a,b|\mM_D,\mM_0,\mM_1,\rho,\sigma,u,v)$, see Eq.~\eqref{AVERAGE_MF_1RSB}, now reads 
\begin{equation}
    \langle (\cdots) \rangle_{1} = \frac{\int \de a~\de b~ g_{\beta,1}(a,b|\mM_D,\mM_0,\mM_1,\rho,\sigma,u,v)}{I_{\beta,1}(\mM_D,\mM_0,\mM_1,\rho,\sigma,u,v)},
\end{equation}
with
\begin{gather}
\label{gbeta_1_ab}
 g_{\beta,1}(a,b|x,y,t,\rho,\sigma,u,v)=
e^{-\beta(a^2+b^2)^2 + \frac{1}{2}(x-t)(a^2+b^2) + \sqrt{y} (a \rho + b \sigma)+\sqrt{t - y} (a u + b v)} \\
I_{\beta,1}(x,y,t,\rho,\sigma,u,v) = \int \de a~\de b~ g_{\beta,1}(a,b|x,y,t,\rho,\sigma,u,v).
\end{gather}
Notice that, as in the RS case, we have removed a factor 2 from the definition of $I_{\beta,1}$ since it cancels out with an equal factor in the numerator of the average $\langle (\cdots) \rangle_{1}$.

The algebraic Eqs.~\eqref{Alg1RSB}, together with the equation obtained with the derivatives in $m$, complete the set of self-consistency equations for the 1RSB parameters. The 1RSB free energy, which has to be computed on the solutions of the self-consistency equations, reads as
\begin{equation}
\begin{split}
    f_{1\text{RSB}} (\beta) = &-\frac{1}{\beta} \Bigg(\log2 - \frac{\mathcal{R}_D - \mathcal{R}_1}{2} + \frac{1}{m} \int \mathscr{D} h \log \int \mathscr{D} u \cosh^m \Xi  \\
    &+\frac{1}{2 m} \int  \mathscr{D} \rho \mathscr{D} \sigma \log \int \mathscr{D} u \mathscr{D} v [2 I_{\beta,1}]^m + \frac{1}{2} s_1(\mA_D,\mA_0,\mA_1,m)  \Bigg),
\end{split}
\end{equation}
where all parameters are the solutions to Eqs.~\eqref{Alg1RSB} and \eqref{1RSBequations}. We notice that the RS equations are easily obtained by putting $m=1$ and taking $q_0=q_1$ and similarly for the other parameters.

\subsection{Simplified 1RSB Ansatz}
A simpler optimization problem, which is worth studying at least at a preliminary level, is the one resulting from a 1RSB ansatz with $q_0=\mR_0=\mM_0=0$. This assumption works for the p-spin model in zero external field, both with spherical variables, see Refs.~\cite{Crisanti92,Castellani05}, and with Ising spin variables, see Ref.~\cite{Gardner85}, so it is reasonable for the present case as well.

The advantage of making this simplified ansatz is a drastic simplification of the self-consistency equations for the remaining 1RSB parameters. Besides having three parameters less to optimize, in this case, the local partition function function $I_{\beta,1}$ does not depend anymore on the auxiliary variable $z$ (or equivalently $\rho,\sigma$), leading to the disappearance of the outer Gaussian integration from the self-consistency equations obtained by the derivative of $f_C$. Moreover, in this case the functions $g_{\beta,1}$ and $I_{\beta,1}$ reduce to the RS integral functions $g_{\beta,0}$ and $I_{\beta,0}$ computed in $\mM_D, \mM_1$ rather than $\mM_D, \mM_0$. Hence, though we are still in a 1RSB ansatz, the average $\langle (\dots) \rangle_0$ will appear in the saddle-point equations. A similar simplification also occurs for the equation that contains the derivative of $f_\tau$ with respect to $\mR_1$: the function $\Xi$ reduces to its RS form, but computed in $\mR_1$, rather than $\mR_0$, i.e. $\Xi= \sqrt{-\mR_1}u$.

Let us report the expression of the self-consistency equations in this case
\begin{subequations} \label{1RSBSimplysystem}
\begin{gather}
    q_1 = \frac{ \int \mathscr{D} u~\cosh^m \Xi \tanh^2 \Xi }{\int \mathscr{D}u \cosh^m \Xi} \\
    q_1 = \frac{\int \mathscr{D}u \mathscr{D}v I_{\beta,0}^m(\mM_D,\mM_1 | u,v) \langle a \rangle_{0}^2 }{\int \mathscr{D}u \mathscr{D}v I_{\beta,0}^m(\mM_D,\mM_1 | u,v) }  \label{1RSBSimplysystemB} \\
    1 =  \frac{\int \mathscr{D}u \mathscr{D}v I_{\beta,0}^m(\mM_D,\mM_1 | u,v) \langle a^2 \rangle_{0} }{\int \mathscr{D}u \mathscr{D}v I_{\beta,0}^m(\mM_D,\mM_1 | u,v) \label{1RSBSimplysystemC}}
\end{gather}
\end{subequations}
with the following algebraic constraints
\begin{subequations} \label{AlgSimpl1RSB}
\begin{gather}
\mathcal{R}_D-\mathcal{M}_D +(m-1)(\mathcal{R}_1-\mathcal{M}_1) q_1 = 1 \\
\mathcal{R}_1-\mathcal{M}_1 +  (\mathcal{R}_D-\mathcal{M}_D) q_1 + (m-2)(\mathcal{R}_1-\mathcal{M}_1)q_1 = 0.
\end{gather}
\end{subequations}
The equation obtained by the vanishing of the action derivative with respect to $m$ can be computed for the present case by putting $\mR_0 = \mM_0 = 0$ in the derivatives computed before and reads as
\begin{equation} \label{mEq}
\begin{split}
    &-\frac{1}{m} \log \int \mathscr{D} u \cosh^m\Xi + \frac{\int \mathscr{D}u \cosh^m\Xi \log \cosh \Xi}{\int \mathscr{D}u \cosh^m\Xi} - \frac{1}{m} \log \int \mathscr{D}u \mathscr{D}v I_{\beta,0}^m  \\
    &  + \frac{ \int \mathscr{D}u \mathscr{D}v I_{\beta,0}^m \log I_{\beta,0}}{ \int \mathscr{D}u \mathscr{D}v I_{\beta,0}^m }  \\
    & - \frac{1}{m} \log \left(1 + m (1-q_1)(\mR_1-\mM_1)\right) + \frac{(1-q_1)(\mR_1 - \mM_1)}{1+m (1-q_1)(\mR_1-\mM_1)} = 0,
\end{split}
\end{equation}
where $q_1$ has been introduced through its expression in terms of the other parameters. The free energy of the model in this simplified 1RSB ansatz reads
\begin{equation} \label{1RSBfreeENsimply}
\begin{split}
    f_{1\text{RSB}} (\beta) &= -\frac{1}{\beta} \Bigg(\log2 - \frac{\mathcal{R}_D - \mathcal{R}_1}{2} + \frac{1}{m} \log \int \mathscr{D} u \cosh^m \Xi  \\
    &+\frac{1}{2 m} \log \int \mathscr{D} u \mathscr{D} v [2 I_{\beta,0}(\mM_D,\mM_1,| u, v)]^m + \frac{1}{2} s_1(\mR_D,\mR_1,\mM_D,\mM_1,m)  \Bigg),
\end{split}
\end{equation}
where, now, we have
\begin{align*}
    s_1 = \frac{m-1}{m}\log(\mathcal{R}_D-\mathcal{M}_D-\mathcal{R}_1+\mathcal{M}_1) +\frac{1}{m}\log[\mathcal{R}_D-\mathcal{M}_D+(m-1)(\mathcal{R}_1-\mathcal{M}_1)].
\end{align*}
By using the algebraic relations \eqref{AlgSimpl1RSB}, the entropic term has the following simpler dependence on $q_1$: 
$$s_1 = \frac{m-1}{m}\log\left(\frac{1}{1-q_1}\right) + \frac{1}{m}\log\left(\frac{1}{1+(m-1)q_1}\right).$$

\section{Numerical Integration}
This section is devoted to the description of the technique used for the numerical integration of the saddle-point self consistency equations which have been obtained for the variational parameters of the model in the previous sections. In particular, we have studied in detail the RS set of Eqs.~\eqref{RSsystem} and the simplified 1RSB set of Eqs.~\eqref{1RSBSimplysystem}, both on Mathematica and by writing dedicated codes in Python.

The numerical solution of the integrals has been performed by means of the Gaussian-Legendre quadrature rule \cite{Barone06,Abramowitz72}, which we briefly explain in the following. Given a function $f:[a,b] \subset \mathbb{R} \rightarrow \mathbb{R}$, the integral of $f$ over its domain can be approximated by
\begin{align}
    \int_a^b \de y f(y)  \approx \frac{b-a}{2} \sum_{i=1}^n w_i f(y_i),
\end{align}
with 
\begin{align*}
    y_i = \left(\frac{b-a}{2}\right) x_i + \left(\frac{b+a}{2}\right).
\end{align*}
In the previous expressions the point $x_i$ is the $i^{\text{th}}$ zero of the Legendre polynomial $P_n(x)$ and $w_i$ is the corresponding weight given by
\begin{align*}
    w_i = \frac{2}{(1-x_i^2)[P'_n(x_i)]^2}.
\end{align*}
Both the values of $x_i$ and $w_i$ are tabulated and can be generated by a specific Python (or Mathematica) routine. The generalization of this technique to two-dimensional integrals is straightforward. The advantage of this technique is that the discretization of the integration domain is very efficient and one obtains a relatively good result already with a small number of points. 

Our integrals are extended between $\pm \infty$, but the integrands are rapidly decreasing functions. We choose symmetric integration domains limited by a parameter $L_g$ in the case of Gaussian integrals and $L_q$ in the case of integrals of the exponential of a 4-degree polynomial, namely the function $g_{\beta,0}$. We have studied the parametrical dependence of integrals on the number of points $n$ and on the quantities $L_g,L_q$ and assessed the values of the parameters such that the results remained stable. The computation of the integrals is particularly demanding in the 1RSB case, due to the presence of nested double integrals, which lead to an increase of the computational complexity of order $O(n^2)$ for each layer. Just to make an example, the discretized version of the integral in the third equation of the simplified 1RSB system (see Eq.~\eqref{1RSBSimplysystemC}) reads as
\begin{eqnarray*}
    I &=& \frac{\sum_{ij}^n w_i w_j e^{-\frac{L_g^2(u_i^2+v_j^2)}{2}} I^m_{\beta,0}(\mM_D,\mM_1|L_g u_i,L_g v_j) \langle (L_q a_k)^2 \rangle_0}{\sum_{ij}^n w_i w_j e^{-\frac{L_g^2(u_i^2+v_j^2)}{2}} I^m_{\beta,0}(\mM_D,\mM_1|L_g u_i,L_g v_j)}, \\
    \langle (L_q a_k)^2 \rangle_0 &=& \frac{\sum_{kl}^n w_k w_l g_{\beta,0}(\mM_D,\mM_1|L_g u_i,L_g v_j, L_q a_k, L_q b_l) (L_q a_k)^2 }{\sum_{kl}^n w_k w_l g_{\beta,0}(\mM_D,\mM_1|L_g u_i,L_g v_j, L_q a_k, L_q b_l) } 
\end{eqnarray*}
where, of course, $I_{\beta,0}$ contains another double integration. During the solution of the saddle point equations at a certain value of the inverse temperature $\beta$, this integral, like the others, has to be computed iteratively many times with respect to tentative values of the 1RSB parameters $\mM_D,\mM_1,m$. Then, the procedure has to be repeated varying the temperature. To speed up this kind of computations the code has been parallelized on GPUs using the Python library PyTorch.

The integration technique of the saddle point equations is based on the optimization of a loss function defined as the sum of the action gradient components squared. Let us briefly describe the technique in general before moving to the case of interest. Suppose we want to find the global minimum of a differentiable cost function $\mathcal{L}(\bm{x})$, depending on $P$ parameters $\{x_i\}_{i=1,\dots,P}$. The Gradient Descent (GD) algorithm is based on the idea that the most efficient way of reaching the minimum of $\mathcal{L}(\bm{x})$ is to follow the opposite direction of its gradient. This can be implemented in following iterative way
\begin{equation} \label{GD}
    \bm{x}_{n+1} = \bm{x}_n - \gamma \nabla \mathcal{L}(\bm{x}_n),
\end{equation}
where the quantity $\gamma$ is the so-called \emph{learning rate}, which defines the step size of the algorithm and the quantity $\gamma \nabla \mathcal{L}(\bm{x}_n)$ is subtracted from $\bm{x}_n$ since we want to move against the gradient. Clearly, this method may encounter some difficulties for functions with many local minima: when one of this minima is reached the gradient of the function is a vector of zeros and the algorithm gets stuck. To overcome the problem, several GD optimizations can be run with random initial conditions: in non-pathological cases the global minimum can be selected \emph{a posteriori}. In order to optimize the loss functions which will be defined in a short while, we used the Adam\footnote{Adam stands for \emph{adaptive moment estimation} and it is usually adopted as a Stochastic Gradient Descent (SGD) algorithm in the context of input-output problem in machine learning. Here, however, we just have to minimize a function with respect to its arguments and we have used it as a simple GD.} optimizer as a GD algorithm with momentum \cite{Kingma14}. The momentum is an additional term to the GD dynamics defined in Eq. \eqref{GD}, which suppresses the oscillations of the gradient, by taking larger steps in the preferred direction of steepest descent.

\subsection{RS Equations}
Consider the set of RS Eqs.~\eqref{RSsystem}. In principle, we have to determine five parameters $q_0,\mR_D,\mR_0,\mM_D$ and $\mM_0$; however, the algebraic constraints \eqref{AlgEqs} can be used to express two of them in terms of the others. For example, we can eliminate $\mR_D$ and $\mR_0$ by writing $\mR_D=\mR_D(\mM_D,q_0)$ and $\mR_0=\mR_0(\mM_0,q_0)$, where
\begin{subequations} \label{RDR0}
\begin{gather}
    \mR_D = \mM_D + \frac{1-2 q_0}{(1-q_0)^2} \label{RD} \\
    \mR_0 = \mM_0 - \frac{q_0}{(1-q_0)^2}. \label{R0}
\end{gather}
\end{subequations}
Incidentally, this expression of the parameters allows us to make an important consistency check on the solution: in order for the RS free energy~\eqref{freeEnRS} to be real valued we need the argument of the logarithm function in the entropic term $s_0$ to be positive definite, i.e.~$\mR_D -\mM_D -\mR_0 +\mM_0 > 0$. When substituting the expressions of $\mR_D$ and $\mR_0$ in terms of the other parameters into the previous condition, we find the simple condition $1-q_0 > 0$, which is always verified at finite temperature, since $q_0\in[0,1]$. We are, then, left with three integral equations in the three parameters $q_0,\mM_D,\mM_0$. The loss function that we need to optimize can be defined as $\mathcal{L}_{\text{RS}} = \sum_i (\partial_{\mathcal{X}_i} A_{\text{RS}})^2$, $\mathcal{X}_i$ being the generic replica parameter, and explicitly reads
\begin{align*}
    \mathcal{L}_{\text{RS}}(q_0,\mM_D,\mM_0, \beta) &= \left(q_0 - \int \mathscr{D}h \tanh^2(\sqrt{-\mathcal{R}_0}h) \right)^2 + \left(q_0 - \int  \mathscr{D} \rho \mathscr{D} \sigma ~\langle a \rangle_{0}^2 \right)^2  \\
    &\quad +  \left(1-\int\mathscr{D} \rho \mathscr{D} \sigma  ~\langle a^2 \rangle_{0}\right)^2,
\end{align*}
where the dependence of the parameters on $\beta$ is implicit and all the integrals are discretized with the procedure described before. 

With respect to standard spin-glass optimization problems, where the Lagrange multipliers conjugated to the overlap variables are usually integrated away at the level of the saddle-point equations \cite{Crisanti92}, here we have the additional difficulty of dealing with unbounded parameters: in fact, if $q_0$ must be in the interval $[0,1]$ for every value of $\beta$, we do not have such strong bounds on $\mM_D$ and $\mM_0$. The only information we have is that, either by looking again at the RS free energy Eq.~\eqref{freeEnRS} or by directly inspecting the equations, the theory is well defined only if $\mR_0 < 0$ and $\mM_0 > 0$. The condition on $\mR_0$ can be used together with Eq.~\eqref{R0} to find an upper bound for $\mM_0$. Eventually, we have
\begin{align} \label{M0bounds}
    0 < \mM_0 < \frac{q_0}{(1-q_0)^2},
\end{align}
which, however, is not so useful in practice a part from the choice of the initial condition selection. Actually, what we learn from Eq.~\eqref{M0bounds} is that as long as $q_0=0$, $\mM_0$ has to vanish too: this is expected to happen at least for low values of $\beta$. If there is a value of $\beta$ from which $q_0$ starts increasing, then, as $q_0 \rightarrow 1$ the upper bound on $\mM_0$ diverges. Nothing can be said, instead, for the definition interval of $\mM_D$.

In order to acquire preliminary knowledge of the parameters region where the global minimum of $\mathcal{L}_{\text{RS}}$ might be located at a certain value of $\beta$, we have visualized the loss landscape, by producing color maps of its projections onto orthogonal planes. From these plots we could clearly identify the paramagnetic solution, which is always present for any value of the temperature. Moreover, for sufficiently high $\beta$ many other minima, though not deep as the paramagnetic one, could be found for non-vanishing values of the parameters $\mM_0$ and $q_0$. However, most of these minima have turned out to be nonphysical, leading to imaginary values of the free energy or other pathological consequences. The absence of boundaries for the parameter $\mM_0$ prevents to thicken the grid over which the loss is computed, but for small intervals of $\mM_0$ values. If the global minimum of the loss has a small basin of attraction, it is very unlikely to be visualized. However, we have managed to exclude some values of $\mM_0$ from the choice of the initial conditions for the GD algorithm.

Starting from high temperature, our algorithm falls into the paramagnetic solution with $q_0=\mM=0=0$, independently of the initial conditions, leading to results which are consistent with the annealed limit. To speed up the search for the optimal parameters, when increasing $\beta$, we always initialize the optimizer for the next step in temperature with the optimized parameters at the previous temperature. However, if one starts with a low value of $\beta$, this procedure may cause the algorithm to remain stuck in the paramagnetic solution, even in presence of a different solution dominating the thermodynamics. This occurrence is typical of first-order phase transitions, where the high temperature solution does not become unstable at the transition (i.e.~a maximum or a saddle in the parameter space), but from being a global minimum it turns into a local minimum. With respect to this, we refer to the section dedicated to the numerical integration of the Ising $p$-spin model, which we have used as a test of the procedure. Therefore, we have tried another strategy: we start from a high value of $\beta$, hoping to fall into a different state than the paramagnetic one and we lower $\beta$ to follow the solution up to the transition point. Having not much information on the value of $\mM_0$ at low temperature, we start the optimization by fixing a high value of $q_0 \in [0,1]$ and randomly choosing the initial condition on $\mM_0$ according to Eq.~\eqref{M0bounds} and to our observations of the loss function. However, no solution can be found at finite $T$ with non-vanishing $q_0$ and $\mM_0$ and a free energy value higher than the value of the RS free energy at the same temperature. Hence, no evidence of a phase transition has been revealed in terms of non-vanishing $q_0$ or $\mM_0$.

\subsection{1RSB Equations}
We now turn to the more complicated case of the set of 1RSB equations \eqref{1RSBSimplysystem}, where we have six parameters to determine: $q_1,\mR_D,\mR_1,\mM_D,\mM_1$ and the breaking parameter $m$. We eliminate $\mR_D$ and $\mR_1$ from the problem, by means of the algebraic constraints \eqref{Alg1RSB}, expressing them as $\mR_D= \mR_D(q_1,m,\mM_D)$ and $\mR_1= \mR_1(q_1,m,\mM_1)$ in the following way
\begin{subequations} \label{RDR1}
\begin{gather}
    \mR_D = \mM_D + \frac{1 + (m-2) q_1}{(1-q_1)[1+(m-1)q_1]} \label{RD1RSB} \\
    \mR_1 = \mM_1 - \frac{q_1}{(1-q_1)[1+(m-1)q_1]}. \label{R1}
\end{gather}
\end{subequations}
As in the RS case, we can verify the consistency of the theory, by using these equations together with the positiveness of the arguments of the two logarithms in the entropic term of the 1RSB free energy~\eqref{1RSBfreeENsimply}: we get two conditions, i.e.~$1-q_1 > 0$ and  $1-(m-1)q_1 > 0$, which are both satisfied for every finite temperature, since $q_1,m\in[0,1]$. Moreover, by using the fact that $\mR_1 < 0$ and $\mM_1>0$, we find a condition on $\mM_1$, which is the 1RSB generalization of Eq.~\eqref{M0bounds} and reads
\begin{align}
    0 < \mM_1 < \frac{q_1}{(1-q_1)[1+(m-1)q_1]}.
\end{align}
As in the RS case, this equation tells us that as long as $q_1=0$, $\mM_1=0$ as well.
Hence in this case we are left with the three integral equations~\eqref{1RSBSimplysystem} plus Eq.~\eqref{mEq}, which by using the expression of $\mR_1$ further simplifies to
\begin{equation} \label{mEq2}
\begin{split}
    &-\log \int \mathscr{D} u \cosh^m\Xi + m \frac{\int \mathscr{D}u \cosh^m\Xi \log \cosh \Xi}{\int \mathscr{D}u \cosh^m\Xi} - \log \int \mathscr{D}u \mathscr{D}v I_{\beta,0}^m  \\
    &  + m \frac{ \int \mathscr{D}u \mathscr{D}v I_{\beta,0}^m \log I_{\beta,0}}{ \int \mathscr{D}u \mathscr{D}v I_{\beta,0}^m }  - \log \left(1 + \frac{m}{1+(m-1)q_1}\right) + m\frac{q_1}{1-q_1} = 0.
\end{split}
\end{equation}
to be solved with respect to $q_1,m,\mM_D$ and $\mM_1$. 

We have adopted two alternative strategies to solve the 1RSB optimization problem, one, more direct, which indeed makes use of Eq.~\eqref{mEq2}, the other one, more subtle, which is based on a ``graphical'' maximization of the 1RSB free energy with respect to $m$. In the first case, the loss function is defined by taking into account all the squared components of the action gradient, expressed in terms of $q_1,m,\mM_D$ and $\mM_1$. We have
\begin{equation}
\begin{split}
    \mathcal{L}_{1\text{RSB}}(q_1,m,\mM_D,\mM_1,\beta) &= \left(q_1 - \frac{ \int \mathscr{D} u~\cosh^m \Xi \tanh^2 \Xi }{\int \mathscr{D}u \cosh^m \Xi} \right)^2 + \left( q_1 - \frac{\int \mathscr{D}u \mathscr{D}v I_{\beta,0}^m \langle a \rangle_{0}^2 }{\int \mathscr{D}u \mathscr{D}v I_{\beta,0}^m }  \right)^2 \\
    &\quad + \left(1 -  \frac{\int \mathscr{D}u \mathscr{D}v I_{\beta,0}^m \langle a^2 \rangle_{0} }{\int \mathscr{D}u \mathscr{D}v I_{\beta,0}^m } \right)^2  + (\partial_m A_{1\text{RSB}})^2 = 0
\end{split}
\end{equation}
where the left hand side of Eq.~\eqref{mEq2} has been denoted as $\partial_m A_{1\text{RSB}}$ for brevity and all the integrals are computed with the Gauss-Legendre quadrature rule implemented in parallel. In the other case, the term  $\partial_m A_{1\text{RSB}}$ drops out from the cost function definition, and optimization with respect to $m$ is performed as follows. At a fixed value of $\beta$, the global minimum of the cost function is found in parallel for different values of $m \in [0,1]$: then, the free energy in Eq.~\eqref{1RSBfreeENsimply} is computed as a function of $m$, i.e. $f_{1\text{RSB}} = f_{1\text{RSB}}(m)$ and its values are sorted. We look for the maximum of the $f_{1\text{RSB}}(m)$ and consider the values of the parameters corresponding to it as the solution of the optimization problem. In order to reduce the error, this procedure can be iterated many times, by using values of $m$ each time closer to the true maximum of $f_{1\text{RSB}}(m)$.

Regarding the dynamics in temperature, we have proceeded as in the RS case, by either increasing $\beta$ starting from the paramagnetic solution or by decreasing $\beta$ starting from many random initializations of the parameters $q_1,\mM_D,\mM_1$ and $m$, which are sampled consistently with their bounds from the regions where the low temperature maps of the loss functions revealed the presence of minima. By increasing $\beta$, we just remain stuck into the paramagnetic solution already found at the RS level and in the annealed limit. By lowering $\beta$, notwithstanding the huge number of attempts with different initial conditions, we are not able to find any good solution besides the paramagnetic one.

\subsection{Test: the Ising p-spin model}
In order to check our numerical integration technique, we have tested the procedure on the 1RSB solution of the Ising $p$-spin model. Let us briefly report some useful result drawn from Ref.~\cite{Gardner85}. 

\begin{figure}[t]
   \begin{minipage}{0.485\textwidth}
     \centering
     \includegraphics[width=\linewidth]{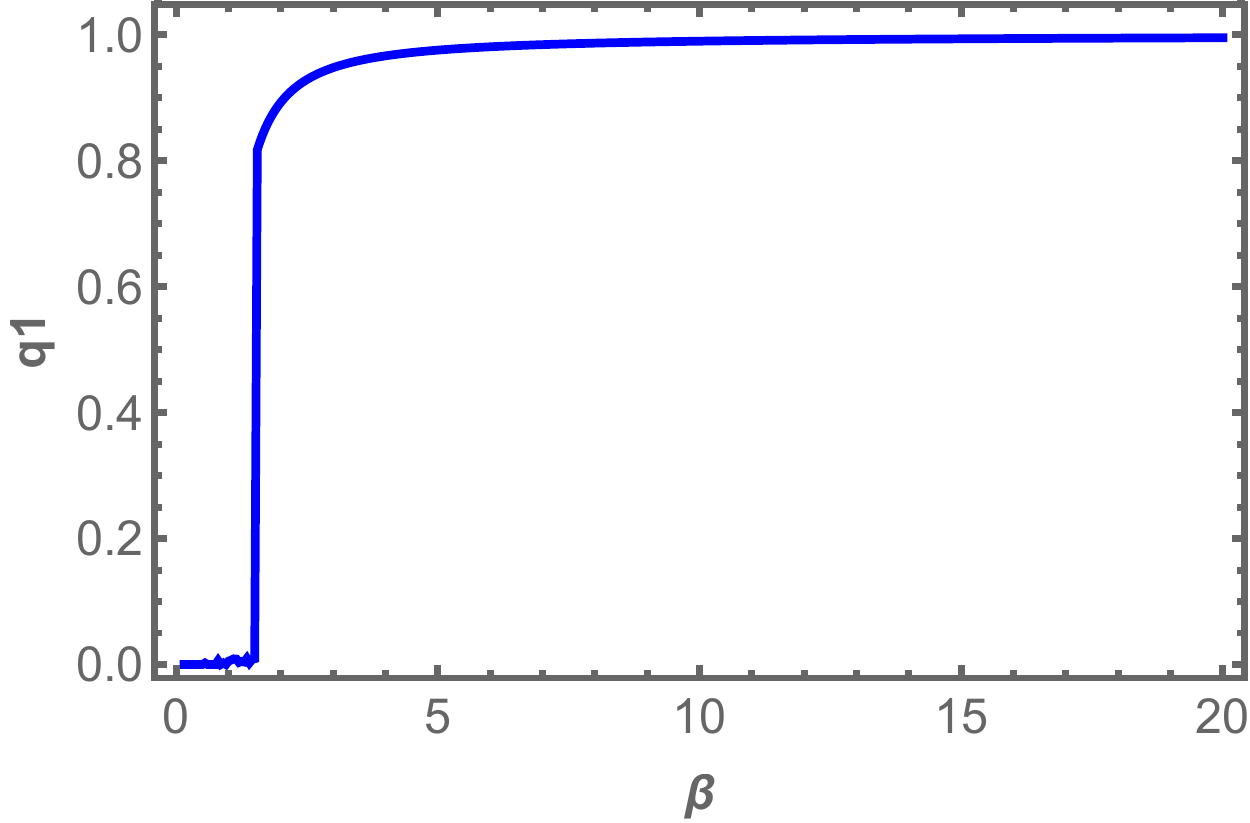}
     \caption{The 1RSB overlap parameter $q_1$ as a function of the inverse temperature. Notice, the jump at $\beta_c= 1.55 \pm 0.025$.} \label{fig:q1_pspin}
   \end{minipage}
   \hfill
   \begin{minipage}{0.485\textwidth}
     \centering
     \includegraphics[width=\linewidth]{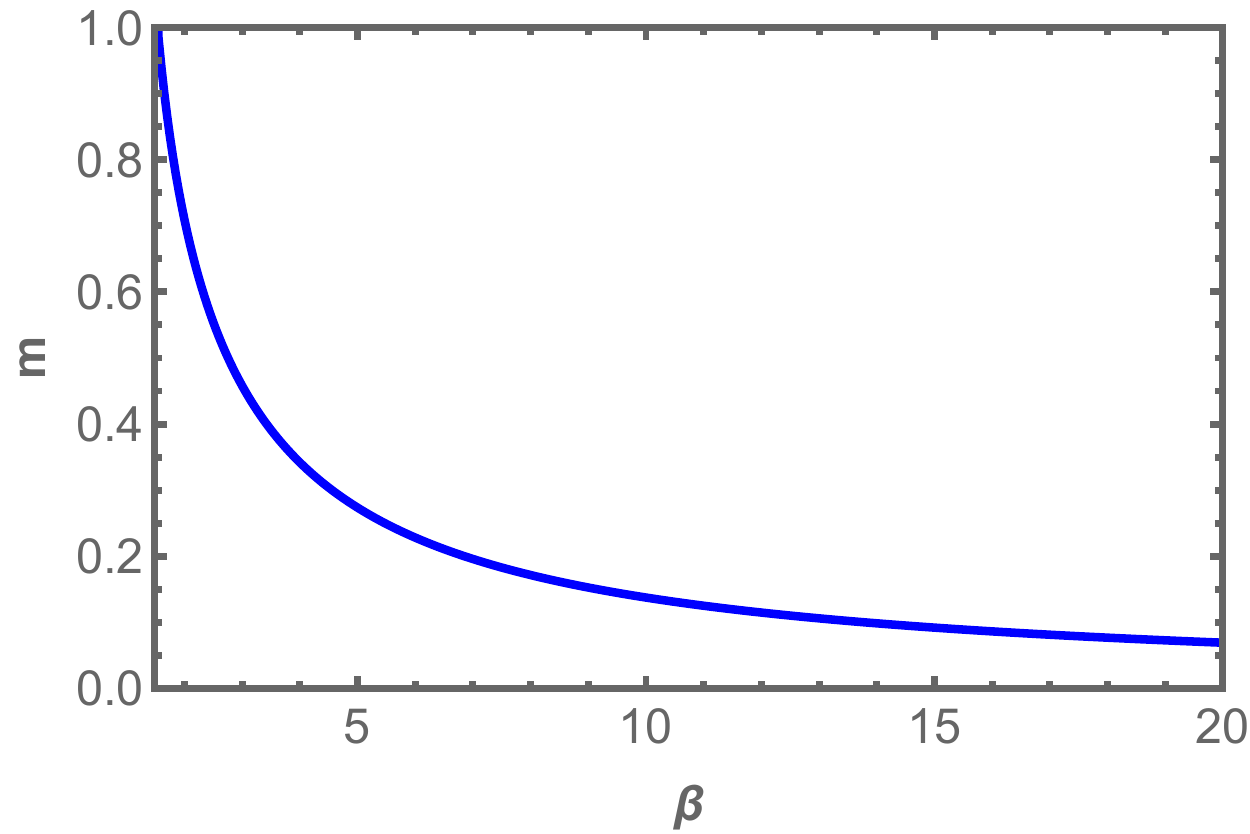}
     \caption{The breaking parameter $m$ as a function of the inverse temperature in the well-behaved region for $\beta>\beta_c$.} \label{fig:m_pspin} 
\hspace{1cm} 
   \end{minipage} 
     \centering
     \includegraphics[width=.7\linewidth]{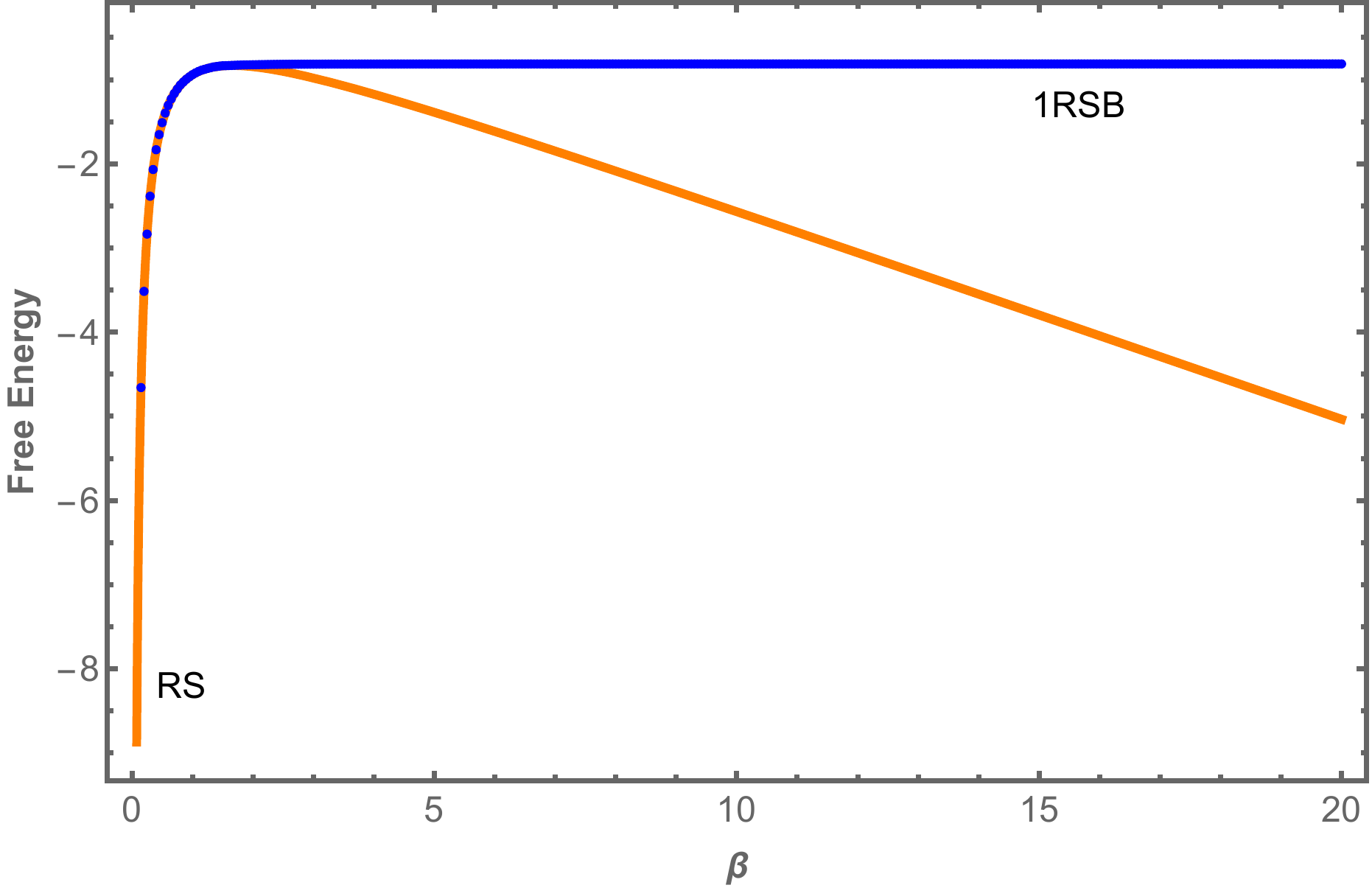}
     \caption{Free energy of the model in the RS (orange, analytical) and 1RSB (blue, numerical) solutions. The thermodynamic free energy is given by the maximum of the two and hence coincides with the RS one in the high temperature region and with the 1RSB one in the low temperature region.} \label{fig:fEen_pspin}
\end{figure}

The temperature at which the RS entropy becomes negative, signaling a thermodynamic anomaly, is $T=1/(2 \sqrt{\log 2})=0.60056\dots$. The critical temperature of the transition to the 1RSB phase is given by the analytical expression
\begin{equation}
    T_c = \frac{1}{2 \sqrt{\log 2}} \left(1+2^{-(p+1)} \sqrt{\frac{\pi}{p (\log 2)^3}}  \right),
\end{equation}
which for the case $p=3$ gives the value $T_c = 0.66712\dots$ to be compared with our numerical integration. 

For an arbitrary value of $p$, the free energy with $q_0$ already set to zero is given by the following expression
\begin{align} \label{freeEN}
\Phi = -\frac{\beta}{4} [1 + (p-1)(1-m)q_1^p - p q_1^{p-1}] - \frac{1}{m\beta} \log \int \msD z ~(2\cosh \Xi )^m,
\end{align}
where in this case 
$$\Xi= z \sqrt{\frac{p}{2}} \beta q_1^{\frac{p-1}{2}}$$
and, as usual, 
$$\msD z = \frac{\de z}{\sqrt{2 \pi}} \exp(z^2/2).$$
For the present case we have just two equations to determine the parameters $q_1$ and $m$ as functions of $\beta$, which read as
\begin{subequations}
\begin{gather}
    q_1 = \frac{\int \msD z~(2\cosh \Xi )^m \tanh^2\Xi}{\int \msD z~(2\cosh \Xi )^m} \\
    \frac{\beta}{4} (p-1) q_1^p + \frac{1}{\beta m^2} \log \int \msD z ~(2\cosh \Xi )^m - \frac{1}{\beta m} \frac{\int \msD z ~(2\cosh \Xi )^m \log 2 \cosh \Xi }{\int \msD z ~(2\cosh \Xi )^m} = 0.\label{mEqpspin}
\end{gather}
\end{subequations}
The second equation is obtained from the derivative in $m$ of the free energy and, in this simple case, can be easily integrated directly. Alternatively, as discussed above, one can solve the first equation parametrically in $m$ and then look for the maximum of the free energy in $m$ ``graphically''. We have tested both procedures and checked that the optimization of the loss functions defined for the two cases in analogy with the previous section yields the same results. Here, we report the results obtained through the optimization of the full loss function by using the \emph{FindRoot} routine on Mathematica. The integrals have been computed both with the Gauss-Legendre quadrature rule and with the \emph{NIntegrate} routine: we have checked the consistency of the values computed in the two ways.

\begin{figure}[t]
     \centering
     \includegraphics[width=\linewidth]{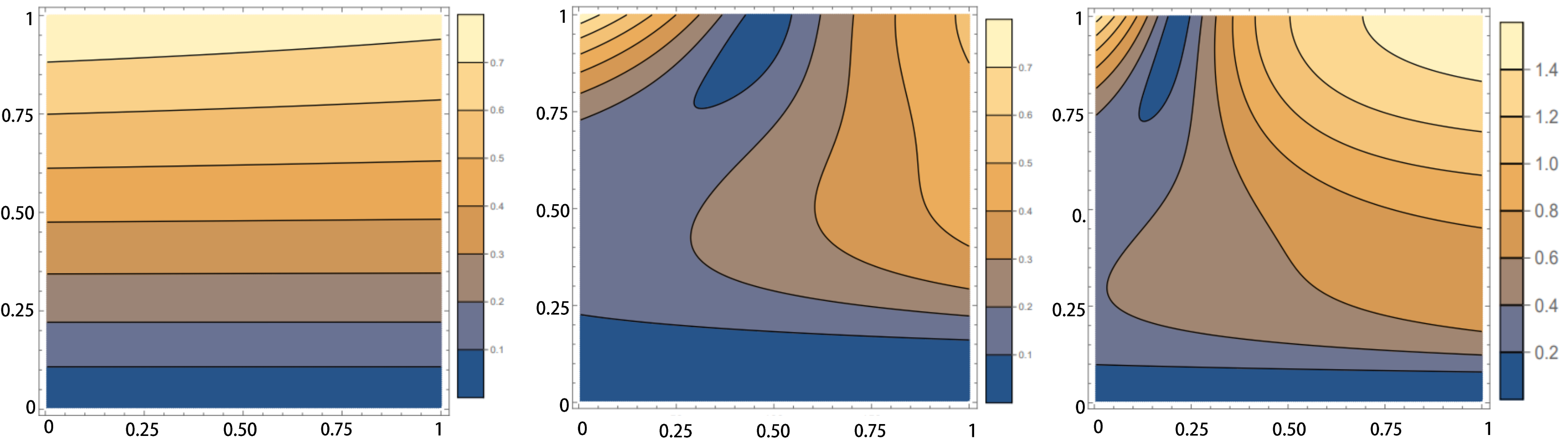}
     \caption{Color map of the $p$-spin loss function at three different values of $\beta$: from left to right  $\beta = 0.5$, $\beta = 3$ and $\beta=6.5$. Vertical axis: $q_1$; horizontal axis $m$; color map: values of the loss increase from blue to light yellow. The two blue valleys correspond to the paramagnetic state (RS solution), with $q_1=0$ and degenerate in $m$ and to the glassy state (1RSB solution) with $q_1 \simeq 1$ and $m$ decreasing when $\beta$ increases. In the left panel, where $\beta < \beta_c$, only the RS solution can be visualized, while in the other two panels the 1RSB solution appears and moves towards $m=0$, when $\beta$ is increased.} \label{fig:Loss_pspin}
\end{figure}

In Figs.~\ref{fig:q1_pspin} and \ref{fig:m_pspin} we display the temperature dependence of the two parameters $q_1$ and $m$ respectively, by plotting them as functions of $\beta$. Data are obtained for the case $p=3$. The jump in $q_1$ occurs at $\beta_c= 1.55 \pm 0.025$, with the uncertainty estimated as half of the $\beta$ spacing ($\delta \beta = 0.05$). This result is in good agreement with the expected critical temperature for the present case. The value of $m$ oscillates for $\beta < \beta_c$, where $q_1=0$, as a sign of the fact that in the high temperature phase the solution is degenerate in $m$. At $\beta_c$ the value of $m$ starts decreasing smoothly from 1 towards zero, see Fig.~\ref{fig:m_pspin}. Eventually, in Fig.~\ref{fig:fEen_pspin} we plot the 1RSB free energy and the RS free energy as functions of $\beta$. The RS free energy is given by the analytical expression $f(\beta) = -\frac{\beta }{4}-\frac{\log (2)}{\beta }$. Notice how at $\beta_c$ there is a bifurcation of the free energy corresponding to the point where the 1RSB free energy dominates the thermodynamics of the model: the physical free energy is the maximum value of the two curves for each value of $\beta$.

It is important to stress once again that starting from the high temperature region, i.e.~a low value of $\beta$, the optimization remains stuck in the RS (or annealed/paramagnetic) solution. This is in line with the fact that the RS solution with $q=0$ remains stable at all temperature values in the $p$-spin model, differently from the SK case. The reason why this happens is that the transition to the 1RSB phase in the case of the Ising $p$-spin model is first order from the point of view of the order parameter, while the SK model is characterized by a continuous order parameter. In other terms, in the $p$-spin case the solution has a jump in the order parameter at the critical temperature to an already existing state, whereas in the SK model the order parameter continuously increases from zero, where the new state starts to exist. In order to obtain the previous results, one can proceed in two ways: either from a very high value of $\beta$ and heating the system, or, when approaching $\beta_c$ from below, by suggesting initial conditions for the loss optimization which are close to the right solution.  

The stability of the RS solution can be clearly visualized in the low temperature color maps of the loss function, which is defined as
\begin{equation}
    \mathcal{L}(q_1,m, \beta) =  \left( q_1 - \frac{\int \msD z~(2\cosh \Xi )^m \tanh^2\Xi}{\int \msD z~(2\cosh \Xi )^m} \right)^2 + (\partial_m \Phi)^2.
\end{equation}
where $\partial_m \Phi$ denotes the left hand side of Eq.~\eqref{mEqpspin}. In Fig.~\ref{fig:Loss_pspin} we display three different plots of the loss function, taken at the following values of the inverse temperature $\beta = 0.5$, $\beta = 3$ and $\beta=6.5$, from left to right. Values of $q_1$ (vertical axis) and $m$ (horizontal axis) are sampled in their definition interval $[0,1]$, with a spacing of $0.005$. Colors correspond to values of the loss function: the gradient of the loss points from blue to yellow . One can clearly identify a valley in the low region of the plots corresponding to the paramagnetic state at $q_1=0$ and degenerate in $m$, which is also present at $\beta<\beta_c$, and a smaller valley in the high region corresponding to the glassy state with a high value of $q_1$, which appears for $\beta > \beta_c$. The glassy state moves to the left of the plot towards lower values of $m$, consistently with the plot in Fig.~\ref{fig:m_pspin}.

\section{Zero Temperature Limit} \label{sec:0temp}
In this section we complete the study of the model, by computing the zero temperature limit of the 1RSB equations. Instead of computing the limit $\beta \rightarrow \infty$ directly on the equations, we first compute the zero temperature limit of the free energy \eqref{1RSBfreeENsimply}, by making scaling hypotheses on the parameters and introducing temperature independent quantities; then, we compute a new set of self-consistency equations for the new parameters, hoping that they turn out to be easier then those at finite temperature. Once the value of these parameters has been found, then the corresponding value of the free energy is the asymptotic value of the 1RSB free energy: if it turns out to be greater than the asymptotic value of the paramagnetic free energy, then we would have evidence of a phase transition at zero temperature.

\subsection{Test: the Ising p-spin model}
\begin{figure}[t]
   \begin{minipage}{0.48\textwidth}
     \centering
     \includegraphics[width=\linewidth]{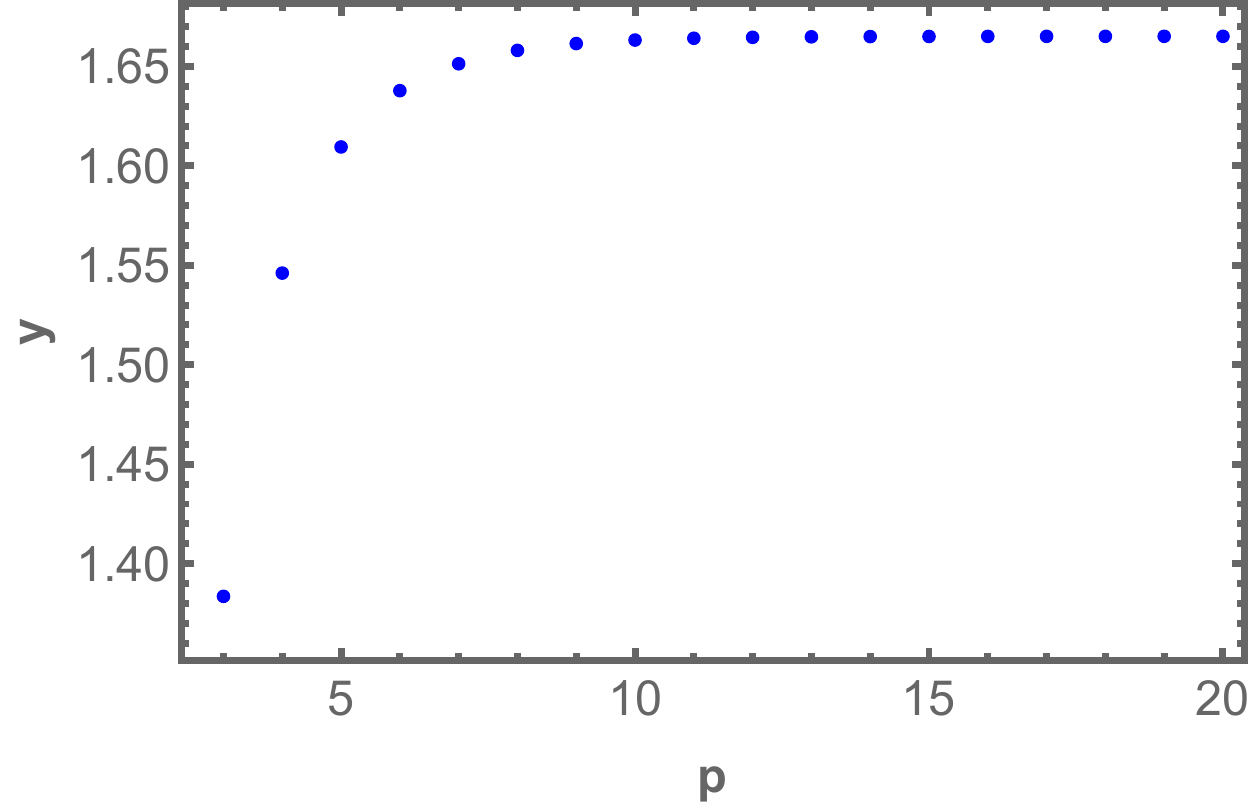}
     \caption{Solution of the self-consistency equation \eqref{yEq1} for $y$ as a function of $p$ in the interval $[3,20]$. The dependence on $p$ becomes weaker as $p$ increases.} \label{fig:y(p)}
   \end{minipage}\hfill
   \begin{minipage}{0.48\textwidth}
     \centering
     \includegraphics[width=\linewidth]{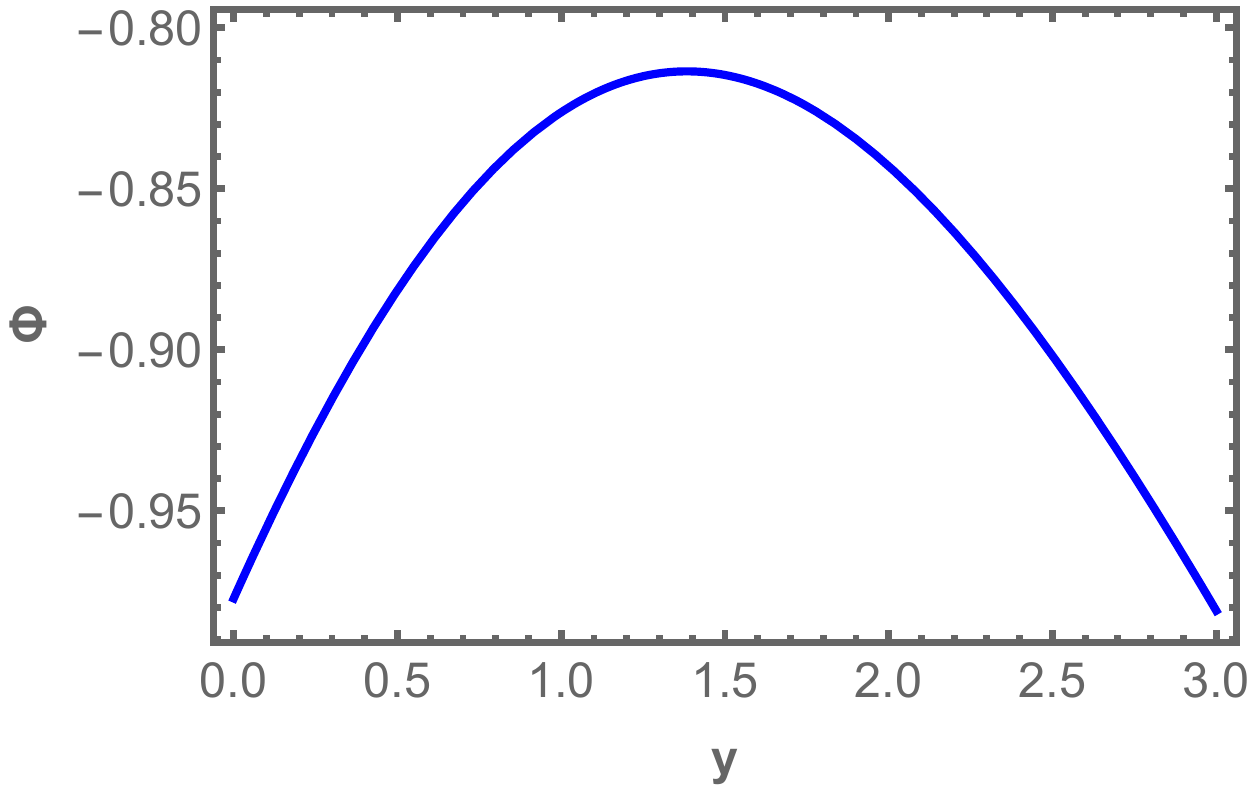}
     \caption{Function $\Phi(y)$ of Eq.~\ref{fig:Phi(y)} in the case $p=3$. The stationary point $y^*=1.38356\dots$ is actually a maximum. We notice that $\Phi(y^*)= -0.813535\dots$.} \label{fig:Phi(y)}
   \end{minipage}
\end{figure}

Let us first report the case of the Ising p-spin in order to gain familiarity with this kind of computations. We take the limit at the leading order in $1/\beta$, thus
\begin{align}
q_1 &= 1 + O(1/\beta) \\
m &= y/\beta + O(1/\beta^2),
\end{align}
which means that $m(T)$ linearly approaches zero when $T \rightarrow 0$, with a slope $y$ to be determined. By direct substitution in the first term of the free energy one finds
\begin{align*}
-\frac{\beta}{4} [1 + (p-1)(1-m)q_1^p - p q_1^{p-1}] = \frac{p-1}{4} y,
\end{align*}
up to terms of order $O(1/\beta^2)$. The integral can be evaluated by considering that
\begin{align*}
- \frac{1}{m\beta} \log \int \msD z ~[2\cosh(z\sqrt{p/2} \beta q^{\frac{p-1}{2}})]^m = - \frac{1}{y} \log  \int \msD z \exp\left[\frac{y}{\beta} \ln (e^{z \beta \sqrt{p/2}} + e^{-z \beta \sqrt{p/2}})\right] 
\end{align*}
and between the two exponentials inside the logarithm the first one dominates for $z>0$, while the second one dominates when $z<0$ in the $\beta \rightarrow \infty$ limit. Thus, we can write
\begin{align*}
- \frac{1}{y} \log  \int \msD z~ \exp\left[\frac{y}{\beta} \log (e^{z \beta \sqrt{p/2}} + e^{-z \beta \sqrt{p/2}})\right]  = - \frac{1}{y} \log  \int \msD z~e^{ y \sqrt{p/2} |z|}
\end{align*}
which can be expressed in terms of an error function as
\begin{align*}
- \frac{1}{y} \log  \int \msD z~e^{ y \sqrt{p/2} |z|} &= - \frac{1}{y} \log \left[e^{p y^2/4}\left(1+ \erf{\frac{\sqrt{p}y}{2}}\right)\right] \nonumber \\
&= -\frac{py}{4} - \frac{1}{y} \log\left( 1+ \erf{\frac{\sqrt{p}y}{2}}\right)
\end{align*}
Thus, at the leading order we get
\begin{align} \label{Free-En1}
\Phi(y) = \frac{1}{y} - \frac{1}{y} \log\left( 1+ \erf{\frac{\sqrt{p}y}{2}}\right)
\end{align}
where $y$ has to be determined by the self-consistency equation $\de \Phi/\de y=0$, i.e.
\begin{align} \label{yEq1}
-\frac{1}{4} + \frac{1}{y^2} \log \left( 1+ \erf{\frac{\sqrt{p}y}{2}}\right) - \frac{1}{y} \sqrt{\frac{p}{\pi}} \frac{e^{-p y^2/4}}{1+ \erf{\frac{\sqrt{p}y}{2}}} = 0.
\end{align}
This equation can be easily solved numerically: Fig.~\ref{fig:y(p)} shows the results for $p \in [3,20]$. For $p=3$, see Fig.~\ref{fig:Phi(y)}, the equation has the solution $y^*=1.38356\dots$, which is a maximum point for the free energy. The value of the free energy in $y^*$ is $\Phi(y^*) = -0.813535\dots$ in perfect agreement with the results of the numerical integration reported in the previous section.

\subsection{Case of interest}
We perform the following scaling ansatzes on the 1RSB parameters of our model
\begin{subequations}
\begin{gather}
    \mR_1 \simeq - r_1 \beta^2 ~~~ r_1 > 0 \\
    \mM_1 \simeq \mu_1 \beta^2 ~~~ \mu_1 > 0 \\
    \mR_D - \mR_1 \simeq \beta r_{-} \\
    \mM_D - \mM_1 \simeq \beta \mu_{-} \\
    m \simeq \frac{y}{\beta}.
\end{gather}
\end{subequations}
The guiding idea is to take the parameters (or the differences between couples of parameters) appearing inside the square roots as of order $O(\beta^2)$ and the others as  $O(\beta)$, apart from $m$, which is supposed to behave as in the $p$-spin case. The first local free energy in Eq.~\eqref{1RSBfreeENsimply} can be simply written by proceeding in analogy to the $p$-spin case and reads as
\begin{equation*}
    f_\tau(r_1,y) = - \frac{r_1}{2} y - \frac{1}{y} \log \left[1 + \erf\left( \frac{\sqrt{r_1}y}{\sqrt{2}} \right) \right].
\end{equation*}
The other local free energy is more difficult to compute in the large-$\beta$ limit. After implementing the scaling hypotheses on the parameters we find
\begin{equation*}
    f_C (\mu_{-}, \mu_1, y) = - \frac{1}{y} \log \int \msD u \msD v \left(\int \de a \de b~e^{\beta F(a,b|\mu_{-},\mu_1, u,v)}\right)^{y/\beta},
\end{equation*}
where the function $F$ corresponds to the argument of the exponent of the function $g_{\beta,0}$, see Eq. \eqref{gbeta0}, computed in $x=\mM_D$ and $y=\mM_1$, and reads
\begin{equation}
    F(a,b | \mu_{-},\mu_1, u,v) = - (a^2+b^2)^2 + \frac{\mu_{-}}{2} (a^2 + b^2) + \sqrt{\mu_1}(au +bv).
\end{equation}
The integrals in $a$ and $b$ can be computed with the saddle point method, since we are in the limit $\beta \rightarrow \infty$. Therefore we have
\begin{equation*}
    f_C (\mu_{-}, \mu_1, m) = - \frac{1}{y} \log \int \msD u \msD v ~e^{ y F(  \mu_{-},\mu_1, u,v  )},
\end{equation*}
where $F( \mu_{-},\mu_1, u,v ) = F(a^*,b^* |\mu_{-},\mu_1, u,v)$ and $a^*,b^*$ are functions of the other parameters, e.g. $a^* = a^*(\mu_{-},\mu_1,u,v)$ and are given by the solution of the coupled equations
\begin{gather*}
    \frac{\partial F}{\partial a} \Big|_{a^*,b^*} = 0 \\
    \frac{\partial F}{\partial b} \Big|_{a^*,b^*} = 0.
\end{gather*}
We do not report here the explicit expression of the function $F(u,v | \mu_{-},\mu_1 )$, since, after $a^*,b^*$ have been substituted, it becomes too cumbersome. Finally the entropic term of the free energy~\eqref{1RSBfreeENsimply} simply reduces to 
\begin{equation*}
    s_1(r_{-},r_1,\mu_{-},\mu_1,y) = - \frac{1}{y} \log \left(1+y \frac{r_1 - \mu_1}{r_{-} - \mu_{-}} \right).
\end{equation*}
Then, the complete free energy is given by
\begin{equation}
    \begin{split}
          f_{1\text{RSB}}(r_{-},r_1,\mu_{-},\mu_1,y) &= \frac{r_{-}}{2} - \frac{r_1}{2} y - \frac{1}{y} \log \left[1 + \erf\left( \frac{\sqrt{r_1}y}{\sqrt{2}} \right) \right] \\
    &\quad - \frac{1}{y} \log \int \msD u \msD v ~e^{ y F(\mu_{-},\mu_1,u,v )}  - \frac{1}{2 y} \log \left(1+y \frac{r_1 - \mu_1}{r_{-} - \mu_{-}} \right).
    \end{split}
\end{equation}
The self-consistency equations for the parameters introduced can be easily derived from the previous free energy by imposing the vanishing of the derivatives. We get the following system of equations
\begin{subequations}
\begin{gather}
    - y - \sqrt{\frac{2}{\pi}} \frac{r_1^{-1/2} e^{-r_1 y^2/2}}{1 + \erf\left(\frac{\sqrt{r_1}y}{\sqrt{2}} \right)} - \frac{1}{r_{-} - \mu_{-} + y (r_1 - \mu_1)} = 0 \\
    \frac{1}{r_{-} - \mu_{-} + y (r_1 - \mu_1)} = - \frac{r_1 - \mu_1}{(r_{-} - \mu_{-})} \\
    - \frac{\int \msD u \msD v~e^{y F} \partial_{\mu_1} F}{\int \msD u \msD v~e^{y F}} + \frac{1}{r_{-} - \mu_{-} + y (r_1 - \mu_1)} = 0 \\
    \frac{\int \msD u \msD v~e^{y F} \partial_{\mu_{-}} F}{\int \msD u \msD v~e^{y F}} + \frac{1}{r_{-} - \mu_{-} + y (r_1 - \mu_1)} = 0, 
\end{gather}
\end{subequations}
to which we have to add the derivative with respect to $y$. Resolution of the equations is in progress.


\newpage

\chapter{A New Mean-Field Theory for the Glassy Random Laser} \label{chap:ML}

In the previous chapter, we have presented a deterministic model with long-range interactions and a topology of the interaction network similar to the mode-locked graph, which can been solved by means of the replica method. It is now time to turn back to our original problem of reaching the analytical solution of the ML 4-phasor model. Inspired by the results of numerical simulations discussed in Chap.~\ref{chap:Univ}, we believe that a mean-field solution for this model may exist, even if most likely of a different kind with respect to the solution already obtained on the fully-connected graph \cite{Antenucci15a,Antenucci15b}. In fact, the model is characterized by a combined effect of quenched disorder due to the random couplings and deterministic dilution induced by the FMC. While in the case of ordered mode-locked graphs, a long-range spatial structure can be identified notwithstanding the dilution \cite{Antenucci15d,Antenucci15e}, we do not expect this to happen in the presence of disordered couplings. What we expect, and indeed what happens as soon as the replica method is applied to the model, is that the heterogeneities induced by the disorder do not simply average out as in the fully connected case, leading to the failure of the standard mean-field Replica Symmetry Breaking theory for spin-glass models. 

On the other hand, the order of the dilution is not such that the model can be defined on a sparse network, e.g.~the Bethe lattice, where the cavity method implemented through message passing algorithms like belief propagation works well \cite{Mezard16}. However, precisely because of the weakness of the dilution, our conjecture, supported by numerical evidence, is that the interaction network is still dense enough to compensate the effect of the heterogeneities and to be compatible with a mean-field approximation, although with a more complicated theory than the standard one. 

In this chapter, after presenting the ML 4-phasor model in connection to the Merit Factor problem and explaining the solution strategy, we report the various steps of the replica computation, which goes along the same lines of the previous chapter, leading to the saddle point equations for the ML 4-phasor model. 

\section{The Model}
For the purpose of defining a new mean-field theory for the glassy random laser, we have developed a technique based on the formal analogy with the Bernasconi model for the MF problem \cite{Bernasconi87}, to which the previous chapter has been entirely devoted. In order to make the discussion more specific, let us recall the model and add some technical details. 

The Hamiltonian function in which we are interested reads
\begin{align} \label{5-Hamilt1}
    \mH[\bm{a}] = - \sum_{\text{FMC}} J_{ijkl} [a_i \overline{a}_j a_k \overline{a}_l + \text{c.c.}],
\end{align}
where $\bm{a}$ is a complex vector on the $N$-sphere and  $J_{ijkl}$ are quenched disordered couplings. The summation is restricted to all those indices which satisfy the FMC in the case of the linear comb
\begin{align}
    |i-j+k-l| = 0.
\end{align}
Due to this constraint on the interacting quadruplets, the Hamiltonian can be written in a more convenient way as follows
\begin{align}
    \mH[\bm{a}] = - \sum_{i<j<k}^N J_{ijk}[a_i \overline{a}_{i+k} a_j \overline{a}_{j+k} + \text{c.c.}],
\end{align}
where we recognize exactly the same structure of the indices of the Bernasconi model. 

Besides the complex variables, the other obvious difference with respect to the case of the previous chapter is the randomness of the couplings, which are independently extracted from the following zero-mean Gaussian distribution
\begin{align}   \label{5-Gaussian}
 P(J_{ijk})= \frac{1}{\sqrt{2 \pi \sigma_J^2}} \exp\left[-\frac{J_{ijk}^2}{2 \sigma_J^2}\right] ~~~~~~~ \sigma_J^2 =  \frac{3! J^2}{2 N^2},
\end{align}
where the scaling of the variance with $N$ ensures the extensivity of the thermodynamic potentials. Moreover the sign in front of the summation in the Hamiltonian is different in the two cases. However, this is not a big deal in the present case, since for random couplings extracted from an unbiased symmetric distribution it makes no real difference whether one has a plus or a minus in the Hamiltonian definition: after averaging the partition function over the Gaussian distribution~\eqref{5-Gaussian}, one is left with the Hamiltonian squared.

\subsection{Strategy of Solution}
Our strategy is to transform the disordered model of
Eq.~\eqref{5-Hamilt1} into a non-disordered one, which is equivalent to the Bernasconi model, though with different variables, and can be solved exactly, by using the associated random unitary model. In order to solve the model we will then need to introduce two averages over the disorder. 

\begin{itemize}
\item \emph{First Average}: this is the average taken over the quenched randomness, which we denote as $\overline{(\cdots)}$. Due to the dilution of the graph, after the average over the Gaussian couplings is performed, one is forced to introduce local matrices $q_i^{\alpha\beta}$. With respect to these variables, our problem simplifies to the study of an ordered model with long-range interactions, which has a Hamiltonian of the kind
\begin{align} \label{5-HamiltMFoverlap}
   \mH = \sum_{k=1}^N  \left(\frac{1}{N}\sum_{i=1}^N q_iq_{i+k}\right)^2 = \sum_{k=1}^N  \mathcal{C}_k^2,  
\end{align}
where we have dropped the dependence on the replica indices, just to make more clear the analogy with the Hamiltonian of the MF problem. Namely, we look for the sequences of $q_i$ for which the correlation $\mathcal{C}_k^2$, summed for all distances $k$, takes the lowest value. In other words, we have a problem which is analogous to the MF problem, but at the level of the local overlaps $q_i$, rather than of the spins. Clearly, if the original variables are phasors $a_k$, their local overlap will be in general a complex quantity, so this is a more general problem with respect to the search for the LABS.

\item \emph{Second Average}: after the first average, one is left with a deterministic model in the local overlaps with long-range interactions. What we can do now is simply to apply all the machinery developed in the previous chapter for the solution of the MF problem, by following Ref.~\cite{Marinari94a}. We will then associate to the model Eq.~\eqref{5-HamiltMFoverlap} the corresponding random unitary model, by replacing the usual Fourier transformation of the variable with generic unitary matrices, over which we will perform a second average. In order to distinguish it from the average over the quenched disorder, we denote this average with $\overline{(\cdots)}^U$. Following this procedure it turns out that, for our problem, one has to introduce the overlap between local overlaps:
\begin{align}
\msQ_{\alpha\beta} = \frac{1}{N} \sum_{i=1}^N q_i^\alpha q_i^\beta.   
\end{align} 
We will refer to this quantity as \emph{superoverlap}. Our main result is to show that for the ML p-spin a
mean-field ansatz for the structure of replica matrices can be done only at the level of $\msQ_{\alpha\beta}$ matrices. Numerical simulations shows clear evidence of a glass transition at low temperatures, so that we are led to assume at low temperatures a replica-symmetry breaking ansatz for $\msQ_{\alpha\beta}$.
\end{itemize}

\section{Average over Disorder}
In order to disentangle the difficulties, we consider the case of non-complex variables, leaving the generalization to phasors for the future. We carry out the computations in parallel for both Ising and spherical spins, up to the point where some ansatz for the solution of the saddle point equations has to be performed. Then, the Hamiltonian we consider is
\begin{align}
    \mH_J(\bm{\sigma}) = - \sum_{i<j<k}^N J_{ijk} \sigma_i\sigma_{i+k}\sigma_{j}\sigma_{j+k}
\end{align}
and the configuration space is given by either
\begin{align}
    \Sigma_N = \left\{ \begin{array}{r}
    \begin{aligned}
     &\{\pm 1\}^N~~~~\text{or} \\ 
     &\mathbb{S}_N = \{\bm{\sigma} : \sum_{i=1}^N \sigma_i^2 = N \}.
     \end{aligned}
\end{array}\right.
\end{align}
We notice that the two cases differ in the number of constraints: for Ising spins, we have $N$ local constraints, whereas spherical spins are locally unbounded, but have to satisfy a global constraint. The partition function of the model can be written as
\begin{align}
    \mZ = \Tr_{\Sigma_N} e^{-\beta \mH_J(\bm{\sigma})},
\end{align}
where the trace is a compact notation for the summation over all possible configurations, which in the case of spherical spins corresponds to an integration over the $N$-sphere, i.e.
\begin{align}
     \Tr_{\Sigma_N} = \left\{ \begin{array}{r}
    \begin{aligned}
     & \prod_{i=1}^N\sum_{\sigma_i=\pm 1}  \\ & \int_{\mathbb{S}_N} \de \bm{\sigma} = \int \prod_{i=1}^n \de\sigma_i~\delta \left(\sum_{i=1}^N \sigma_i^2 - \epsilon N \right),
     \end{aligned}
\end{array}\right.
\end{align}
where $\epsilon$ is a constant which tunes the constraint, i.e.~the radius of the hypersphere (see Chap.~\ref{chap:Condens}). The replicated partition function reads
\begin{align}
    \mZ^n =  \Tr_{\Sigma_N^n} \exp\left[- \beta \sum_{\alpha=1}^n \mH_J(\bm{\sigma}^\alpha) \right]
\end{align}
where $\Sigma_N^n = \bigotimes_{\alpha=1}^n \Sigma_N^\alpha$. In the thermodynamic limit the free energy of the model is given by
\begin{align} \label{5-freeEn1}
    f(\beta) = \lim_{n \rightarrow 0}\lim_{N \rightarrow \infty} - \frac{1}{\beta n N} \log \overline{ \mZ^n},
\end{align}
where the order of the two limits have been exchanged, as is usual in the replica method. 

The average of the replicated partition function over quenched disorder is computed as follows
\begin{align*}
    \overline{\mZ^n} &=  \Tr_{\Sigma_N^n} \overline{\exp \left[\beta \sum_{\alpha=1}^n \sum_{i<j<k}^N J_{ijk} \sigma_i^\alpha \sigma_{i+k}^\alpha \sigma_{j}^\alpha \sigma_{j+k}^\alpha  \right]}  \\
    & = \Tr_{\Sigma_N^n} \prod_{i<j<k}^N \int   \frac{\de J_{ijk}}{\sqrt{2 \pi \sigma _J^2}} \exp\left[ -\frac{J_{ijk}^2}{2 \sigma_J^2} + \beta  J_{ijk} \sum_{\alpha=1}^n \sigma_i^\alpha \sigma_{i+k}^\alpha \sigma_{j}^\alpha \sigma_{j+k}^\alpha  \right]  \\
    & = \Tr_{\Sigma_N^n} \exp\left[ \frac{(\beta J)^2 }{4} \frac{3!}{N^2} \sum_{i<j<k}^N \sum_{\alpha \beta}^n \sigma_i^\alpha \sigma_{i+k}^\alpha \sigma_{j}^\alpha \sigma_{j+k}^\alpha \sigma_i^\beta \sigma_{i+k}^\beta \sigma_{j}^\beta \sigma_{j+k}^\beta \right]  \\
    &= \Tr_{\Sigma_N^n} \exp \left[ \frac{(\beta J)^2 }{4} \sum_{\alpha\beta}^n \sum_{k=1}^N \left(\frac{1}{N} \sum_{i=1}^N \sigma_i^\alpha \sigma_i^\beta \sigma_{i+k}^\alpha \sigma_{i+k}^\beta \right)^2 \right],
\end{align*}
where the last expression is correct up to order $O(1/N)$, since $\frac{3!}{N^2} \sum_{i<j<k} = \frac{1}{N^2} \sum_{ijk} + O(1/N)$, see e.g.~Ref.~\cite{Crisanti92}. We have already reached the point where the standard mean-field computation breaks down for this model: although, as usual, the average over disorder has led to coupled replicas, we are not able to introduce at this point a global overlap between configurations of different replicas. What one can do instead is to change variables from spins to local overlaps $q_i^{\alpha \beta} = \sigma_i^\alpha  \sigma_i^\beta $. In order to predispose the model for a generic unitary transformation, in analogy with the Merit Factor we define the complex overlaps $\mfq_{i}^{\alpha\beta}= \sigma_{2i-1}^\alpha\sigma_{2i-1}^\beta+i \sigma_{2i}^\alpha\sigma_{2i}^\beta $ and multiply the whole partition function by
\begin{align*}
  1 & =
  \prod_{i=1}^{N/2} \prod_{<\alpha \beta>}^n \int \de\mfq_{i}^{\alpha\beta}\de\overline{\mfq}_{i}^{\alpha\beta}~
  \delta\left( \mfq_{i}^{\alpha\beta}-(\sigma_{2i-1}^\alpha\sigma_{2i-1}^\beta+i \sigma_{2i}^\alpha\sigma_{2i}^\beta) \right)
  \delta\left( \overline{\mfq}_{i}^{\alpha\beta}-(\sigma_{2i-1}^\alpha\sigma_{2i-1}^\beta-i \sigma_{2i}^\alpha\sigma_{2i}^\beta)\right)  \\ 
   & = \int \prod_{i=1}^{N/2} \prod_{<\alpha \beta>}^n \left[\de\mfq_i^{\alpha\beta} \de\overline{\mfq}_i^{\alpha\beta}
   \de\lambda_i^{\alpha\beta} \de\overline{\lambda}_i^{\alpha\beta}\right] ~
   \exp\left[\frac{1}{2}\sum_{\alpha \beta}^n \sum_{i=1}^{N/2} \overline{\lambda}_i^{\alpha\beta}[\mfq_{i}^{\alpha\beta}-(\sigma_{2i-1}^\alpha\sigma_{2i-1}^\beta+i \sigma_{2i}^\alpha\sigma_{2i}^\beta)] + \text{c.c.} \right],
\end{align*}
where the integral over the $\lambda_i^{\alpha\beta}$ is between $\pm i \infty$ and the argument of the exponent has been symmetrized in the replica indices. In order to keep a compact notation, we use the symbol $<\alpha \beta>$ in the product to denote how many independent values of the local overlap we have introduced. On one hand, the product has $\alpha < \beta $ terms in the Ising case, where the diagonal terms are fixed by the local constraints and yield the following constant contribution 
\begin{align*}
    \exp \left[ \frac{(\beta J)^2 }{4} \sum_{\alpha}^n \sum_{k=1}^N \left(\frac{1}{N} \sum_{i=1}^N (\sigma_i^\alpha)^2 (\sigma_{i+k}^\alpha)^2 \right)^2 \right] = \exp\left[\frac{(\beta J)^2Nn}{4}\right].
\end{align*}
This term can be dropped from the computation and added eventually to the free energy. On the other hand, the product has $\alpha \leq \beta $ terms in the spherical case, since the diagonal terms are free to vary compatibly with the global constraint. Furthermore, the delta functions implementing the spherical constraint on each replica of the system can be written in terms of the complex overlaps as follows
\begin{align} \label{5-spherConstover}
   \prod_{\alpha=1}^n \delta \left(\sum_{i=1}^N (\sigma_i^\alpha)^2 - \epsilon N \right) = \prod_{\alpha=1}^n\delta\left( \sum_{i=1}^{N/2} (\Re[\mfq_i^{\alpha\alpha}] + \Im[\mfq_i^{\alpha\alpha}]) - \epsilon N \right).
\end{align}
However, in order to keep the notation compact, for the moment this contribution will be left inside the definition of the trace operator for the continuous case. At this point, it is convenient to define an action functional in order to write the partition function in the following way
\begin{align*}
    \overline{\mZ^n} = \Tr_{\Sigma_N^n}  \int \prod_{i=1}^{N/2} \prod_{<\alpha \beta>}^n \left[\de\mfq_i^{\alpha\beta} \de\overline{\mfq}_i^{\alpha\beta}
   \de\lambda_i^{\alpha\beta} \de\overline{\lambda}_i^{\alpha\beta}\right] \exp \left[S(\mfq_i^{\alpha\beta},\overline{\mfq}_i^{\alpha\beta},\lambda_i^{\alpha\beta},\overline{\lambda}_i^{\alpha\beta},\sigma_i^\alpha)\right] ,
\end{align*}
where
\begin{equation} \label{5-action1}
\begin{split}
    S(\mfq_i^{\alpha\beta},\overline{\mfq}_i^{\alpha\beta},\lambda_i^{\alpha\beta},\overline{\lambda}_i^{\alpha\beta},\sigma_i^\alpha) &= \frac{\beta^2J^2}{4} \sum_{\alpha\beta}^n \sum_{k=1}^{N/2} \left( \frac{1}{N}\sum_{i=1}^{N/2} \overline{\mfq}_i^{\alpha\beta} \mfq_{i+k}^{\alpha\beta}\right)^2  \\
    & \quad + \frac{1}{2} \sum_{\alpha\beta}^n \sum_{i=1}^{N/2} \overline{\lambda}_i^{\alpha\beta}[\mfq_{i}^{\alpha\beta}-(\sigma_{2i-1}^\alpha\sigma_{2i-1}^\beta+i \sigma_{2i}^\alpha\sigma_{2i}^\beta)] + \text{c.c.}   
\end{split}
\end{equation}

After averaging over the Gaussian couplings, we have a matrix field theory in the overlap matrices $\mfq_i^{\alpha\beta}$, which do not depend only on the replica indices $\alpha\beta$ but also on the site index $i$. Different indices $i$ and $i+k$ are coupled in the interaction terms. One can define a new Hamiltonian as
\begin{align*}
    \mH = \frac{\beta J^2}{4} \sum_{\alpha\beta}^n \sum_{k=1}^{N/2} \left( \frac{1}{N}\sum_{i=1}^{N} q_i^{\alpha\beta} q_{i+k}^{\alpha\beta}\right)^2.
\end{align*}
which is the equivalent of the Merit Factor Hamiltonian of the local overlaps, while the other terms in the action~\eqref{5-action1} are entropic contributions accounting for the fact that the fundamental variables are the spins. In analogy with the previous chapter, we introduce the generic unitary transform of the complex overlaps as
\begin{align} 
\mQ^{\alpha\beta}_k = \sum_{r=1}^{N/2} U_{kj}~\mfq^{\alpha\beta}_j = \left[ U{\boldsymbol \mfq^{\alpha\beta}} \right]_k, 
\end{align}
where $\bm{U}$ represents a generic $N/2 \times N/2$ matrix of the unitary group. As we already know, the interaction term is diagonalized by this transformation and reads as 
\begin{align*}
    \mH = \frac{\beta J^2}{4} \sum_{\alpha\beta}^n \sum_{k=1}^{N/2} |\mQ_k^{\alpha \beta}|^4.
\end{align*}
We introduce the unitary-tranformed variables in the computation as usual by means of delta functions
\begin{align*}
    1 &= \prod_{<\alpha \beta>}^n \prod_{k=1}^{N/2} \int  \de \mQ^{\alpha\beta}_k \de \overline{\mQ}^{\alpha\beta}_k~\delta\left(\mQ^{\alpha\beta}(k)-[U{\boldsymbol \mfq^{\alpha\beta}}]_k\right)~\delta\left(\overline{\mQ}^{\alpha\beta}(k)-[\overline{U{\boldsymbol \mfq^{\alpha\beta}}}]_k\right) \\
    & = \int  \prod_{<\alpha \beta>}^n \prod_{k=1}^{N/2} \left[ \de \mQ^{\alpha\beta}_k \de \overline{\mQ}^{\alpha\beta}_k  \de \xi^{\alpha\beta}_k \de \overline{\xi}^{\alpha\beta}_k \right] \exp\left[ \frac{1}{2}\sum_{\alpha \beta}^n \sum_{k=1}^{N/2} i \overline{\xi}_k^{\alpha\beta}\left(\mQ^{\alpha\beta}(k)-[U{\boldsymbol \mfq^{\alpha\beta}}]_k\right) + \text{c.c.} \right],
\end{align*}
where, once again, the diagonal terms have been included (excluded) in the continuous (discrete) case and the argument of the exponent has been symmetrized in the replica indexes. In order to lighten the notation, let us introduce the following convention for the symbol of integration over local variables
\begin{align*}
     \mD x = \prod_{<\alpha \beta>}^n \prod_{k=1}^{N/2} \de x_i^{\alpha\beta}.
\end{align*}

At this stage the partition function can be written as 
\begin{align*}
    \overline{\mZ^n} = \Tr_{\Sigma_N^n}  \int \mD\mQ \mD \overline{\mQ} \mD \mfq\mD \overline{\mfq} \mD \xi \mD
    \overline{\xi} \mD \overline{\lambda}\mD \lambda~ \exp \left[ S_U(\mQ, \overline{\mQ}, \mfq,\overline{\mfq},\xi,\overline{\xi},\lambda,\overline{\lambda},\sigma) \right],
\end{align*} 
and the action reads as
\begin{equation} \label{5-action2}
\begin{split}
    S_U(\mQ, \overline{\mQ}, \mfq,\overline{\mfq},\xi,\overline{\xi},\lambda,\overline{\lambda},\sigma) &= \frac{\beta^2J^2}{4} \sum_{\alpha\beta}^n \sum_{k=1}^{N/2} |\mQ_k^{\alpha \beta}|^4  + \frac{1}{2}\sum_{\alpha \beta}^n \sum_{k=1}^{N/2} i \overline{\xi}_k^{\alpha\beta}\left(\mQ^{\alpha\beta}(k)-[U{\boldsymbol \mfq^{\alpha\beta}}]_k\right) + \text{c.c.}\\
    & \quad + \frac{1}{2} \sum_{\alpha\beta}^n \sum_{i=1}^{N/2} \overline{\lambda}_i^{\alpha\beta}[\mfq_{i}^{\alpha\beta}-(\sigma_{2i-1}^\alpha\sigma_{2i-1}^\beta+i \sigma_{2i}^\alpha\sigma_{2i}^\beta)] + \text{c.c.}   
\end{split}
\end{equation}
where the subscript $U$ specifies its
dependence on the specific realization of a random unitary matrix.

\section{Average over Unitary Matrices}
According to Eq.~\eqref{5-action2} the replicated partition function depends on a new source of randomness, that is we have a free energy $f_U(\beta)$ and we aim to compute $\overline{f_U(\beta)}^U$. We can use the fact that $\log \overline{Z^n} = \log\left(1 + (\overline{Z^n}-1) \right)$ and since $(\overline{Z^n}-1) = O(n)$ we can write the free energy in the equivalent form
\begin{align}
    f_U(\beta) = \lim_{n \rightarrow 0}\lim_{N \rightarrow \infty} - \frac{1}{\beta N} \frac{\overline{ \mZ^n}-1}{n}.
\end{align}
This is all very standard, but it allows us to understand that, at variance with the MF problem, in this case it is sufficient to perform the annealed average over the matrices $U$, since, thanks to the average over the couplings, we have already dealt with the problem of integrating the logarithm of the partition function and we are interested in the moments of the partition function. Therefore, the free energy averaged over the unitary group is simply
\begin{align}
    f(\beta) = \lim_{n \rightarrow 0}\lim_{N \rightarrow \infty} - \frac{1}{\beta N} \frac{\overline{\left(\overline{ \mZ^n}\right)}^U-1}{n}.
\end{align}

The computation becomes now very similar to the one performed in the previous chapter. We select the $U$-dependent part of Eq.~\eqref{5-action2}, we introduce auxiliary variables $\Omega_{ik}=i\sum_{\alpha \beta}\xi_k^{\alpha \beta} \overline{\mfq}_i^{\alpha \beta}/2$ and perform the integration on the unitary group as follows
\begin{align*}
    \overline{\exp\left[\sum_{kj}^{N/2} \overline{\Omega}_{kl} U_{kl} + \text{c.c.} \right]}^U &= \int \de U \de U^\dagger \exp \left[ \Tr(\Omega^\dagger U + \text{h.c.})\right] \\
    & = \exp \left[\frac{N}{2} \Tr \mG \left( \frac{\Omega^\dagger \Omega}{N^2} \right) \right] = \exp \left[ \frac{N}{2} \Tr \mG \left( \frac{\Lambda \msQ}{4} \right)\right],
\end{align*}
where we have used the results of Refs.~\cite{Brezin80b, Marinari92}. In particular, we recall that the function $\mG$ is defined as in Eq.~\eqref{MF:G}. Moreover, we have introduced the overlaps
\begin{align}
    \msQ_{\alpha\beta,\gamma\delta} = \frac{1}{N}
\sum_{i=1}^{N/2} \mfq_{i}^{\alpha\beta}
~\overline{\mfq}_{i}^{\gamma\delta}~~~~~\Lambda_{\alpha\beta,\gamma\delta}
  = \frac{1}{N} \sum_{k=1}^{N/2} \xi_{k}^{\alpha\beta}
  ~\overline{\xi}_{k}^{\gamma\delta},
\end{align}
which represent the new global order parameters of the theory: the overlaps between local overlap fields. In order to change variables from the local fields
$\lambda_{k}^{\alpha\beta} $ and $\mfq_{i}^{\gamma\delta}$ to the global
matrices $\msQ_{\alpha\beta,\gamma\delta}$ and
$\Lambda_{\alpha\beta,\gamma\delta}$ we introduce the following
terms in the partition function
\begin{align*}
    1 = \int_{-\infty}^\infty D\msQ~\int_{-i\infty}^{i\infty} D\hat{\msQ}~
    \exp\left[\frac{N}{4} \sum_{\alpha\beta}^n\sum_{\gamma\delta}^n\hat{\msQ}_{\alpha\beta,\gamma\delta}\msQ_{\alpha\beta,\gamma\delta}- \frac{1}{4}\sum_{\alpha\beta}^n\sum_{\gamma\delta}^n\sum_{i=1}^{N/2} \mfq_{i}^{\alpha\beta}  \hat{\msQ}_{\alpha\beta,\gamma\delta} \overline{\mfq}_{i}^{\gamma\delta} \right]
\end{align*}
and, similarly, 
\begin{align*}
    1 =& \int_{-\infty}^\infty~D\Lambda~\int_{-i\infty}^{i\infty}D\hat{\Lambda}~\exp\left[ \frac{N}{4} \sum_{\alpha\beta}^n\sum_{\gamma\delta}^n\hat{\Lambda}_{\alpha\beta,\gamma\delta}\Lambda_{\alpha\beta,\gamma\delta}-\frac{1}{4} \sum_{\alpha\beta}^n\sum_{\gamma\delta}^n\sum_{k=1}^{N/2} \xi_{k}^{\alpha\beta} \hat{\Lambda}_{\alpha\beta,\gamma\delta} \overline{\xi}_k^{\gamma\beta}\right],
\end{align*}
where the integration measures for the global order parameters $\msQ,\Lambda$ and their Lagrange multipliers $\hat{\msQ},\hat{\Lambda}$ read as
\begin{align}
    DX = \prod_{<\alpha\beta>}^n \prod_{<\gamma\delta>}^n \de X_{\alpha\beta,\gamma\delta},
\end{align}
with the usual meaning of the symbol $<\alpha\beta>$. Moreover, let us momentarily denote by $x$ all the local variables of the theory $\{\mQ, \overline{\mQ}, \mfq,\overline{\mfq},\xi,\overline{\xi},\lambda,\overline{\lambda},\sigma \}$ and by $X$ all the global ones $\{ \msQ, \hat{\msQ}, \Lambda, \hat{\Lambda} \}$. With these notations the averaged partition function reads as
\begin{align*}
    \mZ \equiv \overline{\left(\overline{ \mZ^n}\right)}^U = \Tr_{\Sigma_N^n}  \int D \msQ D\hat{\msQ} D\Lambda D\hat{\Lambda}  \mD\mQ \mD \overline{\mQ} \mD \mfq\mD \overline{\mfq} \mD \xi \mD
    \overline{\xi} \mD \overline{\lambda}\mD \lambda ~\exp[S(X,x)],
\end{align*}
where 
\begin{equation} \label{5-action3}
\begin{split}
    S(X,x) &= \frac{\beta^2 J^2}{4}
\sum_{\alpha\beta}^n \sum_{k=1}^{N/2}|\mQ^{\alpha\beta}(k)|^4  +
\frac{1}{2}\sum_{\alpha\beta}^n \sum_{k=1}^{N/2} \left( i \overline{\xi}_{k}^{\alpha\beta}
  \mQ^{\alpha\beta}_k + \text{c.c.} \right) \\ & \quad +\frac{1}{2}\sum_{\alpha\beta}^n \sum_{i=1}^{N/2} \overline{\lambda}_{i}^{\alpha\beta}[\mfq_{i}^{\alpha\beta}-(\sigma_{2i-1}^\alpha\sigma_{2i-1}^\beta+i \sigma_{2i}^\alpha\sigma_{2i}^\beta)] + \textrm{c.c.}  \\
& \quad + \frac{N}{8}~\textrm{Tr} G(\Lambda\msQ) 
 + \frac{N}{4}\sum_{\alpha\beta}\sum_{\gamma\delta}\hat{\Lambda}_{\alpha\beta,\gamma\delta}~\Lambda_{\alpha\beta,\gamma\delta}-\frac{1}{4}\sum_{\alpha\beta}^n\sum_{\gamma\delta}^n\sum_{k=1}^{N/2}
\xi_{k}^{\alpha\beta}~\hat{\Lambda}_{\alpha\beta,\gamma\delta}~\overline{\xi}_{k}^{\gamma\delta}  \\
&\quad + \frac{N}{4}\sum_{\alpha\beta}^n\sum_{\gamma\delta}^n\hat{\msQ}_{\alpha\beta,\gamma\delta}~\msQ_{\alpha\beta,\gamma\delta}-\frac{1}{4}\sum_{\alpha\beta}^n\sum_{\gamma\delta}^n\sum_{i=1}^{N/2}
\mfq_{i}^{\alpha\beta}~\hat{\msQ}_{\alpha\beta,\gamma\delta}~\overline{\mfq}_{i}^{\gamma\delta}.
\end{split}
\end{equation}

We now develop some further manipulations, which will simplify the expression of the partition function. First, the Gaussian integration over the complex matrices $\xi_k^{\alpha\beta}$ can be easily carried out, yielding up to constant terms 
\begin{align*}
    \int  \mD \xi \mD \overline{\xi} \exp \left[ -\frac{1}{4}\sum_{\alpha\beta}^n\sum_{\gamma\delta}^n\sum_{k=1}^{N/2}
\xi_{k}^{\alpha\beta}~\hat{\Lambda}_{\alpha\beta,\gamma\delta}~\overline{\xi}_{k}^{\gamma\delta} + \frac{1}{2}\sum_{\alpha\beta}^n \sum_{k=1}^{N/2} \left( i \overline{\xi}_{k}^{\alpha\beta}
  \mQ^{\alpha\beta}_k + \text{c.c.} \right) \right] = \\ 
  = \exp\left[ -\frac{N}{2} \log \det \hat{\Lambda} + \sum_{\alpha\beta}^n\sum_{\gamma\delta}^n \sum_{k=1}^{N/2} \mQ^{\alpha\beta}_k [\hat{\Lambda}^{-1}]_{\alpha\beta,\gamma\delta}\overline{\mQ}^{\gamma\delta}_k \right].
\end{align*}
Furthermore, we can store all the dependence on the local variables inside the definition of free-energy functions. Let us first consider the variables depending on the local indices of the real space. In order to simplify the dependence on indices it is better to rename the spin variables in a way which makes explicit the fact that they are independent integration variables: $u_i=\sigma_{2i-1}$ and $v_i = \sigma_{2i}$. Then, we define
\begin{equation}
    \begin{split}
      e^{F_q(\hat{\msQ})} = \Tr_{\Sigma_N^n} \int \mD \mfq \mD \overline{\mfq}
    \mD \lambda\mD \overline{\lambda}~\exp \Bigg[&-\frac{1}{4}\sum_{i=1}^{N/2}\sum_{\alpha\beta}^n\sum_{\gamma\delta}^n  \mfq_{i}^{\alpha\beta}~\hat{\msQ}_{\alpha\beta,\gamma\delta}~\overline{\mfq}_{i}^{\gamma\delta}  \\ 
    & +\frac{1}{2}\sum_{\alpha\beta}^n \sum_{i=1}^{N/2} \overline{\lambda}_{i}^{\alpha\beta}[\mfq_{i}^{\alpha\beta}-(u_i^\alpha u_i^\beta+i v_i^\alpha v_i^\beta)] + \textrm{c.c.} \Bigg],  
    \end{split}
\end{equation}
where we will see in a short while that $F_q(\hat{\msQ})$ can be factorized as $\frac{N}{2}f_q(\hat{\msQ})$.

Similarly, the dependence on the local unitary transformed variables can be put in the following free energy, which immediately factorizes in $N/2$ local identical contributions
\begin{equation}
    e^{ \frac{N}{2} f_\mQ(\hat{\Lambda})} = \left\{ \int \prod_{<\alpha,\beta>}^n \de \mQ^{\alpha\beta} \de \overline{\mQ}^{\alpha\beta}
    \exp\left[  \frac{\beta^2J^2}{4} \sum_{\alpha\beta}^n |\mQ^{\alpha\beta}|^4   + \sum_{\alpha\beta}^n\sum_{\gamma\delta}^n  \mQ^{\alpha\beta} [\hat{\Lambda}^{-1}]_{\alpha\beta,\gamma\delta}\overline{\mQ}^{\gamma\delta} \right] \right\}^{\frac{N}{2}}.
\end{equation}
Eventually, we can rewrite the partition function in a very compact expression as follows
\begin{align*}
    \mZ = \int D \msQ D\hat{\msQ} D\Lambda D\hat{\Lambda}  \exp\left[ S(\msQ, \hat{\msQ}, \Lambda, \hat{\Lambda}) \right]
\end{align*}
where 
\begin{equation}
\begin{split}
    S(\msQ, \hat{\msQ}, \Lambda, \hat{\Lambda}) &= \frac{N}{2}\Bigg\{  f_\mQ(\hat{\Lambda})+ f_q(\hat{\msQ}) + \frac{1}{2}\textrm{Tr}~(\hat{\Lambda}\Lambda) + \frac{1}{2}\textrm{Tr}~(\hat{\msQ}\msQ) \\
    &\quad+ \textrm{Tr}~G\left(\frac{\Lambda \msQ}{4} \right) -\log(\det\hat{\Lambda})\Bigg\}
\end{split}
\end{equation}

\subsection{Free Energy of the Local Overlap}
In the discrete case, it is immediate to see that the free energy $F_q(\hat{\msQ})$ corresponds indeed to the sum of $N/2$ independent and identical local free-energies, where the expression of the trace operation is simply
\begin{align*}
    \Tr_{\Sigma_N^n} = \prod_{\alpha=1}^n \prod_{i=1}^{N/2} \sum_{u_i^\alpha=\pm 1} \sum_{v_i^\alpha=\pm 1}.
\end{align*}
Since the exponent in the definition of $F_q(\hat{\msQ})$ is diagonal in the local indices, one can define $N/2$ terms of the kind
\begin{equation}
\begin{split}
    f_\mfq(\hat{\msQ}) &= \log  \prod_{\alpha=1}^n\sum_{u^\alpha=\pm 1} \sum_{v^\alpha=\pm 1} \int \prod_{\alpha < \beta}^n \left[\mfq^{\alpha\beta} \overline{\mfq}^{\alpha\beta} \lambda^{\alpha\beta} \overline{\lambda}^{\alpha\beta} \right] ~  \\ 
    & \quad \times \exp \left[-\frac{1}{4}\sum_{\alpha\beta}^n\sum_{\gamma\delta}^n  \mfq^{\alpha\beta}~\hat{\msQ}_{\alpha\beta,\gamma\delta}~\overline{\mfq}^{\gamma\delta}  
     +\frac{1}{2}\sum_{\alpha\beta}^n \overline{\lambda}^{\alpha\beta}[\mfq^{\alpha\beta}-(u^\alpha u^\beta+i v^\alpha v^\beta)] + \textrm{c.c.} \right]
\end{split}
\end{equation}

In order to show that the relation $F_\mfq(\hat{\msQ}) =
\frac{N}{2} f_\mfq(\hat{\msQ})$ holds also in the continuous case, one has to ``open'' the Dirac delta of the spherical constraint, which we have hidden inside the trace operator
\begin{align}
    \Tr_{\Sigma_N^n} = \int \prod_{\alpha =1}^n \prod_{i=1}^{N/2} \de u_i^\alpha \de v_i^\alpha \prod_{\alpha=1}^n\delta\left( \sum_{i=1}^{N/2} \left[(u_i^\alpha)^2 + (v_i^\alpha)^2 \right] - \epsilon N \right).
\end{align}
Equivalently, the spherical constraint can be written in terms of the local overlaps as in Eq.~\eqref{5-spherConstover}. The operation of passing to the integral representation of a delta function, which in practice amounts to pass from a \emph{microcanonical} (hard) version of the constraint to a \emph{canonical} (soft) one, is harmless only when the interaction network is dense enough (see Chap.~\ref{chap:Condens}). When the graph of interactions is sparse, which is not the case here, the global constraint induce a condensation phenomenon and the equivalence between ensembles breaks down. The opening of the Dirac delta is not harmless and must be handled with much more care. However, in the present case, due to the results of Chap.~\ref{chap:Condens}, we do not have to worry, since a proper localization transition does not take place on the mode-locked graph. Then, by considering the expression of the constraint in the local overlap, we can write
\begin{align*}
    \prod_{\alpha=1}^n &\delta\left(\sum_{i=1}^{N/2} (\Re[\mfq_{i}^{\alpha\alpha}] + \Im[\mfq_{i}^{\alpha\alpha}]) - \epsilon N \right) =\int_{-i\infty}^{i\infty}\prod_{\alpha=1}^n \de h^\alpha \exp\left[ \sum_{\alpha=1}^n h^\alpha \left( \sum_{i=1}^{N/2} (\Re[\mfq_{i}^{\alpha\alpha}] + \Im[\mfq_{i}^{\alpha\alpha}]) - \epsilon N  \right) \right] \nonumber \\
    &\qquad = \int_{-i\infty}^{i\infty}\prod_{\alpha=1}^n \de h^\alpha \exp\left[ - \epsilon N
    \sum_{\alpha=1}^n h^\alpha  + \sum_{i=1}^{N/2} \sum_{\alpha=1}^n
    h^\alpha~(\Re[\mfq_{i}^{\alpha\alpha}] + \Im[\mfq_{i}^{\alpha\alpha}]) \right], \nonumber
\end{align*}
from which we get 
\begin{align}
    \exp\left[F_q(\hat{\msQ}) \right] = \int_{-i\infty}^{i\infty}\prod_{\alpha=1}^n \de h^\alpha~
    e^{- \epsilon N \sum_{\alpha=1}^n h^\alpha }~\mZ[h^\alpha],
\end{align}
where the partition function $\mZ[h^\alpha]$ reads as 
\begin{align*}
\mZ[h^\alpha] =  \int \mD \mfq \mD \overline{\mfq}\mD\lambda \mD\overline{\lambda}\mD u \mD v ~\exp\Bigg[
&-\frac{1}{4}\sum_{i=1}^{N/2}\sum_{\alpha\beta}^n \sum_{\gamma\delta}^n \mfq_{i}^{\alpha\beta}~\hat{\msQ}_{\alpha\beta,\gamma\delta}~\overline{\mfq}_{i}^{\gamma\delta}  + \sum_{i=1}^{N/2} \sum_{\alpha=1}^n h^\alpha~(\Re[\mfq_{i}^{\alpha\alpha}] + \Im[\mfq_{i}^{\alpha\alpha}]) \\
&  + \frac{1}{2}\sum_{\alpha\beta}^n \sum_{i=1}^{N/2}
\overline{\lambda}_{i}^{\alpha\beta} \left[ \mfq_{i}^{\alpha\beta} - (u_i^\alpha u_i^\beta+i v_i^\alpha v_i^\beta) \right] + \text{c.c.}  \Bigg].
\end{align*}
It is now clear that this partition function can be factorized in the product of $N/2$ identical terms, so that we can write
\begin{align}
    F_\mfq(\hat{\msQ}) = \log \int_{-i\infty}^{i\infty}\prod_{\alpha=1}^n \de h^\alpha~\exp \frac{N}{2} \left[ f_q(\hat{\msQ},h) - 2\epsilon \sum_{\alpha=1}^n h^\alpha\right], 
\end{align}
where 
\begin{equation}
\begin{split}
    f_\mfq(\hat{\msQ},h) = &\log \int  \mD \mfq \mD \overline{\mfq}\mD\lambda \mD\overline{\lambda}\mD u \mD v ~\exp\Bigg[
   -\frac{1}{4}\sum_{\alpha\beta}^n \sum_{\gamma\delta}^n \mfq^{\alpha\beta}~\hat{\msQ}_{\alpha\beta,\gamma\delta}~\overline{\mfq}^{\gamma\delta}   \\
    &  + \sum_{\alpha=1}^n h^\alpha(\Re[\mfq^{\alpha\alpha}] + \Im[\mfq^{\alpha\alpha}]) + \frac{1}{2}\sum_{\alpha\beta}^n 
    \overline{\lambda}^{\alpha\beta} \left[ \mfq^{\alpha\beta} - (u^\alpha u^\beta+i v^\alpha v^\beta) \right] + \text{c.c.}  \Bigg].
\end{split}
\end{equation}
Notice that here we have kept for convenience the same notation for the integration measure as before, even if now it has lost the product over the local indices, i.e. $\mathcal{D}x = \prod_{\alpha \leq \beta}^n \de x^{\alpha\beta}$ and equivalently for the spins.

\section{Saddle-Point Equations}
In this section we focus on the case of continuous spherical variables, in which we are mostly interested, since they are closer to phasors, compared to discrete spins. However, the solution of the model has been set up also for the discrete case. By including the result at the end of the previous section, the action of the model can be written as
\begin{equation}
\begin{split}
     S(\hat{\Lambda},\Lambda,\msQ,\hat{\msQ},h)
=\frac{N}{2} &\Big[ f_\mQ(\hat{\Lambda})+ f_\mfq(\hat{\msQ},h) + \textrm{Tr}~G\left(\frac{\Lambda \msQ}{4} \right) + \frac{1}{2}\textrm{Tr}(\hat{\Lambda}\Lambda)  \\
& + \frac{1}{2}\textrm{Tr}~(\hat{\msQ}\msQ) - \textrm{Tr}\log(\hat{\Lambda}) - 2 \epsilon \textrm{Tr}(h) \Big].
\end{split}
\end{equation}

The full set of saddle-point equations for the action reads as: 
\begin{subequations}
\begin{gather}
    \frac{\partial S}{\partial h^\alpha} = \frac{\partial f_\mfq(\hat{\msQ},h)}{\partial h^\alpha} -2\epsilon=0  \\ 
    \frac{\partial S}{\partial \hat{\msQ}_{\alpha\beta,\gamma\delta}} =
    \frac{\partial f_\mfq(\hat{\msQ},h)}{\partial
    \hat{\msQ}_{\alpha\beta,\gamma\delta}} + \frac{1}{2}\msQ_{\alpha\beta,\gamma\delta} =0  \label{eq:saddle-point-1-a} \\
    \frac{\partial S}{\partial \msQ_{\alpha\beta,\gamma\delta}} =
    \hat{\msQ}_{\alpha\beta,\gamma\delta} + \frac{1}{2} [ \Lambda~G'(\Lambda \msQ/4)]_{\alpha\beta,\gamma\delta} =0 \label{eq:saddle-point-1-b} \\ 
    \frac{\partial S}{\partial \Lambda_{\alpha\beta,\gamma\delta}} =
    \hat{\Lambda}_{\alpha\beta,\gamma\delta}+ \frac{1}{2} [ \msQ~G'(\Lambda \msQ/4)]_{\alpha\beta,\gamma\delta}=0  \label{eq:saddle-point-1-c} \\
    \frac{\partial S}{\partial \hat{\Lambda}_{\alpha\beta,\gamma\delta}} = \frac{\partial f_\mQ(\hat{\Lambda})}{\partial \hat{\Lambda}_{\alpha\beta,\gamma\delta}} +\frac{1}{2}\Lambda_{\alpha\beta,\gamma\delta}-[\hat{\Lambda}^{-1}]_{\alpha\beta,\gamma\delta}=0. \label{eq:saddle-point-1-d}
\end{gather}
\end{subequations}
Now, by exploiting the property of the derivative of $\mG$ Eq.~\eqref{G'rel} and following the same procedure as the previous chapter (which assumes commuting matrices), we can eliminate the variable $\Lambda$ and Eqs.~\eqref{eq:saddle-point-1-b} and \eqref{eq:saddle-point-1-c} in favor of the algebraic constraint
\begin{align}
      \msQ \left(\hat{\msQ} -\mM\right) = 1,
  \label{eq:saddle-point-e}
\end{align}
where for convenience we have defined $\mM = \hat{\Lambda}^{-1}/8$ and performed the rescaling $\hat{\msQ} \rightarrow 2 \hat{\msQ}$. Consistently with these redefinitions, the two local free energies can be rewritten as 
\begin{equation}
    f_\mQ(\mM) = \log \int \prod_{\alpha \leq \beta}^n \de \mQ^{\alpha\beta} \de \overline{\mQ}^{\alpha\beta}
    \exp\left[  \frac{\beta^2J^2}{4} \sum_{\alpha\beta}^n |\mQ^{\alpha\beta}|^4   + \sum_{\alpha\beta}^n\sum_{\gamma\delta}^n  \mQ^{\alpha\beta} \mM_{\alpha\beta,\gamma\delta}\overline{\mQ}^{\gamma\delta} \right]
\end{equation}
and 
\begin{equation}
\begin{split}
    f_\mfq(\hat{\msQ},h) = &\log \int  \mD \mfq \mD \overline{\mfq}\mD\lambda \mD\overline{\lambda}\mD u \mD v ~\exp\Bigg[
   -\frac{1}{2}\sum_{\alpha\beta}^n \sum_{\gamma\delta}^n \mfq^{\alpha\beta}~\hat{\msQ}_{\alpha\beta,\gamma\delta}~\overline{\mfq}^{\gamma\delta}  \\
    &  + \sum_{\alpha=1}^n h^\alpha(\Re[\mfq^{\alpha\alpha}] + \Im[\mfq^{\alpha\alpha}]) + \frac{1}{2}\sum_{\alpha\beta}^n 
    \overline{\lambda}^{\alpha\beta} \left[ \mfq^{\alpha\beta} - (u^\alpha u^\beta+i v^\alpha v^\beta) \right] + \text{c.c.}  \Bigg],
\end{split}
\end{equation}
where in the first free energy the local integration variables have been rescaled as $\mQ \rightarrow \mQ/(2\sqrt{2})$. It is worth stressing that all the variable redefinitions performed so far do not affect the theory up to irrelevant constants and a rescaling of the temperature. Hence, the set of saddle-point equations reduces to
\begin{subequations}
\begin{gather}
\frac{\partial f_\mfq(\hat{\msQ},h)}{\partial h^\alpha} - 2\epsilon = 0 \\
\frac{\partial f_\mfq(\hat{\msQ},h)}{\partial \hat{\msQ}_{\alpha\beta,\gamma\delta}} +
\frac{1}{2}\msQ_{\alpha\beta,\gamma\delta} = 0  \\
-\frac{\partial f_\mQ(\mM)}{\partial \mM_{\alpha\beta,\gamma\delta}} + \msQ_{\alpha\beta,\gamma\delta} = 0 \\
\msQ (\hat{\msQ} - \mM )= 1 ,
\end{gather}
\end{subequations}
which can be obtained by extremizing the following reduced action with respect to the matrix elements of $\mM$ and $\hat{\msQ}$:
\begin{equation} \label{MLaction}
    A(\mM,\hat{\msQ},h) = N \left[f_\mQ(\mM) +  2 f_q(\hat{\msQ},h) + \textrm{Tr}\log(\hat{\msQ}-\mM) - ~4\epsilon~\textrm{Tr}(h) \right].
\end{equation}
Notice that, when explicitly computing the derivatives of the free energies, the saddle point equations lead to the physically relevant relations
\begin{gather}
    \langle \text{Re}[\mfq^{\alpha\alpha}] + \text{Im}[\mfq^{\alpha\alpha}] \rangle_{ _{\hat{\msQ}}} = 2\epsilon \\
    \langle \mfq^{\alpha\beta} \overline{\mfq}^{\gamma\delta} \rangle_{ _{\hat{\msQ}}} =  \langle \mQ^{\alpha\beta}\overline{\mQ}^{\gamma\delta}\rangle_{ _\mM} =  \msQ_{\alpha\beta,\gamma\delta}, \label{two-point_corr}
\end{gather}
where the definitions of the averages are intuitively induced by the expression of the local free energies.

\subsection{Symmetries of the Overlap-Overlap Correlations}
The structure of the overlap-overlap matrices -- and with that the whole formalism -- can be simplified and lightened a lot considering the symmetries of the original Hamiltonian under reversal of all spins. Recall that the number of spins in the 4-body interaction term is even. As a consequence the replicated action must be invariant when all spins are flipped in one replica \cite{Altieri16}, namely we need it to be invariant under the transformation
\begin{align} \label{eq:reversal-symmetry}
    \lbrace \mfq^{1\alpha},\mfq^{2\alpha},\ldots,\mfq^{n\alpha}\rbrace~\longrightarrow~\lbrace
-\mfq^{1\alpha},-\mfq^{2\alpha},\ldots,-\mfq^{n\alpha}\rbrace.
\end{align}
The direct consequence of this is that among generic multipoint correlation functions of the kind
\begin{align*}
    \langle \mfq^{\alpha_1\beta_1}~\mfq^{\alpha_2\beta_2}\ldots \mfq^{\alpha_k\beta_k} \rangle,
\end{align*}
only those where each upper index is repeated an even number of times are different from zero. In particular the two point correlations of Eq.~\eqref{two-point_corr} are non-zero only when $\alpha=\gamma$ and $\beta=\delta$
\begin{align}
    \langle \mfq^{\alpha\beta} \overline{\mfq}^{\gamma\delta} \rangle_{ _{\hat{\msQ}}} ~=~ \langle \mfq^{\alpha\beta} \overline{\mfq}^{\gamma\delta} \rangle_{ _{\hat{\msQ}}}~\delta_{\alpha\gamma}~\delta_{\beta\delta} =
\langle \mfq^{\alpha\beta} \overline{\mfq}^{\alpha\beta} \rangle_{ _{\hat{\msQ}}}.
\end{align}
This means that the only non-zero terms of the matrix
$\hat{\msQ}_{\alpha\beta,\gamma\delta}$ are those diagonal with respect to the couple of indices:
\begin{align}
    \hat{\msQ}_{\alpha\beta,\gamma\delta}= \hat{\msQ}_{\alpha\beta}~\delta_{\alpha\gamma}~\delta_{\beta\delta}.
\end{align}
This simple observation greatly simplifies all the mean-field
equations and the matrices appearing therein. The simplified
saddle-point equations are
\begin{subequations} \label{eq:saddle-point-reduced-1-sym}
\begin{gather}
\frac{\partial f_\mQ(\mM)}{\partial \mM_{\alpha\beta}} = \langle \mQ^{\alpha\beta}\overline{\mQ}^{\alpha\beta}\rangle_{ _\mM} = \msQ_{\alpha\beta} \\
-\frac{\partial f_\mfq(\hat{\msQ},h)}{\partial \hat{\msQ}_{\alpha\beta}} =
\frac{1}{2} \langle \mfq^{\alpha\beta} \overline{\mfq}^{\alpha\beta} \rangle_{ _{\hat{\msQ}}} =
\frac{1}{2}\msQ_{\alpha\beta} \\
\frac{\partial f_\mfq(\hat{\msQ},h)}{\partial h^\alpha} = \langle \text{Re}[\mfq^{\alpha\alpha}] + \text{Im}[\mfq^{\alpha\alpha}] \rangle = 2\epsilon \\
\msQ (\hat{\msQ} - \mM) = 1, \label{MLAlgConstr}
\end{gather}
\end{subequations}
where the two local free energies now read:
\begin{equation} \label{freeEnQsimply}
    f_\mQ(\mM) = \log \int \prod_{\alpha \leq \beta}^n \de \mQ^{\alpha\beta} \de \overline{\mQ}^{\alpha\beta}
    \exp\left[  \frac{\beta^2J^2}{4} \sum_{\alpha\beta}^n |\mQ^{\alpha\beta}|^4   + \sum_{\alpha\beta}^n \mM_{\alpha\beta} |\mQ^{\alpha\beta}|^2 \right]
\end{equation}
and 
\begin{equation} \label{freeEnqsimply}
\begin{split}
    f_\mfq(\hat{\msQ},h) = &\log \int  \mD \mfq \mD \overline{\mfq}\mD\lambda \mD\overline{\lambda}\mD u \mD v~\exp\Bigg[
   -\frac{1}{2}\sum_{\alpha\beta}^n \hat{\msQ}_{\alpha\beta} |\mfq^{\alpha\beta}|^2    \\
    & + \sum_{\alpha=1}^n h^\alpha(\Re[\mfq^{\alpha\alpha}] + \Im[\mfq^{\alpha\alpha}]) + \frac{1}{2}\sum_{\alpha\beta}^n 
    \overline{\lambda}^{\alpha\beta} \left[ \mfq^{\alpha\beta} - (u^\alpha u^\beta+i v^\alpha v^\beta) \right] + \text{c.c.}  \Bigg],
\end{split}
\end{equation}


\section{RS Ansatz}
The first step towards a replica-symmetric solution is to bring back the free energy written in the second line of Eq.~\eqref{freeEnqsimply} to the form where spins appear explicitly. We rewind the steps performed, by first integrating over the variables $\bm{\lambda}$ and then proceeding in the following way
\begin{align*}
    f_\mfq(\hat{\msQ},h) &= \log \int \mD \mfq \mD \overline{\mfq} \mD u \mD v \exp\left[ -\frac{1}{2}\sum_{\alpha\beta}^n \hat{\msQ}_{\alpha\beta} |\mfq^{\alpha\beta}|^2 + \sum_{\alpha=1}^n h^\alpha(\Re[\mfq^{\alpha\alpha}] + \Im[\mfq^{\alpha\alpha}]) \right] \\
    & \quad\quad\quad \times \prod_{\alpha \leq \beta} \delta \left( \mfq^{\alpha\beta} - ((u^\alpha u^\beta+i v^\alpha v^\beta) \right) \delta \left( \overline{\mfq}^{\alpha\beta} - ((u^\alpha u^\beta - i v^\alpha v^\beta) \right) \\
    &= \log \int \mD u \mD v \exp\left[ -\frac{1}{2}\sum_{\alpha\beta}^n \hat{\msQ}_{\alpha\beta} \left[(u^\alpha u^\beta)^2+ (v^\alpha v^\beta)^2\right] + \sum_{\alpha=1}^n h^\alpha\left[(u^\alpha)^2 + (v^\alpha)^2 \right]  \right] \\
    &=2 \log \int \prod_{\alpha=1}^n \de \sigma_\alpha \exp\left[ -\frac{1}{2}\sum_{\alpha\beta}^n \sigma_\alpha^2 \hat{\msQ}_{\alpha\beta} \sigma_\beta^2 + \sum_{\alpha=1}^n h^\alpha \sigma_\alpha^2  \right]
\end{align*}
where the integrals in the variables $\bm{u}$ and $\bm{v}$ have been factorized in two identical contributions. In the following, we will refer to this expression of the local free energy in real space as $f_\sigma(\hat{\msQ},h)$, to remind that now the local integration variables are the spins. 

The simplest assumption for the elements of the global order parameters $\hat{\msQ}_{\alpha\beta}$ and $\mM_{\alpha\beta}$ is
the replica-symmetric one:
\begin{align*}
  \hat{\msQ}_{\alpha\beta} & = \hat{q}_D \delta_{\alpha\beta} + \hat{q}_0 (1-\delta_{\alpha\beta})  \\
  \mM_{\alpha\beta} & = \mu_D \delta_{\alpha\beta} + \mu_0 (1-\delta_{\alpha\beta}),
\end{align*}
together with $h_\alpha = h$ for the field. Therefore, we can write
\begin{align*}
    \frac{1}{2} f_\sigma(\hat{\msQ},h) &= \log \int \prod_{\alpha=1}^n \de \sigma_\alpha
  \exp\left[ -\frac{1}{2}(\hat{q}_D-\hat{q}_0) \sum_{\alpha=1}^n \sigma_\alpha^4 -
  \frac{\hat{q}_0}{2}\left( \sum_{\alpha=1}^n \sigma_\alpha^2 \right)^2 + \sum_{\alpha=1}^n h_\alpha \sigma_\alpha^2 \right] \\
  & = \log \int \msD z \prod_{\alpha=1}^n \int  \de \sigma_\alpha
  \exp\left[ -\frac{1}{2}(\hat{q}_D-\hat{q}_0) \sigma_\alpha^4  +  \sqrt{-\hat{q}_0} z \sigma_\alpha^2 +  h_\alpha \sigma_\alpha^2 \right] \\
  & = \log \int \msD z~ \left[ \mZ_0(\hat{q}_D,\hat{q}_0,h,z) \right]^n,
\end{align*}
where $\msD z = e^{-z^2/2}/\sqrt{2\pi}$ and we have defined the local partition function
\begin{gather}
    \mZ_0(\hat{q}_D,\hat{q}_0, h, z) = \int  \de \sigma  f_{0}(\sigma | \hat{q}_D,\hat{q}_0, h, z) \\
    f_{0}(\sigma | \hat{q}_D,\hat{q}_0, h, z) = \exp\left[ -\frac{1}{2}(\hat{q}_D-\hat{q}_0) \sigma^4  +  (\sqrt{-\hat{q}_0} z +  h ) \sigma^2. \right]
\end{gather}
From this finite-$n$ expression, it is easy to find that
\begin{align}
    \lim_{n \rightarrow 0} \frac{1}{n} f_\sigma(\hat{\msQ},h) = 2 \int \msD z~\log \mZ_0(\hat{q}_D,\hat{q}_0, h, z).
\end{align}

The local free energy of the dual space is completely diagonal in the replica indices and can be written as follows
\begin{align*}
     f_\mQ(\mM) &= \log \int \prod_{\alpha \leq \beta}^n \de \mQ^{\alpha\beta} \de \overline{\mQ}^{\alpha\beta}
    \exp\left[  \frac{\beta^2J^2}{4} \sum_{\alpha\beta}^n |\mQ^{\alpha\beta}|^4   + \sum_{\alpha\beta}^n \mM_{\alpha\beta} |\mQ^{\alpha\beta}|^2 \right] \\
    &= \log \int \prod_{\alpha=1}^n \de \mQ^{\alpha\alpha} \de \overline{\mQ}^{\alpha\alpha}  \exp\left[  \frac{\beta^2J^2}{4} \sum_{\alpha=1}^n |\mQ^{\alpha\alpha}|^4   + \mu_D \sum_{\alpha=1}^n |\mQ^{\alpha\alpha}|^2 \right] \\
    &\quad + \log \int \prod_{\alpha < \beta}^n \de  \mQ^{\alpha\beta} \de \overline{\mQ}^{\alpha\beta}
    \exp\left[  \frac{\beta^2J^2}{2} \sum_{\alpha < \beta}^n |\mQ^{\alpha\beta}|^4   + 2 \mu_0 \sum_{\alpha < \beta}^n |\mQ^{\alpha\beta}|^2 \right] \\
    &= n \log \mZ_{\beta} (\mu_D) + \frac{n(n-1)}{2} \log \mZ_{\beta} (\mu_0).
\end{align*}
In the previous expression, the site partition function $\mZ_{\beta,0}$ is defined as
\begin{gather}
    \mZ_{\beta}(\mu) = \int \de x \de \overline{x}~g_{\beta,0}(x | \mu ) \\
    g_{\beta}( x | \mu) = \exp \left[ (2 - \delta_{\mu,\mu_D}) \left(\frac{\beta^2 J^2}{4}|x|^4 + \mu |x|^2\right) \right],
\end{gather}
where the Kronecker delta in the second definition accounts for the factor 2 in the off-diagonal case. Eventually, by taking the limit $n\rightarrow 0$, we get
\begin{equation}
    \lim_{n\rightarrow 0} \frac{1}{n} f_\mQ(\mM) = \log \mZ_{\beta} (\mu_D) - \frac{1}{2} \log \mZ_{\beta} (\mu_0).
\end{equation}
The entropic term in Eq.~\eqref{MLaction} in the limit $n\rightarrow 0$ reads 
\begin{align}
   \lim_{n\rightarrow 0} \frac{1}{n} \log\det(\hat{\msQ}-\mM) =
  \log(\hat{q}_D-\mu_D - (\hat{q}_0-\mu_0)) + \frac{\hat{q}_0-\mu_0}{\hat{q}_D-\mu_D - (\hat{q}_0-\mu_0)},
\end{align}
so that, in conclusion, the RS action is given by
\begin{equation} \label{MLRSaction}
    \begin{split}
     \lim_{n\rightarrow 0} \frac{1}{n} A_{\text{RS}}(\hat{\msQ},\mM,h) &=  4\int \msD z~\log \mZ_0(\hat{q}_D,\hat{q}_0, h , z) +
   \log \mZ_{\beta} (\mu_D) - \frac{1}{2} \log \mZ_{\beta} (\mu_0) \\
  &+ \log(\hat{q}_D-\mu_D - (\hat{q}_0-\mu_0)) + \frac{\hat{q}_0-\mu_0}{\hat{q}_D-\mu_D - (\hat{q}_0-\mu_0)} -4\epsilon h.
    \end{split}
\end{equation}

\subsection{RS Equations}
In this section the self-consistency equations for the RS parameters are derived. Let us start from the computation of the derivatives of the action~\eqref{MLRSaction}, by considering separately its terms. In the following we imply the limit $n\rightarrow 0$, to shorten the notation. We have:
\begin{align*}
    \frac{\partial f_\sigma}{\partial \hat{q}_D}  = 2 \int \msD z \frac{1}{\mZ_0} \partial_{\hat{q}_D} \mZ_0 = 2 \int \msD z \frac{\int \de \sigma \partial_{\hat{q}_D} f_0 }{\int \de \sigma f_0} = - \int \msD z \langle  \sigma^4 \rangle_{0},
\end{align*}
where we have defined the average
\begin{align}
    \langle  (\cdots) \rangle_{0} = \frac{\int \de \sigma f_0(\sigma | \hat{q}_D,\hat{q}_0, h, z ) (\cdots)}{ \mZ_0(\hat{q}_D,\hat{q}_0, h, z) }.
\end{align}
Similarly, we have
\begin{align*}
   \frac{\partial f_\sigma}{\partial \hat{q}_0} = \int \msD z \left[ \langle  \sigma^4 \rangle_{0} - \frac{1}{\sqrt{-\hat{q}_0}} \partial_z \langle  \sigma^2 \rangle_{0} \right] = \int \msD z \left( \langle  \sigma^2 \rangle_{0} \right)^2
\end{align*}
where after integration by parts we have used the fact that $\partial_z \langle  \sigma^2 \rangle_{0} = \sqrt{-\hat{q}_0} \left[\langle  \sigma^4 \rangle_{0} - (\langle  \sigma^2 \rangle_{0})^2 \right]$.
We consider now the free energy of the dual space and define the average
\begin{align}
      \langle  (\cdots) \rangle_{\mu} = \frac{\int \de x \de \overline{x}~ g_{\beta}(x | \mu) (\cdots) }{\mZ_{\beta}(\mu)},
\end{align}
where the subscript $\mu=\{\mu_D,\mu_0\}$ is just a reminder of the $g_\beta$ function argument. It is easy, then, to see that
\begin{gather*}
    \frac{\partial f_\mQ}{\partial \mu_D} = \langle |x|^2 \rangle_{\mu_D} \\
    \frac{\partial f_\mQ}{\partial \mu_0} = - \langle |x|^2 \rangle_{\mu_0}.
\end{gather*}
The derivatives of the entropic term read just like in the previous chapter. Therefore, we can write 
\begin{align*}
    &\frac{\partial A_{\text{RS}}}{\partial \hat{q}_D} = 0  ~~~ \rightarrow ~~~ - 2 \int \msD z \langle  \sigma^4 \rangle_{0} + A = 0 \\
    &\frac{\partial A_{\text{RS}}}{\partial \hat{q}_0} = 0  ~~~ \rightarrow ~~~ 2 \int \msD z \left( \langle  \sigma^2 \rangle_{0} \right)^2 + B = 0 \\
    &\frac{\partial A_{\text{RS}}}{\partial \mu_D} = 0  ~~~ \rightarrow ~~~ \langle |x|^2 \rangle_{\mu_D} - A = 0 \\
    &\frac{\partial A_{\text{RS}}}{\partial \mu_0} = 0  ~~~ \rightarrow ~~~ \langle |x|^2 \rangle_{\mu_0} + B = 0  \\
    &\frac{\partial A_{\text{RS}}}{\partial h} = 0  ~~~ \rightarrow ~~~ \int \msD z \langle  \sigma^2 \rangle_{0} - \epsilon = 0,
\end{align*}
were we have defined
\begin{gather*}
    A = \frac{\hat{q}_D - \mu_D - 2 (\hat{q}_0 - \mu_0) }{[\hat{q}_D - \mu_D - (\hat{q}_0 - \mu_0)]^2} \\
    B =  \frac{\hat{q}_0 - \mu_0}{[\hat{q}_D - \mu_D - (\hat{q}_0 - \mu_0)]^2}.
\end{gather*}

The system of equations can be put in a more familiar form, by exploiting the RS expression of the algebraic constraint \eqref{MLAlgConstr}, which in the $n \rightarrow 0$ limit is given by the set of equations
\begin{subequations}
\begin{gather}
    q_D (\hat{q}_D - \mu_D) - q_0 (\hat{q}_0 -\mu_0) =1 \\
    (q_D - 2q_0)(\hat{q}_0 -\mu_0) + q_0 (\hat{q}_D - \mu_D) = 0.
\end{gather}
\end{subequations}
Thanks to these equations, we can eliminate $\hat{q}_D$ and $\hat{q}_0$ from the saddle-point equations by replacing them with the expressions
\begin{gather} \label{qDhatq0hat}
    \hat{q}_D = \mu_D + \frac{q_D - 2q_0}{(q_0-q_D)^2}  \\
    \hat{q}_0 = \mu_0 - \frac{q_0}{(q_0-q_D)^2},
\end{gather}
which are analogous to Eqs.~\eqref{RDR0}, with the only difference that here $q_D$ is not fixed to 1. By substituting into $A$ and $B$, one finds $A=q_D$ and $B=-q_0$. A further simplification follows by noting that the average $\langle (\cdots) \rangle_{\mu}$ can be rewritten as 
\begin{align}
    \langle (\cdots) \rangle_{\mu} = \frac{\int_0^\infty \de r~r~(\cdots) g_\beta (r | \mu )}{\int_0^\infty \de r~r~g_\beta( r | \mu )},
\end{align}
where we have passed to polar coordinates in the complex integration variables and 
\begin{equation}
 g_\beta ( r | \mu ) = \exp \left[ (2-\delta_{\mu,\mu_D}) \left(\frac{\beta^2 J^2}{4}r^4 + \mu r^2\right) \right].
\end{equation}
Then, we have
\begin{subequations}
    \begin{gather}
        \frac{q_D}{2} = \int \msD z \langle  \sigma^4 \rangle_{0} \\
        \frac{q_0}{2} = \int \msD z \left( \langle  \sigma^2 \rangle_{0} \right)^2 \\
        q_D = \langle r^2 \rangle_{\mu_D} \\
        q_0 = \langle r^2 \rangle_{\mu_0} \\
        \epsilon = \int \msD z \langle  \sigma^2 \rangle_{0}
    \end{gather}
\end{subequations}
where $\langle r^2 \rangle_{\mu}$ can be written in terms of error functions. This is a system in the independent variables $q_D,q_0,\mu_D,\mu_0$ and $h$ and Eqs.~\eqref{qDhatq0hat} must be used instead of $\hat{q}_D$ and $\hat{q}_0$ inside the definition of the function $f_0$. The solution of these equations is in progress.

\section{1RSB ansatz}
In the 1RSB ansatz the overlap matrices have the structure introduced in the previous chapter and Eq.~\eqref{1RSBdecomp} holds. We use the same expression for the free-energy in the direct space as in the previous section, going back to the trace over the spin variables. Then, with simple manipulations we get the following expression
\begin{align}
    \lim_{n \rightarrow 0} \frac{1}{n} f_\sigma(\hat{\msQ},h) = \frac{2}{m} \int \msD z~\log \int \msD y~\mZ_1^m(\hat{q}_D, \hat{q}_0, \hat{q}_1, h, z, y).
\end{align}
where we have defined
\begin{gather}
    \mZ_1(\hat{q}_D, \hat{q}_0, \hat{q}_1, h, z, y) = \int  \de \sigma  f_{1}( \sigma | \hat{q}_D,\hat{q}_0,\hat{q}_1, h, z, y) \\
    f_{1}( \sigma | \hat{q}_D,\hat{q}_0,\hat{q}_1, h, z, y) = \exp\left[ -\frac{1}{2}(\hat{q}_D-\hat{q}_1) \sigma^4  +  (\sqrt{-\hat{q}_0} z + \sqrt{\hat{q}_0-\hat{q}_1}y + h ) \sigma^2 \right]
\end{gather}
as a clear generalization of the RS functions $\mZ_0$ and $f_0$. Similarly, the 1RSB expression of the free-energy in the dual space is given by
\begin{equation}
    \lim_{n\rightarrow 0} \frac{1}{n} f_\mQ(\mM) = \log \mZ_{\beta} (\mu_D) - \frac{m}{2} \log \mZ_{\beta} (\mu_0) + \frac{m-1}{2} \log \mZ_{\beta} (\mu_1) 
\end{equation}
where the local partition function $Z_\beta$ and the function $g_\beta$ are those defined in the previous section. The 1RSB expression of the entropic term is the same as Eq.~\eqref{entr1rsb}, with the matrix $\mA=\hat{\msQ}-\mM$ for the present case.

The self-consistency equations for the 1RSB parameters read as
\begin{subequations}
    \begin{gather}
        \frac{q_D}{2} = \int \msD z \frac{\int \msD y~\mZ_1^m \langle  \sigma^4 \rangle_{1}}{\int \msD y ~\mZ_1^m }  \\
        \frac{q_0}{2} = \int \msD z \frac{\int \msD y~\mZ_1^m \left(\langle  \sigma^2 \rangle_{1}\right)^2 }{\int \msD y ~\mZ_1^m } \\
        \frac{q_1}{2} = \int \msD z \left(\frac{\int \msD y~\mZ_1^m \langle  \sigma^2 \rangle_{1}}{\int \msD y ~\mZ_1^m }\right)^2 \\
        q_D = \langle r^2 \rangle_{\mu_D} \\
        q_0 = \langle r^2 \rangle_{\mu_0} \\
        q_1 = \langle r^2 \rangle_{\mu_1} \\
        \epsilon = \int \msD z \frac{\int \msD y~\mZ_1^m \langle  \sigma^2 \rangle_{1}}{\int \msD y ~\mZ_1^m },
    \end{gather}
\end{subequations}
where 
\begin{align}
    \langle  (\cdots) \rangle_{1} = \frac{\int \de \sigma f_1(\sigma | \hat{q}_D,\hat{q}_0,\hat{q}_1, h, z ) (\cdots)}{ \mZ_1(\hat{q}_D,\hat{q}_0,\hat{q}_1, h, z) }.
\end{align}
Resolution of these equations is in progress.

\newpage

\chapter{Conclusions and Perspectives} \label{chap:disc}
This work finds its place in the statistical mechanical approach to light amplification in disordered media. In particular, it addresses the problem of going beyond the standard mean-field RSB (Replica Symmetry Breaking) theory employed to find the solution of spin-glass models for random lasers, thus improving the theory towards a more realistic description of these optical systems. 

The leading spin-glass model for the study of the glassy lasing transition has been introduced, by connecting its key features with the semiclassical theory of random lasers. In particular, it has been shown that: (\emph{i}) a non-diagonal linear coupling between pairs of cavity modes arises as a result of the interaction with a bath of diffusive modes escaping the system; (\emph{ii}) a 4-body term of interaction accounting for the light-matter interactions emerges in the context of third-order perturbation theory in the mode amplitudes. When the mode dynamics is considered in the slow amplitude basis, where the modes have a definite frequency, as lasing modes approximately have to, both the linear and the nonlinear couplings turn out to be selected by a Frequency Matching Condition (FMC). Moreover, it has been shown that generalizing the results of the Statistical Light-mode Dynamics approach developed by Fischer, Gordon and coworkers leads to a thermodynamic theory for the stationary regime of RLs. The spin-glass (2+4)-phasor Hamiltonian is obtained by taking disordered couplings, where the randomness in the mode-coupling is induced by the randomness in the spatial extension of the modes and by the spatial heterogeneity of the nonlinear optical response.

The standard mean-field theory requires the model to be defined on the complete graph of interactions, where the FMC does not play any role, since it is always satisfied. In this approximation, the model is compatible only with the narrow-bandwidth limit, where the emission spectrum has a width comparable to the broadened linewidth of the single modes. This is the price to pay for the huge simplification that one has in the mean-field fully-connected approximation, which allows to apply in a quite straightforward way the RSB techniques developed for mean-field spin glasses and to derive the phase diagrams described in Chap.~\ref{chap:IntroCap}. However, the regime to which this mean-field solution pertains is very special, therefore preventing the theory from being applied to generic experimental situations. For instance, neglecting the coupling dilution induced by the FMC hinders the reproduction of the central narrowing in random laser empirical spectra. Consequently, it is of great interest to investigate the model on the mode-locked diluted graph of interaction.

So far the most important result that has been found regarding the mode-locked glassy random laser is the evidence of a mixed-order ergodicity breaking phase transition, as revealed by Monte Carlo numerical simulations (see Chap.~\ref{chap:mixedorder}). The joint study of the specific-heat divergence at the critical temperature and of the low temperature behavior of the Parisi overlap distribution function reveals both the first and second-order nature of the transition, which is the typical scenario of a Random First Order Transition. This is a feature which was already predicted on the complete graph of interactions and seems quite solidly preserved in the diluted model. However, in numerical simulations of the Mode-Locked (ML) 4-phasor model preceding the
present thesis work the transition is found not to be compatible with mean-field theory, according to the estimated value of the scaling exponent of the critical region. This exponent $\nu_{\text{eff}}$ appears to be outside the boundaries corresponding to a mean-field universality class. These limits are derived in Chap.~\ref{chap:Univ} through a simple mean-field argument based on the second-order nature of the glass transition, which consists in the divergence of the thermal response at the critical point.

In this work, we have presented new results from numerical simulations of the ML 4-phasor model, showing how the previous results were haunted by strong finite-size effects. Finite size-effects are unavoidable when dealing with simulations of a dense model such as the mode-locked random laser: the number of connections in the graph requires a number of operations which scales as the cube of the system size, thus forbidding the simulation of large enough sizes. In order to reduce these effects, we have developed a simulation strategy based on periodic boundary conditions on the frequency space, for which band-edge modes participate in the same number of interacting quadruplets as the modes in the center of the spectrum. Therefore, a given size of the simulated model with periodic boundary conditions on the frequencies can be regarded as the bulk of a a larger size with free boundaries. By means of this strategy, and by also performing simulations of the original model, but with a larger number of sizes and of disordered samples, we have assessed that the scaling of the critical region is compatible with mean-field theory up to the precision of our analysis. However, the model seem not to be in the the universality class of the Random Energy Model, an feature suggesting that the mode-locked random laser may need a different mean-field solution than its fully connected counterpart. The study of the glass transition has been completed with the analysis of the Parisi overlap probability distribution function, where the use of periodic boundary conditions, has resulted in more pronounced side-peaks. 

Another interesting phenomenon that has been studied in this work is the possibility of a localization -- else termed power condensation -- transition in the mode-locked glassy random laser. In this context, localization is understood as the phenomenon whereby a finite number of modes carries an extensive amount of light intensity, and not in the sense of disorder induced Anderson or many body localization in quantum systems. The presence of localization, as the spherical constraint is tuned above a given threshold, is only theoretically possible in presence of dilution of the interaction network: in the fully-connected case, the high connectivity of the model is sufficient to guarantee the equipartition of the constraint among all degrees of freedom. From the careful finite-size study of the localization order parameter reported in Chap.~\ref{chap:Condens}, we have been able to assess that, although some evidence of incipient localization can be found, the glassy phase of light is not strictly speaking localized. This means that the results of our analysis are not compatible with single light modes carrying an extensive amount of intensity. However, we
have found an anomaly in the finite-size study of the participation ratio, whose size dependence is compatible with the presence of high power modes, carrying an intensity $|a_k|^2 \sim N^{1-\Psi/2}$, with $\Psi > 0$. We stress that in presence of power condensation those
modes would have intensity $|a_k|^2 \sim N$, i.e.~$\Psi = 0$. Moreover, the study of the spectral entropy has revealed that the low temperature phase of the model is characterized by the breaking of equipartition. We have termed ``pseudo-localization'' the transition to this hybrid phase, where the light intensity is not completely localized and at the same time is not equipartitioned among the modes.

One of the most relevant aspects of the picture revealed by the numerical results presented in this work is that the critical temperature of the glass and of the pseudo-localization transitions is the same within the statistical uncertainty. This occurrence makes the mode-locked random laser a very interesting problem where ergodicity breaking manifests itself in a twofold way: replica-symmetry breaking, which is typical of quenched disordered systems, and localization, which has mostly been studied in the context of quantum many-body systems. The opportunity given by this model is to study both transitions at the same time, opening the way to more general studies for arbitrary nonlinearities and degrees of coupling dilution.

Supported by the numerical evidence that the mode-locked random laser is, indeed, a mean-field model, we have approached its solution with analytical techniques. The similarity of the mode-locking dilution rule with the kind of correlations in the Hamiltonian of the Bernasconi model has led us to perform a preliminary study of the Merit Factor problem. Though the presence of a low temperature glass phenomenology is suggested by numerical studies of the finite-size Hamiltonian, within the accuracy of our analysis, the model does not exhibit any transition at finite temperature. The presence of a transition has been investigated through the replica method applied to the model in the space where the spin variables are mapped by a random unitary matrix. In Chap.~\ref{chap:MF}, the model has been solved in the annealed limit, in the replica symmetric ansatz and with one step of replica symmetry breaking. The self-consistency equations for the free parameters have been deeply studied with different integration techniques and computational tools: the only solution revealed at finite temperature is the paramagnetic one. The study of the zero temperature limit of the 1RSB free energy and self-consistency equations is still in progress. Other important information may come from the study of the stability of the RS solution, which helps to distinguish the kind of replica symmetry breaking possibly characterizing the low temperature phase.

The solution of the mode-locked random laser has been addressed in Chap.~\ref{chap:ML}, where the replica technique developed for the Bernasconi model has been adapted to the case of interest. In order to simplify the computation, we have first considered real spherical variables, eliminating the technicality of dealing with the phases at an initial step of investigation. In this case, after the average over disorder, we pass to a generalized Fourier space by transforming the local overlaps with a random unitary matrix. The major difficulty of defining a global order parameter for the model and finding closed equations to determine it as function of temperature has been successfully addressed, with the introduction of a new order parameter, a \emph{superoverlap}, which is a measure of the correlations among local overlaps. The analysis has been completed up to the formal derivation of the 1RSB self-consistency equations for the order parameters of the model. 

Future developments of the work presented in the present thesis are being considered. In the remaining part of this chapter, the most relevant directions of research that we aim to follow are discussed.

A first natural continuation of the work presented in Chap. \ref{chap:MF} is to investigate the reason why a glassy phase at finite temperature has not been found for the Merit Factor problem. This may be due either to some technical issues of the computation procedure, which we are still carefully inspecting, or to the fact that the transition occurs at zero temperature. While the study of the zero-temperature saddle-point equations is in progress, we also aim to compute the stability of the RS solution. In fact, if the RS solution turns out to be unstable below a certain value of the temperature, this may be taken as an indication that the correct low-temperature solution of the model should be looked for with a FRSB ansatz. However, the more reasonable scenario seems to be that there is no transition at all in the model with random unitary matrices. It may be that the the mapping of the deterministic model onto a disordered model with unitary matrices in not correct: the mapping to a a model with two random orthogonal transformations of the spin variables seems to be more reliable and will be deeply studied in future. On top of that, numerical studies of the original deterministic model and of the models with random orthogonal and unitary transformations are in progress. 

Regarding the Mode-Locked 4-phasor model, though the saddle point equations have not been studied in detail yet, the replica approach based on random unitary matrices may suffer the same problems as in the Merit Factor problem. Probably, in this case as well a double orthogonal transformation is needed to reach the correct solution. However, the approach developed in this work proved useful at least to deal with the dependence on the site indices due to the deterministic dilution rule. Similar computations will be performed on the model with the introduction of orthogonal matrices. Once the study of the 1RSB saddle-point equations is completed and a phase diagram for the model is obtained, a straightforward extension of the new mean-field theory is to include the phases of the modes and determine the role played by the phase locking in the diluted case. Then, we plan to generalize the computation to the case of a non-zero mean coupling distribution, in order to study the effect of ferromagnetic alignment with respect to the disorder of the couplings. 

Another analytical approach to the study of the Mode-Locked 4-phasor model, which may yield a useful basis for comparison with the replica computation presented in this work (or variations on the theme), implies performing an expansion in the coupling magnitude, which, given the density of the mode-locked graph, still has to decrease as a power of the size also in the diluted model. The small coupling expansion can also be interpreted as a high temperature expansion, the well-known Plefka/Geroges-Yedida expansion \cite{Plefka82,Georges91}. A useful reference for this approach is Ref. \cite{Gradenigo20b}, where the second-order truncation of the expansion is performed on the $p$-spin model, precisely to make a comparison with the replica computation. In the case of the ML 4-phasor model, one has to perform the expansion keeping the implementation of the Frequency Matching Condition. Eventually, the third order truncation of the expansion, by including the Onsager reaction term, yields the TAP (Thouless-Anderson-Palmer) free energy, paving the way to a TAP analysis of the model.

As discussed in the first part of this work, the numerical approach gives the opportunity to gain physical insight on the mode-locked diluted models and to bridge the theory with the experiments. Work is in progress regarding the numerical simulation of the ML 4-phasor model with a continuous frequency distribution and considering also the relaxation dynamics towards equilibrium \cite{Trinca22,Trinca23}. Extracting the frequencies uniformly in the interval $[0,1]$ allows for a more realistic description of the frequency profile of a random laser with respect to the frequency comb, which actually applies to the case of standard closed cavity lasers. Moreover, experimental measurements RL spectra are always affected by a transient of non-equilibrium dynamics, which equilibrium data collected for the results presented in this work do not take into account. One important study which is in progress in this context is the test of the correspondence between IFO (Intensity Fluctuation Overlap) and Parisi overlap distribution functions in the case of the mode-locked diluted interaction graph. In fact, this correspondence has been analytically proved only in the mean-field fully-connected theory and it would be of great importance to check its validity also in the diluted case, which is closer to experimental RLs. 

Another direction of investigation regards the addition of the 2-body term in the study of the mode-locked diluted model, which for simplicity has been discarded in the whole thesis. For instance, the simulations of the complete (2+4)-phasor model, on the mode-locked graph, should be able, in principle, to show traces of FRSB, if this feature of the solution on the fully-connected graph survives in the diluted model. Of course, in this case the effect of FRSB will have to be carefully disentangled from the finite size effects on the overlap distribution function (see Chap~\ref{chap:Univ}). Simulations in presence of the linear term can be used also to study the inclusion of a gain profile to the dynamics, which so far has been considered in the ordered case only. Furthermore, once a code is written to simulate a gain profile, it could be fed with random laser gain curves measured in experiments.

\newpage

\begin{appendices}

\chapter{Integration over the Unitary Group} \label{app:UG}
In this Appendix we derive the result contained in Eq.~\eqref{MF:AverDis} with two different techniques, by considering elementary cases, which were already discussed in Refs.~\cite{Brezin80b,Marinari94a}. Our aim is just to provide an explanation for the specific function $\mG$ appearing in Eq.~\eqref{MF:G}. 

The key idea is to reduce the difficulty of the integration over the Haar measure of the unitary group in Eq.~\eqref{ExtFieldProblem}, by passing to a scalar problem, where the integral can be performed with the saddle point method in the large-$N$ limit. In fact, as pointed out in Ref.~\cite{Brezin80b}, the general matrix source problem can not be solved with this technique, since the Lagrange multiplier, which implements the unitary constraint in the integration measure, is itself a matrix. Thus, after carrying out the integration over the unitary group, one is left with $N^2$ coupled variables and the saddle point method is not applicable. For this reason, in order to deal with the general problem, Brezin and Gross developed in Ref.~\cite{Brezin80b} an alternative procedure based on the study of the equations of motion in the large-$N$ limit.

\subsubsection{Equations of motion}
Let us start by reviewing the simple example provided in Ref.~\cite{Brezin80b}. Consider the case of a unitary $N$-dimensional vector $\bm{u}\bm{u}^\dagger = 1$ in an external complex vector field $\bm{a}$. In this case, the problem simplifies to the computation of the following integral
\begin{align} \label{Ex1}
    Z = K \int \prod_{i=1}^N \de u_i \de \overline{u}_i  \delta\left( \sum_{i=1}^N |u_i|^2 - 1 \right) e^{N \left[\sum_{i=1}^N (u_i \overline{a}_i + \overline{u}_i a_i) \right]},
\end{align}
where $K$ is a constant and the delta function restricts the integration to the space of unitary vectors. First, we notice that the partition function $Z$ satisfies the following equation:
\begin{align} \label{ZEqmotion}
    \sum_{i=1}^N \frac{\partial^2 Z}{\partial a_i \partial \overline{a}_j} = N^2 Z.
\end{align}
This can be checked by computing the derivatives of $Z$ with respect to the components of the external field:
\begin{align*}
    \frac{\partial Z}{\partial \overline{a}_i} = K \int \prod_{i=1}^N \de u_i \de \overline{u}_j  \delta\left( \sum_{i=1}^N |u_i|^2 - 1 \right) N u_i e^{N \left[ \sum_{i=1}^N (u_i \overline{a}_i + \overline{u}_i a_i) \right]}
\end{align*}
and 
\begin{align*}
    \frac{\partial^2 Z}{\partial a_j \partial \overline{a}_i} = K \int \prod_{i=1}^N \de u_i \de \overline{u}_j  \delta\left( \sum_{i=1}^N |u_i|^2 - 1 \right) N^2 u_i \overline{u}_j e^{N \left(\sum_{i=1}^N u_i \overline{a}_i + \overline{u}_i a_i \right)}.
\end{align*}
Considering the case $i=j$ and summing over $i$, one immediately finds Eq.~\eqref{ZEqmotion} due to the constraint. 

We now make an important remark: in order for the theory to be invariant under unitary transformations, the dependence of $Z$ on the external field can only be mediated by its modulus. This can be seen as follows: after $\bm{u} \rightarrow U\bm{u}$ with $UU^\dagger = I$, the action in Eq.~\eqref{Ex1} reads $N\left[\bm{u}(U\bm{a}^\dagger) + (U^\dagger \bm{a}) \bm{u}^\dagger  \right]$, and the only way to make $U$ disappear from the theory is that the dependence is on $\bm{a}\bm{a}^\dagger$. Then, if we denote $\lambda=\bm{a} \bm{a}^\dagger$, it must be $Z=Z(\lambda)$. Given this remark, Eq.~\eqref{ZEqmotion} takes the form of an ordinary differential equation:
\begin{align} \label{PartialZ}
    \lambda Z''(\lambda) + N Z'(\lambda) = N^2 Z(\lambda),
\end{align}
where the apex denotes derivatives with respect to $\lambda$. This can easily be proved by repeatedly using the chain rule to obtain
\begin{align*}
    \frac{\partial^2 Z}{\partial a_j \partial \overline{a}_i} = \frac{\de Z}{\de \lambda^2 } a_i \overline{a}_j + \frac{\de Z}{\de \lambda} \delta_{ij}
\end{align*}
and, again, taking $i=j$ and summing over $i$. By now imposing the fact that $\log Z$ is proportional to $N$, we look for solutions of Eq.~\eqref{PartialZ} of the kind $Z=\exp[N \mG (\lambda)]$. By plugging this ansatz into Eq.~\eqref{PartialZ} we find an equation for $\mG(\lambda)$, which in the large-$N$ limit reduces to
\begin{align} \label{ODEforG}
    \lambda (\mG'(\lambda))^2 + \mG'(\lambda) = 1.
\end{align}
The solution of this differential equation with initial condition $\mG(0)=0$ is 
\begin{align}
    \mG(\lambda) = - 1 + \sqrt{1 + 4\lambda} - \log\left[\frac{1}{2} + \frac{1}{2}\sqrt{1+4\lambda}  \right],
\end{align}
which, apart from irrelevant constant factors, corresponds to the formula in Eq.~\eqref{MF:G}. 

The general solution of the external field problem posed in Eq.~\eqref{ExtFieldProblem} is a generalization of this procedure and leads to a result which is much more complicated. However, the final expression of the partition function involves double trace operators, which for large $N$ are irrelevant in the replica computation carried out in Chap.~\ref{chap:MF}: by taking into account only terms with a single trace operation, the general result contained in Ref.~\cite{Brezin80b} is a very simple generalization of the scalar case. If we denote by $A$ the external matrix field, we have
\begin{align}
    Z = \exp \left[ N \Tr \mG(A A^\dagger)  \right],
\end{align}
which corresponds to Eq.~\eqref{MF:AverDis}, with $A=\Omega/N$.

\subsubsection{Saddle-point computation}
The other example we deal with is taken from Ref.~\cite{Marinari94a}. Consider the case in which $\Omega$ has only one element different form zero which is extensive in $N$, say $\Omega_{11}=\omega N /2 $, so that the trace operation reduces to 
\begin{equation}
    \Tr(\Omega^{\dagger} U+\text{h.c.}) = \overline{\omega}\frac{N}{2}u_{11} + \text{c.c.}
\end{equation}
In this case, we have to compute the integral
\begin{equation}
Z = \int \prod_{i,j=1}^{N/2} \de u_{ij} \de \overline{u}_{ij} \prod_{i=1}^{N/2} \delta\left(\sum_{j=1}^{N/2}|u_{ij}|^2 -1 \right) e^{\overline{\omega}\frac{N}{2}u_{11} + \text{c.c.}}
\end{equation}
where the integration over the Haar measure of the unitary group has been opened with $N/2$ global constraints, one for each line in order to implement the unitary condition $U U^{\dagger}=I$. Here, we are considering the case of $N/2 \times N/2$ matrices only to be coherent with the treatment of Chap.~\ref{chap:MF}, but the final result will be the same of the previous example a part from an overlall factor $1/2$. The integration over the elements of every line but the first one gives a constant. Hence, one is left with
\begin{equation}
\begin{split}
Z &= K\int \prod_{j=1}^{N/2} \de u_{1j} \de \overline{u}_{1j} \delta\left(\sum_{j=1}^{N/2}|u_{1j}|^2 -1 \right) e^{\overline{\omega}\frac{N}{2}u_{11} + \text{c.c.}} \\
&= K\int \prod_{j=1}^{N/2} \de x_j \de \overline{x}_j \delta\left(\sum_{j=1}^{N/2}|x_j|^2 -1 \right) e^{\overline{\omega}\frac{N}{2}x_1 + \text{c.c.}},
\end{split}
\end{equation}
where in the second line we have just renamed the integration variables and dropped a redundant index, i.e. $u_{1j} \rightarrow x_j$. By comparing this integral with the one in Eq.~\eqref{Ex1}, we can see that this is a special case of the previous example.

The integration in $x_1$ can be isolated as follows
\begin{equation*}
\begin{split}
Z &=K \int \de x_1 \de \overline{x}_1~e^{\overline{\omega}\frac{N}{2}x_1 + \text{c.c.}} \int \prod_{j=2}^{N/2} \de x_j \de \overline{x}_j \delta\left(\sum_{j=2}^{N/2}|x_j|^2 -(|x_1|^2-1) \right) \\
&= K \int \de x \de \overline{x}~e^{\overline{\omega}\frac{N}{2}x + \text{c.c.}} \int \prod_{j=1}^{N/2-1} \de x_j \de \overline{x}_j \delta\left(\sum_{j=1}^{N/2-1}|x_j|^2 -(|x|^2-1) \right).
\end{split}
\end{equation*}
The $N/2-1$-dimensional integral can be performed easily by passing to the real and imaginary parts of the variables $x_j=a_j+i b_j$ and using the property of the delta of a function:
\begin{equation*}
\begin{split}
\int \prod_{j=1}^{N/2-1} \de x_j \de \overline{x}_j \delta\left(\sum_{j=1}^{N/2-1}|x_j|^2 -(|x|^2-1) \right) &\propto \int  \prod_{j=1}^{N/2-1} \de a_j \de b_j \delta\left(\sum_{j=1}^{N/2-1}(a_j^2+b_j^2) -(|x|^2-1) \right) \\
&= \int \prod_{i=1}^{2(N/2-1)} \de t_i  \delta\left(\sum_{j=1}^{2(N/2-1)}t_i^2 -(|x|^2-1) \right) \\
& \propto \int_0^\infty \de r~r^{2(N/2-1)-1} \delta\left(r^2 -(|x|^2-1) \right) \\
&= \int_0^\infty \de r~r^{2(N/2-1)-1} \frac{\delta(r-\sqrt{1-|x|^2})}{2\sqrt{1-|x|^2}} \\
&\approx (1-|x|^2)^{N/2},
\end{split}
\end{equation*}
where the last step holds in the large-$N$ limit. Hence, the integral we aim to compute boils down to
\begin{equation}
Z = \int \de x \de \overline{x}~ e^{\overline{\omega}\frac{N}{2}x + \text{c.c.}} (1-|x|^2)^{N/2} = \int \de x \de \overline{x}~ \exp \frac{N}{2} \left[\log(1-|x|^2) + \overline{\omega}x + \omega \overline{x} \right], 
\end{equation}
which is a one-dimensional integral that can be solved with the saddle point method in the large-$N$ limit. 

By passing to real and imaginary part in $x=a+ib$ and $\omega=v+iw$, we have
\begin{equation}
Z = \int \de a\de b~ \exp \frac{N}{2} \left[\log(1-a^2-b^2)+ 2va+2wb \right]  \approx \exp \frac{N}{2} \mathcal{G}(a^*,b^*,v,w) .
\end{equation}
where
\begin{equation}
\mathcal{G}(a,b,v,w)=\log(1-a^2-b^2)+ 2va+2wb
\end{equation}
and $(a^*,b^*)$ is its maximum. The saddle point equations are
\begin{subequations}
\begin{equation}
v a^2 + v b^2 +a -v =0
\end{equation}
\begin{equation}
w a^2 + w a^2 +b -w =0
\end{equation}
\end{subequations}
and the solution that maximizes $\mathcal{G}$ is
\begin{equation}
(a^*,b^*)=\left(\frac{-v-v\sqrt{1+4v^2+4w^2}}{2(v^2+w^2)},\frac{-w-w\sqrt{1+4v^2+4w^2}}{2(v^2+w^2)} \right).
\end{equation}
Calculating $\mathcal{G}$ in $(a^*,b^*)$ one finds after some algebra
\begin{equation}
\mathcal{G}(|\omega|)= \mathcal{G}(a^*,b^*,v,w)= -1 + \sqrt{1+4|\omega|^2} - \log\left[\frac{1}{2}+\frac{1}{2}\sqrt{1+4|\omega|^2}\right].
\end{equation}
This result corresponds precisely to Eq.~\eqref{MF:G}, taking into account that $\Omega_{11}=\omega \frac{N}{2} $. It is interesting to note that the dependence of $\mG$ in the modulus of the external source $\omega$. This is a consequence of the invariance of the theory under unitary transformations.

\newpage

\chapter{RS computations for the MF problem} \label{app:RSMF}
In this Appendix, we report all the Replica Symmetric (RS) computations for the Random Unitary model of the MF problem. In the first part, the RS action is derived, by implementing the ansatz \eqref{RSansatz} in the action of the model. In order to simplify the calculation, the three terms by which the action \eqref{Action2} is comprised are treated separately. Then, self-consistency equations for the RS parameters are obtained, by imposing the stationarity of the RS action. 

\section{RS action}

\subsubsection{Local free energy $\bm{f_\tau}$}
Consider first the free energy in the spin variables. It is convenient to use the notation $\Tr_{\bm{\tau}} = \prod_{a=1}^n\left[\sum_{ \{\tau^a \}}\right] $, where now $\bm{\tau}$ denotes a vector in the replica space. By plugging the RS ansatz in place of the matrix $\mathcal{R}$, we have
\begin{equation*}
\begin{split}
f_\tau(\mR_D,\mR_0) &= \log \Tr_{\bm{\tau}}  e^{-\frac{1}{2}\sum_{ab}^n \tau^a[\mR_D\delta_{ab}+ \mR_0(1-\delta_{ab})]\overline{\tau}^b} \\
&=\log e^{-(\mR_D-\mR_0)n} \Tr_{\bm{\tau}} e^{-\frac{1}{2}\mR_0|\sum_a^n \tau^a|^2} \\
&= - n(\mR_D-\mR_0) + \log \Tr_{\bm{\tau}} \int\frac{\de h \de \overline{h}}{4\pi} e^{-|h|^2/2 + \sqrt{-\frac{\mR_0}{4}}\sum_a^n (\tau^a \overline{h} + \overline{\tau}^a h)}  \\
&= - n(\mR_D-\mR_0) + \log \int\frac{\de h \de \overline{h}}{4\pi} e^{-|h|^2/2} \Tr_{\bm{\tau}} e^{\sqrt{-\mR_0}\sum_a^n \Re(\tau^a\overline{h})},
\end{split}
\end{equation*}
where in the second step a Hubbard-Stratonovich transformation has been used to decouple the square. In order to take the trace over the spins, let us pass to the real and imaginary parts of the complex variables, by defining $\tau^a=\rho^a+i\sigma^a$ and $h=h_R + i h_I$. Since $\Re(\tau^a\overline{h})=\rho^a h_R + \sigma^a h_I$, we have
\begin{equation*}
\begin{split}
\Tr_{\bm{\tau}} e^{\sqrt{-\mR_0}\sum_a^n \Re(\tau^a\overline{h})} &= \prod_{a=1}^n \sum_{\{\rho^a,\sigma^a\}} e^{\sqrt{-\mR_0}\sum_a^n (\rho^a h_R + \sigma^a h_I) } \\
&= \left( \sum_{\rho=\pm 1} e^{\sqrt{-\mR_0} \rho h_R} \sum_{\sigma = \pm 1} e^{\sqrt{-\mR_0}\sigma h_I} \right)^n \\
&= 4^n \cosh^n(\sqrt{-\mR_0} h_R)\cosh^n(\sqrt{-\mR_0} h_I).
\end{split}
\end{equation*}
Now, we consider the integration over the complex Gaussian variable $h$. It is easy to see that the two integrals in the real and imaginary parts of $h$ can be factorized in two identical contributions, leading to
\begin{align}
    f_\tau(\mR_D,\mR_0) = n\log4 -n (\mR_D-\mR_0) + 2 \log \int \frac{\de h}{\sqrt{2\pi}} e^{-h^2/2} \cosh^n(\sqrt{-\mR_0} h),
\end{align}
which is the expression of the local free energy at finite $n$. We now have to take the limit  $\lim_{n \rightarrow 0} f_\tau/n$, i.e. we only have to keep $\mathcal{O}(n)$ terms in the expression of $f_\tau$. The term containing the Gaussian integral in the expression above can be treated as follows
\begin{align*}
    \log \int \frac{\de h}{\sqrt{2\pi}} e^{-h^2/2} \cosh^n(\sqrt{-\mR_0} h) &= \log \int \frac{\de h}{\sqrt{2\pi}} e^{-h^2/2} \exp \left[n \cosh (\sqrt{-\mR_0} h) \right] \\
    &\approx \log \int \frac{\de h}{\sqrt{2\pi}} e^{-h^2/2} \left[ 1 + n  \cosh (\sqrt{-\mR_0} h) \right] \\
    &=  \log \left[1 + n \int \frac{\de h}{\sqrt{2\pi}} e^{-h^2/2} \cosh (\sqrt{-\mR_0} h)  \right] \\
    &\approx n \int \frac{\de h}{\sqrt{2\pi}} e^{-h^2/2} \cosh (\sqrt{-\mR_0} h).
\end{align*}
These are all standard manipulations, which are true at first order in the limit $n\rightarrow 0$ and will be repeatedly used in the following. Eventually, by redefining $\lim_{n \rightarrow 0} f_\tau/n \rightarrow f_\tau$ for convenience, we have
\begin{equation} \label{ftauRS-app}
f_\tau(\mR_D,\mR_0)= \log4 - \mR_D + \mR_0 + 2\int\frac{\de h}{\sqrt{2\pi}} e^{-h^2/2} \log \cosh (\sqrt{-\mR_0}h),
\end{equation} 
This expression resembles the RS free energy of standard spin-glass models, such as the SK model, see e.g.~Ref.~\cite{Mezard87}, except for the fact that it does not depend explicitly on the temperature: in this model the temperature appears only in the local free energy $f_C$.

\subsubsection{Local free energy $\bm{f_C}$}
Let us now turn to the local free energy that is computed in the generalized Fourier space. By using the RS ansatz on the matrix $\mathcal{M}$, we proceed as follows
\begin{equation*}
\begin{split}
f_C(\mM_D,\mM_0)&=\log \int \prod_{a=1}^n \de C^a \de \overline{C}^a e^{-\beta \sum_a^n |C^a|^4 + \frac{1}{2}(\mM_D-\mM_0)\sum_a^n|C^a|^2} e^{\frac{1}{2}\mM_0 |\sum_a^nC^a|^2} \\
&=\log \int \frac{\de z \de \overline{z}}{4\pi} e^{-|z|^2/2} \prod_{a=1}^n \int \de C^a \de \overline{C}^a e^{-\beta |C^a|^4 + \frac{1}{2}(\mM_D-\mM_0)|C^a|^2 + \sqrt{\frac{\mM_0}{4}} (C^a \overline{z}+\overline{C}^a z)} \\
&=\log \int \frac{\de z \de \overline{z}}{4\pi} e^{-|z|^2/2} \left[\int \de C \de \overline{C} e^{-\beta |C|^4 + \frac{1}{2}(\mM_D-\mM_0)|C|^2 + \sqrt{\mM_0} \Re(C \overline{z})}\right]^n ,
\end{split}
\end{equation*}
where analogously to the previous case the square of the sum in the exponential has been decoupled by introducing a complex Gaussian integration on the auxiliary variable $z$ and the dependence on $n$ has been factorized. Here, at variance with the free energy $f_\tau$ it is convenient to keep the complex formalism, which is more compact. By using the definition \eqref{Ibeta0}, the finite-$n$ expression of the second free energy can be compactly written as 
\begin{align}
    f_C(\mM_D,\mM_0) = \log \int \frac{\de z \de \overline{z}}{4\pi} e^{-|z|^2/2} I_{\beta,0}^n(\mM_D,\mM_0,z).
\end{align}
By expanding linearly in the limit  $n\rightarrow0$, using the same manipulations of the previous section, and replacing $\lim_{n \rightarrow 0} f_C/n \rightarrow f_C$, we finally get
\begin{equation} \label{fcRS-app}
f_C(\mM_D, \mM_0)= \int \frac{\de z \de \overline{z}}{4\pi} e^{-|z|^2/2} \log I_{\beta,0}(\mM_D,\mM_0,z).
\end{equation}

\subsubsection{Entropic term}
In order to rewrite the entropic term in the action, note that for a generic RS matrix $\mathcal{A}$ the following relation holds
\begin{equation*}
\det \mA = (\mathcal{A}_D-\mathcal{A}_0)^{n-1}(\mathcal{A}_D+(n-1)\mathcal{A}_0) ,
\end{equation*}
as a consequence of the fact that $A$ has only two kind of eigenvalues: $\mathcal{A}_D-\mathcal{A}_0$ with degeneracy $n-1$ and $\mathcal{A}_D+(n-1)\mathcal{A}_0$ with degeneracy $1$. Thus, in the present case we have
\begin{equation}
\begin{split}
\Tr\log(\mathcal{R}-\mathcal{M}) &= \log \det (\mathcal{R}-\mathcal{M}) \\
&= (n-1)\log(\mathcal{R}_D -\mathcal{M}_D- \mathcal{R}_0 + \mathcal{M}_0) \\
&\quad + \log(\mathcal{R}_D -\mathcal{M}_D- \mathcal{R}_0 + \mathcal{M}_0 +n(\mathcal{R}_0 - \mathcal{M}_0 )),
\end{split}
\end{equation}
and by carefully taking limit the $n \rightarrow 0$ up to order $O(n)$, we find
\begin{equation} \label{TrLogRS}
\lim_{n \rightarrow 0} \frac{1}{n} \Tr\log(\mathcal{R}-\mathcal{M}) = \log(\mathcal{R}_D -\mathcal{M}_D- \mathcal{R}_0 + \mathcal{M}_0 ) + \frac{\mathcal{R}_0 - \mathcal{M}_0}{\mathcal{R}_D -\mathcal{M}_D- \mathcal{R}_0 + \mathcal{M}_0}.
\end{equation}
For convenience, let us define a function 
$s_0(\mR_D, \mR_0, \mM_D, \mM_0)$ equal to the right hand side of the previous equation,
so that $\lim_{n \rightarrow 0} \Tr\log(\mathcal{R}-\mathcal{M})/n = s_0$. 

By considering together the three expressions Eqs.~\eqref{ftauRS-app},~\eqref{fcRS-app} and \eqref{TrLogRS} the expression of the RS action \eqref{RSaction} reads
\begin{equation*}
\begin{split}
 \lim_{n \rightarrow 0} \frac{1}{n} A_{\text{RS}} &= \log 2 - \frac{\mathcal{R}_D - \mathcal{R}_0}{2} +  \int\frac{\de h}{\sqrt{2\pi}} e^{-h^2/2} \log \cosh (\sqrt{-\mathcal{R}_0}h) \\
&\quad + \frac{1}{2}\int \frac{\de z \de \overline{z}}{4\pi} e^{-|z|^2/2} \log I_{\beta,0}(\mM_D, \mM_0, z) \\
&\quad + \frac{1}{2}  s_0(\mR_D, \mR_0, \mM_D, \mM_0),
\end{split}
\end{equation*}
where the overall factor $1/2$ in Eq.~\eqref{RSaction} has been taken into account.

\section{RS equations}
Let us consider separately the various terms, as in the previous section, and compute the derivatives of the action. 

The derivative of the local free energy $f_\tau$~\eqref{ftauRS-app} with respect to $\mR_D$ gives
\begin{align} \label{DevftauRSD}
    \frac{\partial f_\tau }{\partial \mR_D}  = - 1.
\end{align}
The derivative of $f_\tau$ with respect to $\mR_0$ is less immediate. In order to obtain a simplified expression, we proceed as follows 
\begin{align*}
\frac{\partial f_\tau}{\partial \mathcal{R}_0}  &= 1 - 2 \frac{1}{2 \sqrt{-\mathcal{R}_0}} \int\frac{\de h}{\sqrt{2\pi} }e^{-h^2/2} h \tanh(\sqrt{-\mathcal{R}_0}h) \\
&= 1 - \frac{1}{\sqrt{-\mathcal{R}_0}} \int\frac{\de h}{\sqrt{2\pi}} e^{-h^2/2} \frac{\partial}{\partial h}\tanh(\sqrt{-\mathcal{R}_0}h) \\
&= 1 - \int\frac{\de h}{\sqrt{2\pi} }e^{-h^2/2} \left[ 1-\tanh^2(\sqrt{-\mathcal{R}_0}h) \right],
\end{align*}
where an integration by parts has been performed and the resulting boundary term neglected\footnote{It may be a superfluous remark, but notice that, even if apparently the sign has not changed after the integration by parts, in fact, it has changed, since $\partial_h(e^{-h^2/2}) = - e^{-h^2/2} h $. This kind of integration by parts will be used several times in the following.}. Since the Gaussian integral is normalized to $1$, we finally get 
\begin{align} \label{DevftauRS0}
\frac{\partial f_\tau}{\partial \mathcal{R}_0}   =\int\frac{\de h}{\sqrt{2\pi} }e^{-h^2/2} \tanh^2(\sqrt{-\mathcal{R}_0}h).
\end{align}

Consider now the local free energy $f_C$ in Eq.~\eqref{fcRS-app}. For the following computations we will go on working with complex numbers, instead of passing to their real and imaginary parts. This is done only to keep our notation more compact, and, in fact, some step will be purely formal. The derivative of $f_C$ with respect to $\mM_D$ gives
\begin{align*}
    \frac{\partial  f_C}{\partial \mathcal{M}_D}  &= \int \frac{\de z \de \overline{z}}{4\pi} e^{-|z|^2/2} \partial_{\mM_D} \log I_{\beta,0}(\mM_D,\mM_0, z) \\
    &= \int \frac{\de z \de \overline{z}}{4\pi} e^{-|z|^2/2} \frac{\int \de C\de \overline{C}~ g_{\beta,0}(C | \mM_D,\mM_0, z) |C|^2 /2 }{I_{\beta,0}(\mM_D,\mM_0, z)}
\end{align*}
and, by using the average defined in Eq.~\eqref{RSaver}, we get
\begin{align} \label{DevfcRSD}
     \frac{\partial f_C }{\partial \mathcal{M}_D}   = \frac{1}{2} \int \frac{\de z \de \overline{z}}{4\pi} e^{-|z|^2/2}  \langle |C^2| \rangle_{0}. 
\end{align}
As for the other free energy, the derivative of $f_C$ with respect to the off-diagonal element $\mM_0$ requires some non-trivial step. We have
\begin{align*}
\frac{\partial f_C}{\partial\mathcal{M}_0}   &= \int \frac{\de z \de \overline{z}}{4\pi} e^{-|z|^2/2} \partial_{\mM_0} \log I_{\beta,0}(\mM_D,\mM_0, z) \\
&= \int \frac{\de z \de \overline{z}}{4\pi} e^{-|z|^2/2}  \frac{\int \de C \de \overline{C} ~ g_{\beta,0}(C | \mM_D,\mM_0, z) (-|C|^2/2 + (\mathcal{M}_0)^{-1/2} \Re(C \overline{z})/2)}{I_{\beta,0}(\mM_D,\mM_0, z)} \\
&= - \frac{1}{2} \int \frac{\de z \de \overline{z}}{4\pi} e^{-|z|^2/2} \langle |C|^2 \rangle_{0} + \frac{1}{2\sqrt{\mathcal{M}_0}} \int \frac{\de z \de \overline{z}}{4\pi} e^{-|z|^2/2} \frac{\overline{z}}{2}\langle C \rangle_{0}  \\ 
& \quad +\frac{1}{2\sqrt{\mathcal{M}_0}} \int \frac{\de z \de \overline{z}}{4\pi} e^{-|z|^2/2} \frac{z}{2} \langle \overline{C} \rangle_{0} \\
&= - \frac{1}{2}\int \frac{\de z \de \overline{z}}{4\pi} e^{-|z|^2/2} \langle |C|^2 \rangle_{0}+  \frac{1}{\sqrt{\mathcal{M}_0}} \int \frac{\de z \de \overline{z}}{4\pi} e^{-|z|^2/2} \frac{\overline{z}}{2} \langle C \rangle_{0},
\end{align*}
where the definition of the average $\langle (\cdots) \rangle_{0}$ has been used, together with the fact that in the next-to-last step the third integral can be cast into the second one, by simultaneously changing variables to $C \leftrightarrow \overline{C}$ and $z \leftrightarrow \overline{z}$. This gives twice the contribution of the second integral. We now formally integrate by parts in $z$, and, therefore, we can write
\begin{align} \label{DevfcRSstep}
  \frac{\partial f_C}{\partial\mathcal{M}_0}  =  - \frac{1}{2} \int \frac{\de z \de \overline{z}}{4\pi} e^{-|z|^2/2} \langle |C|^2 \rangle_{0} +  \frac{1}{\sqrt{\mathcal{M}_0}} \int \frac{\de z \de \overline{z}}{4\pi} e^{-|z|^2/2} \frac{\partial}{\partial z}\langle C \rangle_{0}, 
\end{align}
where we used the fact that $\partial_z(e^{-z\overline{z}/2}) = -e^{-z\overline{z}/2}  \overline{z} /2$. The derivative of the expectation value of $C$ is computed as follows
\begin{align*}
    \frac{1}{\sqrt{\mathcal{M}_0}} \frac{\partial}{\partial z}\langle C \rangle_{0} &=  \frac{1}{\sqrt{\mathcal{M}_0}}  \frac{\int \de C\de \overline{C}~ g_{\beta,0}(C |\mM_D,\mM_0, z)  C \sqrt{\mM_0} \frac{\overline{C}}{2} }{I_{\beta,0}(\mM_D,\mM_0, z)} \\
    &\quad - \frac{1}{\sqrt{\mathcal{M}_0}} \frac{\int \de C\de \overline{C}~ g_{\beta,0}(C | \mM_D,\mM_0, z)  C \int \de C\de \overline{C}~ g_{\beta,0}(C | \mM_D,\mM_0, z)  \sqrt{\mM_0} \frac{\overline{C}}{2}}{\left(I_{\beta,0}(\mM_D,\mM_0, z) \right)^2} \\
    &= \frac{1}{2} \langle |C|^2 \rangle_{0} - \frac{1}{2} |\langle C \rangle_{0}|^2,
\end{align*}
where in the last step the linearity of complex conjugation has been used to pass from the integral of the complex conjugate to the complex conjugate of the integral. Once integrated over the Gaussian variable $z$, the first of the two terms in the last step is exactly equal to the first term of Eq.~\eqref{DevfcRSstep}, so that the final result is
\begin{align} \label{DevfcRS0}
     \frac{\partial }{\partial\mathcal{M}_0} f_C = - \frac{1}{2} \int \frac{\de z \de \overline{z}}{4\pi} e^{-|z|^2/2} |\langle C \rangle_{0}|^2.
\end{align}

Eventually, we are able to write the self-consistency equations for the RS parameters, by considering together Eqs.~\eqref{DevfcRSD},~\eqref{DevfcRS0},~\eqref{DevftauRSD} and~\eqref{DevftauRS0}. We have
\begin{align*}
    &\frac{\partial A_{\text{RS}}}{\partial \mR_D}=0 ~~~~\rightarrow~~~~  -1 + \frac{\partial s_0}{\partial \mR_D}  = 0 \\
    &\frac{\partial A_{\text{RS}}}{\partial \mM_D}=0 ~~~~\rightarrow~~~~  \frac{1}{2} \int \frac{\de z \de \overline{z}}{4\pi} e^{-|z|^2/2}  \langle |C^2| \rangle_{0} + \frac{\partial s_0}{\partial \mM_D} = 0 \\
    &\frac{\partial A_{\text{RS}}}{\partial \mR_0}=0 ~~~~\rightarrow~~~~ \int\frac{\de h}{\sqrt{2\pi} }e^{-h^2/2} \tanh^2(\sqrt{-\mathcal{R}_0}h) + \frac{\partial s_0}{\partial \mR_0} = 0 \\
    &\frac{\partial A_{\text{RS}}}{\partial \mM_0} =0 ~~~~\rightarrow~~~~ -\frac{1}{2} \int \frac{\de z \de \overline{z}}{4\pi} e^{-|z|^2/2}  |\langle C \rangle_{0}|^2 + \frac{\partial s_0}{\partial \mM_0} = 0, 
\end{align*}
where the derivatives of the entropic term defined in Eq.~\eqref{TrLogRS} are easy to compute and read as
\begin{gather*}
    \frac{\partial s_0}{\partial \mR_D} = - \frac{\partial s_0}{ \partial \mM_D} =  \frac{\mathcal{R}_D-\mathcal{M}_D-2( \mathcal{R}_0 - \mathcal{M}_0) }{(\mathcal{R}_D -\mathcal{M}_D- \mathcal{R}_0 + \mathcal{M}_0)^2}  \\
    \frac{\partial s_0}{\partial \mR_0} = - \frac{\partial s_0}{ \partial \mM_0} = \frac{\mathcal{R}_0-\mathcal{M}_0}{(\mathcal{R}_D -\mathcal{M}_D- \mathcal{R}_0 + \mathcal{M}_0)^2}.
\end{gather*}

The previous set of equations can be simplified as follows. The first equation in the set, which contains the derivative with respect to the diagonal element $\mR_D$, fixes the condition
\begin{equation}
(\mathcal{R}_D -\mathcal{M}_D - \mathcal{R}_0 + \mathcal{M}_0)^2=\mathcal{R}_D -\mathcal{M}_D- 2 (\mathcal{R}_0 - \mathcal{M}_0),
\end{equation}
which can be eliminated after substitution in all the other equations of the set, leading to 
\begin{gather*}
   \frac{1}{2} \int \frac{\de z \de \overline{z}}{4\pi} e^{-|z|^2/2}  \langle |C^2| \rangle_{0} - 1 = 0 \\
    \int\frac{\de h}{\sqrt{2\pi} }e^{-h^2/2} \tanh^2(\sqrt{-\mathcal{R}_0}h) + \frac{\mathcal{R}_0-\mathcal{M}_0}{\mathcal{R}_D -\mathcal{M}_D- 2 (\mathcal{R}_0 - \mathcal{M}_0)} = 0  \\
    - \frac{1}{2} \int \frac{\de z \de \overline{z}}{4\pi} e^{-|z|^2/2}  |\langle C \rangle_{0}|^2 - \frac{\mathcal{R}_0-\mathcal{M}_0}{\mathcal{R}_D -\mathcal{M}_D- 2 (\mathcal{R}_0 - \mathcal{M}_0)} = 0. 
\end{gather*}

A further simplification comes from the RS expression of the algebraic relation \eqref{AlgConstr}, which defines the overlap in terms of the other variables. By considering the product of two RS matrices, we get a system of only two independent equations
\begin{gather*}
\mathcal{R}_D-\mathcal{M}_D + (n-1)(\mathcal{R}_0-\mathcal{M}_0)q_0=1 \\
(\mathcal{R}_0-\mathcal{M}_0)(1+(n-2)q_0)+ (\mathcal{R}_D-\mathcal{M}_D)q_0=0 ,
\end{gather*}
which, always at order $O(n)$, simplifies to
\begin{subequations} \label{AlgEqs-app}
\begin{gather}
    \mathcal{R}_D-\mathcal{M}_D -(\mathcal{R}_0-\mathcal{M}_0)q_0=1    \\
    (\mathcal{R}_0-\mathcal{M}_0)(1-2q_0)+ (\mathcal{R}_D-\mathcal{M}_D)q_0=0. 
\end{gather}
\end{subequations}
If one isolates $q_0$ from the second equation, one finds
\begin{equation} \label{q0}
q_0= \frac{\mathcal{R}_0-\mathcal{M}_0}{\mathcal{R}_D-\mathcal{M}_D-2 (\mathcal{R}_0 - \mathcal{M}_0)} .
\end{equation}
By substituting this expression of $q_0$ in the self-consistency equations, we find
\begin{gather*}
    q_0 = \int\frac{\de h}{\sqrt{2\pi} }e^{-h^2/2} \tanh^2(\sqrt{-\mathcal{R}_0}h)  \\
    q_0 = \frac{1}{2} \int \frac{\de z \de \overline{z}}{4\pi} e^{-|z|^2/2}  |\langle C \rangle_{0}|^2  \\
    1 = \frac{1}{2} \int \frac{\de z \de \overline{z}}{4\pi} e^{-|z|^2/2}  \langle |C^2| \rangle_{0}  
\end{gather*}
This set of equations is completed by the two algebraic relations among the RS parameters, Eqs.~\eqref{AlgEqs-app}.

\newpage

\chapter{1RSB computations for the MF problem} \label{app:1RSBMF}
In this Appendix, the 1RSB computations for the Random Unitary model of the MF problem are reported by following the same scheme of Appendix \ref{app:RSMF}. First the 1RSB action is derived, by implementing the ansatz \eqref{1RSBdecomp} on all the order parameters of the theory; then self-consistency equations are obtained by imposing the stationarity of the action.

\section{1RSB action}
\subsubsection{1RSB local free energy $\bm{f_\tau}$}
Let us consider the free energy $f_\tau$ in Eq.~\eqref{localFreeEntau}. We perform the 1RSB ansatz on the matrix $\mR$, which leads to the introduction of three parameters $\mR_D, \mR_0$ and $\mR_1$. In the following, we avoid writing the dependence of $f_\tau$ on these parameters  explicitly. By using the same notation of the previous Appendix for the trace over the replicated spin variables $\bm{\tau}$, we have
\begin{align*}
f_\tau &= \Tr_{\bm{\tau}} e^{-\frac{1}{2}\mathcal{R}_0\sum_{ab}^n\tau^a\overline{\tau}^b -\frac{1}{2}(\mathcal{R}_1-\mathcal{R}_0)\sum_k^{n/m}\sum_{ab}^m \tau^a\overline{\tau}^b - \frac{1}{2}(\mathcal{R}_D-\mathcal{R}_1) \sum_a^n |\tau_a|^2} \\
&=-n(\mathcal{R}_D-\mathcal{R}_1) + \log \Tr_{\bm{\tau}}e^{-\frac{1}{2}\mathcal{R}_0 \left|\sum_{a}^n\tau^a\right|^2}e^{\frac{1}{2}(\mathcal{R}_0-\mathcal{R}_1)\sum_k^{n/m} \left|\sum_{a}^m \tau^a\right|^2}.
\end{align*}
The first exponential in the trace operator is of the same kind of the one already encountered in the RS computation and can be decoupled by introducing only one auxiliary Gaussian variable with a Hubbard-Stratonovich transformation 
\begin{equation*}
e^{-\mathcal{R}_0 \left|\sum_{a}^n\tau^a\right|^2}= \int \frac{\de h \de \overline{h}}{4\pi} e^{-|h|^2/2}e^{\sqrt{-\mathcal{R}_0} \sum_a^n \Re(\tau^a \overline{h})}.
\end{equation*}
On the other hand, the second exponential requires the introduction of $n/m$ auxiliary Gaussian variables, one for each diagonal block:
\begin{equation*}
e^{\frac{1}{2}(\mathcal{R}_0-\mathcal{R}_1)\sum_k^{n/m} \left|\sum_{a}^m \tau^a\right|^2} = \prod_k^{n/m} \int \frac{\de u_k \de \overline{u}_k}{4\pi} e^{-|u_k|^2/2} e^{\sqrt{(\mathcal{R}_0 - \mathcal{R}_1)} \sum_a^m \Re(\tau^a \overline{u}_k) }.
\end{equation*}
In order to take the trace over the complex spin variables, as in the previous Appendix, we pass to the real and imaginary parts both of the spins $\tau^a=\rho^a+i\sigma^a$ and of the auxiliary Gaussian fields $h=h_R+ih_I$ and $u_k=u_k^R+iu_k^I$. Moreover, we use the fact that $\sum_a^n = \sum_k^{n/m}\sum_a^m$ and $\prod_a^n = \prod_k^{n/m}\prod_a^m$. Consider the two exponential functions depending on the the parameters $\mR_0$ and $\mR_1$ in the previous expressions. We can write their product as follows:
\begin{align*}
\prod_k^{n/m} \prod_a^m \sum_{\{\tau^a\}} & e^{\sqrt{-\mathcal{R}_0}\Re(\tau^a \overline{h})}e^{\sqrt{\mathcal{R}_0 - \mathcal{R}_1} \Re(\tau^a \overline{u}_k)} =  \prod_k^{n/m} \prod_a^m \sum_{\{\rho^a,\sigma^a\}}  e^{\sqrt{-\mathcal{R}_0}(\rho^a h_R + \sigma^a h_I )} \times \\
& \qquad \qquad \qquad \qquad \qquad \qquad \qquad \qquad \times e^{\sqrt{\mathcal{R}_0 - \mathcal{R}_1} (\rho^a u_k^R + \sigma^a u_k^I)} \\
&=  \prod_k^{n/m} \prod_a^m \sum_{\rho^a =\pm 1} e^{\rho^a(\sqrt{-\mathcal{R}_0} h_R + \sqrt{\mathcal{R}_0 - \mathcal{R}_1} u_k^R )}\sum_{\sigma^a =\pm 1} e^{\sigma^a(\sqrt{-\mathcal{R}_0} h_I + \sqrt{\mathcal{R}_0 - \mathcal{R}_1} u_k^I )} \\
&= 4^n \prod_k^{n/m} \cosh^m(\sqrt{-\mathcal{R}_0} h_R + \sqrt{\mathcal{R}_0 - \mathcal{R}_1} u_k^R ) \times \\
&\qquad \times \cosh^m(\sqrt{-\mathcal{R}_0} h_I + \sqrt{\mathcal{R}_0 - \mathcal{R}_1} u_k^I ) \\
&= 4^n \prod_k^{n/m} \cosh^m\Xi(\mR_0,\mR_1, h_R,u_k^R) \cosh^m\Xi(\mR_0,\mR_1, h_I,u_k^I),
\end{align*}
where the function $\Xi$ is defined as the argument of the $\cosh$ function (see Eq.~\eqref{xifunc}).

When the final expression of the previous sequence is integrated over the Gaussian variables it factorizes in two equivalent contributions, one containing an integral in $h_R$ and $n/m$ identical integrals in $u_k^R$, the other one containing an integral in $h_I$ and $n/m$ identical integrals in $u_k^I$. By changing integration variables the two contributions give a square, which taken out of the $\log$, leads to
\begin{align}
    f_\tau = n \log 4 -n(\mathcal{R}_D-\mathcal{R}_1) + 2 \log \int \mathscr{D} h \left (\int \mathscr{D} u \cosh^m \Xi \right)^{\frac{n}{m}},
\end{align}
where the compact notation for the Gaussian integration measure has been adopted. This is the expression of $f_\tau$ in the 1RSB ansatz at finite $n$. After carefully taking the limit $n \rightarrow 0$ and replacing as usual $\lim_{n \rightarrow 0} f_\tau /n \rightarrow f_\tau$, we find the $O(n)$ expression
\begin{align} \label{ftau1RSB}
    f_\tau(\mR_D,\mR_0,\mR_1,m) = \log 4 - (\mathcal{R}_D-\mathcal{R}_1) + \frac{2}{m} \int \mathscr{D} h  \log \int \mathscr{D} u \cosh^m\Xi(\mR_0,\mR_1, h,u)
\end{align}
Once again, this expression resembles the 1RSB free energy of the SK model, even if here the dependence on temperature is only implicit in the parameters.

\subsubsection{1RSB local free energy $\bm{f_C}$}
We perform the 1RSB ansatz on the matrix $\mM$ and plug it into the free energy $f_C$ in Eq.~\eqref{localFreeEntau}, so that we get
\begin{align*}
f_C &=\log \int \prod_{a}^n \de C^a \de \overline{C}^a e^{-\beta \sum_a^n |C^a|^4 +\frac{1}{2}\mathcal{M}_0 \sum_{ab}^n C^a \overline{C}^b  + \frac{1}{2} (\mathcal{M}_D-\mathcal{M}_1)\sum_a^n|C^a|^2}  \\
&\qquad \qquad \times e^{\frac{1}{2} (\mathcal{M}_1-\mathcal{M}_0)\sum_k^{n/m}\sum_{ab}^m C^a \overline{C}^b} \\
&=\log \int \prod_a^n \de C^a \de \overline{C}^a e^{-\beta \sum_a^n |C^a|^4 +\frac{1}{2} (\mathcal{M}_D-\mathcal{M}_1)\sum_a^n|C^a|^2} e^{ \frac{\mathcal{M}_0}{2} |\sum_a^n C^a|^2} \\
&\qquad \qquad \times e^{ \frac{1}{2}(\mathcal{M}_1-\mathcal{M}_0)\sum_k^{n/m}|\sum_a^m C^a|^2}. 
\end{align*}
As for the other local free energy, we decouple the squares introducing auxiliary Gaussian variables through the following relations
\begin{equation*}
e^{ \frac{\mathcal{M}_0}{2} |\sum_a^n C^a|^2}=\int \frac{\de z \de \overline{z}}{4\pi} e^{-|z|^2/2} e^{\sqrt{\mathcal{M}_0}\sum_a^n \Re(C^a\overline{z})}
\end{equation*}
and 
\begin{equation*}
e^{ \frac{1}{2}(\mathcal{M}_1-\mathcal{M}_0)\sum_k^{n/m}|\sum_a^m C^a|^2} = \int  \prod_k^{n/m} \left[ \frac{\de w_k \de \overline{w}_k}{4\pi} e^{-|w_k|^2/2} \right] e^{\sqrt{\mathcal{M}_1-\mathcal{M}_0}\sum_k^{n/m} \sum_a^m \Re(C^a\overline{w}_k)}.
\end{equation*}
After reorganizing the terms in the free energy and factorizing identical contributions, one finds
\begin{align} \label{fC1RSBfiniten}
    f_C &= \log \int \frac{\de z \de \overline{z}}{4\pi} e^{-|z|^2/2} \left[ \int \frac{\de w \de \overline{w}}{4\pi} e^{-|w|^2/2} I^m_{\beta,1}(\mM_D,\mM_0,\mM_1, z,w) \right]^\frac{n}{m},
\end{align}
where Eq.~\eqref{Ibeta1} has been used. Eq.~\eqref{fC1RSBfiniten} is the finite-$n$ expression of the local free-energy $f_C$. By expanding linearly in $n$ and replacing $\lim_{n \rightarrow 0} f_C / n \rightarrow f_C$, we find
\begin{gather} 
    f_C(\mM_D,\mM_0, \mM_1,m) = \frac{1}{m} \int \mathscr{D}[z\overline{z}] \log \int \mathscr{D}[w\overline{w}] I^m_{\beta,1}(\mM_D,\mM_0,\mM_1 , z,w) \label{fC1RSB}\\
     \mathscr{D}[z\overline{z}] = \frac{\de z \de \overline{z}}{4\pi} e^{-|z|^2/2},  \label{GaussianComplex}
\end{gather}
where to shorten the notation, we have introduced the symbol above for the complex Gaussian integration measure. Incidentally, we notice that, if $z=\sigma + i\rho$, then $\mathscr{D}[z\overline{z}]=\mathscr{D}\sigma \mathscr{D} \rho$, taking into account the usual factor $2$ coming from the Jacobian of the transformation.

\subsubsection{1RSB entropic term}
A useful property of 1RSB matrices, which is of great advantage in writing entropic contributions to the action, is that a given matrix $\mA$ of this kind has only three kinds of eigenvalues
\begin{gather*}
a_1=\mA_D-\mA_1 \qquad d_1=n-\frac{n}{m} \\
a_2=\mA_D+(m-1)\mA_1-m\mA_0 \qquad d_2=\frac{n}{m}-1 \\
a_3=\mA_D+(m-1)\mA_1+(n-m)\mA_0 \qquad d_3=1,
\end{gather*}
where $d_1,d_2$ and $d_3$ are the degeneracies, see e.g.~\cite{Castellani05}. Hence, the determinant of a generic 1RSB matrix is given by
\begin{equation*}
\det \mA=(\mA_D-\mA_1)^{n-\frac{n}{m}}(\mA_D+(m-1)\mA_1-m\mA_0)^{\frac{n}{m}-1}(\mA_D+(m-1)\mA_1+(n-m)\mA_0).
\end{equation*}
Let us put $\mA=\mR-\mM$. Then, the entropic term in the action~\eqref{Action2} is given by $\log \det \mA$, which at order $O(n)$ reads as
\begin{equation} \label{1RSBentr}
\begin{split}
  \lim_{n \rightarrow 0} \Tr\log(\mA)/n &=\frac{m-1}{m}\log(\mA_D-\mA_1)  +\frac{1}{m}\log[\mA_D+(m-1)\mA_1 - m \mA_0] \\ 
   &\quad +\frac{\mA_0}{\mA_D+(m-1)\mA_1 - m \mA_0}.
\end{split}
\end{equation}
Analogously to the RS case, we define a function $s_1(\mA_D,\mA_0,\mA_1,m)$, in which we store the expression of the 1RSB entropic term reported in the right hand side of the previous equation.

The 1RSB expression of the action~\eqref{Action2}, can be written collecting the results of Eqs.~\eqref{ftau1RSB}, \eqref{fC1RSB} and~\eqref{1RSBentr}. We have
\begin{equation*} 
\begin{split}
\lim_{n\rightarrow0} \frac{1}{n} A_{1\text{RSB}} &= \log 2 - \frac{\mathcal{R}_D-\mathcal{R}_1}{2} + \frac{1}{m} \int \mathscr{D}h \log  \int \mathscr{D}u \cosh^m \Xi  \\
&\quad +\frac{1}{2 m}\int \mathscr{D}[z\overline{z}] \log \int \mathscr{D}[w\overline{w}]~I^m_{\beta,1}(\mM_D,\mM_0,\mM_1 , z,w) \\
&\quad + \frac{1}{2} s_1(\mA_D,\mA_0,\mA_1,m),
\end{split}
\end{equation*}
where we recall that the functions $\Xi$ and $I_{\beta,1}$ have been defined respectively in Eqs.~\eqref{xifunc} and~\eqref{Ibeta1}, and the overall factor $1/2$ in Eq.~\eqref{Action2} has been taken into account.

\section{1RSB equations}
Let us consider the local free energy $f_\tau$ in Eq.~\eqref{ftau1RSB}, which is the simplest one and resembles to the paradigmatic case of the SK model. The derivative in $\mR_D$ is immediate and leads to the same equation as in the RS case, see Eq.~\eqref{DevfcRSD}. The derivatives with respect to $\mR_0$ and $\mR_1$ need some preliminary remark to be computed more easily. Notice that, given the definition of the function $\Xi$ in Eq.~\eqref{xifunc}, the following relations hold
\begin{gather*}
    \partial_{\mR_0} \Xi = -\frac{1}{2\sqrt{-\mR_0}} h + \frac{1}{2\sqrt{\mR_0-\mR_1}} u, \\
    \partial_{\mR_1} \Xi = - \frac{1}{2\sqrt{\mR_0-\mR_1}} u, \\
    \partial_h \Xi = \sqrt{-\mR_0}, ~~~~  \partial_u \Xi = \sqrt{\mR_0-\mR_1}.
\end{gather*}
As a consequence of the last line of relations, the derivative of any function of $\Xi$ in $h$ or $u$ is equal up to the coefficient $c_{h,u}=\partial_{h,u}\Xi = \{\sqrt{-\mR_0}, \sqrt{\mR_0-\mR_1} \}$. In particular, this applies to the following derivatives compactly written for both the derivative in $h$ and in $u$
\begin{gather} 
   \partial_{h,u} \cosh^m \Xi = m~c_{h,u} \cosh^m\Xi \tanh\Xi \label{UsefulReldevftau1} \\
   \partial_{h,u} (\cosh^m \Xi \tanh \Xi) =    c_{h,u}  \cosh^m\Xi \left[ 1 + (m-1) \tanh^2\Xi \right], \label{UsefulReldevftau2}
\end{gather}
which will appear in the computations. In the previous expressions, the power $m$ is restored after the derivative, by multiplying and dividing by $\cosh\Xi$, which also leads to the presence of $\tanh\Xi$.

With the help of the previous relations, the derivative of $f_\tau$ with respect to $\mR_0$ is computed as follows:
\begin{align*}
    \frac{\partial f_\tau}{\partial \mathcal{R}_0}  &= \frac{2}{m} \int \mathscr{D} h~\frac{\int \mathscr{D} u ~ \partial_{\mR_0} \cosh^m\Xi }{\int \mathscr{D} u~\cosh^{m}\Xi}       \\
    &= - \frac{1}{ \sqrt{-\mR_0}} \int \mathscr{D} h h \frac{ \int \mathscr{D} u~\cosh^{m}\Xi \tanh \Xi }{\int \mathscr{D}u \cosh^m \Xi} + \frac{1}{\sqrt{\mR_0-\mR_1}} \int \mathscr{D} h  \frac{ \int \mathscr{D} u~u \cosh^{m}\Xi \tanh \Xi }{\int \mathscr{D}u \cosh^m \Xi} \\
    &= -\frac{1}{ \sqrt{-\mR_0}} \left[ \int \mathscr{D} h \frac{ \int \mathscr{D} u\partial_h \left(\cosh^m \Xi \tanh \Xi\right) }{\int \mathscr{D}u \cosh^m \Xi} - \int \mathscr{D} h \frac{ \int \mathscr{D} u\cosh^m \Xi \tanh \Xi \int \mathscr{D} u \partial_h \cosh^m\Xi }{\left(\int \mathscr{D}u \cosh^m \Xi\right)^2} \right] \\
    &\quad+ \frac{1}{\sqrt{\mR_0-\mR_1}} \int \mathscr{D}h \frac{ \int \mathscr{D} u~ \partial_u \left(\cosh^m \Xi \tanh \Xi \right) }{\int \mathscr{D}u \cosh^m \Xi}
\end{align*}
where the first term has been integrated by parts in $h$ (and the derivative distributed on the numerator and denominator of the integrand) and the second one in $u$. It is then clear, that the first and third term yield an equal and opposite contributions, as a consequence of Eq.~\eqref{UsefulReldevftau2}. We are therefore left only with the second term in the previous expression, which after using Eq.~\eqref{UsefulReldevftau1}, leads to
\begin{align} \label{ftaudev01RSB}
     \frac{\partial f_\tau}{\partial \mathcal{R}_0}  = m \int \mathscr{D} h \left( \frac{ \int \mathscr{D} u~\cosh^m \Xi \tanh \Xi }{\int \mathscr{D}u \cosh^m \Xi}\right)^2.
\end{align}
The derivative of $f_\tau$ with respect to $\mR_1$ is slightly easier: we have
\begin{align*}
     \frac{\partial f_\tau}{\partial \mathcal{R}_1}  &= 1 - \frac{2}{m} \int \mathscr{D} h~\frac{\int \mathscr{D} u ~ \partial_{\mR_1} \cosh^m\Xi }{\int \mathscr{D} u~\cosh^{m}\Xi} \\
     &= 1 - \frac{1}{\sqrt{\mR_0 -\mR_1}} \int \mathscr{D} h  \frac{\int \mathscr{D}u~u \cosh^m\Xi \tanh\Xi }{\int \mathscr{D}u \cosh^m\Xi} \\
     &=1 - \frac{1}{\sqrt{\mR_0 -\mR_1}} \int \mathscr{D} h \frac{\int \mathscr{D}u~\partial_u(\cosh^m\Xi \tanh\Xi) }{\int \mathscr{D}u \cosh^m\Xi} \\
     &= 1 - \int \mathscr{D} h \frac{\int \mathscr{D}u~\cosh^m\Xi [1 + (m-1) \tanh^2\Xi]}{\int \mathscr{D}u \cosh^m\Xi}.
\end{align*}
Since the first term in the numerator of the integrand is equal to the denominator, it simplifies and cancels out the additive 1. Eventually, we find
\begin{align} \label{ftaudev11RSB}
    \frac{\partial  f_\tau}{\partial \mathcal{R}_1} =  - (m-1) \int \mathscr{D} h \frac{\int \mathscr{D}u~\cosh^m\Xi \tanh^2\Xi}{\int \mathscr{D}u \cosh^m\Xi}.
\end{align}

Let us now compute the derivatives of the local free energy $f_C$. To simplify the procedure we have developed a similar technology with respect to the other free energy. In fact, as mentioned before, there is a certain symmetry between the two free energies, and one could already guess the final result for the derivatives of $f_C$. The symmetry relays on the fact that the role of the function $\cosh$ is played by the integral function $I_{\beta,1}$, and, the argument of the exponential function $g_1$ defined in Eq. \eqref{g1beta} is the counterpart of $\Xi$.

We split the computation, by first considering the derivatives of $I_{\beta,1}^m$ in the 1RSB parameters, in order to have some results ready for use when computing the derivatives of $f_C$. For the derivative in $\mM_D$ we have
\begin{align*}
    \partial_{\mM_D} I_{\beta,1}^m &= m I_{\beta,1}^{m-1} \int \de C \de\overline{C}~\partial_{\mM_D} g_{\beta,1} = m I_{\beta,1}^m \frac{\int \de C \de\overline{C} g_{\beta,1} \frac{|C|^2}{2}}{ I_{\beta,1}} = \frac{m}{2} I_{\beta,1}^m \langle |C|^2 \rangle_{1},
\end{align*}
for the derivative in $\mM_0$
\begin{align*}
    \partial_{\mM_0} I_{\beta,1}^m &= m I_{\beta,1}^m \frac{\int \de C \de\overline{C} g_{\beta,1} \left[\frac{1}{2\sqrt{\mM_0}} \Re{(C\overline{z})} - \frac{1}{2\sqrt{\mM_1-\mM_0}} \Re{(C\overline{w})} \right]}{I_{\beta,1}} \\
    &= \frac{m}{2\sqrt{\mM_0}} I_{\beta,1}^m \langle \Re{(C\overline{z})} \rangle_{1} - \frac{m}{2\sqrt{\mM_1-\mM_0}} \langle \Re{(C\overline{w})} \rangle_{1},
\end{align*}
and similarly for the derivative in $\mM_1$:
\begin{align*}
    \partial_{\mM_1} I_{\beta,1}^m = \frac{m}{2} I_{\beta,1}^m \langle |C|^2 \rangle_{1} + \frac{m}{2\sqrt{\mM_1-\mM_0}} I_{\beta,1}^m \langle \Re{(C\overline{w})} \rangle_{1}
\end{align*}

Given these preliminary results, we can compute the derivatives of the local free energy. The case of the diagonal element $\mM_D$ is the simplest one: we immediately find
\begin{align} \label{fCdevD1RSB}
    \frac{\partial f_C}{\partial \mathcal{M}_D}    & = \frac{1}{2 m} \int \mathscr{D}[z\overline{z}] \frac{\int \mathscr{D}[w \overline{w}] I_{\beta,1}^m \langle |C|^2 \rangle_{1} }{\int \mathscr{D}[w \overline{w}] I_{\beta,1}^m }.
\end{align}
For the derivative with respect to $\mM_0$, we proceed as follows
\begin{align*}
    \frac{\partial f_C}{\partial \mathcal{M}_0}  & = \frac{1}{m} \int \mathscr{D}[z\overline{z}] \frac{\int \mathscr{D}[w \overline{w}] \partial_{\mM_0} I_{\beta,1}^m   }{\int \mathscr{D}[w \overline{w}] I_{\beta,1}^m } \\
    & = \frac{1}{2\sqrt{\mM_0}} \left[ \int \mathscr{D}[z\overline{z}] \frac{\overline{z}}{2}  \frac{\int \mathscr{D}[w \overline{w}] I_{\beta,1}^m \langle C \rangle_{1} }{\int \mathscr{D}[w \overline{w}] I_{\beta,1}^m } + \int \mathscr{D}[z\overline{z}] \frac{z}{2}  \frac{\int \mathscr{D}[w \overline{w}] I_{\beta,1}^m \langle \overline{C} \rangle_{1} }{\int \mathscr{D}[w \overline{w}] I_{\beta,1}^m } \right] \\
    &\quad - \frac{1}{2\sqrt{\mM_1 - \mM_0}} \left[ \int \mathscr{D}[z\overline{z}]  \frac{\int \mathscr{D}[w \overline{w}] \frac{\overline{w}}{2} I_{\beta,1}^m \langle C \rangle_{1} }{\int \mathscr{D}[w \overline{w}] I_{\beta,1}^m } + \int \mathscr{D}[z\overline{z}]  \frac{\int \mathscr{D}[w \overline{w}]  \frac{w}{2} I_{\beta,1}^m \langle \overline{C} \rangle_{1} }{\int \mathscr{D}[w \overline{w}] I_{\beta,1}^m } \right]
\end{align*}
Similarly to the RS case, the two couples of terms inside square brackets are equal, after changing variables o $C \leftrightarrow \overline{C}$, $z \leftrightarrow \overline{z}$ and $w \leftrightarrow \overline{w}$, hence we have
\begin{align*}
    \frac{\partial f_C }{\partial \mathcal{M}_0} & = \frac{1}{\sqrt{\mM_0}} \int \mathscr{D}[z\overline{z}] \frac{\overline{z}}{2}  \frac{\int \mathscr{D}[w \overline{w}] I_{\beta,1}^m \langle C \rangle_{1} }{\int \mathscr{D}[w \overline{w}] I_{\beta,1}^m } - \frac{1}{\sqrt{\mM_1 - \mM_0}} \int \mathscr{D}[z\overline{z}]  \frac{\int \mathscr{D}[w \overline{w}] \frac{\overline{w}}{2} I_{\beta,1}^m \langle C \rangle_{1} }{\int \mathscr{D}[w \overline{w}] I_{\beta,1}^m } \\
    &= \frac{1}{\sqrt{\mM_0}} \left[ \int \mathscr{D}[z\overline{z}]   \frac{\int \mathscr{D}[w \overline{w}] \partial_z \left(I_{\beta,1}^m \langle C \rangle_{1} \right)}{\int \mathscr{D}[w \overline{w}] I_{\beta,1}^m }  - \int \mathscr{D}[z\overline{z}]   \frac{\int \mathscr{D}[w \overline{w}] I_{\beta,1}^m \langle C \rangle_{1} \int \mathscr{D}[w \overline{w}] \partial_z I_{\beta,1}^m }{\left(\int \mathscr{D}[w \overline{w}] I_{\beta,1}^m \right)^2 }     \right] \\
    &\quad - \frac{1}{\sqrt{\mM_1 - \mM_0}} \int \mathscr{D}[z\overline{z}]  \frac{\int \mathscr{D}[w \overline{w}] \partial_w \left(I_{\beta,1}^m \langle C \rangle_{1}\right) }{\int \mathscr{D}[w \overline{w}] I_{\beta,1}^m }
\end{align*}
where formal integration by parts has been performed in $z$ for the first term (the derivative has been already distributed) and in $w$ for the second one. Notice the perfect symmetry with the corresponding computation performed on the other free energy. We now have to compute the derivatives in $z$ and $w$, which correspond to the derivatives in $h,u$ in the case of $f_\tau$. Since, in analogy with the case of the function $\Xi$, the dependence of the function $g_{\beta,1}$ on $z$ and $w$ is of the same kind, both the derivatives in $z$ and $w$ of more complex objects which have $g_{\beta,1}$ as an argument yield the same result a part from a coefficient, coming from the derivative of the exponent argument in the expression of $g_{\beta,1}$, i.e.~$c_{z,w} = \{ \sqrt{\mM_0}, \sqrt{\mM_1-\mM_0} \}$. Therefore, we can compactly write
\begin{gather*}
    \partial_{z,w} I_{\beta,1}^m = \frac{m c_{z,w}}{2} \langle \overline{C} \rangle_{1} \\
    \partial_{z,w} \left(I_{\beta,1}^m \langle C \rangle_{g_1} \right) = \frac{c_{z,w}}{2} I_{\beta,1}^m \left(\langle |C|^2 \rangle_{1} + |\langle C \rangle_{1}|^2 \right).
\end{gather*}
By considering this result, the only term that survives in the derivative of $f_C$ with respect to $\mM_0$ is second one in the last step of our computation. This leads to the following result
\begin{align} \label{fCdev01RSB}
    \frac{\partial f_C}{\partial \mathcal{M}_0}  = -\frac{m}{2} \int \mathscr{D}[z\overline{z}]   \left| \frac{\int \mathscr{D}[w \overline{w}] I_{\beta,1}^m \langle C \rangle_{1} }{\int \mathscr{D}[w \overline{w}] I_{\beta,1}^m  } \right|^2
\end{align}
With an analogous computation, the derivative of $f_C$ with respect to $\mM_1$ yields the result
\begin{align} \label{fCdev11RSB}
    \frac{\partial f_C}{\partial \mathcal{M}_1}  = \frac{m-1}{2} \int \mathscr{D}[z\overline{z}]  \frac{\int \mathscr{D}[w \overline{w}] I_{\beta,1}^m |\langle C \rangle_{1}|^2 }{\int \mathscr{D}[w \overline{w}] I_{\beta,1}^m  } 
\end{align}
A nice remark, which catches the eye by looking at Eqs.~\eqref{fCdevD1RSB}, \eqref{fCdev01RSB} and \eqref{fCdev11RSB}, is that there is a correspondence between the position of the parameter in a 1RSB (or more in general RSB) matrix and the level of integration at which the square modulus appears in the derivatives of the free energy with respect to that parameter: the more internal is the parameter position, the more internal is the level where the square modulus appears. This hierarchical correspondence can be seen also in the derivatives of the free energy $f_\tau$ in Eqs.~\eqref{ftaudev01RSB} and \eqref{ftaudev11RSB}; in this case the derivative with respect to diagonal term $\mR_D$ is one, coherently with the fact that the square modulus of Ising spins is one.

Now that we have computed all the derivatives of the free energies in the action~\eqref{1RSBaction}, we can write the self-consistency equations, by assembling all the partial results:
\begin{align*}
    &\frac{\partial A_{\text{1RSB}}}{\partial \mR_D}=0 ~~~~\rightarrow~~~~  -1 + \frac{\partial s_1}{\partial \mR_D}  = 0 \\
    &\frac{\partial A_{\text{1RSB}}}{\partial \mR_0}=0 ~~~~\rightarrow~~~~   m \int \mathscr{D} h \left( \frac{ \int \mathscr{D} u~\cosh^m \Xi \tanh \Xi }{\int \mathscr{D}u \cosh^m \Xi}\right)^2 + \frac{\partial s_1}{\partial \mR_0}  = 0  \\
    &\frac{\partial A_{\text{1RSB}}}{\partial \mR_1}=0 ~~~~\rightarrow~~~~   - (m-1) \int \mathscr{D} h \frac{\int \mathscr{D}u~\cosh^m\Xi \tanh^2\Xi}{\int \mathscr{D}u \cosh^m\Xi}  + \frac{\partial s_1}{\partial \mR_1}  = 0  \\
    &\frac{\partial A_{\text{1RSB}}}{\partial \mM_D} =0 ~~~~\rightarrow~~~~  \frac{1}{2 m} \int \mathscr{D}[z\overline{z}] \frac{\int \mathscr{D}[w \overline{w}] I_{\beta,1}^m \langle |C|^2 \rangle_{1} }{\int \mathscr{D}[w \overline{w}] I_{\beta,1}^m }  + \frac{\partial s_1}{\partial \mM_D}  = 0  \\
    &\frac{\partial A_{\text{1RSB}}}{\partial \mM_0}=0 ~~~~\rightarrow~~~~   -\frac{m}{2} \int \mathscr{D}[z\overline{z}]   \left| \frac{\int \mathscr{D}[w \overline{w}] I_{\beta,1}^m \langle C \rangle_{1} }{\int \mathscr{D}[w \overline{w}] I_{\beta,1}^m  } \right|^2  + \frac{\partial s_1}{\partial \mM_0}  = 0  \\
    &\frac{\partial A_{\text{1RSB}}}{\partial \mM_1} =0 ~~~~\rightarrow~~~~  \frac{m-1}{2} \int \mathscr{D}[z\overline{z}]  \frac{\int \mathscr{D}[w \overline{w}] I_{\beta,1}^m |\langle C \rangle_{1}|^2 }{\int \mathscr{D}[w \overline{w}] I_{\beta,1}^m  }   + \frac{\partial s_1}{\partial \mM_1}  = 0  
\end{align*}
where, by using the shorthand notation $\mA=\mR-\mM$, the derivatives of the entropic term read
\begin{align*}
    \frac{\partial s_1}{\partial \mathcal{R}_D} = -\frac{\partial s_1}{\partial \mathcal{M}_D} &= \frac{\mathcal{A}_0}{(\mathcal{A}_D+(m-1)\mathcal{A}_1-m\mathcal{A}_0)^2} 
     -\frac{1}{m}\frac{1}{\mathcal{A}_D+(m-1)\mathcal{A}_1-m\mathcal{A}_0} \\ &\quad -\frac{m-1}{m} \frac{1}{\mathcal{A}_D-\mathcal{A}_1} \\
    \frac{\partial s_1}{\partial \mathcal{R}_0}  = - \frac{\partial s_1}{\partial \mathcal{M}_0} &= -m\frac{\mathcal{A}_0}{(\mathcal{A}_D+(m-1)\mathcal{A}_1-m\mathcal{A}_0)^2} \\
    \frac{\partial s_1}{\partial \mathcal{R}_1}= - \frac{\partial s_1}{\partial \mathcal{M}_1} &= (m-1)\Bigg[ \frac{\mathcal{A}_0}{(\mathcal{A}_D+(m-1)\mathcal{A}_1-m\mathcal{A}_0)^2}
    -\frac{1}{m}\frac{1}{\mathcal{A}_D+(m-1)\mathcal{A}_1-m\mathcal{A}_0} \\
    &\quad -\frac{1}{m} \frac{1}{\mathcal{A}_D-\mathcal{A}_1}\Bigg].
\end{align*}
In order write a more compact set of equations, it is convenient to use the algebric constraint Eq.~\eqref{AlgConstr}, which has to be written in the 1RSB ansatz for this purpose. To visualize the constraint in the 1RSB case, let us write it in matrix form, for the simple case of $n=4$ and $m=2$:
\begin{equation}
\begin{pmatrix}
\mathcal{A}_D & \mathcal{A}_1 & \mathcal{A}_0 & \mathcal{A}_0   \\
\mathcal{A}_1 & \mathcal{A}_D & \mathcal{A}_0 & \mathcal{A}_0   \\
\mathcal{A}_0 & \mathcal{A}_0 & \mathcal{A}_D & \mathcal{A}_1   \\
\mathcal{A}_0 & \mathcal{A}_0 & \mathcal{A}_1 & \mathcal{A}_D   
\end{pmatrix}  
\begin{pmatrix}
1 & q_1 & q_0 & q_0   \\
q_1 & 1 & q_0 & q_0   \\
q_0 & q_0 & 1 & q_1   \\
q_0 & q_0 & q_1 & 1   
\end{pmatrix} =
\begin{pmatrix}
1 & 0  & 0 & 0  \\
0 &1  & 0 & 0  \\
0 & 0  & 1 & 0  \\
0 & 0 & 0 & 1  \\
\end{pmatrix}
\end{equation}
We find the following three independent equations, by considering the product of the first line of $A$ respectively with the first, the second and the $(m+1)$-th column of $\mQ$
\begin{gather*}
\mathcal{A}_D +(m-1)\mathcal{A}_1 q_1 + (n-m)\mathcal{A}_0 q_0 =1 \\
\mathcal{A}_1 + \mathcal{A}_D q_1 + (m-2)\mathcal{A}_1 q_1 + (n-m)\mathcal{A}_0 q_0 =0 \\
\mathcal{A}_D q_0 + (m-1)\mathcal{A}_1 q_0 + \mathcal{A}_0 + (m-1) \mathcal{A}_0 q_1 + (n-2m)\mathcal{A}_0 q_0 = 0,    
\end{gather*}
which in the limit $n \rightarrow 0$ reduce to
\begin{subequations} \label{Alg1RSB-app}
\begin{gather}
     \mathcal{A}_D +(m-1)\mathcal{A}_1 q_1 - m\mathcal{A}_0 q_0 =1   \\
      \mathcal{A}_1 + \mathcal{A}_D q_1 + (m-2)\mathcal{A}_1 q_1 - m\mathcal{A}_0 q_0 =0   \\
      \mathcal{A}_D q_0 + (m-1)\mathcal{A}_1 q_0 + \mathcal{A}_0 + (m-1) \mathcal{A}_0 q_1 - 2m \mathcal{A}_0 q_0 = 0. 
\end{gather}
\end{subequations}
Note that from the first two equations one finds
\begin{equation*} 
q_1= 1-\frac{1}{\mathcal{A}_D-\mathcal{A}_1}.
\end{equation*}
Moreover, by subtracting the first equation to the third one and using the expression of $q_1$, one finds the following expression for $q_0$:
\begin{equation*} 
q_0= 1 - \frac{1}{m}\frac{1}{\mathcal{A}_D+(m-1)\mathcal{A}_1-m\mathcal{A}_0} -\frac{m-1}{m} \frac{1}{\mathcal{A}_D-\mathcal{A}_1}.
\end{equation*}
Now, we can proceed as in the RS case: the first equation of the set, which contains the derivatives in the diagonal element $\mR_D$, can be eliminated after using it in all the other equations. Then, by using the definitions of $q_0$ and $q_1$, after some algebra, we finally get to the nicer and more familiar set of equations
\begin{gather*}
    q_0 = \int \mathscr{D} h \left( \frac{ \int \mathscr{D} u~\cosh^m \Xi \tanh \Xi }{\int \mathscr{D}u \cosh^m \Xi}\right)^2 \\
    q_1 = \int \mathscr{D} h \frac{\int \mathscr{D}u~\cosh^m\Xi \tanh^2\Xi}{\int \mathscr{D}u \cosh^m\Xi} \\
    1 = \frac{1}{2} \int \mathscr{D}[z\overline{z}] \frac{\int \mathscr{D}[w \overline{w}] I_{\beta,1}^m \langle |C|^2 \rangle_{1} }{\int \mathscr{D}[w \overline{w}] I_{\beta,1}^m }  \\
    q_0 = \frac{1}{2} \int \mathscr{D}[z\overline{z}]   \left| \frac{\int \mathscr{D}[w \overline{w}] I_{\beta,1}^m \langle C \rangle_{1} }{\int \mathscr{D}[w \overline{w}] I_{\beta,1}^m  } \right|^2  \\
    q_1 = \frac{1}{2} \int \mathscr{D}[z\overline{z}]  \frac{\int \mathscr{D}[w \overline{w}] I_{\beta,1}^m |\langle C \rangle_{1}|^2 }{\int \mathscr{D}[w \overline{w}] I_{\beta,1}^m  }, 
\end{gather*}
which has to be completed by the three algebraic relations~\eqref{Alg1RSB-app} among the 1RSB parameters.

\end{appendices}

\backmatter
\cleardoublepage
\phantomsection

\addcontentsline{toc}{chapter}{Bibliography}

\printbibliography

\end{document}